\documentclass[twocolumn]{aastex631}

\shorttitle{SOMA V - Clustered Protostars}
\shortauthors{Telkamp et al.}
\graphicspath{{./}{figures/}}
\usepackage{multirow}

\begin{document}

\title{The SOFIA Massive (SOMA) Star Formation Survey. V.\\
Clustered Protostars}

\correspondingauthor{Zoie Telkamp}
\email{zrt7qc@virginia.edu}
\author[0000-0001-6465-9590]{Zoie Telkamp}
\affiliation{Dept. of Astronomy, University of Virginia, Charlottesville, Virginia 22904, USA}

\correspondingauthor{Rub\'{e}n Fedriani}
\email{fedriani@iaa.es}
\author[0000-0003-4040-4934]{Rub\'{e}n Fedriani}
\affil{Instituto de Astrof\'{i}sica de Andaluc\'{i}a, CSIC, Glorieta de la Astronom\'{i}a s/n, E-18008 Granada, Spain}

\author[0000-0002-3389-9142]{Jonathan C. Tan}
\affiliation{Dept. of Astronomy, University of Virginia, Charlottesville, Virginia 22904, USA}
\affiliation{Dept. of Space, Earth \& Environment, Chalmers University of Technology, 412 93 Gothenburg, Sweden}

\author[0000-0003-1964-970X]{Chi-Yan Law}
\affiliation{Osservatorio Astrofisico di Arcetri, Largo Enrico Fermi, 5, 50125 Firenze FI, Italy}
\author[0000-0001-7511-0034]{Yichen Zhang}
\affiliation{Department of Astronomy, Shanghai Jiao Tong University, 800 Dongchuan Rd., Minhang, Shanghai 200240, China}

\author[0000-0002-9912-5705]{Adele Plunkett}
\affiliation{National Radio Astronomy Observatory, 520 Edgemont Road, Charlottesville, VA 22903, USA}

\author[0009-0005-0394-3754]{Samuel Crowe}
\affiliation{Dept. of Astronomy, University of Virginia, Charlottesville, Virginia 22904, USA}

\author[0000-0001-8227-2816]{Yao-Lun Yang}
\affiliation{Star and Planet Formation Laboratory, RIKEN Cluster for Pioneering Research, Wako, Saitama 351-0198, Japan}

\author[0000-0001-7378-4430]{James M. De Buizer}
\affiliation{Carl Sagan Center for Research, SETI Institute, Mountain View, CA, USA}

\author[0000-0003-3315-5626]{Maria T. Beltran}
\affiliation{INAF-Osservatorio Astrofisico di Arcetri, Largo E. Fermi 5, 50125 Firenze, Italy}

\author[0000-0001-6551-6444]{M\'{e}lisse Bonfand}
\affiliation{Dept. of Astronomy, University of Virginia, Charlottesville, Virginia 22904, USA}

\author[0000-0001-9857-1853]{Ryan Boyden}
\affiliation{Dept. of Astronomy, University of Virginia, Charlottesville, Virginia 22904, USA}

\author[0000-0001-5551-9502]{Giuliana Cosentino}
\affiliation{European Southern Observatory, Karl-Schwarzschild-Strasse 2, D-85748 Garching, Germany}

\author[0000-0003-1602-6849]{Prasanta Gorai}
\affiliation{Rosseland Centre for Solar Physics, Institute of Theoretical Astrophysics, Sem Salands vei 13, 0371 Oslo, Norway}

\author[0000-0001-6159-2394]{Mengyao Liu}
\affiliation{Dept. of Astronomy, University of Virginia, Charlottesville, Virginia 22904, USA}

\author[0000-0001-8596-1756]{Viviana Rosero}
\affiliation{Division of Physics, Mathematics, and Astronomy, California Institute of Technology, Pasadena, CA 91125, USA}

\author[0000-0003-4402-6475]{Kotomi Taniguchi}
\affiliation{National Astronomical Observatory of Japan, National Institutes of Natural Sciences, 2-21-1 Osawa, Mitaka, Tokyo 181-8588, Japan}

\author[0000-0002-6907-0926]{Kei E. I. Tanaka}
\affiliation{Department of Earth and Planetary Sciences, Institute of Science Tokyo, Meguro, Tokyo, 152-8551, Japan}

\author[0000-0003-0090-9137]{Tatiana M. Rodr\'iguez}
\affiliation{I. Physikalisches Institut, Universit\"at zu K\"oln, 50937 K\"oln, Germany}



\begin{abstract}
We present $\sim8-40\,\mu$m SOFIA-FORCAST images of seven regions of ``clustered" star formation as part of the SOFIA Massive (SOMA) Star Formation Survey. We identify a total of 34 protostar candidates and build their spectral energy distributions (SEDs). We fit these SEDs with a grid of radiative transfer models based on the Turbulent Core Accretion (TCA) theory to derive key protostellar properties, including initial core mass, $M_c$, clump environment mass surface density, $\Sigma_{\rm cl}$, and current protostellar mass, $m_*$. We also carry out empirical graybody (GB) estimation of $\Sigma_{\rm cl}$, which allows a case of restricted SED fitting within the TCA model grid. We also release version 2.0 of the open-source Python package \emph{sedcreator}, designed to automate the aperture photometry and SED building and fitting process for sources in clustered environments, where flux contamination from close neighbors typically complicates the process. Using these updated methods, SED fitting yields values of $M_c\sim30-200\:M_{\odot}$, $\Sigma_{\text{cl,SED}}\sim0.1-3\:{\rm{g\:cm}}^{-2}$, and $m_*\sim4-50\:M_{\odot}$. The graybody fitting yields smaller values of $\Sigma_{\text{cl,GB}}\lesssim1\:{\rm{g\:cm}}^{-2}$. From these results, we do not find evidence for a critical $\Sigma_{\rm{cl}}$ needed to form massive ($\gtrsim 8\:M_\odot$) stars. However, we do find tentative evidence for a dearth of the most massive ($m_*\gtrsim30\:M_\odot$) protostars in the clustered regions suggesting a potential impact of environment on the stellar initial mass function.
\end{abstract}


\section{Introduction} \label{sect:introduction}

Due to their enormous luminosities, massive stars impact a vast range of scales and processes, from the reionization of the universe, to galaxy evolution, to the formation of star clusters and planets. Despite their importance, no consensus has been reached on the formation mechanisms of massive stars. Theories range from extensions of the standard core accretion model that describes low-mass star formation, e.g., the Turbulent Core Accretion (TCA) model \citep{mckee2002,mckee2003}, to competitive accretion models that form massive stars in the centers of crowded protoclusters of low-mass protostars \citep[e.g.,][]{bonnell1998,wang2010,grudic2022}. 

The SOFIA Massive (SOMA) Star Formation Survey (PI: Tan) aims to test theoretical models of massive star formation by characterizing a sample of $\ga$ 50 high- and intermediate-mass protostars spanning a range of environments and evolutionary stages. These objects have been observed with the SOFIA-Faint Object infraRed CAmera for the SOFIA Telescope (FORCAST) instrument \citep{herter2018} from $\sim$ 10 to 40\,$\mu$m. Paper I of the survey \citep{debuizer2017} presented the first 8 sources, most of which are massive protostars. Paper II \citep{liu2019} presented 7 especially luminous sources, which are some of the most massive protostars in the survey. Paper III \citep{liu2020} presented 14 intermediate-mass sources. Paper IV \citep{Fedriani_2023} analyzed 11 sources that are characterized as relatively ``isolated'' based on their $37\,\mu$m images and presented \verb+sedcreator+, a Python package designed to automate the aperture photometry and SED fitting process for protostars. Paper IV also re-analyzed the sources from Papers I-III using this package to produce a sample of 40 uniformly analyzed massive protostars. The results placed constraints on the environmental conditions needed for massive star formation. However, while the SOMA I-IV samples explored sources across a wide range of masses, they excluded  sources in highly clustered environments.

Here, in SOMA Paper V, we present 7 regions of ``clustered'' protostars, defined as MIR sources exhibiting radio emission and surrounded by several other MIR sources within $\sim$60\arcsec. Altogether, we detect and analyze 34 sources within these crowded regions, significantly increasing the size of the SOMA sample. With these new results, we examine whether the trends observed in SOMA I-IV continue to hold in these new regions and discuss the constraints this places on massive star formation theories. This is a critical step in understanding how environment shapes the massive star formation process.

As in SOMA Papers I-IV, we measure fluxes from infrared (IR) images obtained with Spitzer, SOFIA, and Herschel facilities to construct the spectral energy distribution (SED) of each protostar. We fit these SEDs with the \citet[][hereafter ZT18]{zhang2018} radiative transfer (RT) models to constrain the properties of the protostars. Due to the difficulty in separating the fluxes of sources in these clustered regions, new methods are needed to independently analyze the sources within them. Thus, we also present version 2.0 of \verb+sedcreator+, which includes new and updated tools to explore massive star formation in these environments. This updated package enables the simultaneous analysis of multiple sources in a star-forming region, yielding an individual SED for each source while taking into account nearby sources.





In \S2 of this paper, we describe the observations and data we use for our analysis. In \S3, we discuss the source detection process we have developed based off of the Astrodendro Python package \citep{Robitaille2019}, along with the changes made to the flux calculation and SED fitting methods in \verb+sedcreator+ to enable analysis of multiple sources in a region. In \S4, we present the MIR to FIR images and results of the SED fitting. In \S5, we discuss these results and examine trends in massive star forming environments, and in \S6, we provide a summary of the results from the entire SOMA survey to date.

\section{Observations} \label{sect:observations}

We carried out observations of the seven massive star-forming regions using the Stratospheric Observatory for Infrared Astronomy (SOFIA\footnote{SOFIA is jointly operated by the Universities Space Research Association, Inc. (USRA), under NASA contract NAS2-97001, and the Deutsches SOFIA Institute (DSI) under DLR contract 50 OK 0901 to the University of Stuttgart.}) and the Faint Object infraRed CAmera for the SOFIA Telescope \citep[FORCAST][]{herter2018} instrument. We used four filters, centered on 7.7, 19.7, 31.5, and 37.1\,$\mu$m. We estimate these 7 regions to possess 34 protostars, using the \verb+Astrodendro+ Python package \citep{Robitaille2019} to identify a list of peaks that we condense using methods described in \S3.

We followed the methods of Papers I, II, III, and IV to perform the photometric and astrometric calibrations. We estimate the calibration error of the SOFIA observations to be between $\sim3\% - 7\%$. The astrometric precision for the SOFIA 7\,$\mu$m image is about 0\farcs1, while the value for longer wavelength SOFIA images is about 0\farcs4 (see Paper I for further details). We used pipeline-reduced and calibrated data from the SOFIA archive. When available, we supplemented the SOFIA observations with publicly-available images of Spitzer/IRAC \citep{werner2004,fazio2004} at 3.6, 4.5, 5.8, and 8.0\,$\mu$m from the Spitzer Heritage Archive, and Herschel/PACS and SPIRE \citep{griffin2010} at 70, 160, 250, 350, and 500\,$\mu$m from the ESA Herschel Science Archive.


\begin{deluxetable*}{lllcccccc}
\tablecaption{SOFIA FORCAST Observations: Observation Dates \& Exposure Times (seconds)\label{tab:sofia_obs}}
\tablehead{
\colhead{Source} & \colhead{R.A.(J2000)} & \colhead{Decl.(J2000)} & \colhead{$d$ (kpc)} & \colhead{Obs. Date} & \colhead{7.7$\,{\rm \mu m}$} & \colhead{19.7$\,{\rm \mu m}$} & \colhead{31.5$\,{\rm \mu m}$} & \colhead{37.1$\,{\rm \mu m}$}
}
\startdata
AFGL 5180 & 06$^h$08$^m$53$\fs$3 & $+$21$\arcdeg$38$\arcmin$28$\farcs$7 
& 1.76 & 2015 Jan 29 & 268 & $\cdots$ & $\cdots$ & 270  \\
G018.67$+$00.03 & 18$^h$24$^m$53$\fs$8 & $-$12$\arcdeg$39$\arcmin$18$\farcs$7 & 10.8 & 2018 Aug 22 & 346 & 810 & 509 & 1263 \\
G28.37$+$00.07 & 18$^h$42$^m$53$\fs$2 & $-$04$\arcdeg$00$\arcmin$08$\farcs$0 & 5.00 & 2018 Sep 06 & 283 & $\cdots$ & 1338 & 773 \\
G030.76$+$00.20 & 18$^h$46$^m$44$\fs$6 & $-$01$\arcdeg$50$\arcmin$29$\farcs$8 & 6.4 & 2018 Sep 10 & 329 & 911 & 597 & 1062 \\
G058.77$+$00.65 & 19$^h$38$^m$48$\fs$8 & $+$23$\arcdeg$09$\arcmin$39$\farcs$3 & 3.30 & 2018 Aug 29 & $\cdots$ & $\cdots$ & $\cdots$ & 1235 \\
IRAS 20343$+$4129 & 20$^h$36$^m$07$\fs$2 & $+$41$\arcdeg$39$\arcmin$51$\farcs$3 & 1.40 & 2018 Aug 22 & 303 & 852 & 812 & 1387 \\
W3 IRS5 & 02$^h$25$^m$40$\fs$7 & $+$62$\arcdeg$05$\arcmin$52$\farcs$6 & 1.83 & 2015 May 30 & 116 & 113 & 121 & 113 \\
\enddata
\tablecomments{
The source positions listed here are the same as the positions of the
primary source for each region in Figs. \ref{fig:AFGL5180}, \ref{fig:G18.67}, \ref{fig:G28.37}, \ref{fig:G30.76}, \ref{fig:G58.77} , \ref{fig:IRAS20343+4129}, \ref{fig:W3}. Source distances are from the literature, discussed below (see Sect.\,\ref{sect:sources}).}
\end{deluxetable*}

\section{Methods}\label{sect:methods}
Here we describe the new and updated methods we use to identify sources, measure fluxes, and build and fit SEDs in each of the 7 regions in our sample.

\subsection{Source Detection}\label{sect:source_detection} 

In Papers I-IV of the SOMA survey, source positions were generally derived from radio continuum peaks. However, while visual inspection of each of the SOMA V regions reveals multiple sources, many have not yet been identified in radio continuum data. In addition, given the crowded nature of these regions, flux contamination from neighbors complicates the process of identifying individual sources in the IR images. To address this, we define a method of automatically generating source lists for these clustered regions. We perform this method using the SOFIA 37.1~$\mu$m images, with this emission expected to trace warm dust in the inner regions of protostellar envelopes \citep[e.g.,][]{zhang2018}. We selected this wavelength because SOFIA's higher angular resolution compared to Herschel $70\:{\rm \mu m}$ images aids in distinguishing the sources from one another. Also, using the longest FORCAST wavelength allows us to pierce deeper into the thick gas and dust of star-forming regions. The process works as follows.

First, we apply a median filter with a kernel of 20\arcsec\ to the 37.1 $\mu$m image. We then subtract the resulting median-filtered image from the original image. This 
makes the sources stand out more sharply against the local background. Next, we run the \verb+Astrodendro+ Python package \citep{Robitaille2019} \citep[see][for a description of the original dendrogram algorithm]{Rosolowsky2008} on the resulting image to identify structures above a given brightness threshold and size. We set the minimum size of a source to 3 times that of the beam and the minimum peak height to 6 times the noise. We selected these values after testing multiple combinations and seeing which combination consistently produced reasonable source guesses (determined by eye) for all 7 regions. 
Since the noise can vary significantly across a single SOFIA image, determining the value from one patch would not be representative of the entire image. To account for this, we cover the image in circular apertures of radius 3.5\arcsec, chosen to be comparable to the beam size of the 37.1 $\mu$m images, and calculate their enclosed fluxes. Next, we take the median absolute deviation (MAD) of these values and use this as the noise estimate. We take the MAD instead of the standard deviation to minimize the impact of source flux on the calculation. 

After obtaining the source candidates from \verb+Astrodendro+, we perform a down-selection step. The minimum aperture size we use in the SOMA survey is 5\arcsec, based on the beam size of the Herschel 70 $\mu$m images. As such, we do not allow sources to be closer than 10\arcsec\ to one another, acknowledging that this could lead to some sources being ignored in highly clustered regions. To ensure that this condition is met, we iterate through the source list and build up a new ``down-selected'' catalog of sources that fulfill this requirement. We start by adding the brightest source to the new catalog. We then move on to the second brightest source in the original list and add it to the new catalog if it is at least 10\arcsec\ from the brightest source. We continue to iterate through the remaining original sources, only adding a source to the new catalog if it is farther than 10\arcsec\ from each of the sources already in the new catalog. After this step is complete, we have our final source catalog and can begin the SED analysis process.


\subsection{Sedcreator Version 2.0}\label{sect:sedfluxer}


To analyze SEDs of sources in crowded regions, we present \verb+sedcreator+ version 2.0, an updated version of the open-source Python package presented in \cite{Fedriani_2023} and hosted in both GitHub\footnote{\url{https://github.com/fedriani/sedcreator}} and PyPi\footnote{\url{https://pypi.org/project/sedcreator/}} (the documentation can be accessed at this URL \url{https://sedcreator.readthedocs.io/}). The main functionalities of \verb+sedcreator+ are partitioned into SedFluxer and SedFitter. SedFluxer provides tools to measure fluxes and build SEDs from an image, using a number of functions from \citet{photutils2020}. It also presents an algorithm to automatically determine the aperture radius for a source. SedFitter fits an SED using a massive star formation radiative transfer model grid by \cite{zhang2018}. Version 2.0 presents new and updated methods for SedFluxer to analyze protostars in crowded regions. Instead of analyzing each source in isolation, version 2.0 takes in a list of source coordinates in a given region and performs the aperture selection process and flux calculations for each source while considering positions and flux contributions from its neighbors. In the following sections, we describe the updated methods introduced in \verb+sedcreator+ version 2.0.

\subsubsection{Optimal Aperture Algorithm}\label{sect:opt_rad}

Paper IV presents an ``optimal aperture algorithm" to determine the aperture size of an isolated extended source in an unbiased and reproducible way. The goal of this algorithm is to find a radius that encompasses most of the source's flux in an image, while minimizing the background contribution. To achieve this, it estimates the point at which the gradient of the background-subtracted flux as a function of aperture radius drops below a certain value (see Paper IV for details and illustrative plots). If used in a crowded region, this algorithm would overestimate aperture sizes due to flux contributions from neighboring sources. Here, we present an updated version of the algorithm designed to work in crowded regions. In the case where only one source is present, the algorithm works as before. In the case where multiple source positions are inputted, the new algorithm determines each radius while considering all neighboring sources in an iterative process until a solution is converged upon for each one. 

This process works as follows. The algorithm takes in an image and a list of source coordinates. Each source is assigned an initial aperture radius of 5\arcsec, the smallest aperture size used in the SOMA survey. Starting with the primary source, the algorithm increases each source's aperture until the gradient of its enclosed background-subtracted flux reaches a defined threshold (as described in SOMA IV). An aperture's growth is stopped early if it reaches the central coordinate of another source in the list, regardless of whether the gradient threshold condition has been met. When two or more apertures overlap with one another, each pixel in the region of overlap is assigned to the closest source and excluded from the flux calculations for all other sources. After each source's aperture has been updated, the algorithm repeats this process, starting at the primary source and cycling through the list again. At the end of each iteration, the algorithm calculates how much each source's aperture changed from the previous iteration. As soon as none of the apertures are changing by more than 1\%, the algorithm stops iterating and returns the final apertures, along with maps specifying which source each pixel has been assigned to. In the situation where convergence is not reached after a user-defined number of iterations, the algorithm outputs the apertures corresponding to the iteration with the lowest average percent change.

\begin{figure*}
\begin{center}
    \includegraphics[scale=0.6]{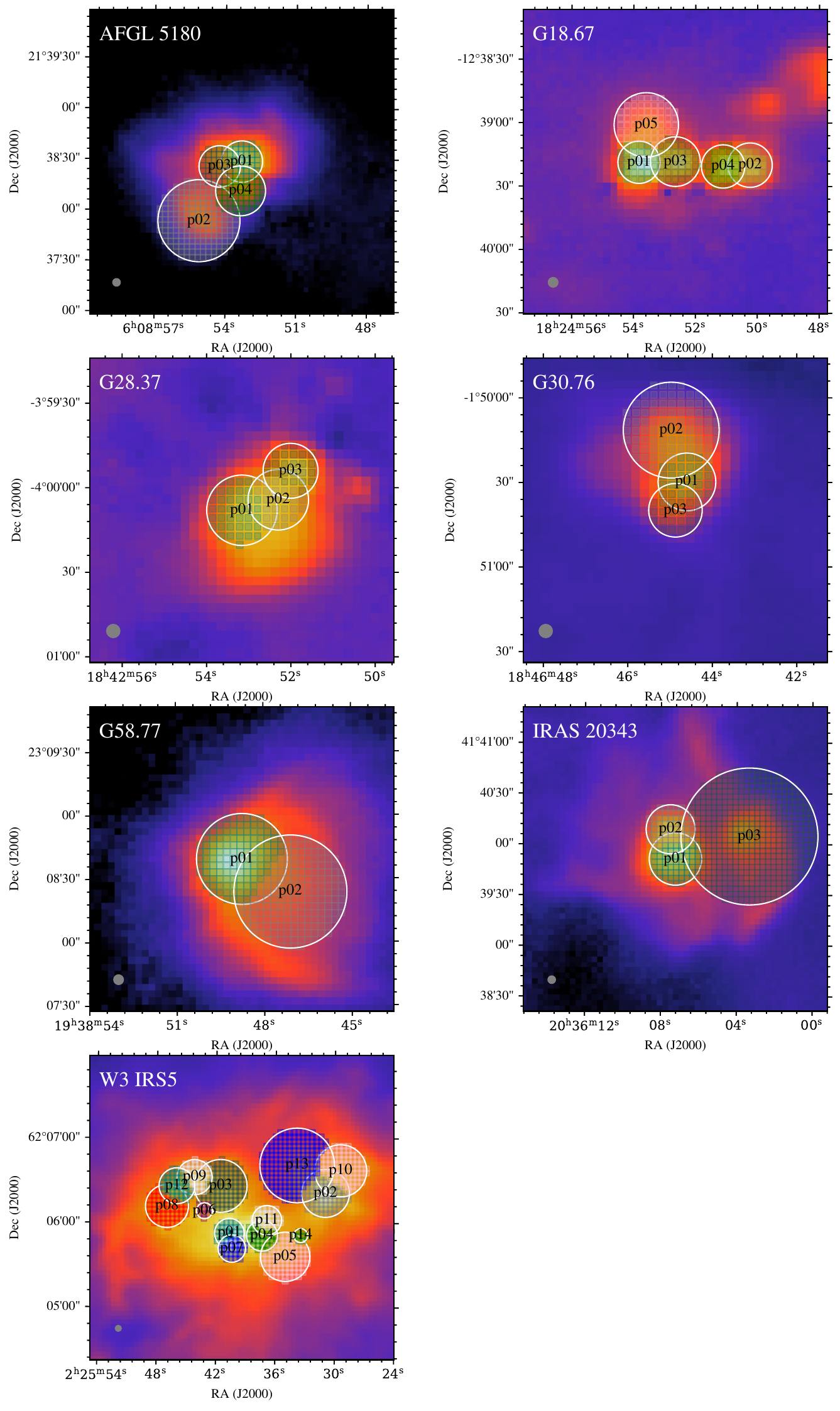}
\end{center}
\caption{Herschel 70 $\mu$m images for each SOMA V region. The color map for each image is normalized by the peak flux in that image. The white circles denote the apertures used for the fiducial photometry. Pixels are colored to show which aperture each pixel is assigned to. Gray circles in the lower left show the resolution of each image. The black crosses in all panels denote the positions of the sources as described in Section \ref{sect:methods}.}
\label{fig:masks}
\end{figure*}

\subsubsection{Flux Measurements}
\label{sect:flux}
Following the fiducial ``Fixed Aperture" method described in Papers I-IV, we determine the aperture radius of a given source using its Herschel 70 $\mu$m image, which is near the peak of the SED, and apply this radius at all wavelengths. The source flux at each wavelength is then obtained through aperture photometry using \verb+sedcreator+'s SedFluxer tool (see Paper IV for details), with several updates. First, the assignment map that the optimal aperture algorithm outputs is used to determine which pixels are assigned to other sources and are thus excluded from this source's flux measurement. In addition, version 2.0 modifies the error estimation process. The previous method and updates are summarized below.

As in Paper IV, for wavelengths $\lambda >100\:\mu$m, which are expected to be potentially impacted by confusion with the cold clump environment, we use the background intensity itself as the measure of the error estimate. 
At shorter wavelengths, Paper IV introduced a method of error estimation via measurement of the fluctuations of the flux within a region of the annulus with an equivalent area to the main aperture. As such, an annulus around the aperture (extending to 2 times the aperture radius) is split into three sectors, and the standard deviation of their fluxes (after aliasing to avoid bias; see Paper IV) is taken as the fluctuation estimate. In crowded regions, any pixels that belong to a neighboring source's aperture will be excluded from this calculation. In particularly crowded regions, this could lead to most of the pixels of a given annulus sector being masked out. To address this issue, if any sector of the annulus has 80\% or more of its area masked out, we do not include this sector in the fluctuation calculations. If this is true for two or more sectors, we return to the background error estimation method used at long wavelengths. 
i.e., taking the median of the entire annulus (other than those that are masked out). In addition, if 50\% or more of the total annulus is masked out, we default to this background estimation method. 
Finally, as in SOMA I-IV, an additional assumed flux calibration error of 10\% of the background-subtracted flux is added in quadrature to yield a total flux error estimate.

\subsubsection{SED Fitting}

After constructing the SEDs for each source, we fit them with a grid of radiative transfer models from ZT18, based on the TCA model of massive star formation. In this model, the initial conditions of massive star formation are gravitationally bound prestellar ``cores'' embedded in a larger-scale ``clump'' of ambient gas. This core is internally supported by pressure from both turbulence and magnetic fields. When the core collapses, it is assumed to form a single rotationally-supported disk that yields a single star or small N multiple via disk fragmentation. However, in the ZT18 grid of RT models, only the case of single star formation from a core is considered, i.e., the limiting case where a single star dominates the luminosity. The ZT18 grid is constructed from two primary initial condition variables that determine the evolutionary history of the protostar: the initial mass of the core ($M_c$) and the mass surface density of the clump in which the core is embedded ($\Sigma_{\rm cl}$). The current protostellar mass ($m_*$) indicates where the protostar is in its evolutionary timeline. Together, $M_c$, $\Sigma_{\rm cl}$, and $m_*$ are the main intrinsic properties that define the model grid. In total there are 432 different physical models with different values of these three parameters.
Two secondary properties are the inclination angle of the outflow axis to the line of sight ($\theta_{\text{view}}$), which is sampled with 20 different values, and the level of foreground extinction ($A_V$), which is explored
up to a maximum value of $A_V=1000$~mag. To complete the fitting, we use \verb+sedcreator+'s SedFitter tool, which fits the ZT grid of $432 \times 20 = 8640$ models to the data, minimizing each model's $\chi^2$ function over $A_V$ to make the process more efficient. This returns the model parameters and corresponding $\chi^2$ values, in order of ``best'' to ``worst'' fit according to its corresponding $\chi^2$ value.

In analysis of the SOMA V regions we find that there are relatively faint, noisy sources being fit with the SED models. Primarily from considering these sources, we make the following updates to the definition of ``good'' model fits compared to SOMA IV. First, when carrying out the normalization of the reduced $\chi^2$ by the total number of data points, $N$, we do not count a data point if it is only used as an upper limit to a given model. This mainly applies to the short wavelength data, i.e., $\lambda<8\:{\rm \mu m}$, which are treated as being upper limits on the models, since the models do not include contributions from transiently heated small dust grains or from PAH emission features. However, note that if a given RT model predicts a flux that is greater than an upper limit data point, then its $\chi^2$ contribution is counted and the data point is counted in the sum of $N$. As in SOMA I-IV, we only consider models that satisfy $R_\mathrm{core}<2R_\mathrm{aper}$, where $R_\mathrm{core}$ is the core radius of the RT model and $R_\mathrm{aper}$ is the aperture radius found in the images using the above methods. This ensures that only physically consistent core radius are considered.

Then, there are three cases we consider to define a set of ``good'' models, depending on the value of the minimum value of $\chi^2$, i.e., $\chi^2_{\rm min}$:\\

\noindent (i) If $\chi^2_{\rm min}<1$, we average over all models with $\chi^2<2$ (this is similar to the method of SOMA IV).\\


\noindent (ii) If $1\leq \chi_{\rm min}^2\leq 2$, then we average over all models with $\chi^2<2$, but also require that at least 10 physical models, i.e., with different combinations of $M_c$, $\Sigma_{\rm cl}$, and $m_*$. This is designed to make sure there is a reasonable level of sampling over these parameters.\\


\noindent (iii) If $\chi^2_{\rm min}>2$, we average over the 10 best physical models, i.e., with different combinations of $M_c$, $\Sigma_{\rm cl}$, and $m_*$, with the worst of these defining a limiting value $\chi^2_{\rm lim}$. Then for these models, all viewing angles that yield $\chi^2<\chi^2_{\rm lim}$ are included.


\subsection{Empirical Derivation of $\Sigma_{\rm cl}$}

To further constrain the parameters derived from the SED fitting process, we empirically calculate the local $\Sigma_{\rm cl}$ value around each source using the Herschel 70, 160, 250, 350, and 500\,$\mu$m images. The background fluxes are measured as described in Section \ref{sect:flux}, using an annular region around each source that excludes flux from other source apertures. We fit the resulting background SED for each source at these wavelengths with a single-temperature graybody distribution represented by the following equation:
\begin{equation}
    I_{\nu} \simeq B_{\nu}(1-e^{-\tau_{\nu}}) = B_{\nu}(1-e^{-\Sigma_{\rm cl}\kappa_{\nu}}),
\end{equation}
where $I_{\nu}$ is the intensity observed at a given frequency, $B_{\nu}$ is the Planck function, $\tau_{\nu}$ is the dust optical depth, and $\kappa_{\nu}$ is the dust opacity. 
For consistency, we use the same opacities as ZT18 used when constructing the grid of RT models. These values come from the opacity law defined by \citet{Whitney2003}. We use the curve\_fit function of the SciPy optimize Python package to perform the graybody fitting. We assign a 10\% error to the background flux values. We then use the covariance matrix returned by curve\_fit to obtain uncertainties on the resulting $\Sigma_{\rm cl}$ values.

\section{Results}\label{sect:results}

We first (in \S\ref{sect:measured_sed_results}) describe some general results of the SED fitting of the SOMA V sources and then (in \S\ref{sect:sources}) present summaries of individual regions. Then, in the following section, \S\ref{sect:discussion}, we describe the results of analysis of the overall SOMA I-V sample.

\subsection{Source Images and SEDs}\label{sect:measured_sed_results}

Overall, in the seven regions of clustered massive star formation, we have identified and characterized a total of 34 sources. The algorithmically-obtained apertures and pixel assignment maps for these sources are shown on top of the Herschel 70 $\mu$m images in Figure \ref{fig:masks}. 

Multi-wavelength SOFIA-FORCAST images from 7.7 to 37~$\rm \mu m$ and archival Spitzer-IRAC 8~$\rm \mu m$ and Herschel-PACS 70~$\rm \mu m$ images (when available) for each region, including identified sources and their apertures, are presented in Figures~\ref{fig:AFGL5180} to \ref{fig:W3}. The aperture radii and fluxes obtained for each source are listed in Table~\ref{tab:fluxes}.

To build the SED for each source, we consider fluxes, when available, from Spitzer-IRAC images at 3.6, 4.5, 5.8, and 8.0 $\mu$m, SOFIA-FORCAST images at 7.7, 19.1, 31.5, and 37.1 $\mu$m, and Herschel-PACS/SPIRE 70, 160, 250, 350, and 500 $\mu$m. 
As described in Section \ref{sect:methods}, the fluxes are obtained through aperture photometry using the SedFluxer functionality of \verb+sedcreator+ version 2.0, where we have updated the package to enable usage in crowded regions. 

Figure \ref{fig:sed1} shows the measured SEDs for each source in this sample, with the good model fits overlaid. For 24 of the 34 SOMA V sources, the best fitting model has $\chi^2<1$. Three sources have a best fit with $1<\chi^2<2$. The following seven sources have relatively poor fits for their best models: G30.76 p01 ($\chi_{\rm min}^2=3.1$); IRAS20343 p01 ($\chi_{\rm min}^2=5.9$); W3 p01 ($\chi_{\rm min}^2=20.98$); W3 p06 ($\chi_{\rm min}^2=2.3$); W3 p09 ($\chi_{\rm min}^2=3.3$); W3 p11 ($\chi_{\rm min}^2=2.4$); and W3 p12 ($\chi_{\rm min}^2=2.7$). Thus, most of the SOMA V SEDs are reasonably well fit with the RT models for massive protostars forming via TCA.

The three primary physical parameters constrained by the SED fitting process are the initial core mass ($M_c$), mass surface density of the surrounding clump environment ($\Sigma_{\rm cl}$), and current protostellar mass ($m_*$). Figure \ref{fig:sed_2D_results_soma_v} shows the good model distributions in the $\Sigma_{\rm cl}$--$M_c$, $m_*$--$M_c$, and $m_*$--$\Sigma_{\rm cl}$ planes for each source.
This figure thus displays the degeneracies that arise from deriving protostellar properties via SED fitting alone.





Table \ref{tab:best_models} lists the properties of the best five models and the average and dispersion of all the good models. When an empirical $\Sigma_{\rm cl,GB}$ is available via graybody fitting, only models with the closest (in logarithmic space) $\Sigma_{\rm cl}$ value sampled by the grid are included in this averaging.
Averages are calculated as in SOMA IV, where all good models are equally weighted, including models with the same intrinsic physical properties but different viewing angles (however, note that only the best value of $A_V$ is considered for these). 



\subsection{Region summaries and identified sources}\label{sect:sources}


\subsubsection{AFGL 5180}

AFGL~5180 (also known as G188.95+0.89 or S252) is a massive star-forming region located at a distance of $1.76\pm0.11\:$kpc \citep{oh2010}. AFGL 5180 is part of the Gemini OB1 star-forming complex \citep{Zucker2020} and has been shown to be the source of bright Class II 6.7 GHz methanol maser emission \citep{Menten1991,Goedhart2014}.
\cite{Mutie2021} reported eight 1.3 mm sources in this region.
AFGL 5180 also hosts multiple outflows, identified in $^{12}\text{CO}(2-1)$ line emission by \cite{Mutie2021}. \cite{Crowe2024} studied this region using high-resolution near-infrared (NIR) imaging data and identified at least three outflows, consistent with multiple driving sources.

\begin{figure*}
\includegraphics[width=1.0\textwidth]{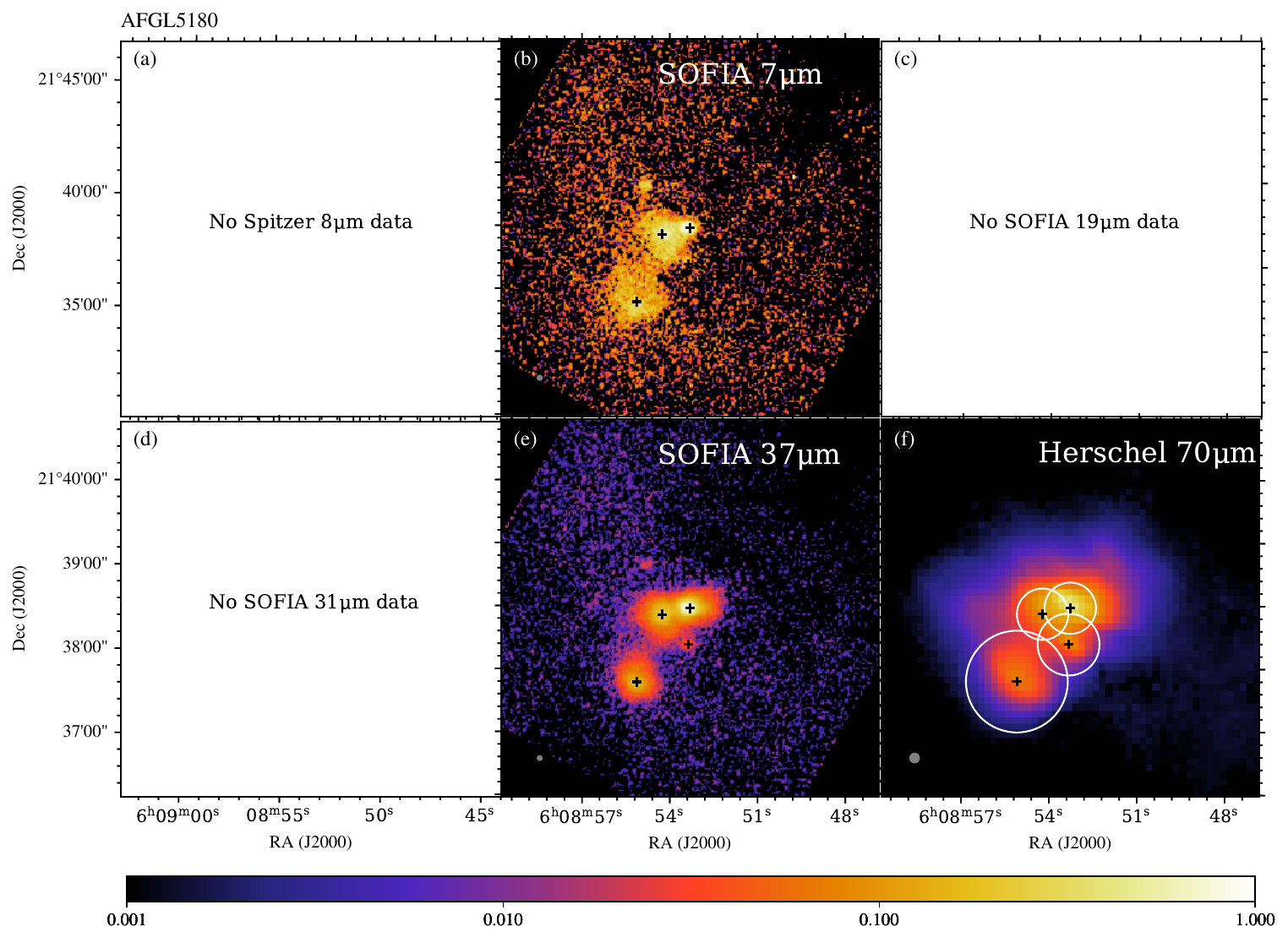}
\caption{Multiwavelength images of AFGL 5180 with the facility and wavelength given in the upper right of each panel. The color map indicates the relative flux intensity compared to that of the peak flux in each image panel. The white circles shown in (f) denote the apertures used for the fiducial photometry. Gray circles in the lower left show the resolution of each image. The black crosses in all panels denote the positions of the sources as described in Section \ref{sect:methods}.\label{fig:AFGL5180}}
\end{figure*}

Figure \ref{fig:AFGL5180} shows the multiwavelength data for AFGL~5180. SOFIA $19.7\:{\rm \mu m}$ and $31.5\:{\rm \mu m}$ data were not available for this region. Four sources were found using the detection algorithm operating on the 37~$\rm \mu m$ image. We note that there is a relatively faint $37\:{\rm \mu m}$ source seen to the north of the main cluster that does not meet the threshold to be identified by the algorithm. While this source is visible at $7.7\:{\rm \mu m}$, it does not stand out at $70\:{\rm \mu m}$. 

As described below, the SED fitting results imply that p01, p02 and p03 are massive protostars with $m_*\sim 12-16\:M_\odot$, while p04 has $\sim 5\:M_\odot$. Protostar p01, with $m*\simeq 12^{+8}_{-5}\:M_\odot$ is the primary outflow driving source of S4/mm1 discussed in the NIR study by \citet{Crowe2024}. Protostar p04 is associated with AFGL 5180 S region discussed by \citet{Crowe2024}. Protostars p02 and p03 were not specifically discussed by \citet{Crowe2024}. However, p02 is co-located with a bright NIR star apparent in their HST/WFC3-IR images. It is possible that this star is already formed and locally heating dust of a molecular cloud to give appearance of a protstellar SED. Similarly, we note that p03 coincides with the NIR reflection nebula that is likely tracing the outflow cavity of p01. Thus, even though the source appears quite spatially concentrated at $37\:{\rm \mu m}$, it is possible that this region is externally heated and masquerading as a protostar. The absence of significant mm emission from this region \citep[see][]{Mutie2021,Crowe2024} tends to favor this latter interpretation. These examples in AFGL 5180 highlight the caveats of protostellar source characterization from MIR - FIR imaging and SED fitting. In the particular case of p01, our inferred source luminosity and protostellar mass may be underestimates, i.e., if the fluxes from p03 should be included in this source.


\subsubsection{G18.67+00.03}

\begin{figure*}
\includegraphics[width=1.0\textwidth]{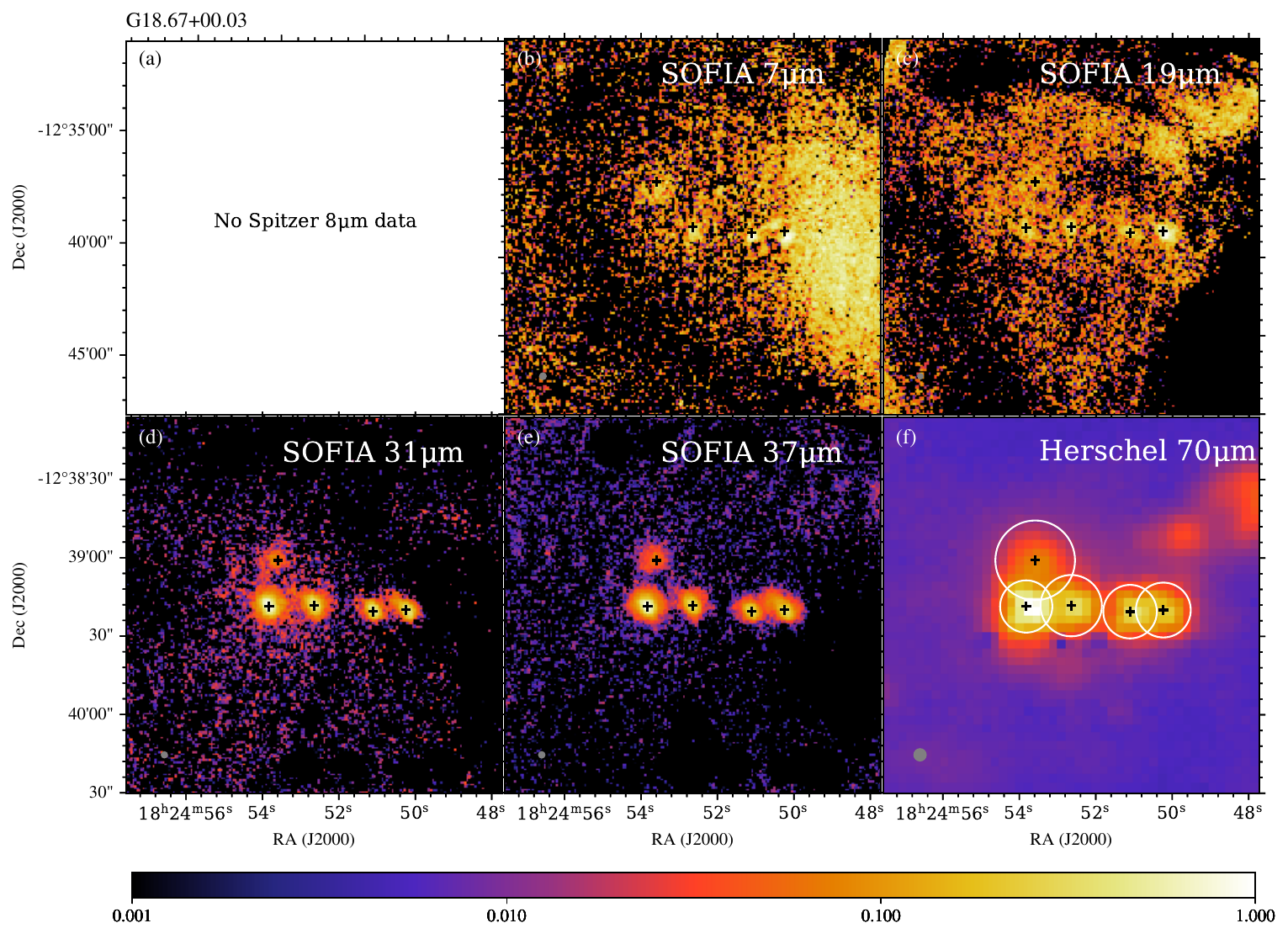}
\caption{Multiwavelength images of G18.67+0.03, following the format of Figure~\ref{fig:AFGL5180}. \label{fig:G18.67}}
\end{figure*}

G18.67+00.03 is a massive protocluster located at a distance of 10.8~kpc \citep{Cyganowski_2012} and identified using GLIMPSE 4.5 $\mu$m images from \cite{Cyganowski_2008}. \cite{Cyganowski_2012} detected four compact 1.3~mm continuum sources in this region, three of which are known 44 GHz Class I $\rm CH_3OH$ maser sources \citep{Cyganowski_2009}. Two of these sources also correspond to 6.7~GHz Class II $\rm CH_3OH$ masers. Species including SO, DCN, $\rm CH_3OH$, $\rm H_2CO$, and CO isotopes have all been detected toward these sources. Each of the three maser sources exhibits both redshifted and blueshifted $^{13}\text{CO}(2-1)$ emission offset from one another and centered on a compact source, implying the presence of bipolar molecular outflows \citep{Cyganowski_2012}. 

Figure \ref{fig:G18.67} shows multiwavelength data of G18.67+00.03, where five sources are identified. Four of these correspond to the 1.3 mm continuum sources of \cite{Cyganowski_2012}, and an additional fainter source has been found slightly to the north of these. All five sources are in the Hi-GAL catalog \citep{Molinari_2016}. 

From our SED fitting results we find that p01 - p03 in this region are protostars with $m_*\sim 20\:M_\odot$, while p04 has $\sim18\:M_\odot$ and p05 has $\sim14\:M_\odot$. Thus, all five sources are massive protostars. We note the striking linearity of the apparent spatial distribution of p01 to p04. We also note that given the relatively large distance to this region, the projected separation between the protostars, i.e., $\sim 10\arcsec$ corresponds to relatively large scales of about 0.5~pc.

\subsubsection{G28.37+00.07}

\begin{figure*}
\includegraphics[width=1.0\textwidth]{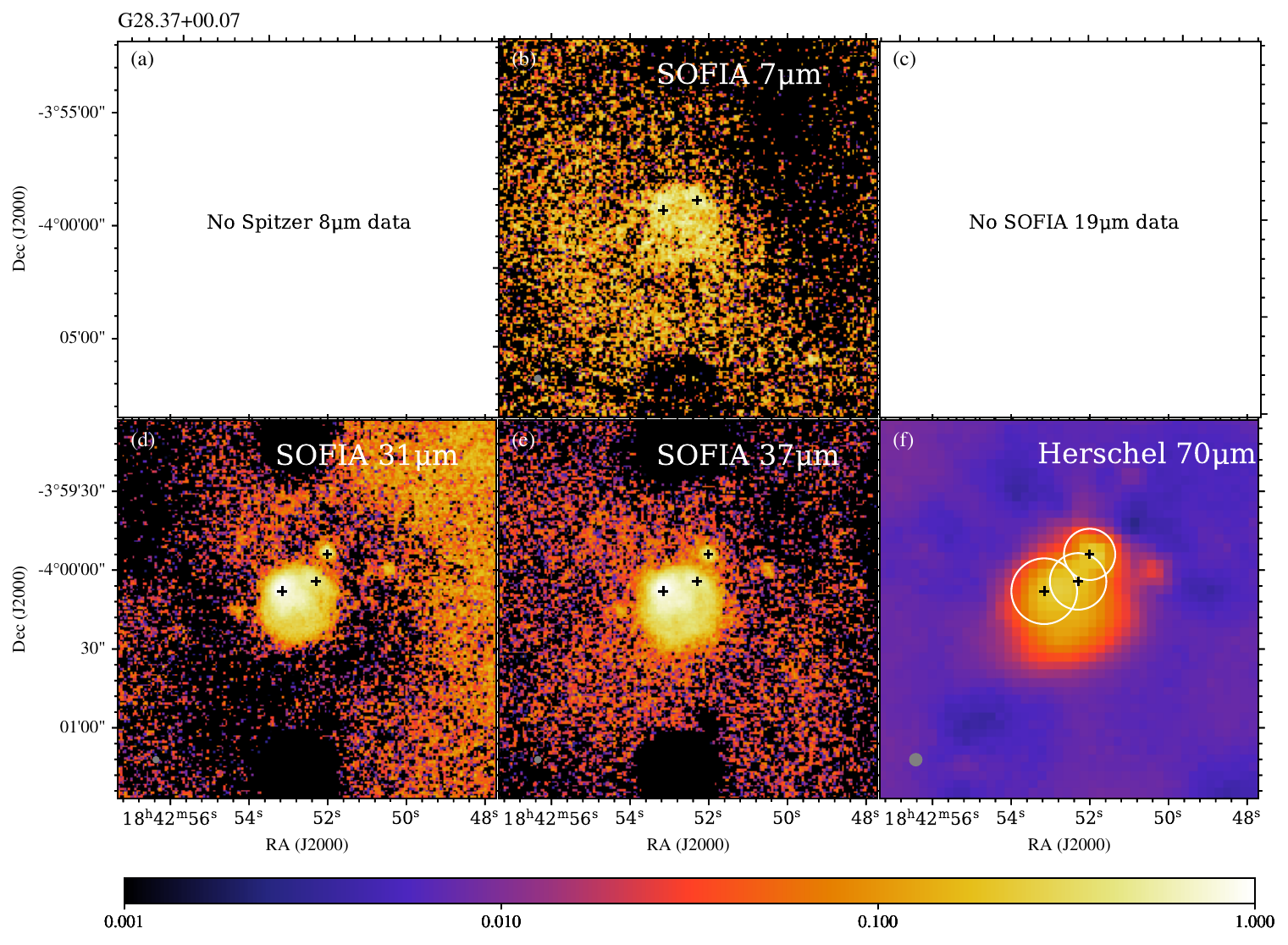}
\caption{Multiwavelength images of G28.37+00.07, following the format of Figure \ref{fig:AFGL5180}. \label{fig:G28.37}}
\end{figure*}

G28.37+00.07 \citep[also known as Cp23 in][]{moser2020} is located in IRDC G28.37+00.07 \citep[also known as IRDC C in the sample of][]{butler2009,butler2012}, which is one of the most massive IRDCs in the Galaxy. The source has a kinematic distance of about 5 kpc \citep{Simon2006}. 
From near-IR (NIR) and MIR extinction maps, the mass of IRDC G28.37+00.07 is estimated to be $68,300\:M_\odot$ \citep{Kainulainen2013}. \citet{Lim2014} used sub-mm dust emission observed by Herschel to estimate its mass as 72,000 $M_\odot$. G28.37+00.07 is one of the most crowded regions in the IRDC, with four sources in the Hi-GAL catalog. In the previous SED analysis of the region by \citet{moser2020}, G28.37+00.07 was unresolved and was modeled as a single source with $m_*=8\:M_\odot$, an initial core mass of $M_c\simeq 300\:M_\odot$, and a current core envelope mass of $M_{\rm env}=300\:M_\odot$. 

Figure \ref{fig:G28.37} shows the multiwavelength data for G28.37+00.07. We note that SOFIA $19.7 \mu$m data were available, but contained negative flux artifacts and so were not used in the analysis. Three sources were detected for this region, two of which correspond to sources in the Hi-GAL catalog. In Figure \ref{fig:G28.37}, we can see a fainter source to the right of the other three, but this source did not meet our detection threshold. 

From our new SED fitting results we find that p01 has a mass of $m_*\sim 14\:M_\odot$. The source p02 is estimated to have $m_*\sim 5\:M_\odot$, while p03 is estimated to have $m_*\sim 8.5\:M_\odot$. We note that the uncertainties in these masses are relatively large, so that these estimates are consistent with that of \citet{moser2020}.

\subsubsection{G30.76+00.20}

\begin{figure*}
\includegraphics[width=1.0\textwidth]{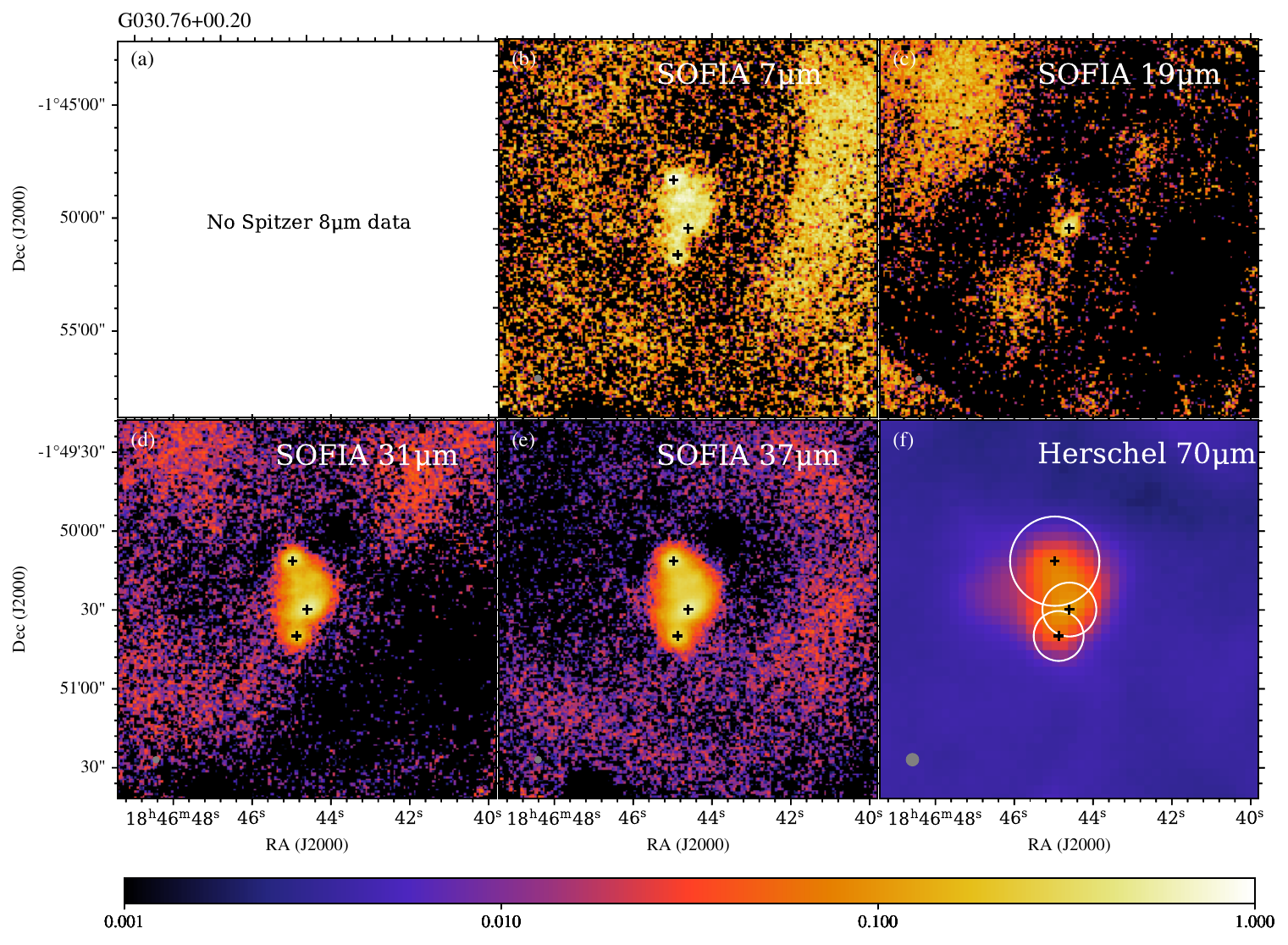}
\caption{Multiwavelength images of G30.76+00.20, following the format of Figure \ref{fig:AFGL5180}.\label{fig:G30.76}}
\end{figure*}

G030.76+00.20 is a massive star-forming region located at a distance of 6.4 kpc \citep{Veneziani2013}. A FIR source was first reported at this location in the Herschel Hi-GAL catalog \citep{Molinari_2016}.
Figure \ref{fig:G30.76} shows the multiwavelength data available for this region. Three sources were identified, each with $m_*\sim 10-15\:M_\odot$.


\subsubsection{G58.77+00.65}

\begin{figure*}
\includegraphics[width=1.0\textwidth]{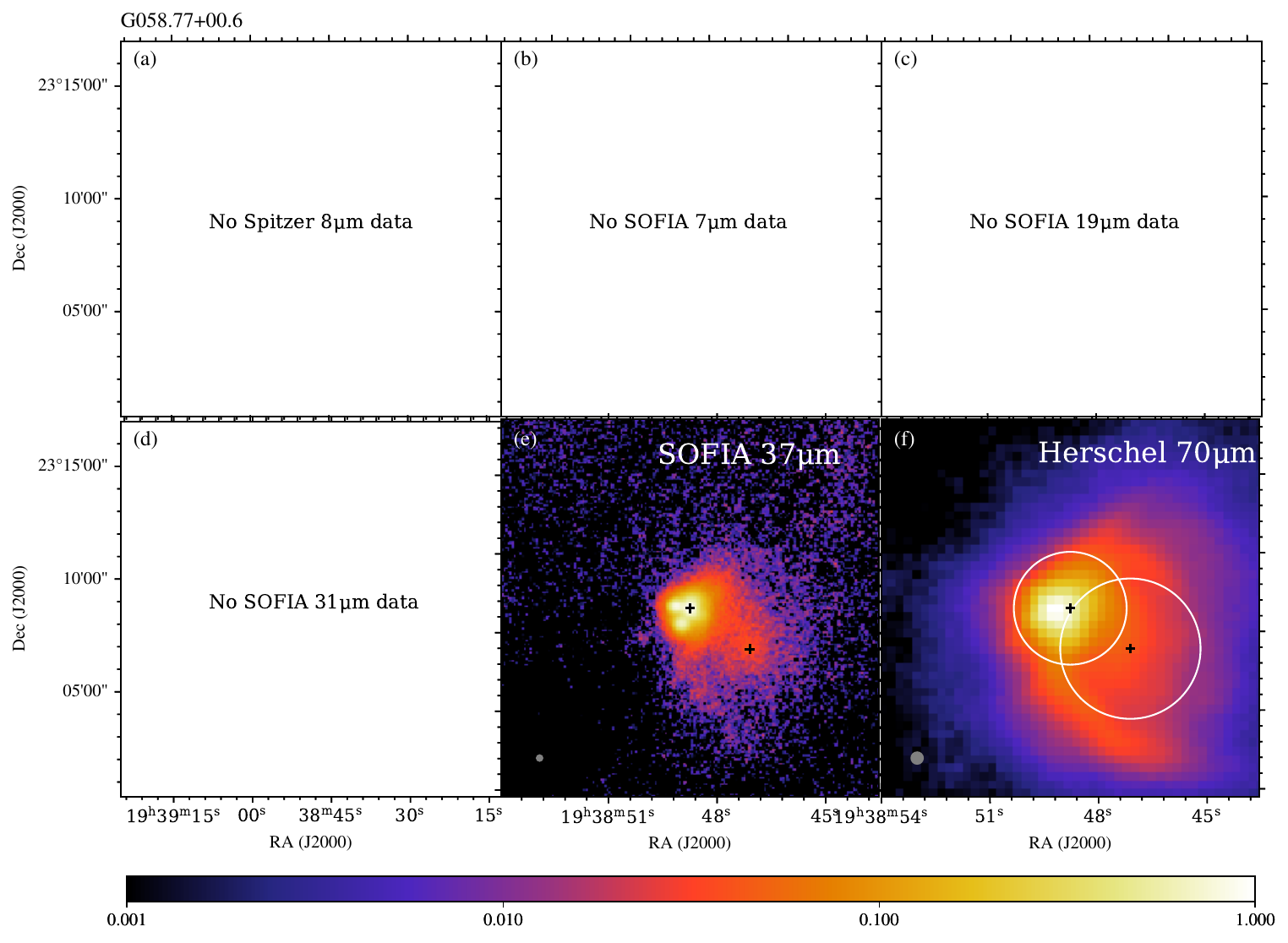}
\caption{Multiwavelength images of G58.77+00.65, following the format of Figure \ref{fig:AFGL5180}. \label{fig:G58.77}}
\end{figure*}

G58.77+00.65 was first reported in the CORNISH survey of ultra-compact H II region candidates \citep{cornish}. This region is located at a distance of 3.3~kpc \citep{mege2021}.  Multiwavelength data for G58.77+00.65 are shown in Figure~\ref{fig:G58.77}. Unfortunately, SOFIA $7.7$, $19.7$, and $31.5$ $\mu$m data were not available for this region. Four sources were originally detected in this region, three of which were located within a projected circular radius of $\sim6\arcsec$, which may indicate a very high protostellar number density. However, for our SED fitting two of these were removed during the source down selection process, which ensures that no two sources are within 10\arcsec\ of one another. The brightest of the remaining sources corresponds with a Hi-GAL-identified source \citep{Molinari_2016}. Since this source aperture also contains the removed sources, its flux measurements may be overestimates of the flux from the primary source in the region.
With this caveat in mind, our SED fitting of p01 indicates $m_*\sim13\:M_\odot$. The other source that is resolved in the region, i.e., p02, is inferred to have $m_*\sim 11\:M_\odot$.

\subsubsection{IRAS 20343+4129}

\begin{figure*}
\includegraphics[width=1.0\textwidth]{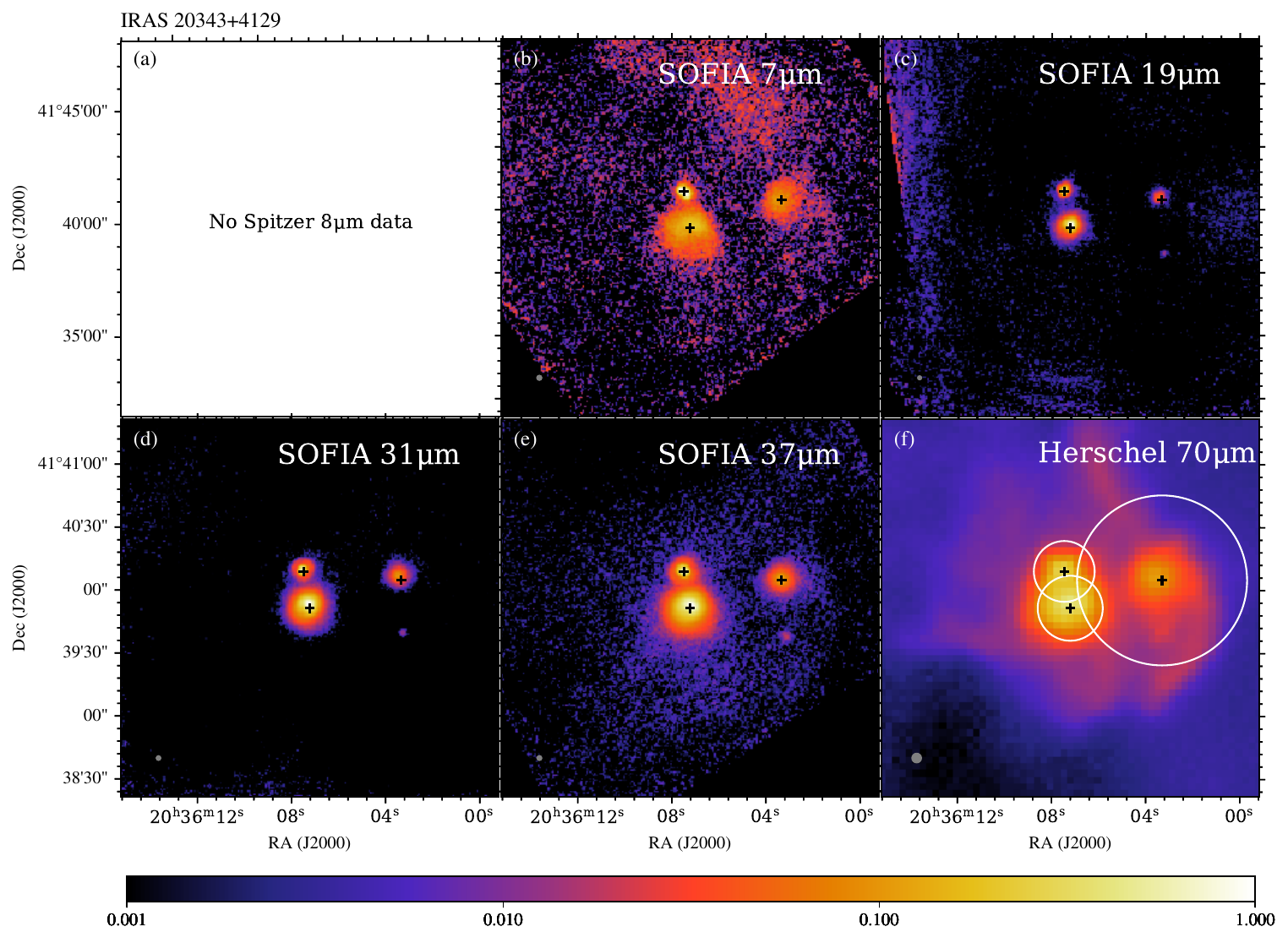}
\caption{Multiwavelength images of IRAS 20343+4129, following the format of Figure \ref{fig:AFGL5180}. \label{fig:IRAS20343+4129}}
\end{figure*}

IRAS 20343+4129 was first reported in the \cite{sridharan2002} list of 
high-mass protostellar candidates, which listed its distance as 1.4~kpc.
This region hosts two massive molecular outflows \citep{beuther2002b} and three K-band continuum YSOs exhibiting circumstellar $\text{H}_2$ emission \citep{Kumar2002}. \cite{Campbell_2008} found two of these sources to have mid-IR counterparts (IRS 1 and IRS 3). The brighter of the two CO outflows is located between these two sources and is oriented north-south, with red and blue lobes that are extended east-west \citep{Campbell_2008}. \cite{Kumar2002} suggest that the extended $H_2$ emission in the east-west direction towards IRS 1 is attributed to a circumstellar disk perpendicular to the outflow axis orientation. \cite{Palau2007} detected a 1.3~mm dust peak and a CO(2-1) peak that coincide with IRS2 and argued that the source is a low- or intermediate-mass YSO. 

Figure \ref{fig:IRAS20343+4129} shows some of the data available for IRAS 20343+4129. Three sources were detected using the methods described in \S\ref{sect:methods}. The fainter source seen to the south of p03 did not meet our detection threshold. From SED fitting we find that p01 and p02 can both be fit by protostellar RT models with $m_*\sim 10\:M_\odot$. The third source, p03, which we note has a larger aperture, can be fit with a similarly massive model.

\subsubsection{W3 IRS5}

\begin{figure*}
\includegraphics[width=1.0\textwidth]{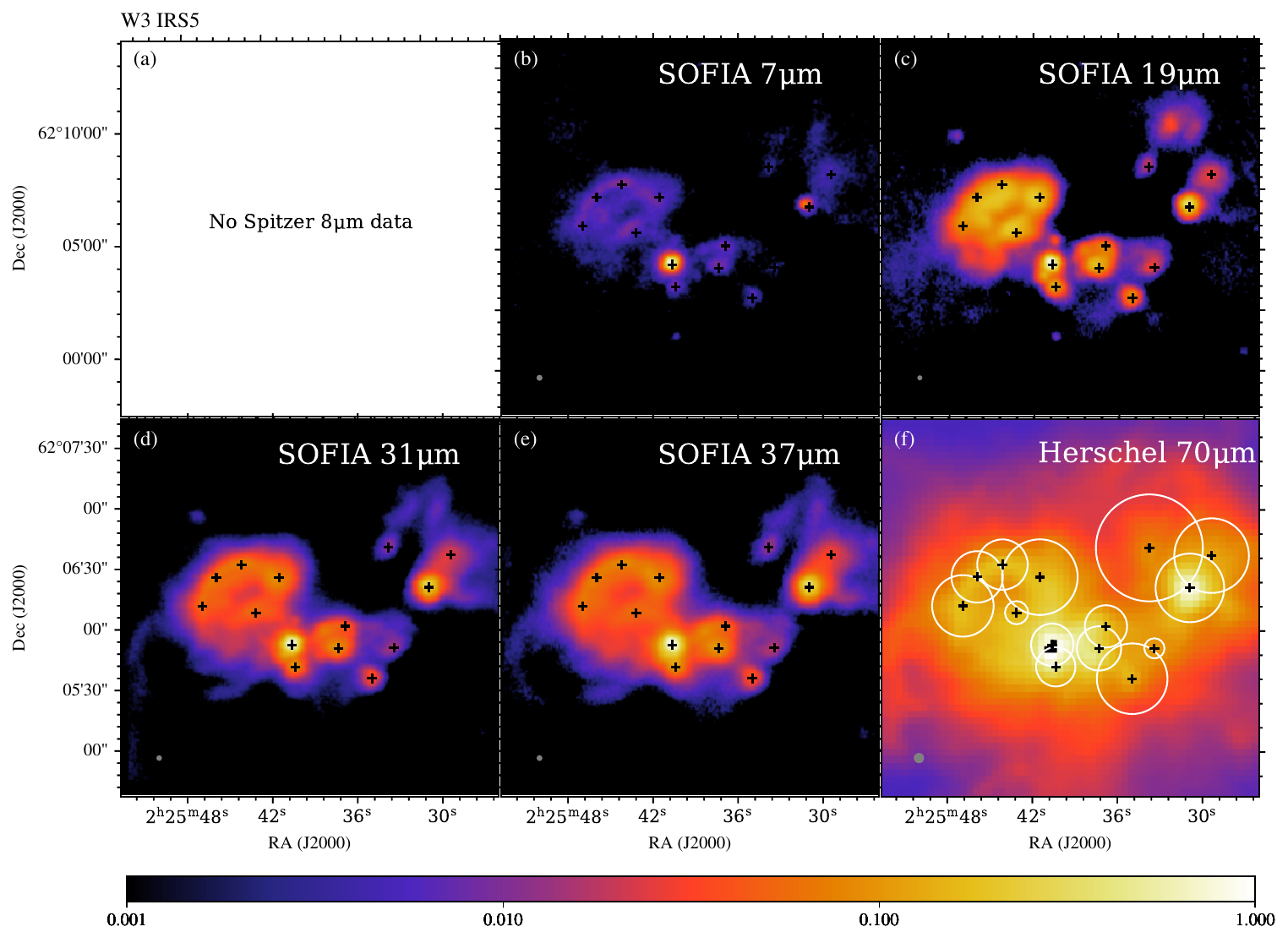}
\caption{Multiwavelength images of W3 IRS5, following the format of Figure \ref{fig:AFGL5180}. \label{fig:W3}}
\end{figure*}

W3 IRS5 is a massive star-forming region within the Perseus arm located $1.83\pm0.14\:$kpc away \citep{Imai}. \cite{1995ApJ...454..831C} estimated its total luminosity to be $2\times10^5\:L_{\odot}$. It has a high stellar projected number density of $10^4\: {\rm pc}^{-2}$, estimated by 
\cite{2008hsf1.book..264M} using near-IR observations. Using 1.4~mm observations with the Plateau de Bure Interferometer (PdBI), \cite{Rodon} estimated the protostellar number density to be $>10^6\:{\rm pc}^{-3}$. The close and dense nature of W3 IRS5 make it an interesting region in which to study massive star formation in a clustered environment \citep{Wang2013}. From (sub)millimeter observations, W3 IRS5 is also known to have a physically and chemically complex molecular structure \citep{Wang2013}. Bipolar outflows have been detected using CO \citep[e.g.,][]{1992ApJ...386..604M,Ridge2001} and SiO \citep[e.g.,][]{Gibb2007ASO,Rodon}, likely driven by infrared sources NIR1 and NIR2 \citep{Megeath2005ApJ...622L.141M}. NIR1 and NIR2 are coincident with two hypercompact H II regions detected by \cite{vanderTak2005} in the mid-IR. 

Figure \ref{fig:W3} shows our SOFIA multiwavelength data of W3 IRS5. Fourteen protostar candidates have been identified using our source detection algorithm. The primary source, p01, is located near the center of the region and has an inferred $m_*\sim 10 - 20\:M_\odot$, depending on whether or not restricted SED fitting is carried out using the empirical estimate of $\Sigma_{\rm cl,GB}$. However, we note that this is a source for which the best-fit model has a very poor $\chi^\sim 15$. From inspection of the fits, this may indicate difficulties in resolving the true FIR fluxes near the peak of the SED. The other 13 sources are generally fit quite well by the RT models. The next three sources, p02 - p04 have $m_*\sim 20\:M_\odot$. Among the remaining sources, are all good candidates for being massive protostars with estimations of $m_*\gtrsim 8\:M_\odot$. We caution that not all of these sources show concentrated morphologies in their MIR-FIR emission, so there may be some confusion with externally heated structures. Similarly, there are some relatively faint, but concentrated sources that we see in the SOFIA-FORCAST images, but which do not enter our source list, either because they fail to meet the detection threshold at $37\:{\rm \mu m}$ or because they are too close to other sources for resolution of their FIR emission.

\section{Global SOMA Survey Results}\label{sect:discussion}

Here we discuss the results from the 34 protostellar candidate sources identified in the 7 regions presented in this paper, along with the sources presented in Papers I-IV of the SOMA survey, but re-analyzed with the updated definitions of ``good'' SED model fits described above.



\subsection{Graybody-derived $\Sigma_{\rm cl}$}

Here we examine results of the graybody-derived values of mass surface density, i.e., $\Sigma_{\rm cl,GB}$ (see Sect. 3.3). This is a more direct, empirical measure of the mass surface density of the clump environment compared to that inferred from the SED fitting of the protostellar models, i.e., $\Sigma_{\rm cl,SED}$. Our goals are thus to see how well these compare, but also to examine results of ``restricted SED fitting'', when $\Sigma_{\rm cl,SED}$ is forced to be the value in the ZT18 model grid that is closest to the observed $\Sigma_{\rm cl,GB}$.

\begin{figure*}[!htb]
\includegraphics[width=0.45\textwidth]{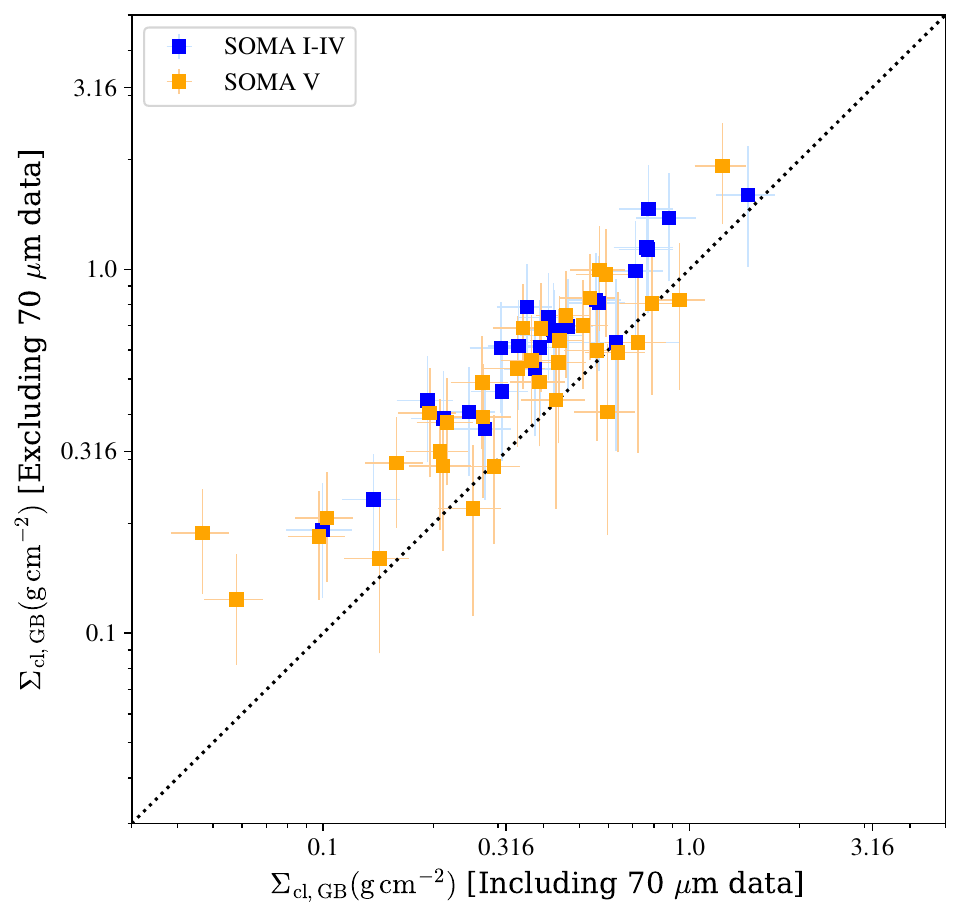}
\includegraphics[width=0.45\textwidth]{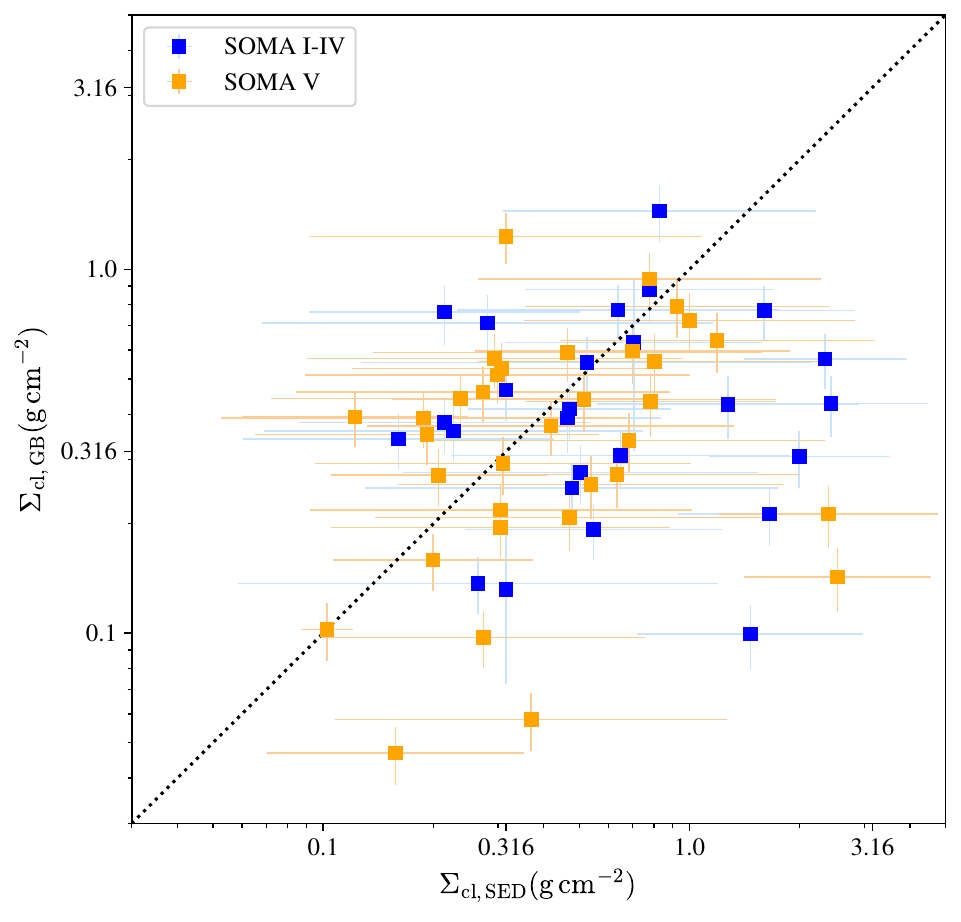}
\caption{{\it (a) Left:} Comparison of $\Sigma_{\rm cl,GB}$ measured only with Herschel 160 to 500~$\rm \mu m$ fluxes versus that measured with Herschel 70 to 500~$\rm \mu m$ fluxes.
%
{\it (b) Right:} Comparison of $\Sigma_{\rm cl,GB}$ (measured using Herschel 70 to 500~$\rm \mu m$ fluxes) versus $\Sigma_{\rm cl,SED}$.
}
\label{fig:GBsigma}
\end{figure*}

We first examine the sensitivity of the measurement of $\Sigma_{\rm cl,GB}$ to whether or not the Herschel $70 {\rm \mu m}$ flux values are included. Figure \ref{fig:GBsigma}a shows that $\Sigma_{\rm cl,GB}$ increases by about a factor of two when we remove the $70\:{\rm \mu m}$ fluxes from the graybody fitting process. Note, the emission at $70\:{\rm \mu m}$ could be dominated by warmer dust that is not well represented by a single temperature graybody model \citep[e.g., see][]{Guzman2015}. However, from our tests on comparing mass surface density measurements of protostellar cores (denoted by $\Sigma_{\rm c}$) defined by the remaining envelope mass, we find that more accurate results are obtained by including the $70 {\rm \mu m}$ flux (see Appendix~A).
Thus we retain this method as our fiducial case, but acknowledge that systematic uncertainties, perhaps up to a factor of two, result from the approximation of the emitting material as a single temperature graybody. 

In addition, it is important to note that the derived $\Sigma_{\rm cl,GB}$ values depend on the choice of the opacity law. If we instead use the opacity values from \cite{Ossenkopf1994} for grains with a thin ice mantle (as used by \citet{Lim2016} in the study of cold IRDC environments), we obtain $\Sigma_{\rm cl,GB}$ values that are factors of about three times smaller for each source. Thus, the choice of dust opacity may be the dominant source of systematic uncertainty in the measurement of $\Sigma_{\rm cl,GB}$.

Figure \ref{fig:GBsigma}b compares $\Sigma_{\rm cl,GB}$ values with those derived from SED fitting, $\Sigma_{\rm cl,SED}$, i.e., the average of good models. Note, these include all the sources of this paper (SOMA V), but only those sources of SOMA I - IV that have Herschel data allowing the $\Sigma_{\rm cl,GB}$ measurement. We see that $\Sigma_{\rm cl,SED}$ values are typically in reasonable agreement with $\Sigma_{\rm cl,GB}$, but with a tendency to be, on average, modestly larger by within a factor of two. However, given the potential systematic uncertainties in $\Sigma_{\rm cl,GB}$, discussed above, and in $\Sigma_{\rm cl,SED}$, we conclude that their agreement is satisfactory, i.e., within expectations. One can also see from Fig.~\ref{fig:GBsigma}b that the individual uncertainties in $\Sigma_{\rm cl,SED}$, which are the dispersion of the ``good'' model fits, are generally larger than the intrinsic uncertainties in $\Sigma_{\rm cl,GB}$ (i.e., ignoring systematic uncertainties in assumptions of single temperature graybody and choice of dust opacities). This motivates the case of ``restricted SED fitting'' in which we set $\Sigma_{\rm cl,SED}$ to be the value in the ZT18 model grid that is closest to $\Sigma_{\rm cl,GB}$ of a given source.


\subsection{The Environmental Conditions for Massive Star Formation}

\begin{figure*}[!htb]
\includegraphics[width=0.36\textwidth]{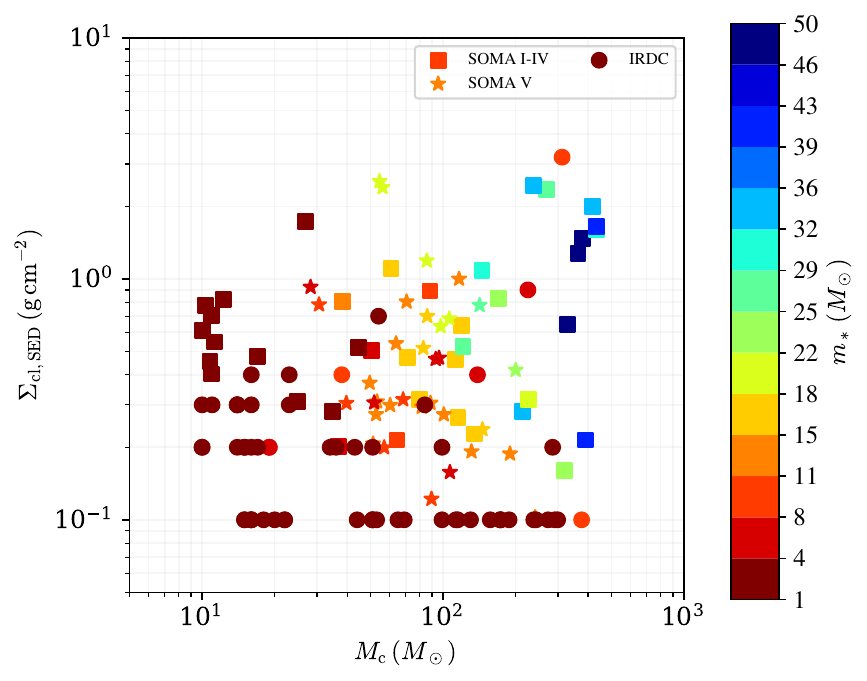}
\includegraphics[width=0.3\textwidth]{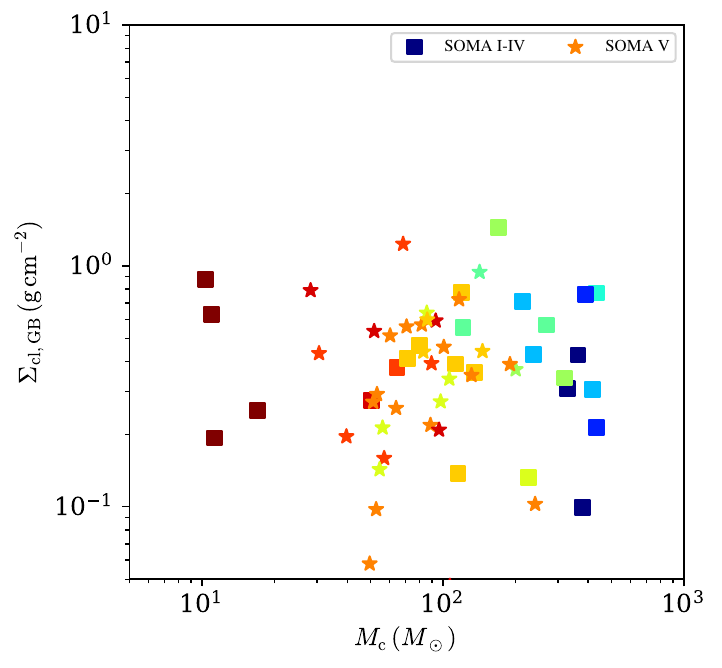}
\includegraphics[width=0.3\textwidth]{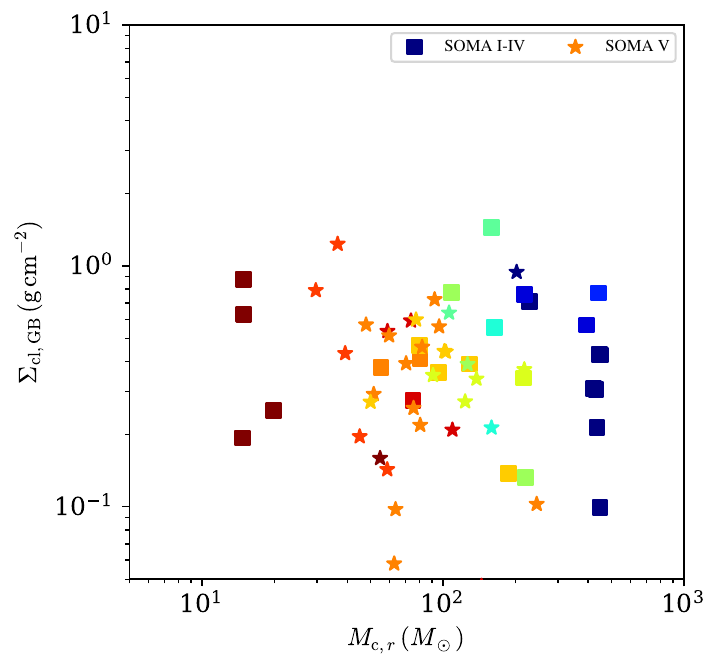}
\caption{{\it (a) Left panel:} Mass surface density of the clump environment ($\Sigma_{\rm cl}$) versus initial mass of the core ($M_c$) for SOMA I-V sources and the IRDC samples. Each data point is the average of good model fits. Each point is also color-coded with the current mass of the protostar ($m_*$). {\it (b) Middle panel:} Same as (a), but $\Sigma_\mathrm{cl}$ values are derived from the graybody fitting for the 60 SOMA I-V sources with available Herschel 70-500 $\mu m$ data (see text).
{\it (c) Right panel:} As (b), but $M_c$ and $m_*$ have been calculated with restricted SED fitting, i.e., using only models with a $\Sigma_\mathrm{cl}$ closest to the graybody fitting-derived value.
\label{fig:masses}}
\end{figure*}

\begin{figure*}[!htb]
\includegraphics[width=0.5\textwidth]{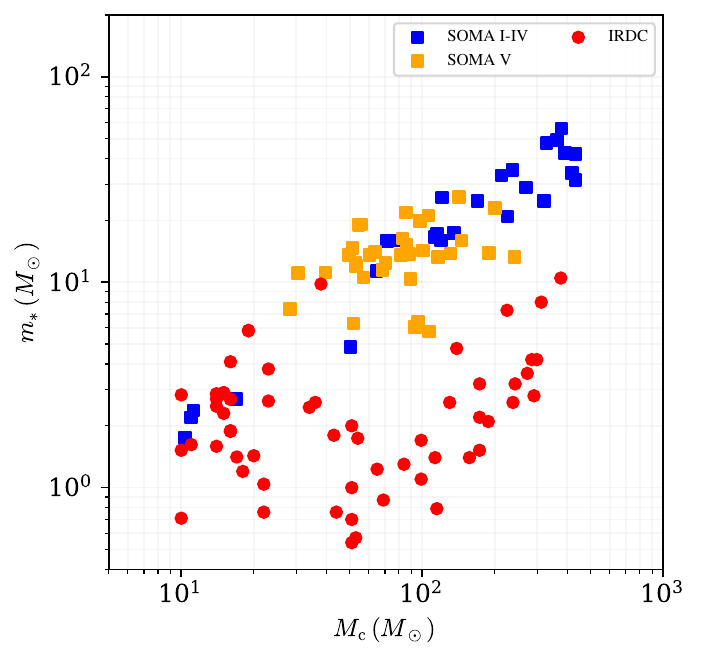}
\includegraphics[width=0.5\textwidth]{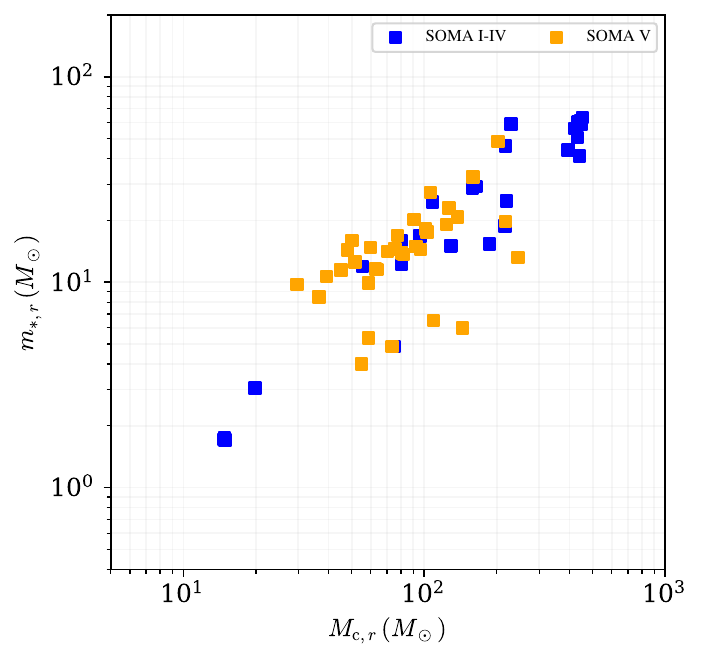}
\caption{{\it (a) Left panel:} Current protostellar mass ($m_*$) versus initial core mass ($M_c$) for SOMA I-V sources and the IRDC samples. Each data point is the average of good model fits. {\it (b) Right panel:} Same as (a), but $M_c$ and $m_*$ have been calculated with restricted SED fitting, i.e., using only models with a $\Sigma_\mathrm{cl}$ closest to the graybody fitting-derived value. 
\label{fig:mstar_Mc}}
\end{figure*}

Figure \ref{fig:masses}(a) shows clump environment mass surface density as derived from SED fitting, $\Sigma_{\rm cl,SED}$, versus initial core mass, $M_c$, showing the average of ``good'' models. The SOMA I-IV and V samples are plotted, as well as samples of protostars identified in infrared dark clouds (IRDCs) \citep{moser2020,liu2021}. The data points are also color coded by the SED-derived current protostellar mass, $m_*$. Figure~\ref{fig:masses}(b) plots the equivalent results, but for $\Sigma_{\rm cl,GB}$ and showing only those SOMA sources with Herschel data. Figure~\ref{fig:masses}(c) is the same as (b), but showing the results of restricted SED fitting where $\Sigma_{\rm cl,SED}$ has been set equal to a value in the ZT18 grid closest to $\Sigma_{\rm cl,GB}$. Figure \ref{fig:mstar_Mc}(a) shows $m_*$ versus $M_c$ for the SOMA I-V and IRDC samples, while Figure \ref{fig:mstar_Mc}(b) shows the results of the $\Sigma_{\rm cl,GB}$-restricted fitting.

We see that the initial core mass of the SOMA V sample ranges from $M_c \sim 30-200\:M_{\odot}$, i.e., a somewhat narrower range than that covered by the ZT18 model grid ($10-480\:M_{\odot}$) and that found in the SOMA I-IV and IRDC samples. The SOMA V sample shows a broad range of $\Sigma_{\rm cl,SED}$ from $\sim 0.1-3 \text{g cm}^{-2}$, i.e., the full range covered by the ZT18 models. However, as also shown in Fig.~\ref{fig:GBsigma}, we note that $\Sigma_{\rm cl,GB}$ occupies a systematically lower range that extends up to $\sim 1.5\:{\rm g\:cm}^{-2}$.



The current protostellar mass of the SOMA V sources ranges from $m_* \sim 4-40\:M_{\odot}$. As with initial core mass, this appears to be a somewhat narrower range than found in the SOMA I-IV plus IRDC samples. In the overall sample, we see that IRDC sources tend to represent earlier evolutionary stage protostars, i.e., with $m_*\sim 0.01 - 0.1 M_c$, while the SOMA sources typically have $m_* \simeq 0.1-0.3 M_c$.

\begin{figure*}[!htb]
\includegraphics[width=0.33\textwidth]{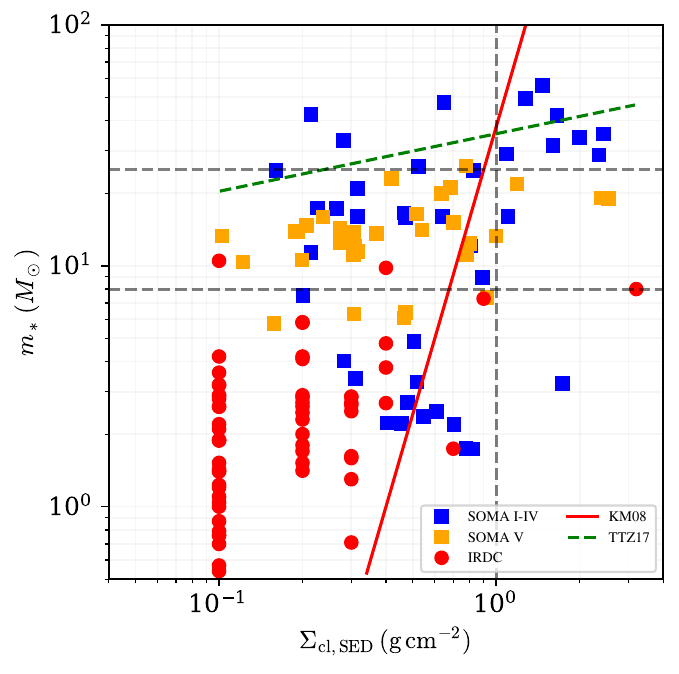}
\includegraphics[width=0.33\textwidth]{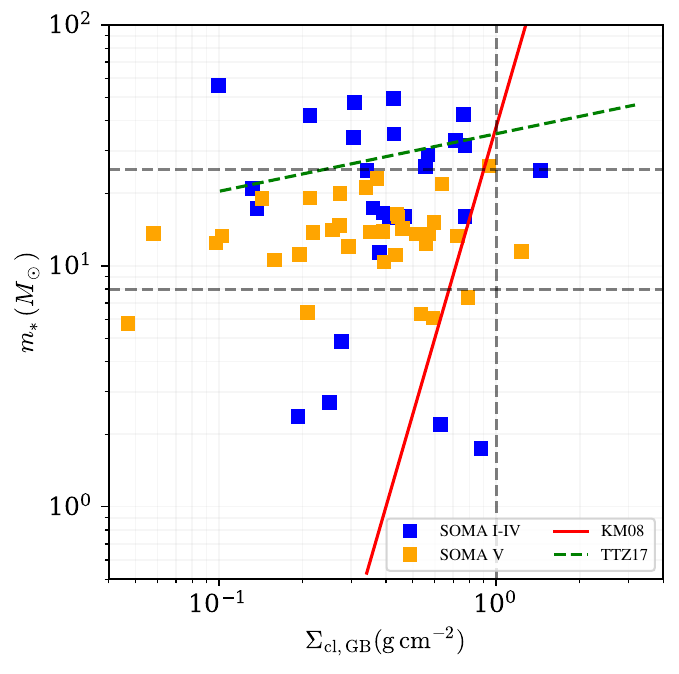}
\includegraphics[width=0.33\textwidth]{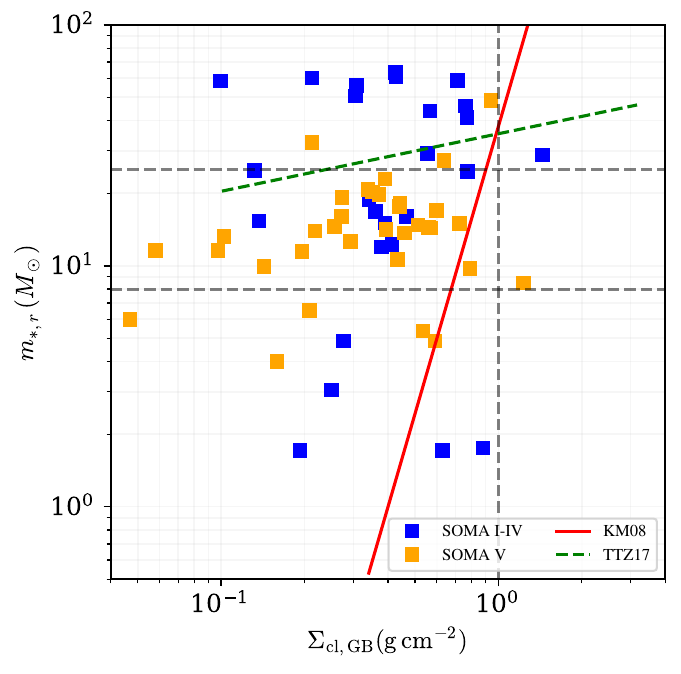}
\caption{{\it (a) Left panel:} Current protostellar mass ($m_\mathrm{*}$) versus clump environment mass surface density ($\Sigma_\mathrm{cl}$) for the SOMA I-V sources and IRDC protostars (see the text). Lines indicating reference values of $m_*=8$ and $25\:M_\odot$ and $\Sigma_{\rm cl}=1\:{\rm g\:cm}^{-2}$ (see the text) are highlighted. The red solid line shows the fiducial prediction of \citet{krumholz2008} (assuming their parameter values of $\delta=1$ and $T_b=10$\,K) for the minimum $\Sigma_{\rm cl}$ needed to form a star of given mass $m_*$. The green-dashed line shows the results for the final stellar mass formed from $100\:M_\odot$ prestellar cores as a function of $\Sigma_{\rm cl}$ \citep{tanaka2017}.
{\it (b) Middle panel:} $m_\mathrm{*}$ versus $\Sigma_\mathrm{cl}$ derived from the graybody fitting for the 60 SOMA I-V sources with available Herschel 70-500 $\mu m$ data (see the text).
{\it (c) Right panel:} As (b), but $m_*$ has been calculated using only models with a $\Sigma_\mathrm{cl}$ closest to the graybody fitting-derived value.
\label{fig:kmplot}}
\end{figure*}

Figure \ref{fig:kmplot}(a) shows the current protostellar mass $m_*$ versus clump environment mass surface density $\Sigma_{\rm cl}$, as derived from the SED fitting of ``good'' models. 
Figure \ref{fig:kmplot}(b) shows the equivalent results, but with $\Sigma_{\rm cl}$ now derived empirically from the graybody fitting around the source. Figure \ref{fig:kmplot}(c) is the same as (b), but with restricted SED fitting using the graybody fitting results to constrain $\Sigma_{\text{cl,SED}}$. 

Some models of massive star formation predict that in order to form massive protostars, a minimum value of $\Sigma_{\text{cl}}$ is required. For example, \citet{krumholz2008} (hereafter KM08) predicted that in order for high-mass stars to form, $\Sigma_{\text{cl}}$ must be $\gtrsim 1 \text{g cm}^{-2}$, with these conditions being needed so that a surrounding population of low-mass protostars has high enough accretion rates and luminosities to heat up the massive core, increase the Jeans mass, and prevent its fragmentation. The red solid line in each panel of Figure~\ref{fig:kmplot} shows the minimum $\Sigma_{\rm cl}$ needed to form a star of a given mass $m_*$ under the KM08 condition. Under this condition, all massive protostars should lie to the right of this line. 
As already noted in the SOMA IV paper, the derived properties of the SOMA and IRDC protostars are distributed broadly across a range of $\Sigma_{\rm cl}$ values that extend far below the KM08 limit. The results for $\Sigma_{\rm cl,GB}$ extend to even smaller values, so that relatively few sources are above the KM08 limit. Thus, even with the possibility of factors of several systematic uncertainties in $\Sigma_{\rm cl}$ from graybody fitting (see above), these results do not indicate any evidence for a threshold $\Sigma_{\rm cl}$ being needed for massive star formation. An alternative mechanism of preventing fragmentation via magnetic field support (requiring $\gtrsim$ 0.1 mG field strengths) has been proposed by \citet{butler2012}, with there being increasing observational evidence for such dynamically important $B-$fields present in early-stage \citep[e.g.,][]{2024ApJ...967..157L} and late-stage \citep[e.g.,][]{Wang2014,2024A&A...686A.281B} protostellar cores.


The TCA model of \cite{mckee2002,mckee2003} predicts that higher $\Sigma_{\text{cl}}$ leads to higher pressures in a self-gravitating clump and thus higher prestellar core densities. 
This leads to more efficient star formation from a core experiencing internal feedback processes from the protostar \citep{tanaka2017}. For example, a $M_c=100\:M_\odot$ core forms only a $\sim 20\:M_\odot$ star in a $\Sigma_{\rm cl}\sim 0.1 \: {\rm g\:cm}^{-2}$ environment, but a $\sim 50\:M_\odot$ star in a $\sim 3\:{\rm g\:cm}^{-2}$ environment. This example result for a $100\:M_\odot$ core is shown by the green dashed line in the panels of Fig.~\ref{fig:kmplot}. 

The SOMA IV paper found tentative evidence for the upper envelope of the points in the $m_*-\Sigma_{\rm cl}$ diagram following the shape of this TCA internal feedback prediction. The equivalent results of our analysis including the SOMA V sources are shown in Fig.~\ref{fig:kmplot}a. Since the SOMA I-IV sources define the most massive end of the distribution, this result still holds: i.e., the most massive, $\gtrsim 30\:M_\odot$ protostars tend to require $\Sigma_{\rm cl}\gtrsim 1\:{\rm g\:cm}^{-2}$ and the upper envelope could be bounded by the protostars expected to form from $\sim 200\:M_\odot$ cores. However, when we consider the results of $\Sigma_{\rm cl,GB}$ shown in Fig.~\ref{fig:kmplot}b and c, this trend becomes less clear. There are examples of $\sim 50\:M_\odot$ protostars forming in clump environments of only $\sim 0.1\:{\rm g\:cm}^{-2}$ and little hint of an increasing upper envelope of $m_*$ with increasing $\Sigma_{\rm cl}$. In the context of the TCA models, which are being fit to these sources, this requires the initial core masses in the low-$\Sigma_{\rm cl}$ environments to be systematically more massive.
A related point is that with lower values of $\Sigma_{\rm cl}$ from graybody fitting, the restricted SED fitting results with these conditions tend to increase the derived values of $m_*$. This is explained by the fact that lower $\Sigma_{\rm cl}$ values imply lower accretion rates, so that a larger $m_*$ is needed to produce a given bolometric luminosity. 





\begin{figure*}[!htb]
\includegraphics[width=0.48\textwidth]{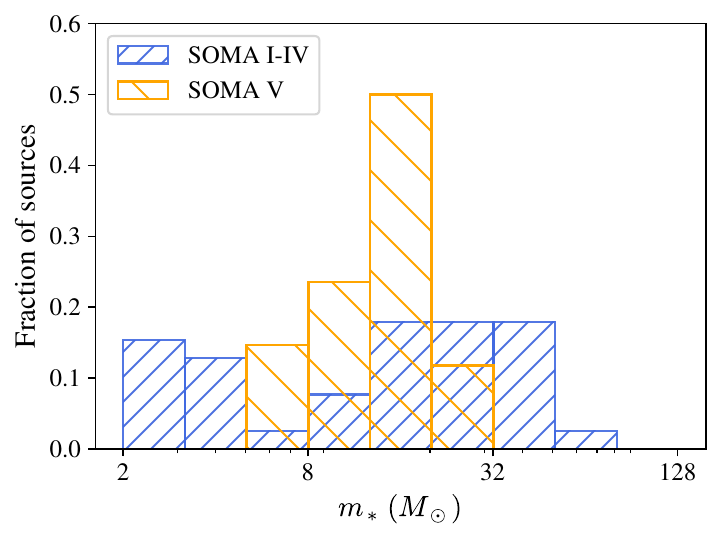}
\includegraphics[width=0.48\textwidth]{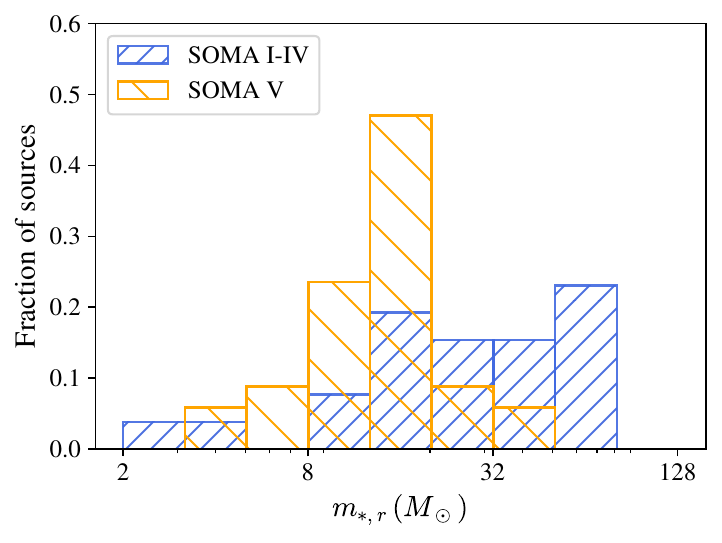}
\caption{{\it (a) Left panel:} Current protostellar mass ($m_*$) distribution for SOMA I-IV sources (blue) and SOMA V sources (orange). 
{\it (b) Right panel:} As (a), but for restricted SED fitting, i.e., only considering models with a $\Sigma_\mathrm{cl}$ closest to the graybody-derived value. 
\label{fig:mass_hists}}
\end{figure*}

In Figure \ref{fig:kmplot}, we also notice that most of the SOMA V sources have $m_* \lesssim 25$ $M_{\odot}$ (marked by the upper gray dashed line). 
Figure \ref{fig:mass_hists} compares the mass distribution for the SOMA I-IV sources to the distribution for the SOMA V sources, before and after restricting the range of good models based on the graybody-derived $\Sigma_{\rm cl}$ values. Again, we see that SOMA V sources are somewhat lacking from the highest $m_*$ bins. To assess the significance in the variation between these two distributions, we performed a Kolmogorov–Smirnov test to see if these distributions are consistent with each other. However, given the relatively small sample sizes, this yielded only marginally significant $p$-values of 0.025 (for Figure \ref{fig:mass_hists}a) and 0.003 (for Figure \ref{fig:mass_hists}b).

Nevertheless, these results could indicate an environmental dependence on the initial mass function of stars at the high-mass end, i.e., the most massive stars ($\gtrsim30\:M_\odot$) do not form in clustered environments. The five SOMA sources with the most massive values of $m_*$ are: G45.12+0.13 and G309.92+0.48 (presented in SOMA II) and AFGL2591, G32.03+0.05N, and G25.40-0.14 (presented in SOMA IV). After examining their SOFIA-FORCAST images, we note that these tend to be relatively isolated protostars.

Potential physical reasons for such a dependence of the IMF on environment could be related to increased competition for gas among stars that are forming in relatively close proximity. Such competition might occur at the pre-stellar core stage or at the later protostellar stage \citep[e.g.,][]{2010ApJ...725..134P}.

However, we also caution that our above observational result may be influenced by methodological differences in how flux estimation and SED fitting is done in crowded regions. As described above, our method of defining source apertures in crowded regions is limited by the presence of neighboring sources. This could cause a systematic underestimation in flux (both via limiting the area and by overestimating the background), and thus mass, of the most massive sources. On the other hand, we note that it seems likely that in some of the SOMA V regions we are actually overestimating fluxes since we cannot resolve sources that are closer than 10\arcsec\ apart, e.g., in G58.77+00.65. To more definitely establish if there is an environmental dependence of maximum protostellar mass will require independent measurements of mass, e.g., dynamical mass measurements via studies of accretion disk kinematics.


\section{Summary and Conclusions}\label{sect:conclusions}

In this paper, the fifth in the SOFIA Massive (SOMA) Star Formation Survey, we have analyzed 34 protostar candidates in 7 regions of relatively clustered massive star formation. 
We have also presented version 2.0 of \verb+sedcreator+, a Python package designed to facilitate the flux measurement and SED fitting process that was introduced by \citet{Fedriani_2023} (SOMA Paper IV). 
In this new version of the package, we introduce a source detection method using the dendrogram algorithm on median-filtered $37\:{\rm \mu m}$ images.
We also present an updated tool to automatically determine the aperture radii of sources in isolated or crowded regions so that this process can be done efficiently and uniformly. In addition, we have updated the flux calculation process to account for the presence of neighboring sources. We have presented an improved method of defining ``good'' SED model fits, which improves results, especially for faint sources. We have also developed a method for empirical clump mass surface density estimation via graybody fitting, which can then be used for ``restricted SED fitting''. Here we summarize our main findings:

\begin{enumerate}
    \item The SEDs of these 34 sources are generally well fit with the ZT18 models. We report the averaged properties of ``good'' model fits and their associated dispersions to account for the degeneracies present in the SED fitting process.
    \item These sources span a wide range of environments and properties. After fitting the SEDs with the ZT18 RT models, we estimate initial core masses spanning $M_c \sim 30-200\:M_{\odot}$, clump mass surface densities spanning the entire model grid range from $\Sigma_{\text{cl}} \sim 0.1-3  \text{ g cm}^{-2}$, and current protostellar masses from $m_* \sim 4-40\:M_{\odot}$.
    \item The distribution of SOMA protostars in the $m_*$ versus $\Sigma_{\text{cl}}$ plane provides insight into the conditions required for massive star formation to occur. Consistent with the SOMA I-IV results, massive protostars in the SOMA V sample appear to cover the entire sampled range of $\Sigma_{\text{cl}}$, which is inconsistent with the prediction that massive stars require a minimum of $\Sigma_{\text{cl}} \sim$ 1 ${\rm g\:cm}^{-2}$ to form. In addition to doubling the number of massive protostars forming in low-$\Sigma_{\text{cl}}$ environments, SOMA V also shows that this trend continues to hold in crowded regions. 
    \item We note tentative evidence for an environmental dependence on the initial mass function of stars at the high-mass end, i.e., the most massive stars ($\gtrsim30\:M_\odot$) do not appear to form in the most clustered environments. However, further investigation of this result is needed, likely requiring independent, dynamical estimates of protostellar masses.    
\end{enumerate}

\noindent
{\it Acknowledgements:} R.F. acknowledges support from the grants Juan de la Cierva FJC2021-046802-I, PID2020-114461GB-I00, PID2023-146295NB-I00, and from the Severo Ochoa grant CEX2021-001131-S funded by MCIN/AEI/ 10.13039/501100011033 and by ``European Union NextGenerationEU/PRTR''. J.C.T. acknowledges support from USRA-SOFIA grant 09$\_$0085, NSF grants AST-2009674 and AST-2206450, ERC Advanced Grant 788829 (MSTAR), and the CCA Sabbatical Visiting Researcher program. G.C. acknowledges support from the ESO Fellowship Program and funding from the Swedish Research Council (VR Grant; Project: 2021-05589).

\appendix 
\restartappendixnumbering

\section{Testing Graybody $\Sigma$ Measurements}

\begin{figure*}[!htb]
\includegraphics[width=0.48\textwidth]{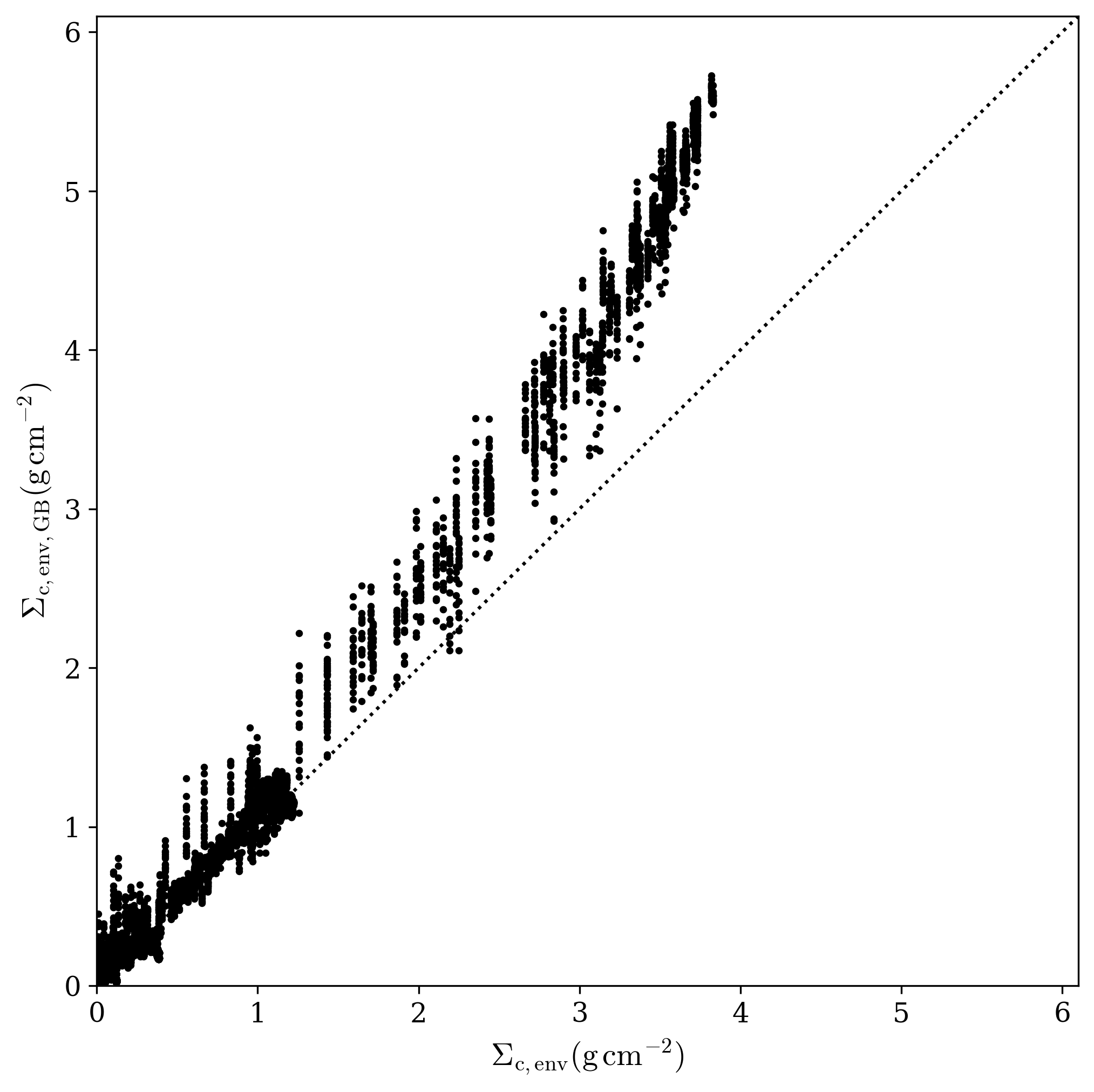}
\includegraphics[width=0.48\textwidth]{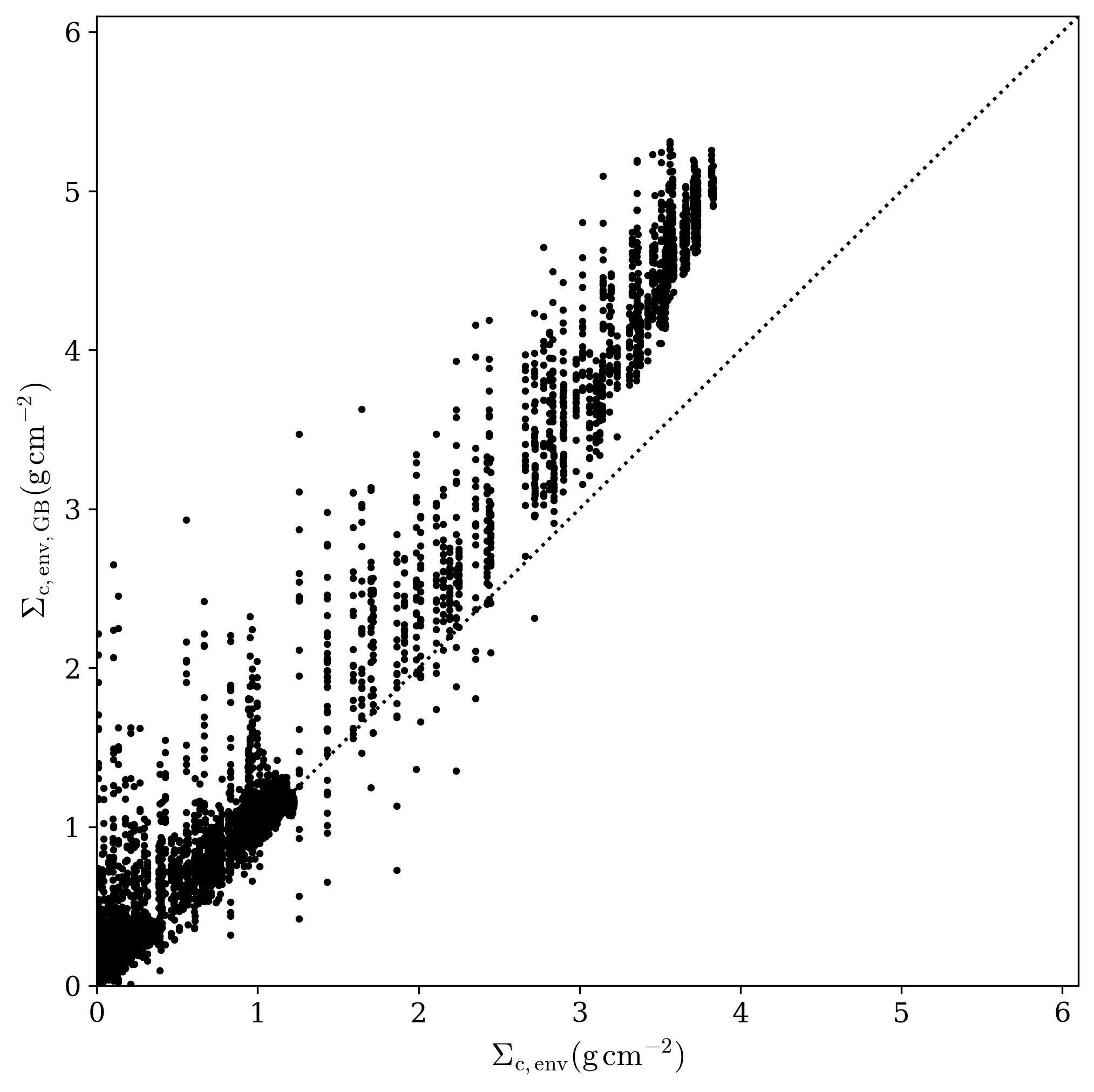}
\caption{{\it (a) Left panel:} Clump environment mass surface density derived from graybody fitting to Herschel 70-500 $\mu$m background flux values ($\Sigma_\mathrm{c, env, GB}$) vs. theoretical $\Sigma_\mathrm{env}$ for the ZT18 grid of Radiative transfer models.
{\it (b) Right panel:} As (a), but excluding the Herschel 70 $\mu$ m data from the fitting.
\label{fig:appendixSigmas}}
\end{figure*}

Here we apply the graybody fitting method described in Section \ref{sect:methods} to the ZT18 grid of radiative transfer models to derive $\Sigma_{\rm c,env,GB}$, i.e., the mass surface density of the protostellar core envelope, which can then be compared with the actually value of the model. To accomplish this, we first constructed an SED of the predicted fluxes at the Herschel 70, 160, 250, 350, and 500 $\mu$m bands for each model in the grid. We then fit the graybody distribution given in Section \ref{sect:methods} to each of these SEDs to derive a corresponding $\Sigma_{\rm c,env,GB}$ value. Next, we repeated this process, but excluded the 70 $\mu$m data from the graybody fitting. We then computed the actual $\Sigma_{\rm c,env}$ for the protostellar cores defined by the remaining envelope mass ($M_{\rm c,env}$) for each of the models via:
\begin{equation}
    \Sigma_{\rm c,env} = M_{\rm c,env} / (\pi R_c^2),
\end{equation}
where $R_c$ is the radius of the core.
Figure \ref{fig:appendixSigmas} shows the results of this comparison when the $70 \mu$m data is included (a) and excluded (b). From this figure, we can see that including the $70 \mu$m data leads to better agreement between $\Sigma_{\rm c,env,GB}$ and $\Sigma_{\rm c,env}$. In particular, there is less scatter in the derived values of $\Sigma_{\rm c,env,GB}$ for a given input $\Sigma_{\rm c,env}$. However, we do note a systematic trend for $\Sigma_{\rm c,env,GB}$ to overestimate the true value, especially for the highest mass surface densities. Nevertheless, we expect that the impact for interpretation of the SOMA sources of this effect is modest, since most values of $\Sigma_{\rm cl,GB}$ are $\lesssim 1\:{\rm g\:cm}^{-2}$.

\section{SOMA V SED Fits}
Here we present the results of the SED fitting for the 34 sources in the SOMA V sample obtained using Sedcreator (version 2.0). Tables \ref{tab:fluxes} and \ref{tab:best_models} present the measured fluxes and best models. Figure \ref{fig:sed1} contains the source SEDs and ``good" model fits (see main text for details), and Figure \ref{fig:sed_2D_results_soma_v} presents the good model distributions in the $\Sigma_{\rm cl}-M_c$, $m_*-M_c$, and $m_*-\Sigma_{\rm cl}$ planes.

\renewcommand{\thefigure}{A\arabic{figure}}
\begin{figure*}[!htb]
\includegraphics[width=0.5\textwidth]{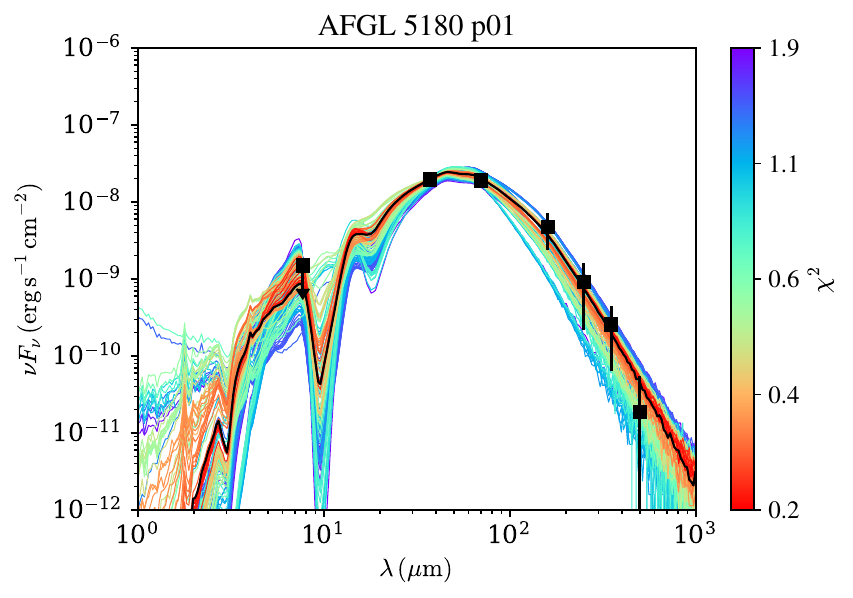}
\includegraphics[width=0.5\textwidth]{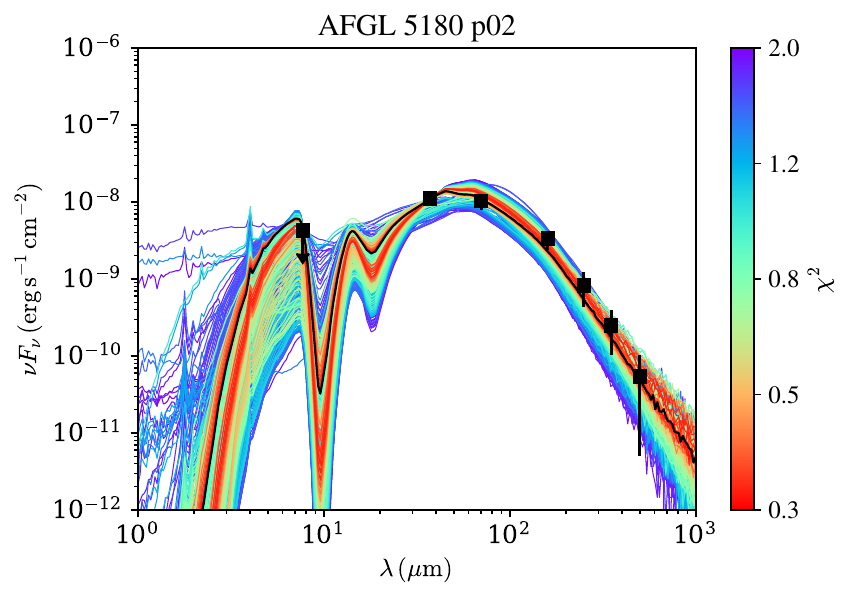}
\includegraphics[width=0.5\textwidth]{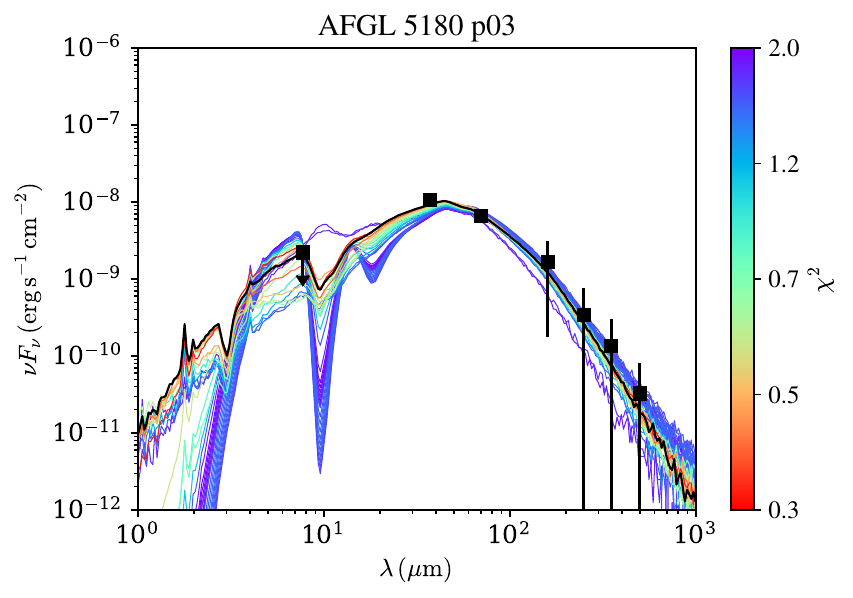}
\includegraphics[width=0.5\textwidth]{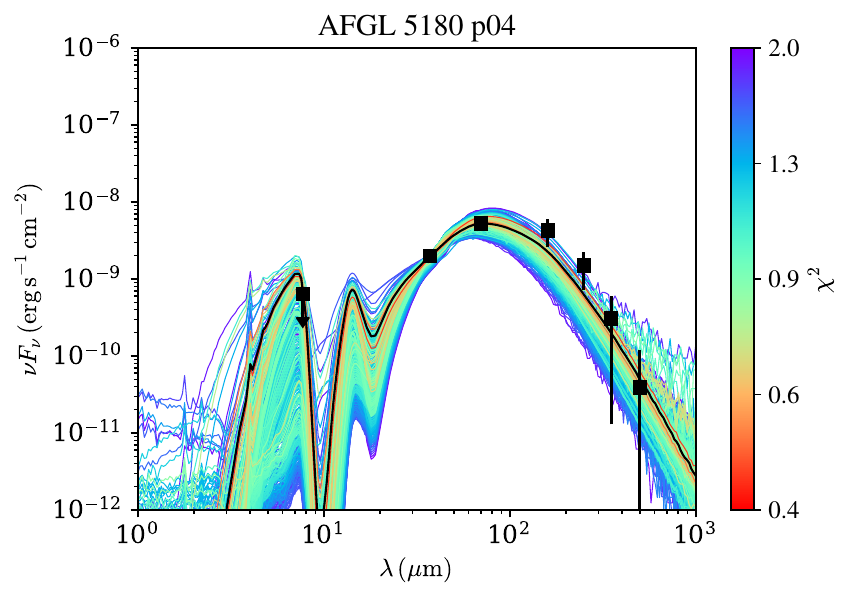}
\includegraphics[width=0.5\textwidth]{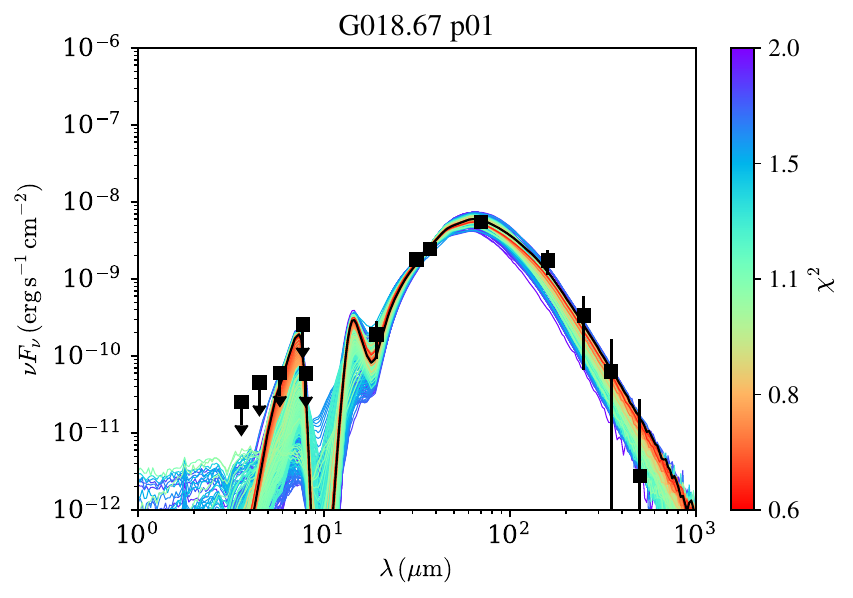}
\includegraphics[width=0.5\textwidth]{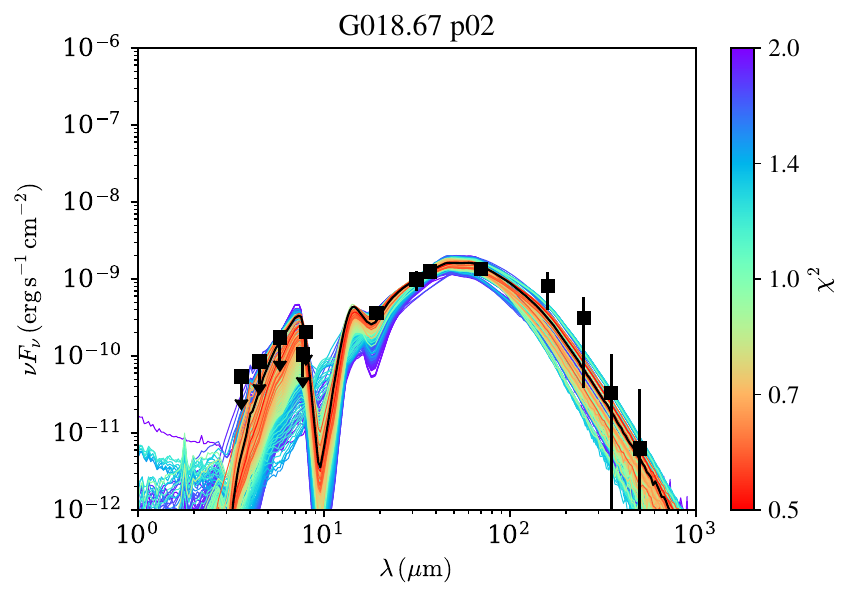}
\caption{
SOMA V sources analyzed with Sedcreator. Protostellar model fitting to the fixed aperture, background-subtracted SED data using the ZT18 model grid. For each source, the best fitting protostar model is shown with a black line, while all other ``good'' model fits (see text) are shown with colored lines (red to blue with increasing $\chi^2$). Flux values are those from Table\,\ref{tab:fluxes}. Note that the data at $\lesssim8\,{\rm \mu m}$ are treated as upper limits (see text). The resulting model parameters are listed in Table\,\ref{tab:best_models}. \label{fig:sed1}}
\end{figure*}

\renewcommand{\thefigure}{A\arabic{figure}} 
\addtocounter{figure}{-1}
\begin{figure*}[!htb]
\includegraphics[width=0.5\textwidth]{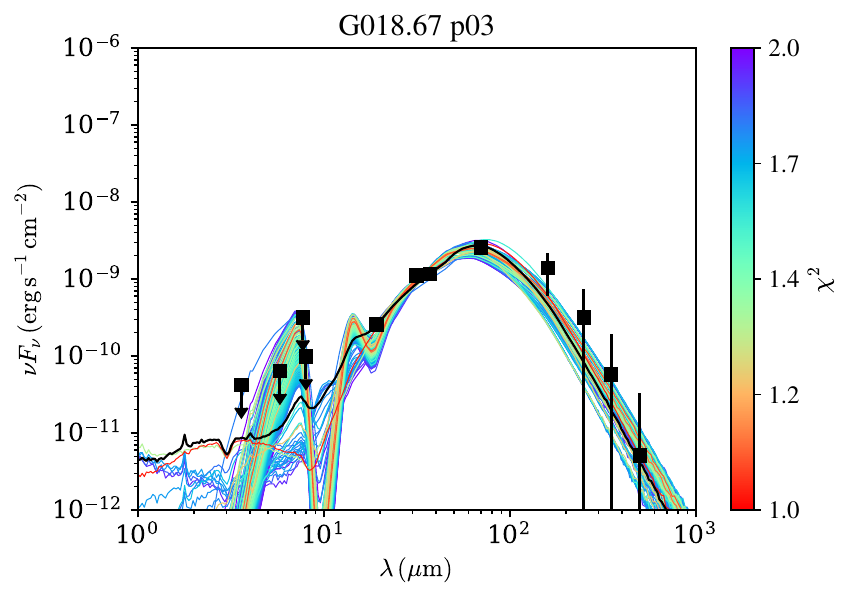}
\includegraphics[width=0.5\textwidth]{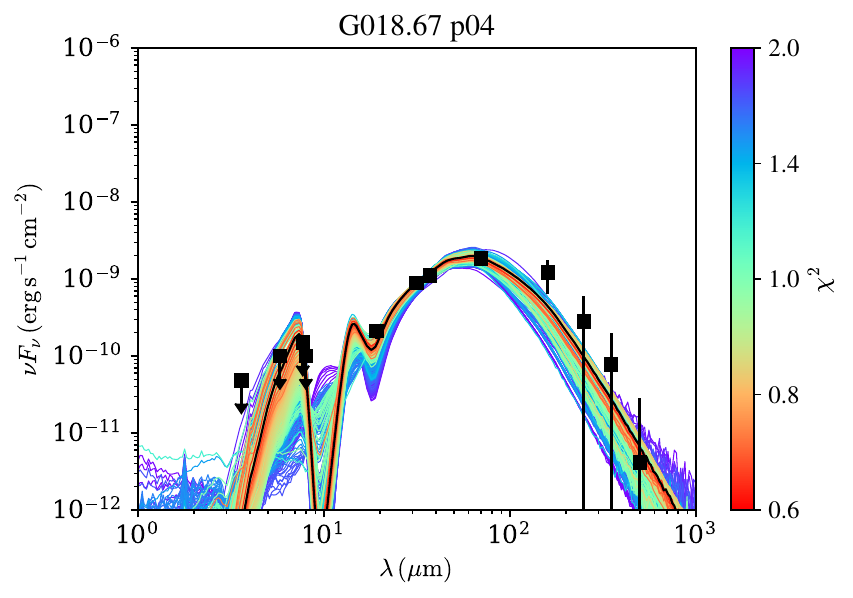}
\includegraphics[width=0.5\textwidth]{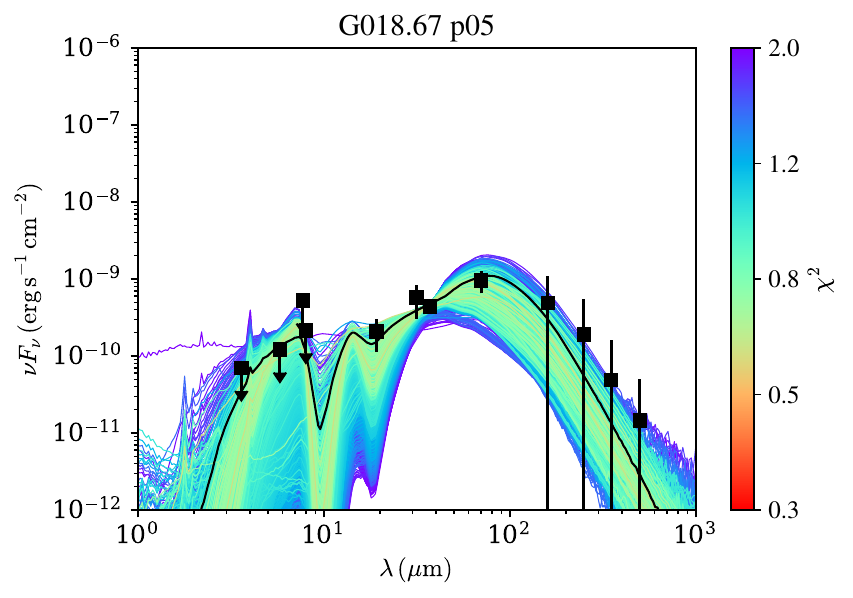}
\includegraphics[width=0.5\textwidth]{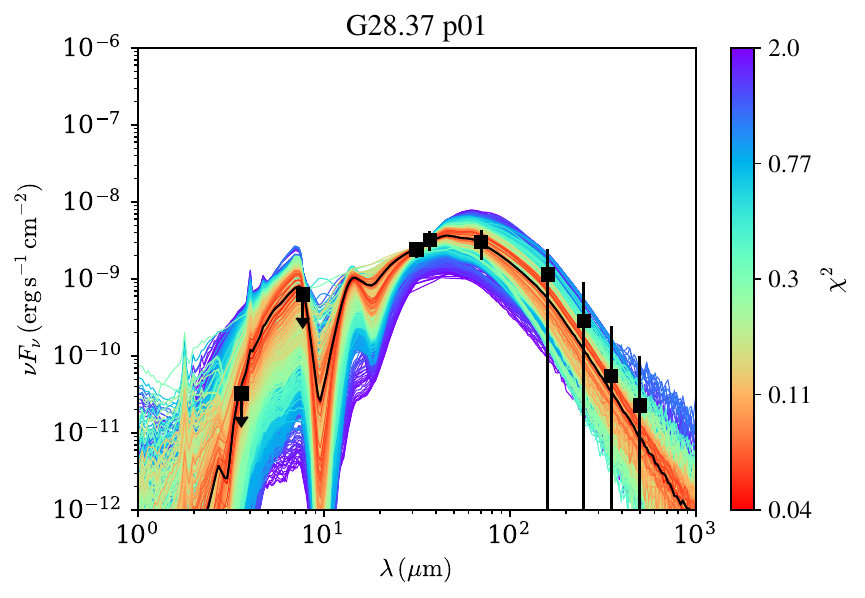}
\includegraphics[width=0.5\textwidth]{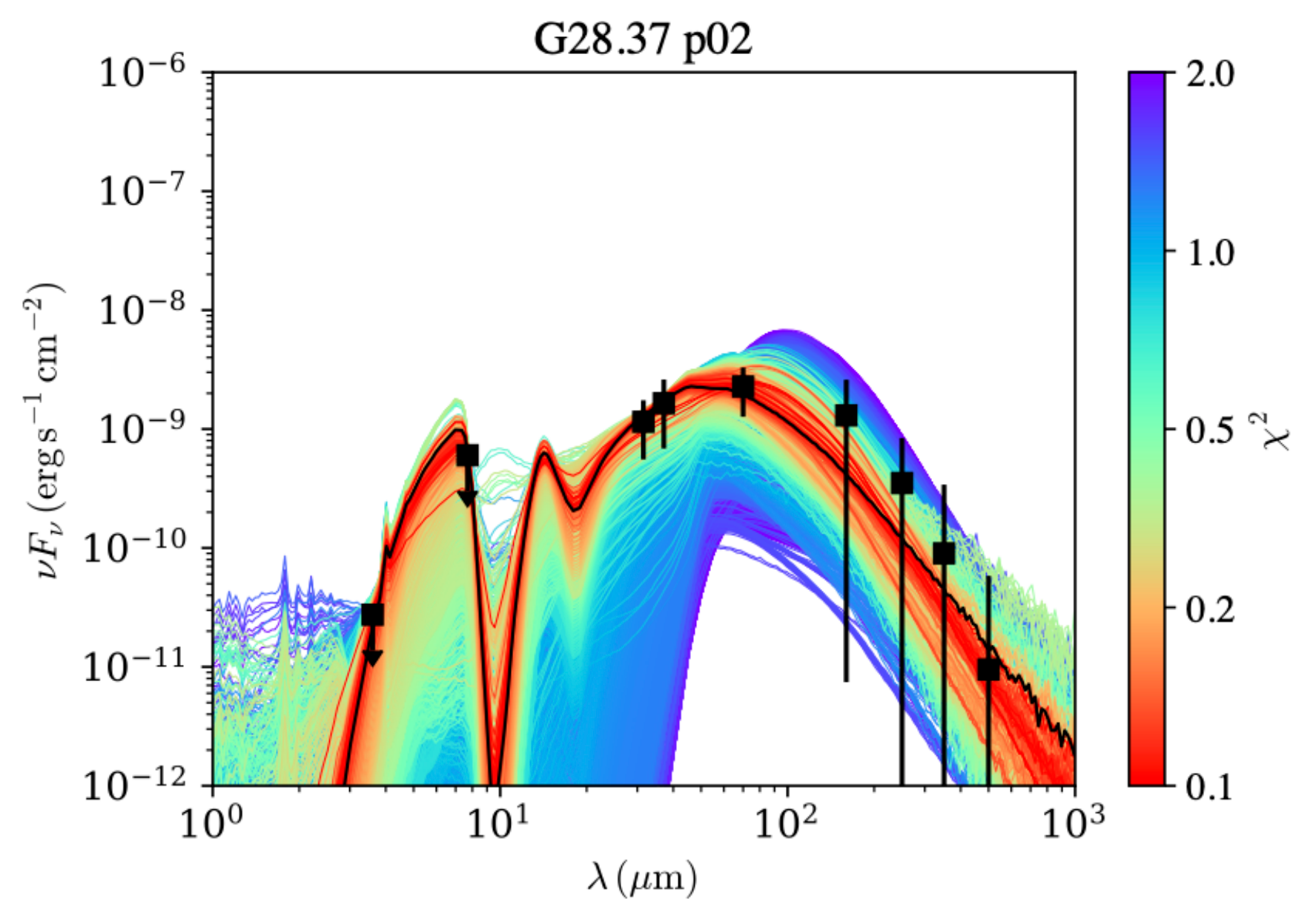}
\includegraphics[width=0.5\textwidth]{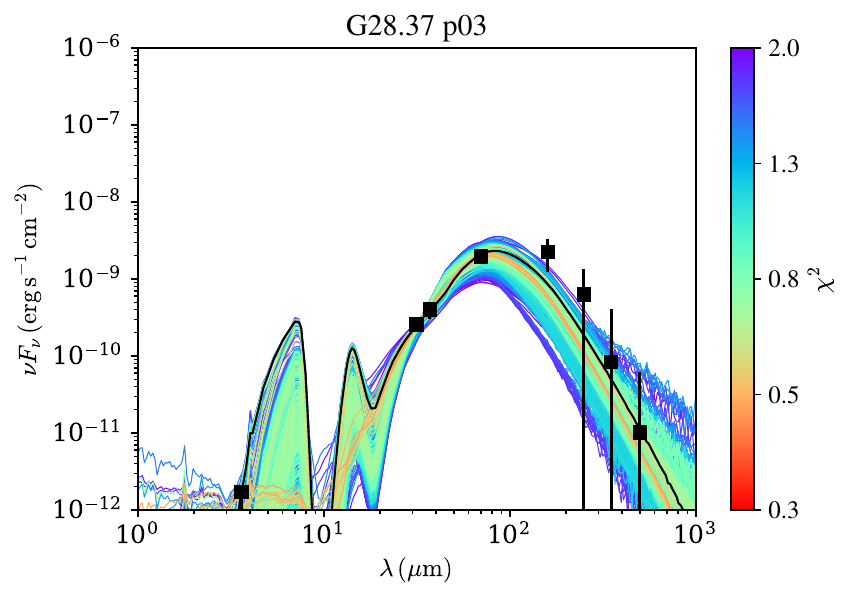}
    \caption{(Continued.)}
\end{figure*}

\renewcommand{\thefigure}{A\arabic{figure}}
\addtocounter{figure}{-1}
\begin{figure*}[!htb]
\includegraphics[width=0.5\textwidth]{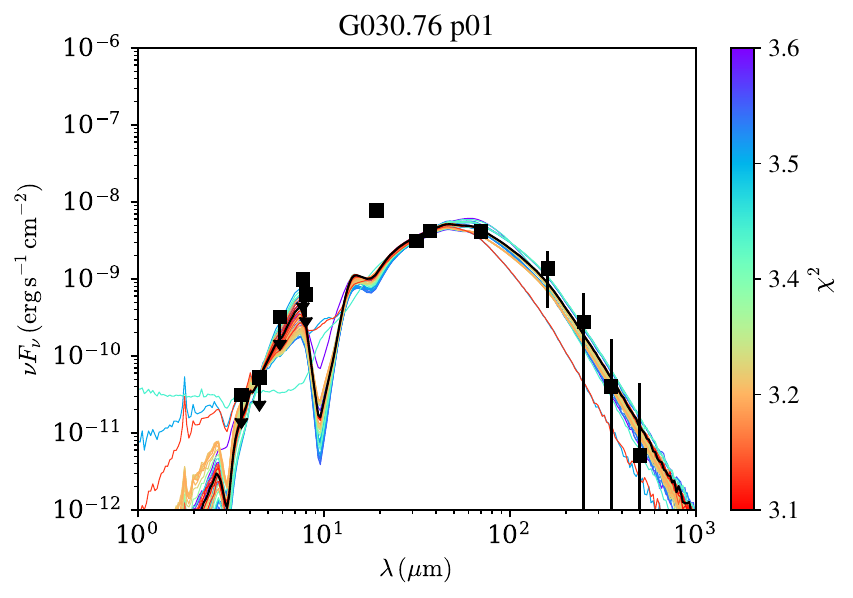}
\includegraphics[width=0.5\textwidth]{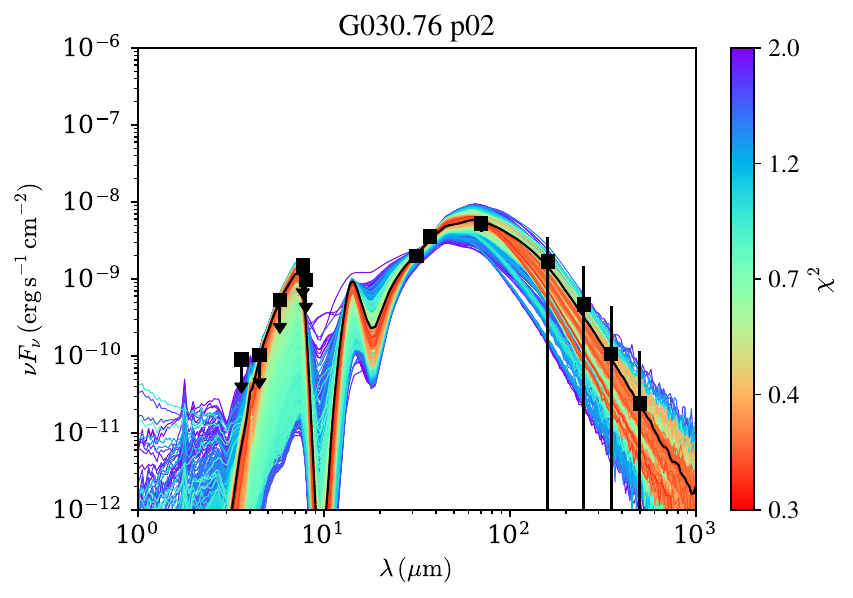}
\includegraphics[width=0.5\textwidth]{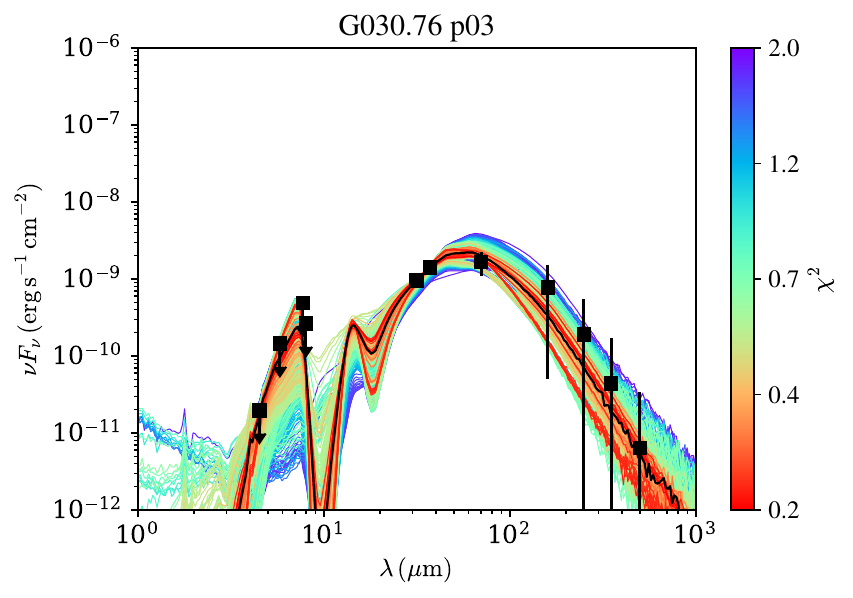}
\includegraphics[width=0.5\textwidth]{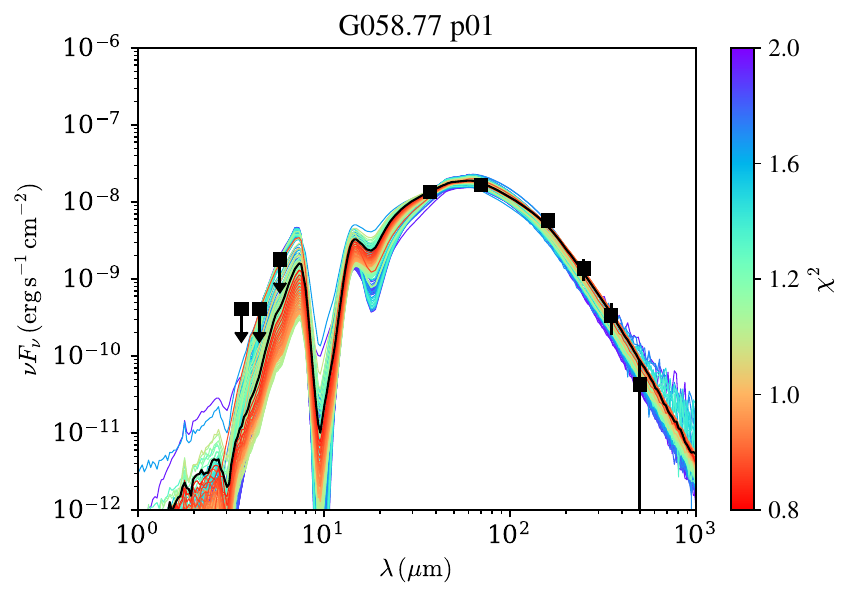}
\includegraphics[width=0.5\textwidth]{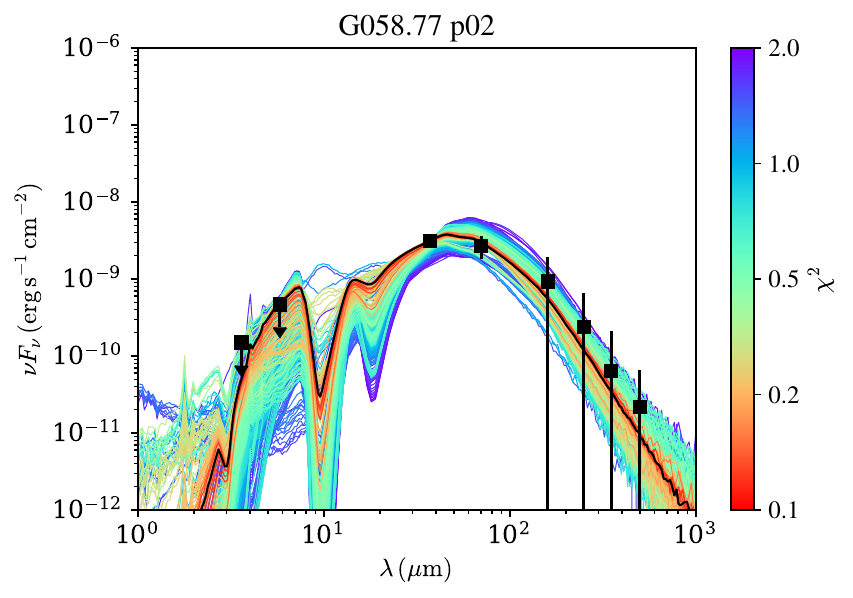}
\includegraphics[width=0.5\textwidth]{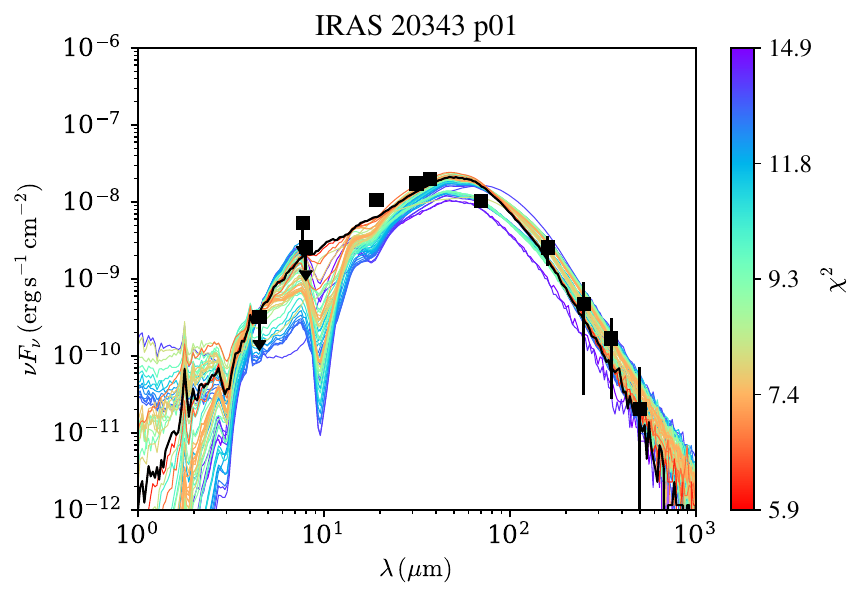}
    \caption{(Continued.)}
\end{figure*}

\renewcommand{\thefigure}{A\arabic{figure}}
\addtocounter{figure}{-1}
\begin{figure*}[!htb]
\includegraphics[width=0.5\textwidth]{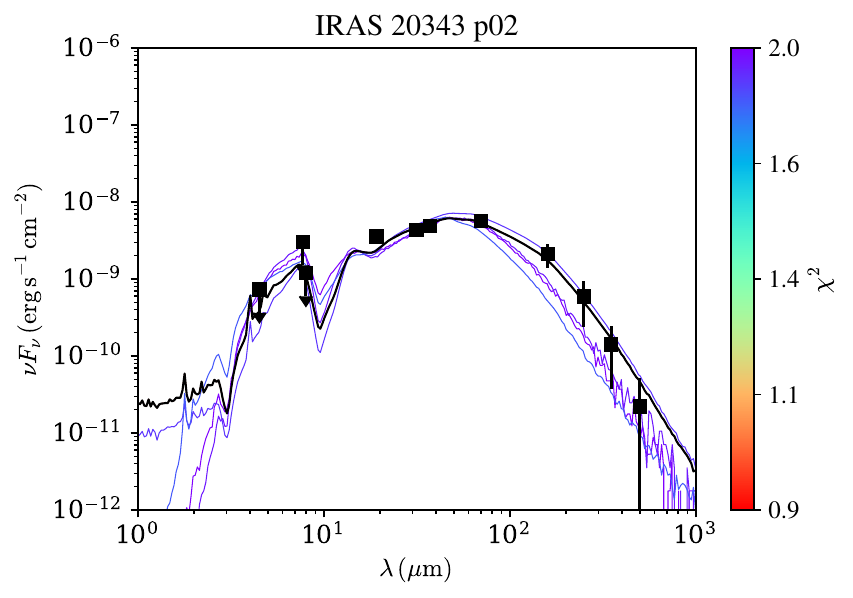}
\includegraphics[width=0.5\textwidth]{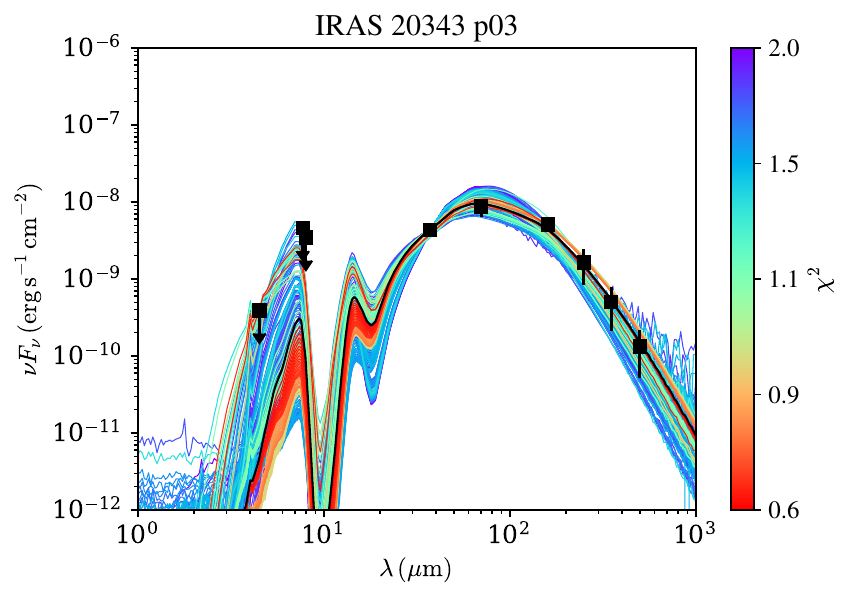}
\includegraphics[width=0.5\textwidth]{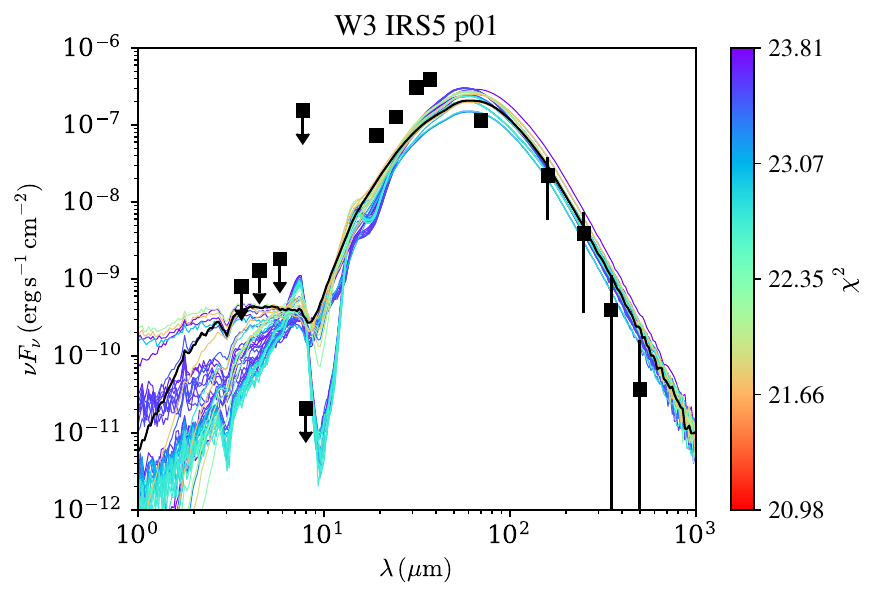}
\includegraphics[width=0.5\textwidth]{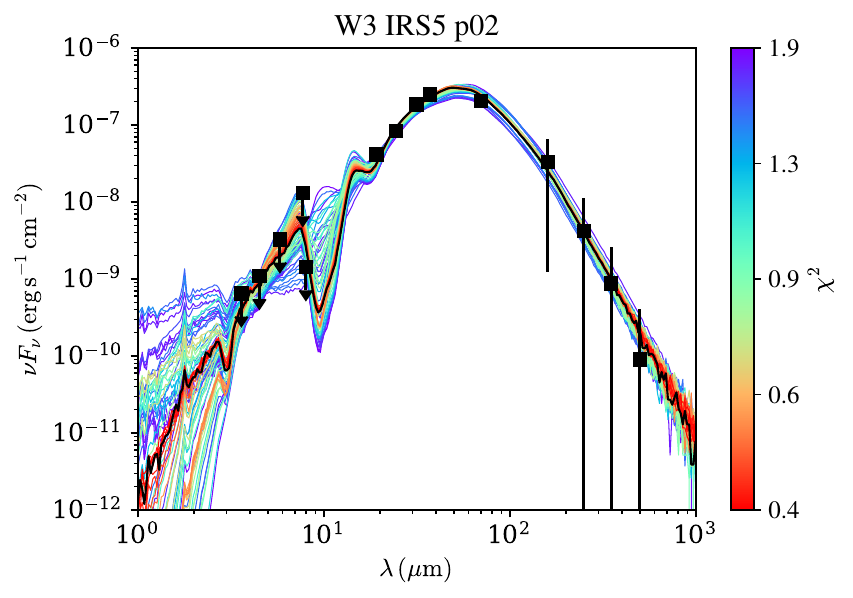}
\includegraphics[width=0.5\textwidth]{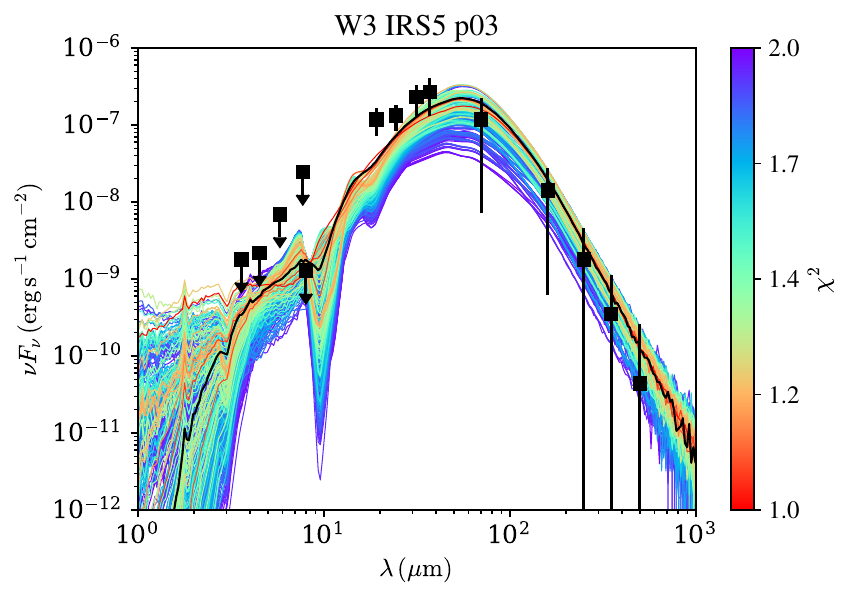}
\includegraphics[width=0.5\textwidth]{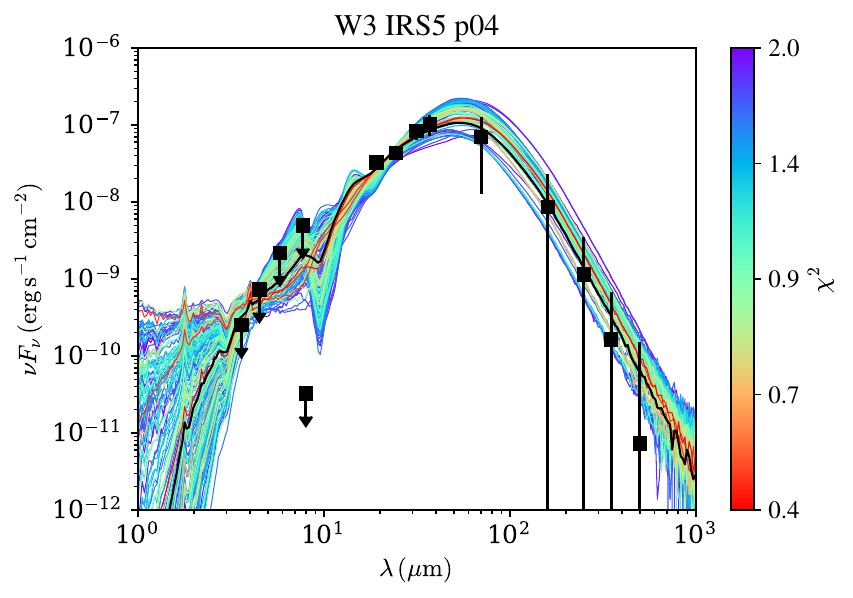}
    \caption{(Continued.)}
\end{figure*}

\renewcommand{\thefigure}{A\arabic{figure}}
\addtocounter{figure}{-1}
\begin{figure*}[!htb]
\includegraphics[width=0.5\textwidth]{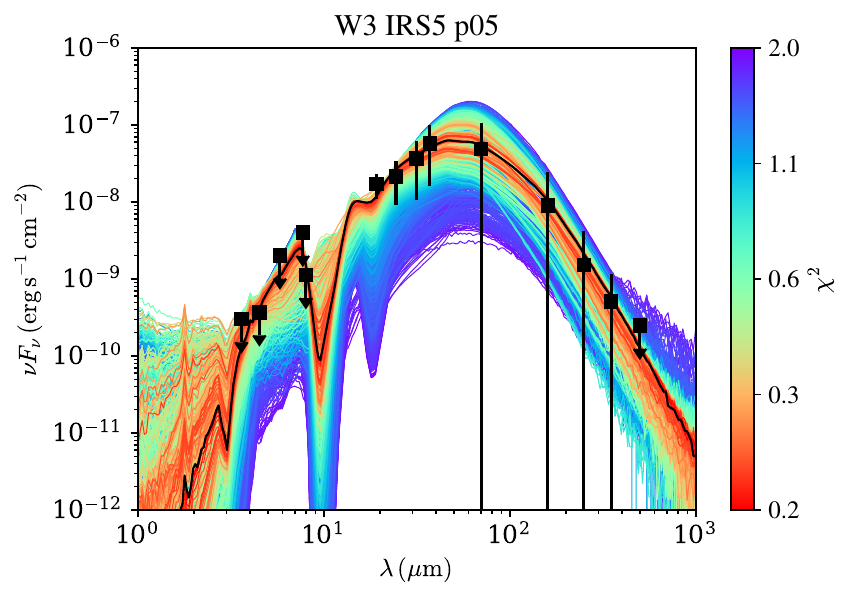}
\includegraphics[width=0.5\textwidth]{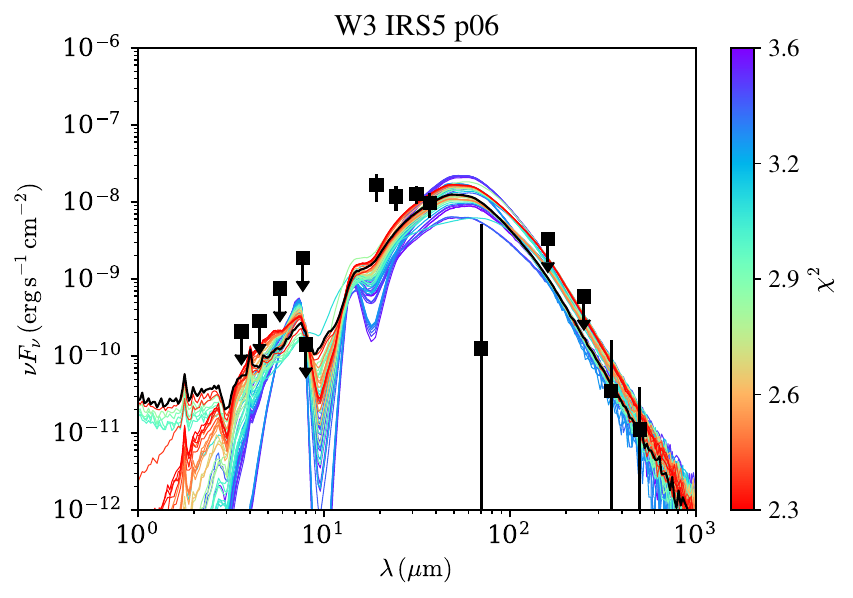}
\includegraphics[width=0.5\textwidth]{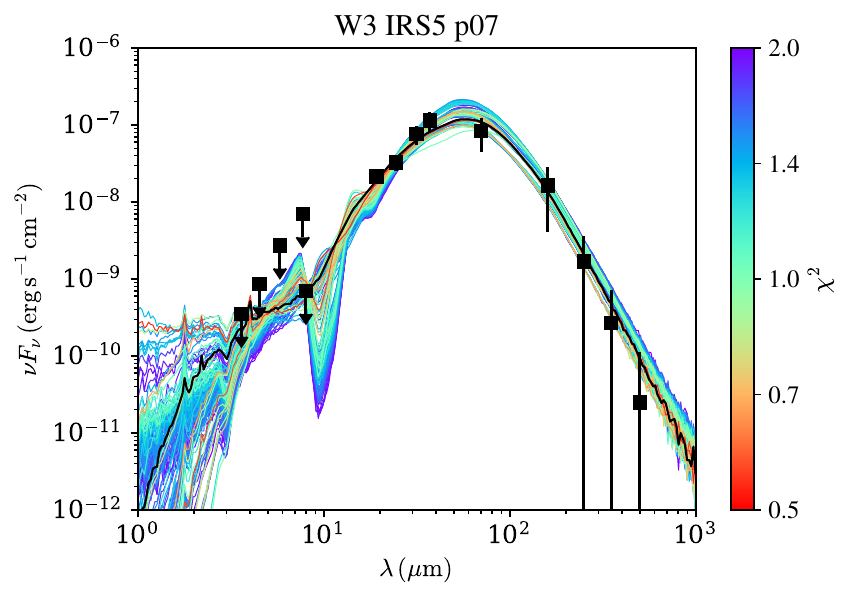}
\includegraphics[width=0.5\textwidth]{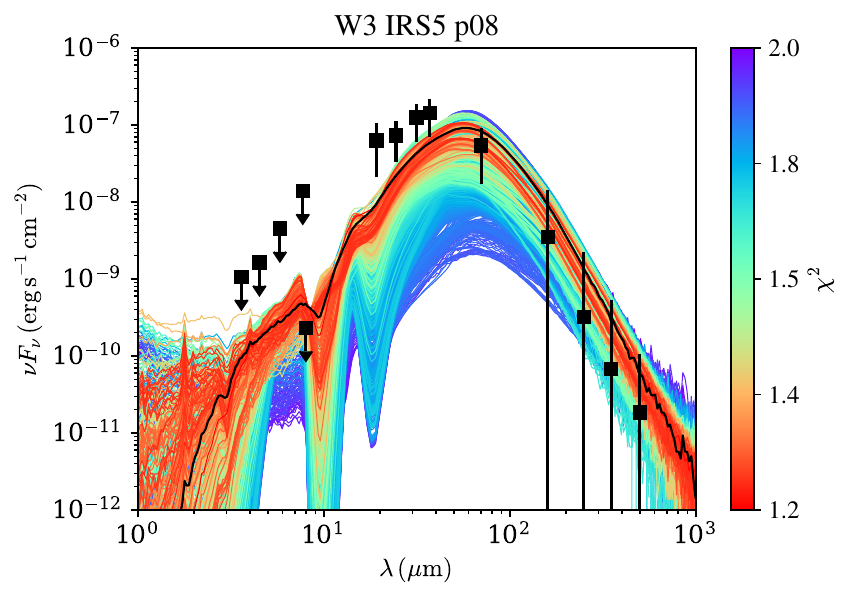}
\includegraphics[width=0.5\textwidth]{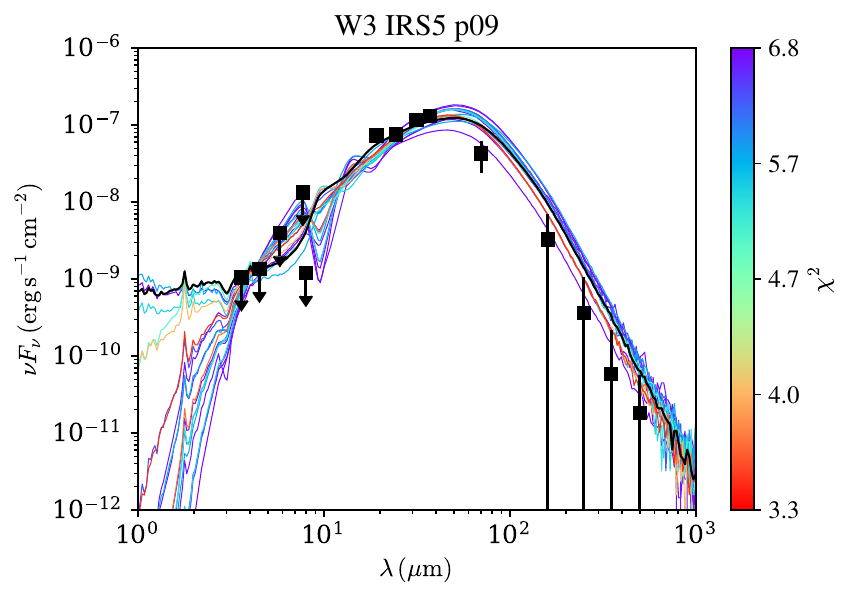}
\includegraphics[width=0.5\textwidth]{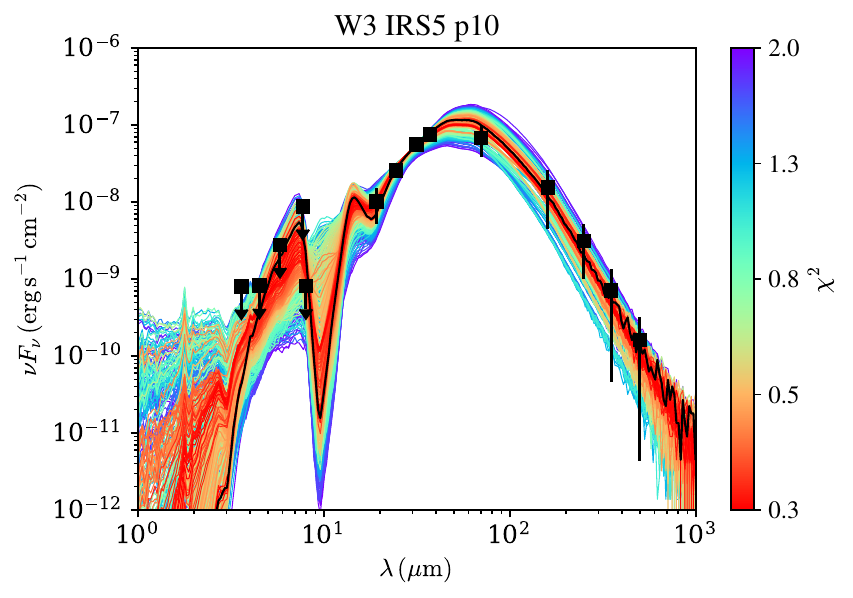}
    \caption{(Continued.)}
\end{figure*}

\renewcommand{\thefigure}{A\arabic{figure}}
\addtocounter{figure}{-1}
\begin{figure*}[!htb]
\includegraphics[width=0.5\textwidth]{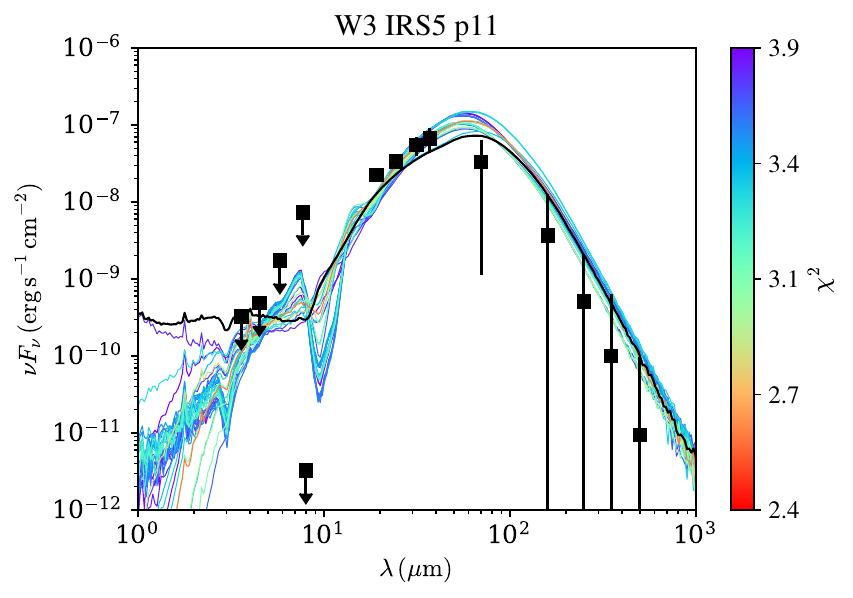}
\includegraphics[width=0.5\textwidth]{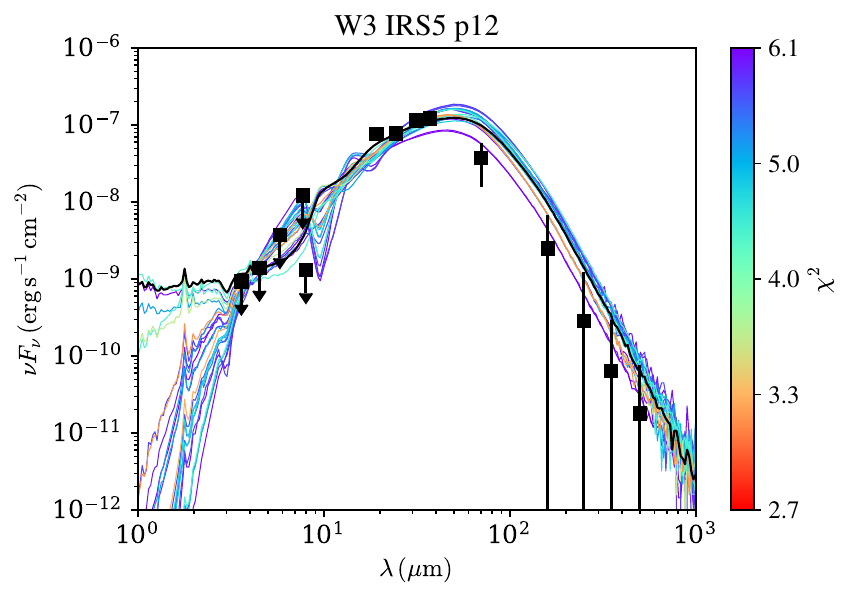}
\includegraphics[width=0.5\textwidth]{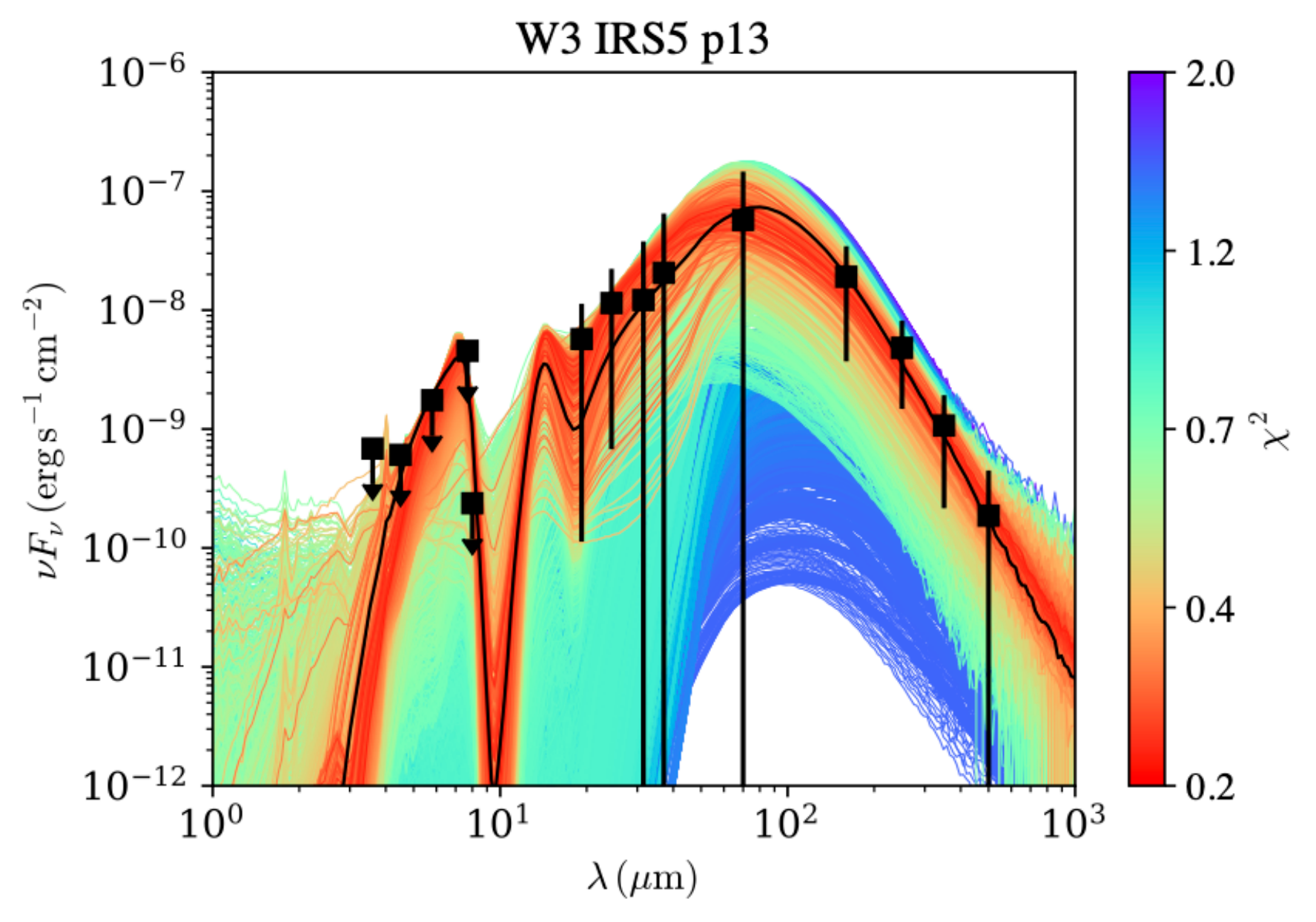}
\includegraphics[width=0.5\textwidth]{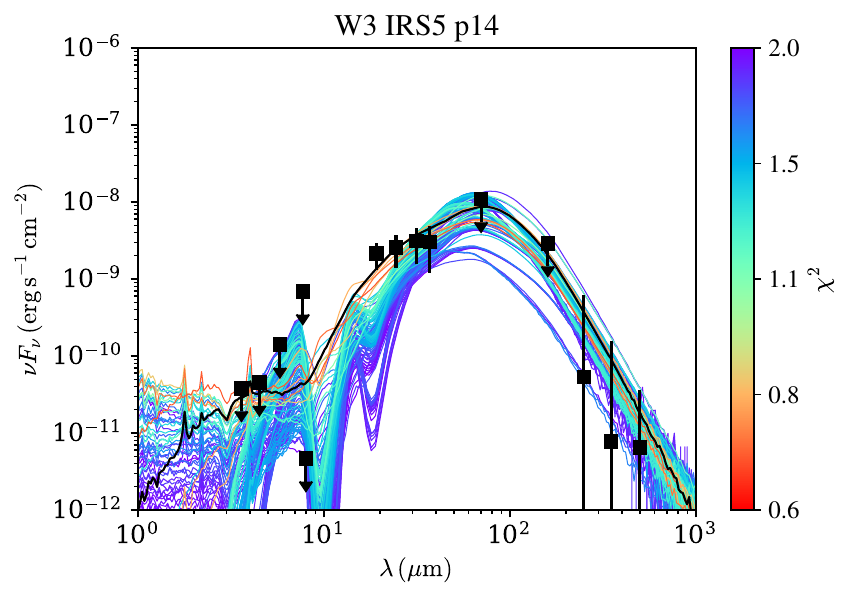}
    \caption{(Continued.)}
\end{figure*}

\renewcommand{\thefigure}{A\arabic{figure}}
\begin{figure*}[!htb]
\includegraphics[width=1.0\textwidth]{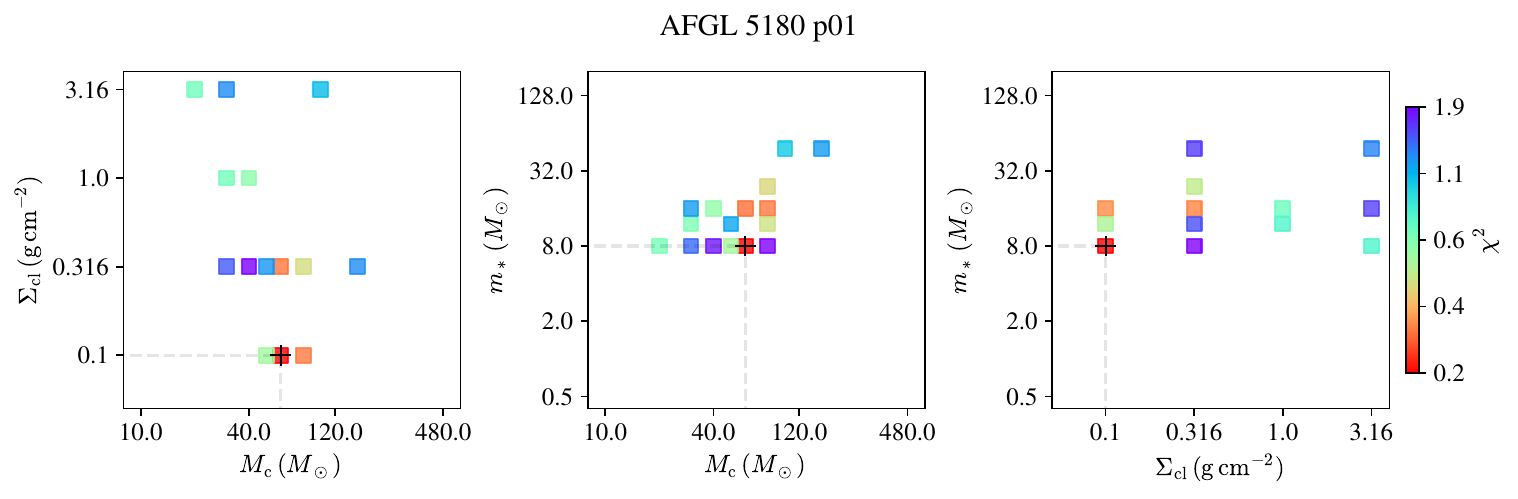}
\includegraphics[width=1.0\textwidth]{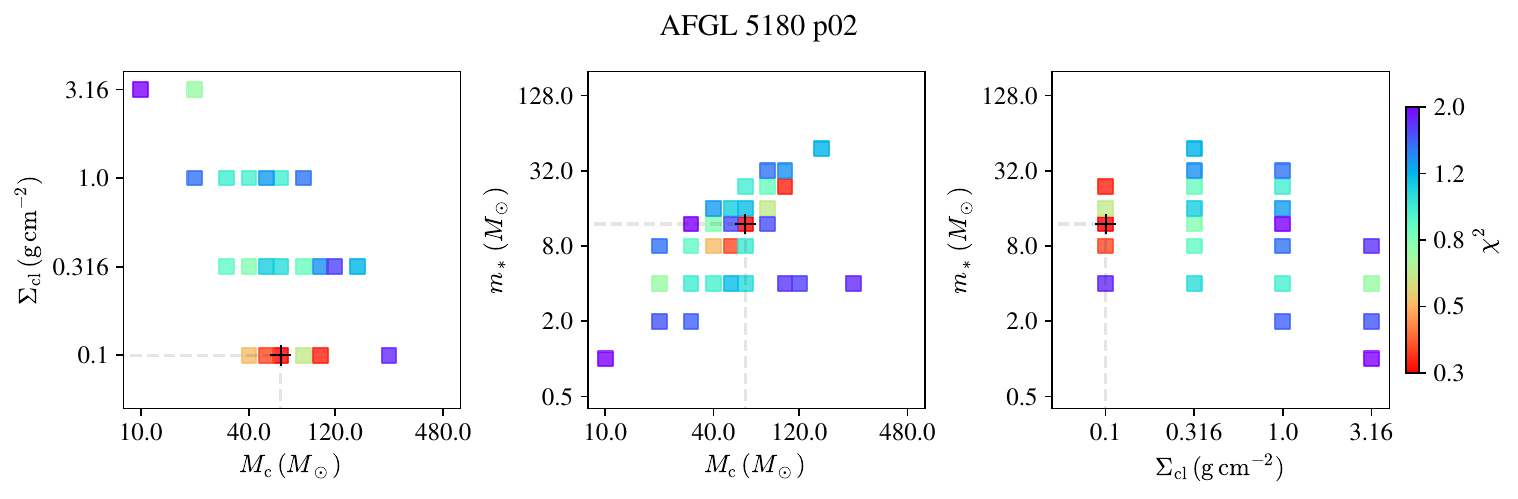}
\includegraphics[width=1.0\textwidth]{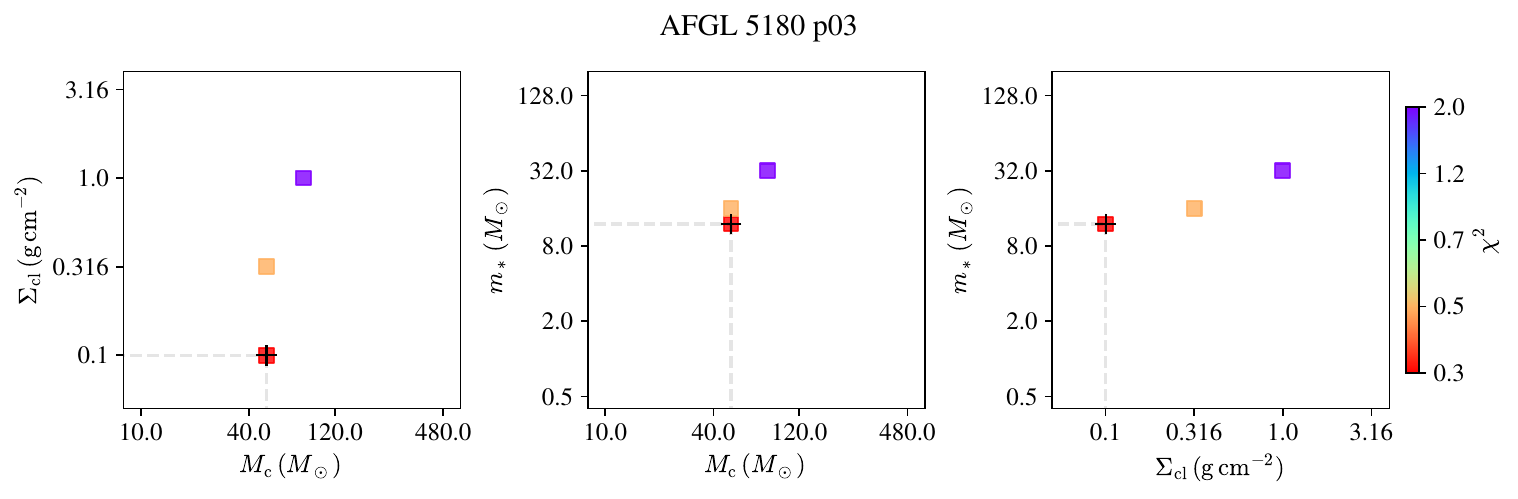}
\includegraphics[width=1.0\textwidth]{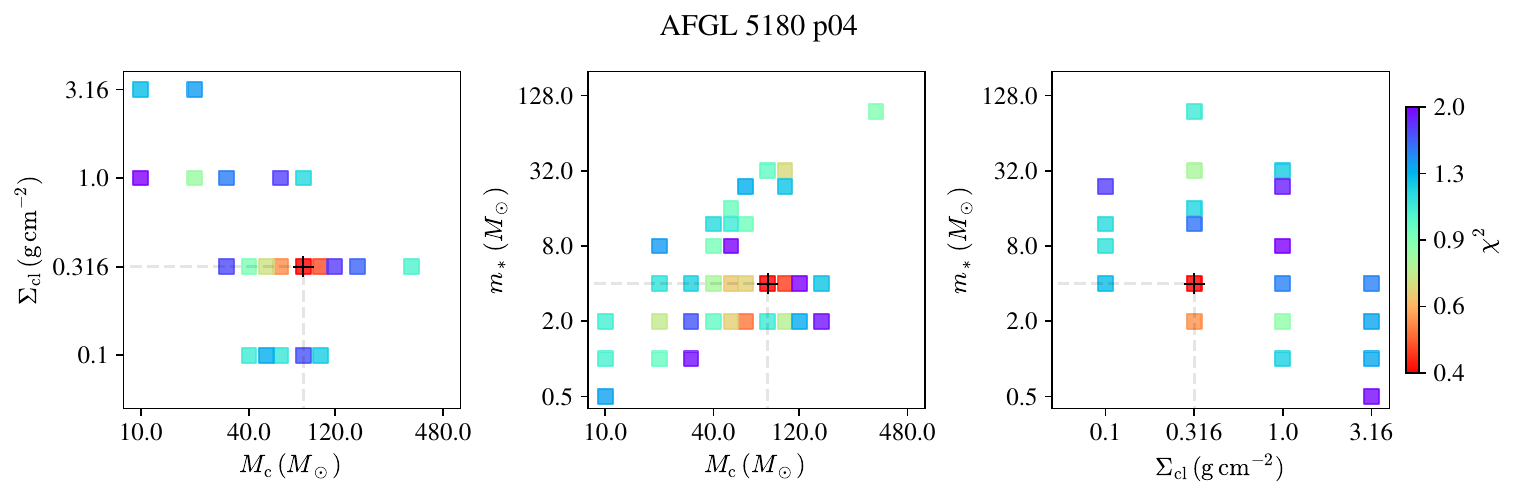}
\caption{$\Sigma_{\rm cl}$ vs. $M_c$ (left column), $m_*$ vs.  $M_c$ (center column), and $m_*$ vs. $\Sigma_{\rm cl}$ (right column) for the ``good" model fits for each source, color-coded by $\chi^2$ value. The black cross distinguishes the best-fit model.
}
\label{fig:sed_2D_results_soma_v}
\end{figure*}

\renewcommand{\thefigure}{A\arabic{figure}}
\addtocounter{figure}{-1}
\begin{figure*}[!htb]
\includegraphics[width=1.0\textwidth]{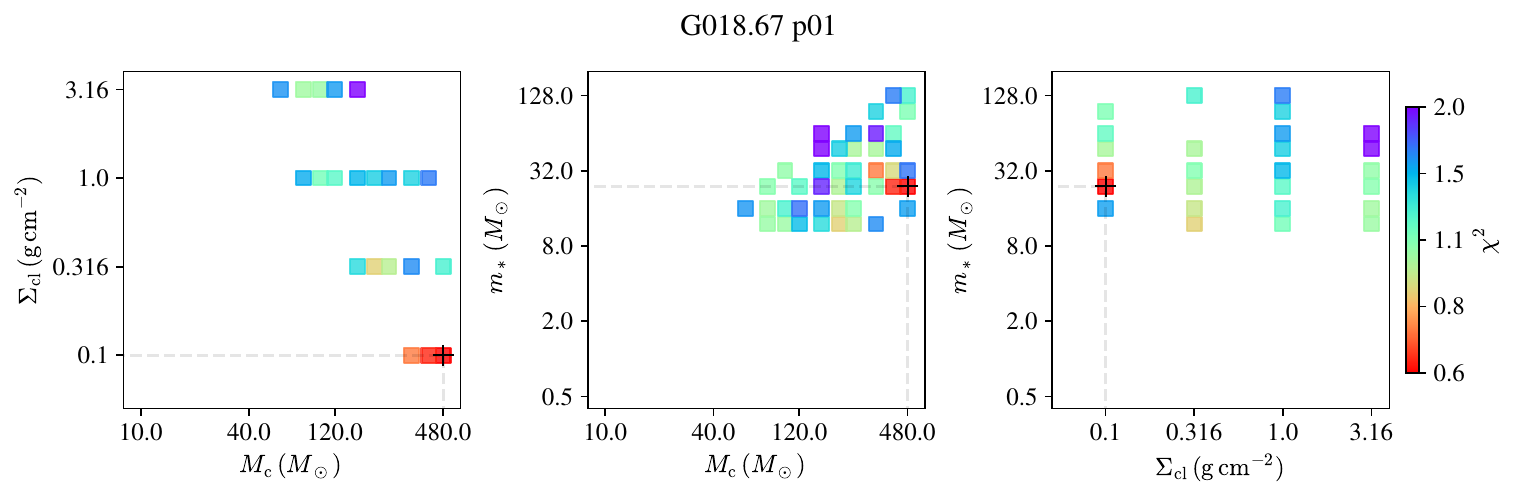}
\includegraphics[width=1.0\textwidth]{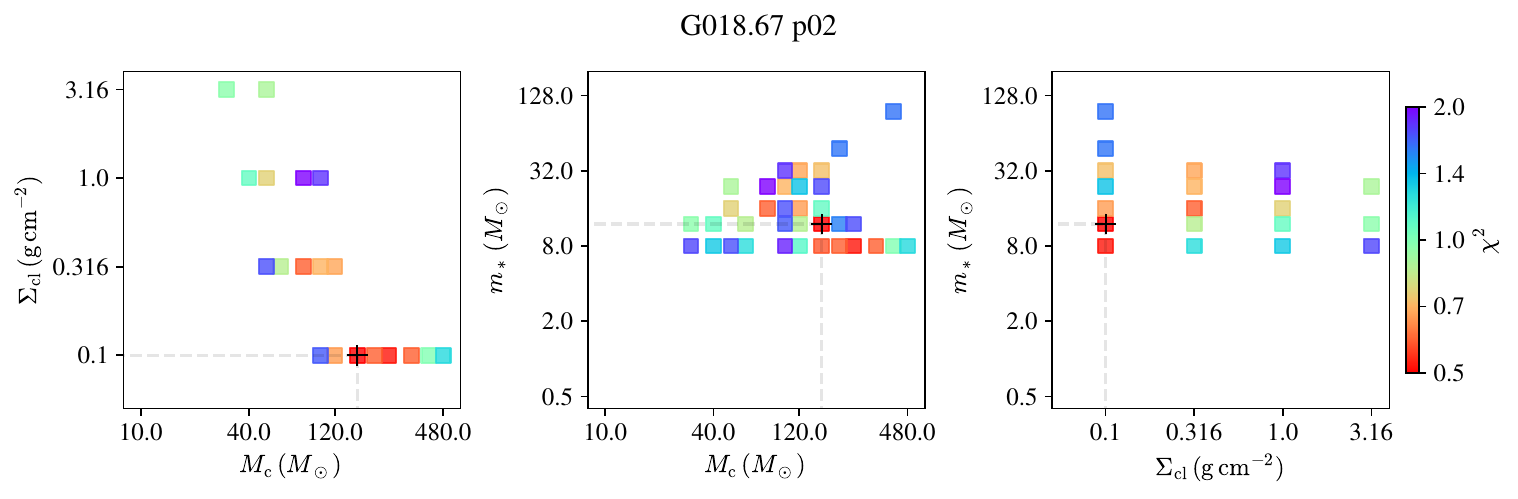}
\includegraphics[width=1.0\textwidth]{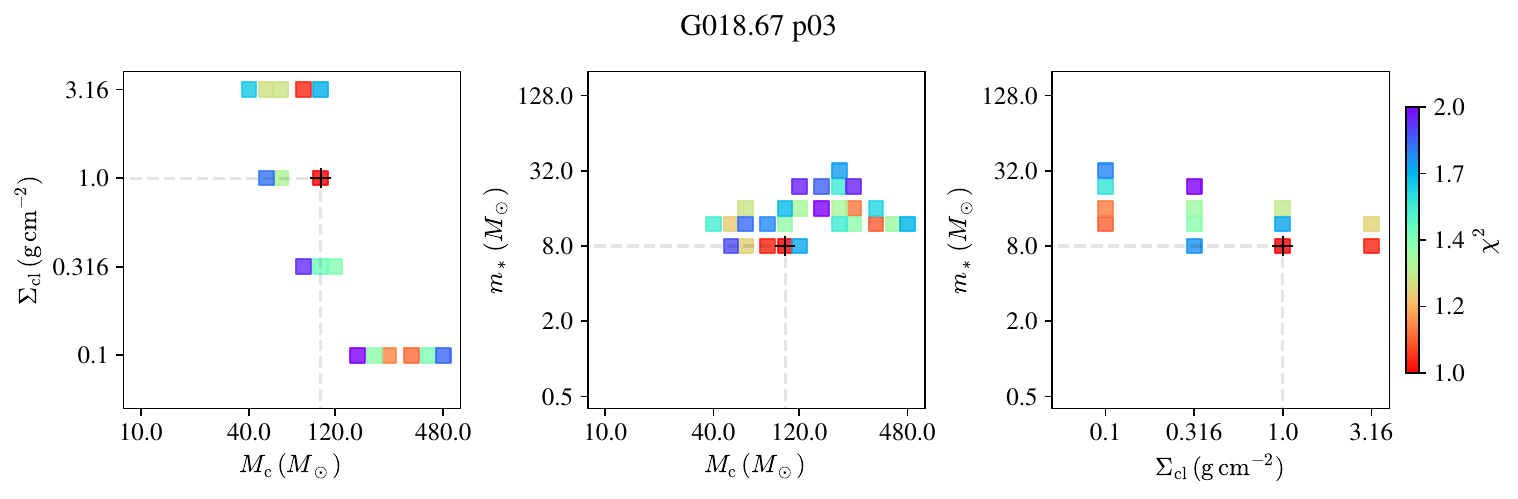}
\includegraphics[width=1.0\textwidth]{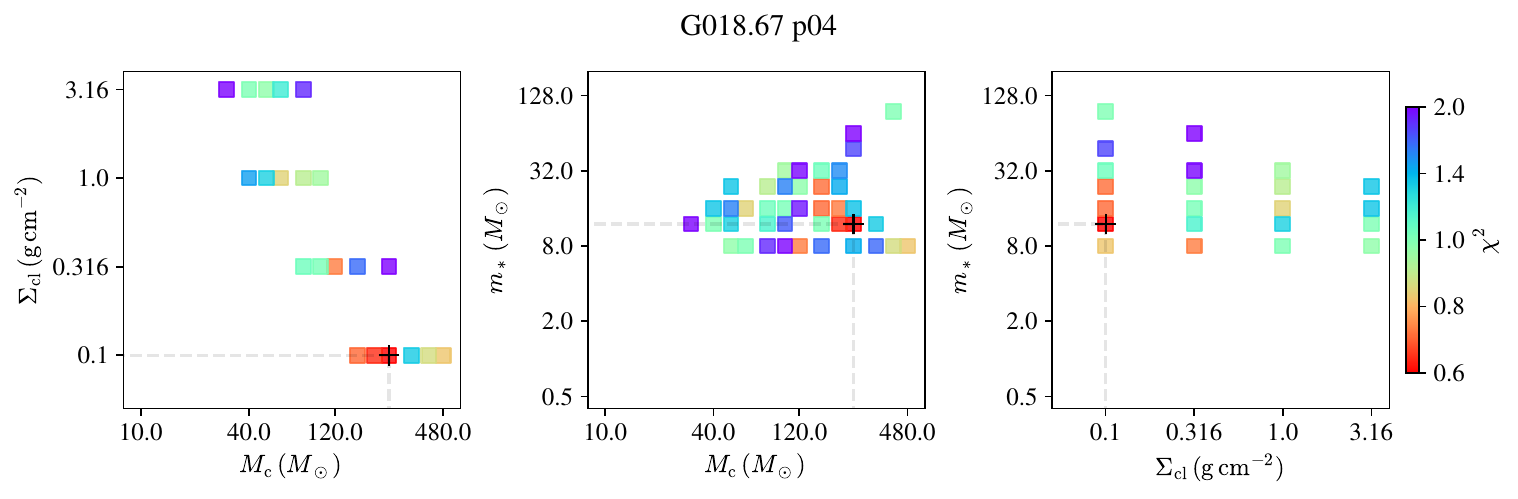}
\caption{(Continued.)}
\end{figure*}

\renewcommand{\thefigure}{A\arabic{figure}}
\addtocounter{figure}{-1}
\begin{figure*}[!htb]
\includegraphics[width=1.0\textwidth]{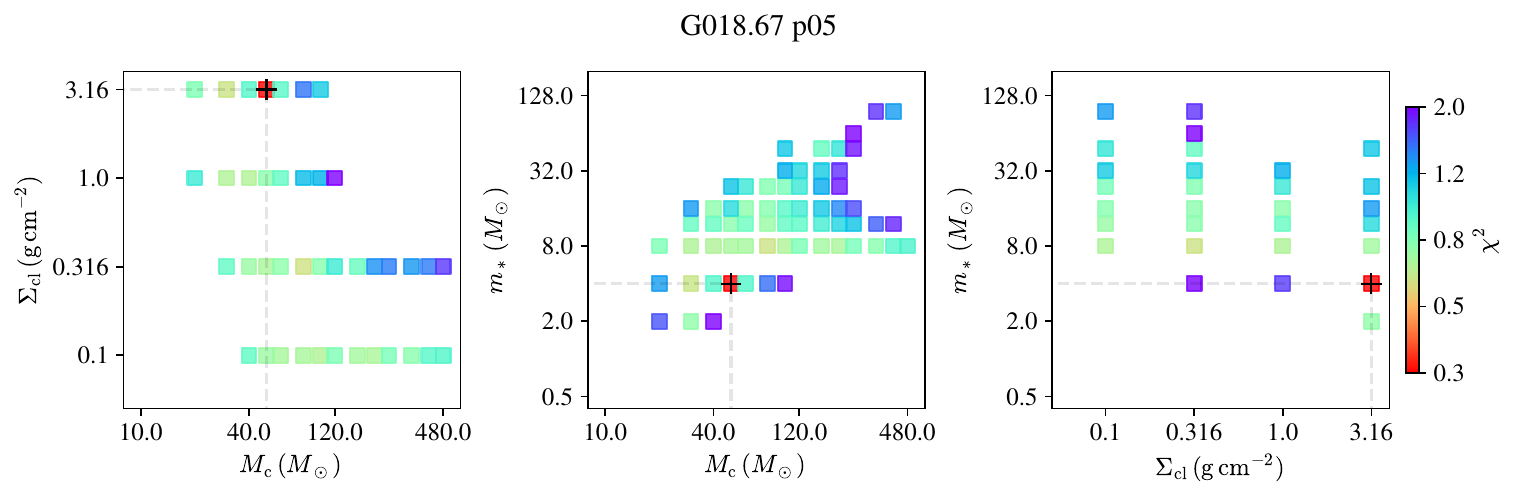}
\includegraphics[width=1.0\textwidth]{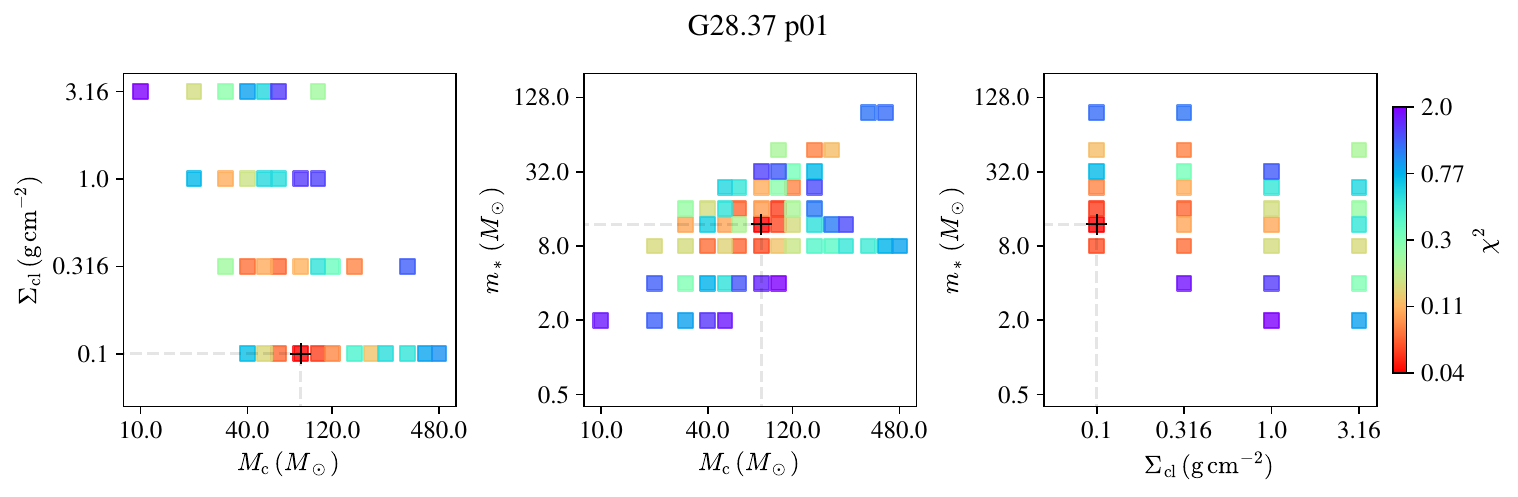}
\includegraphics[width=1.0\textwidth]{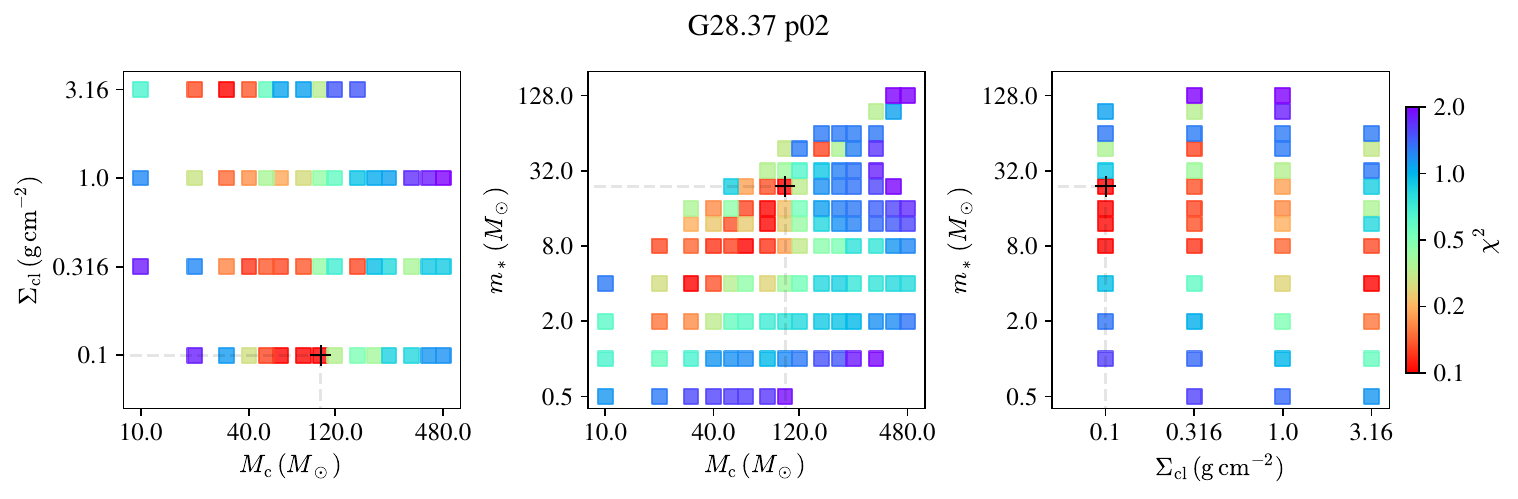}
\includegraphics[width=1.0\textwidth]{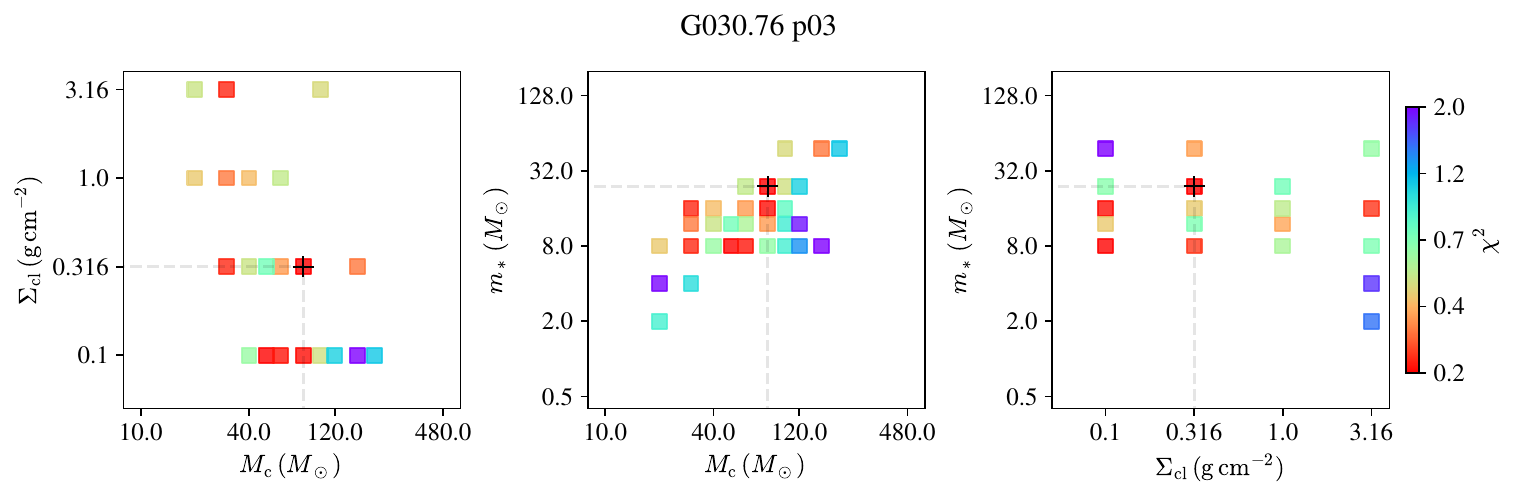}
\caption{(Continued.)}
\end{figure*}

\renewcommand{\thefigure}{A\arabic{figure}}
\addtocounter{figure}{-1}
\begin{figure*}[!htb]
\includegraphics[width=1.0\textwidth]{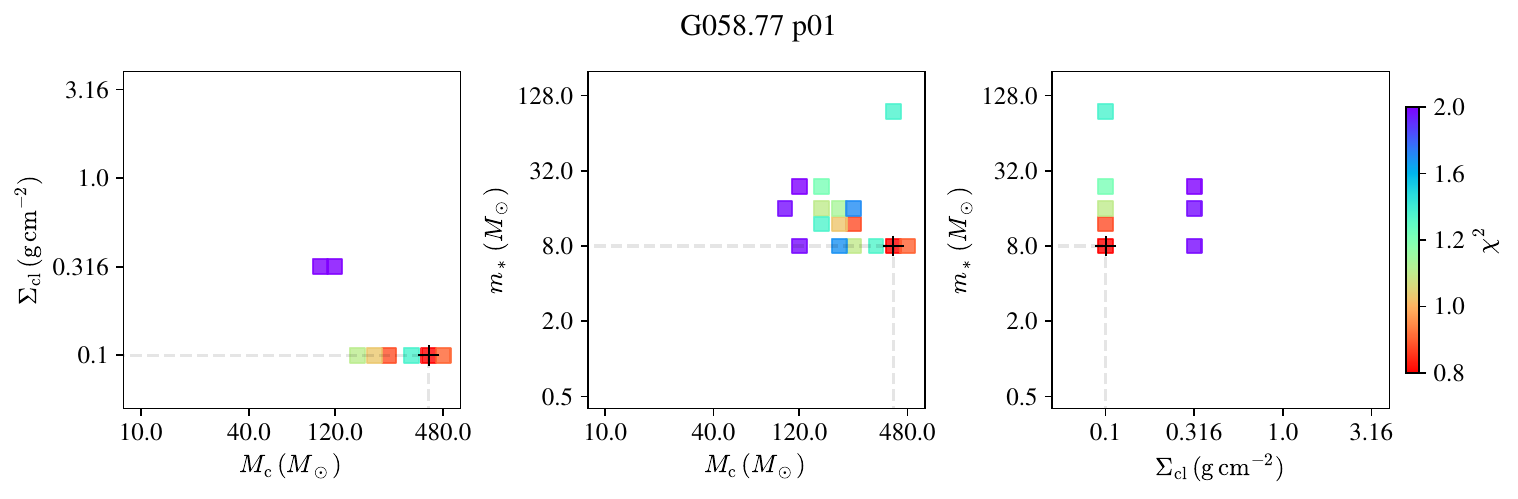}
\includegraphics[width=1.0\textwidth]{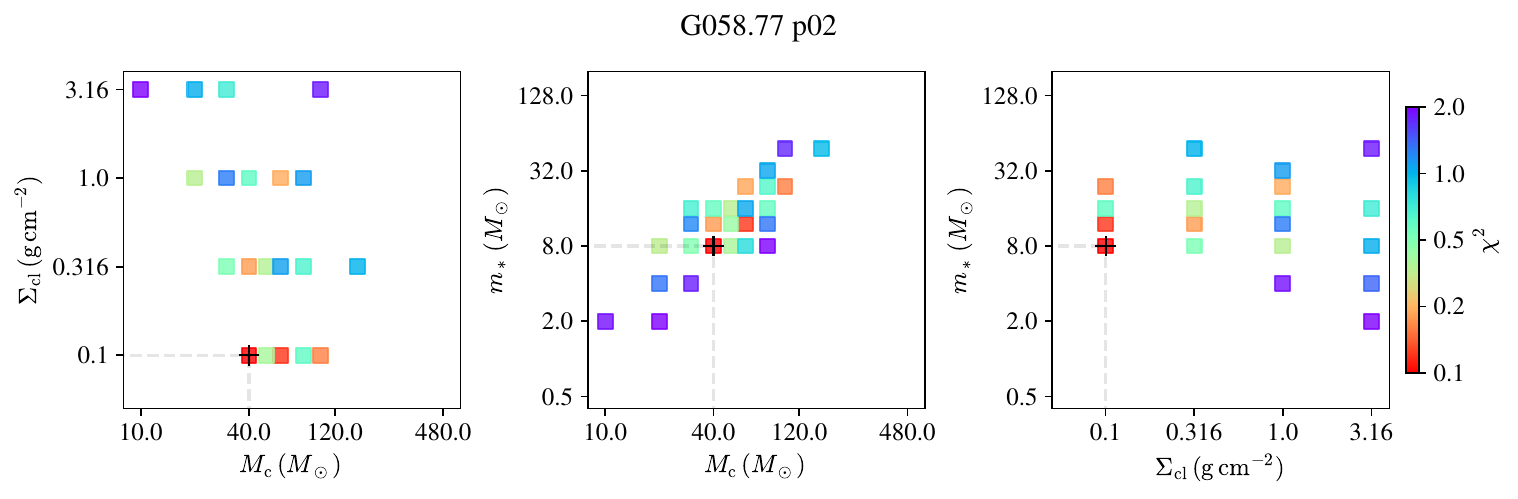}
\includegraphics[width=1.0\textwidth]{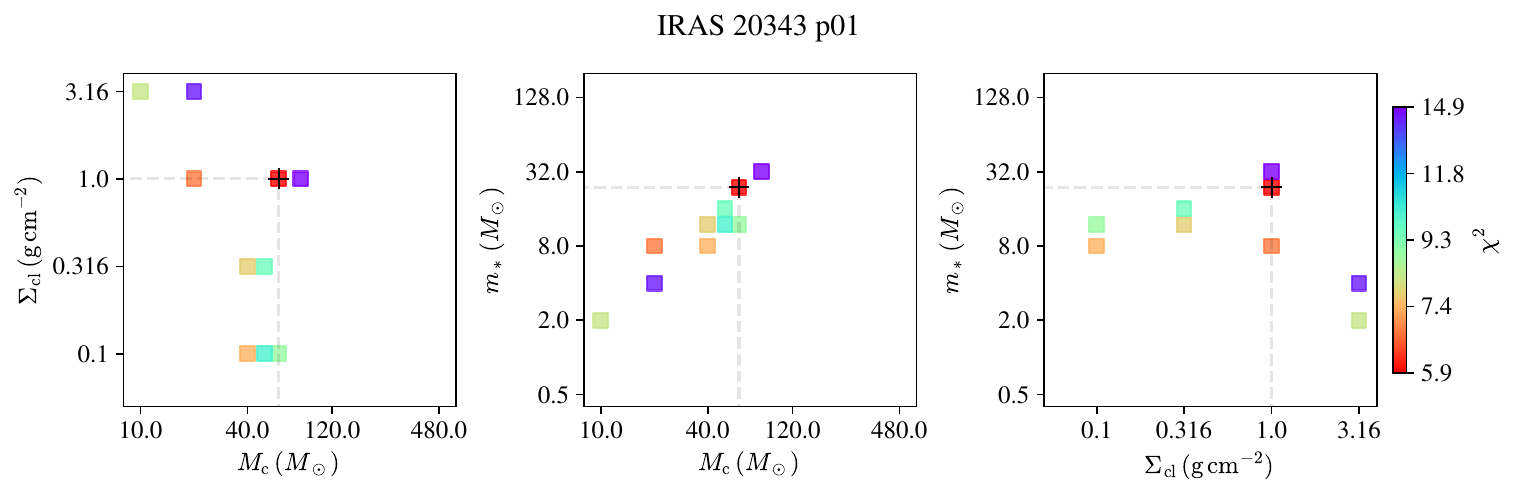}
\includegraphics[width=1.0\textwidth]{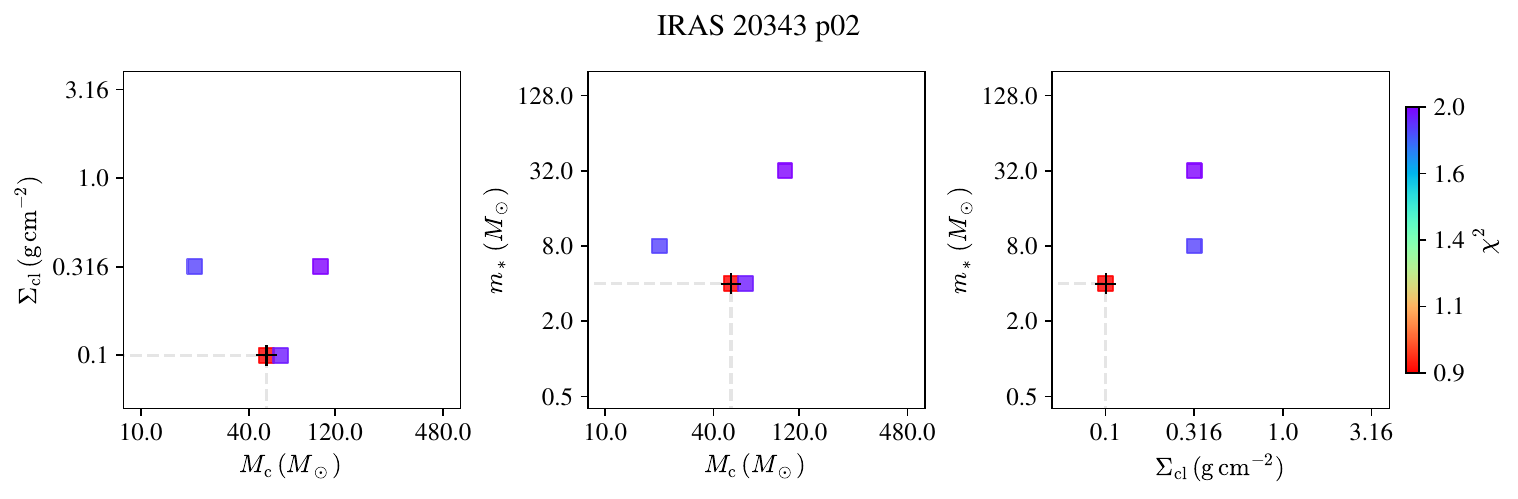}
\caption{(Continued.)}
\end{figure*}

\renewcommand{\thefigure}{A\arabic{figure}}
\addtocounter{figure}{-1}
\begin{figure*}[!htb]
\includegraphics[width=1.0\textwidth]{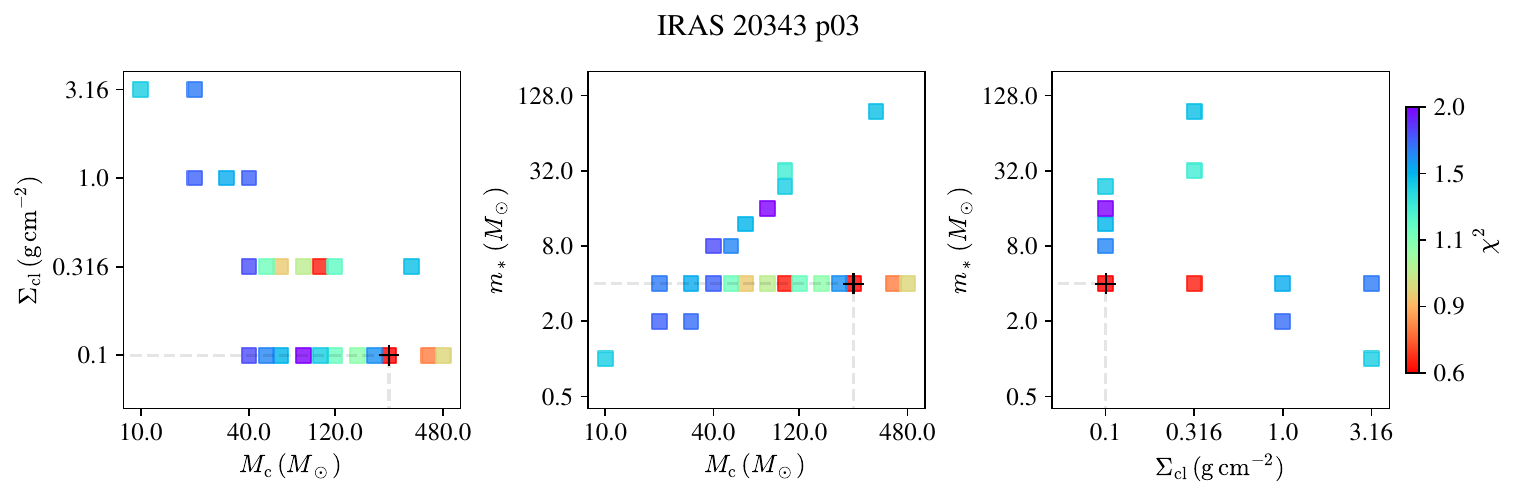}
\includegraphics[width=1.0\textwidth]{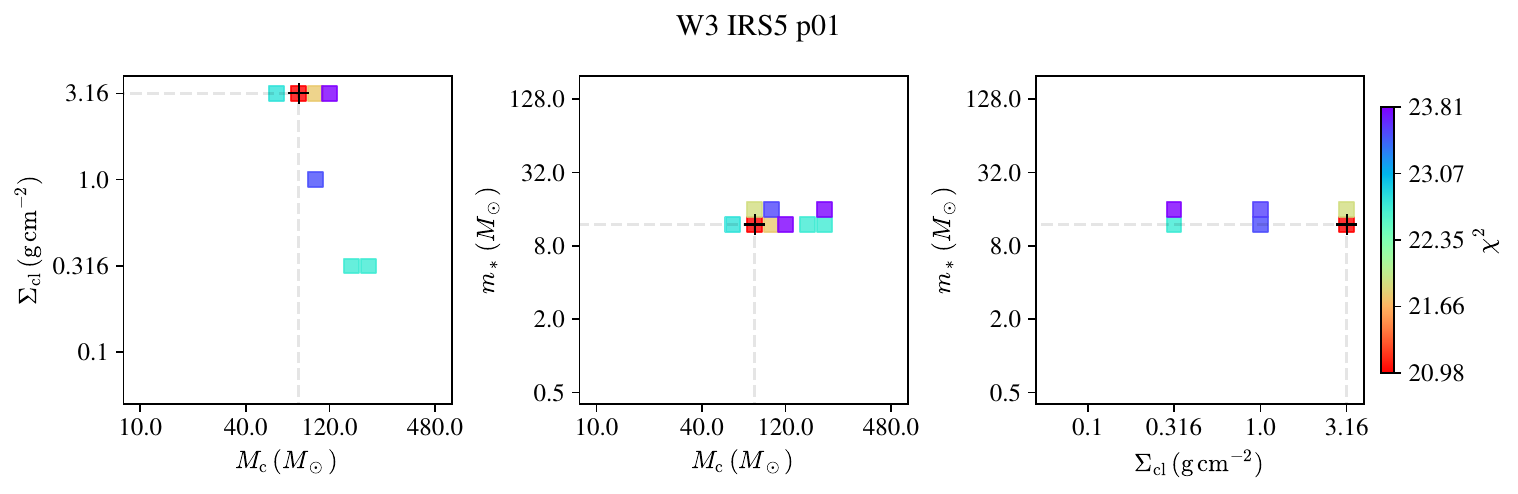}
\includegraphics[width=1.0\textwidth]{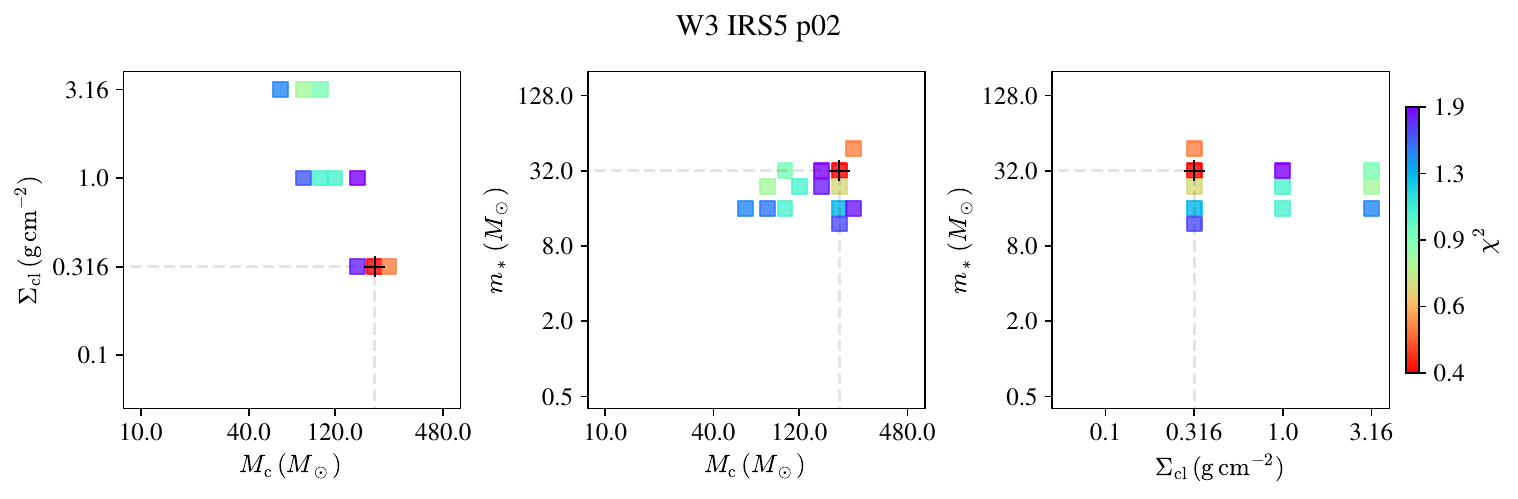}
\includegraphics[width=1.0\textwidth]{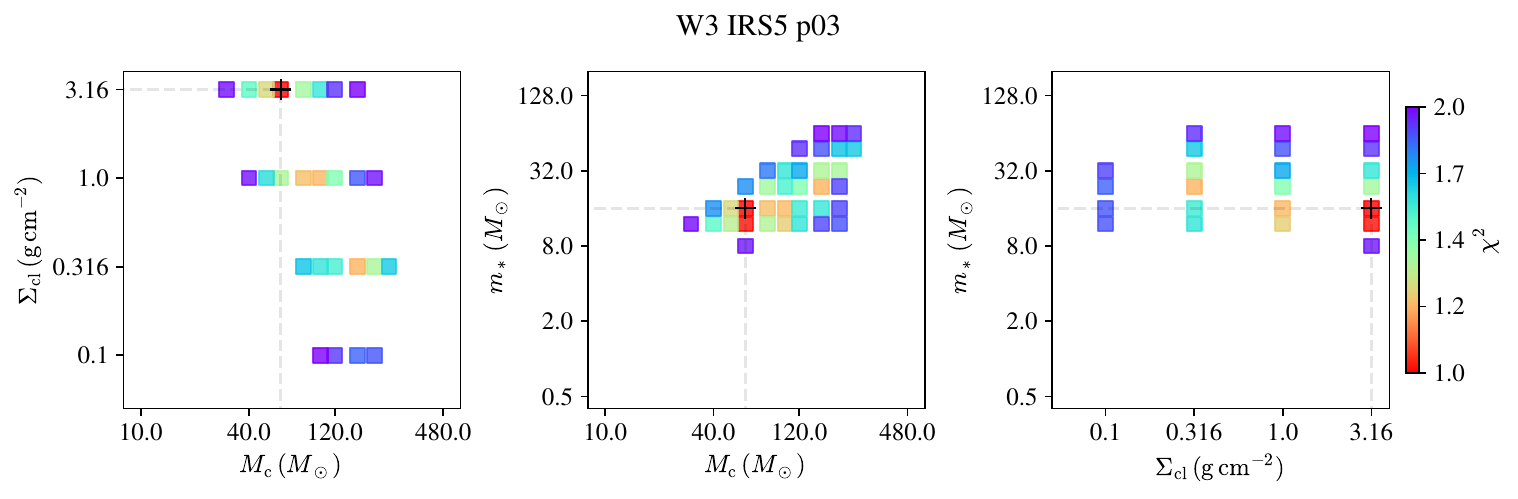}
\caption{(Continued.)}
\end{figure*}

\renewcommand{\thefigure}{A\arabic{figure}}
\addtocounter{figure}{-1}
\begin{figure*}[!htb]
\includegraphics[width=1.0\textwidth]{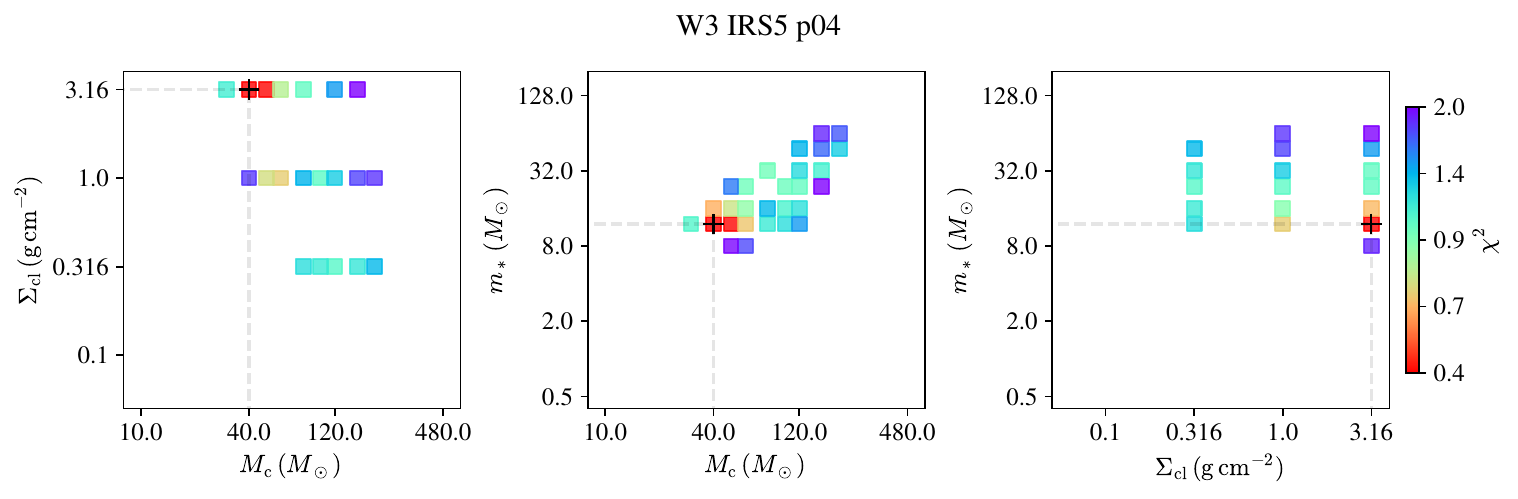}
\includegraphics[width=1.0\textwidth]{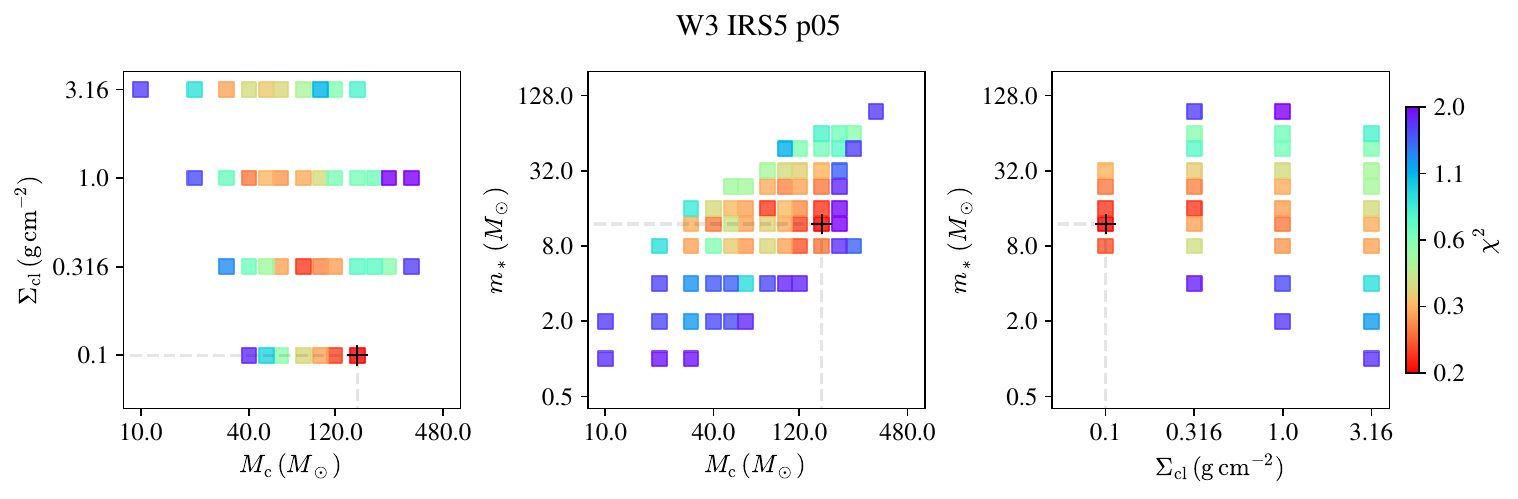}
\includegraphics[width=1.0\textwidth]{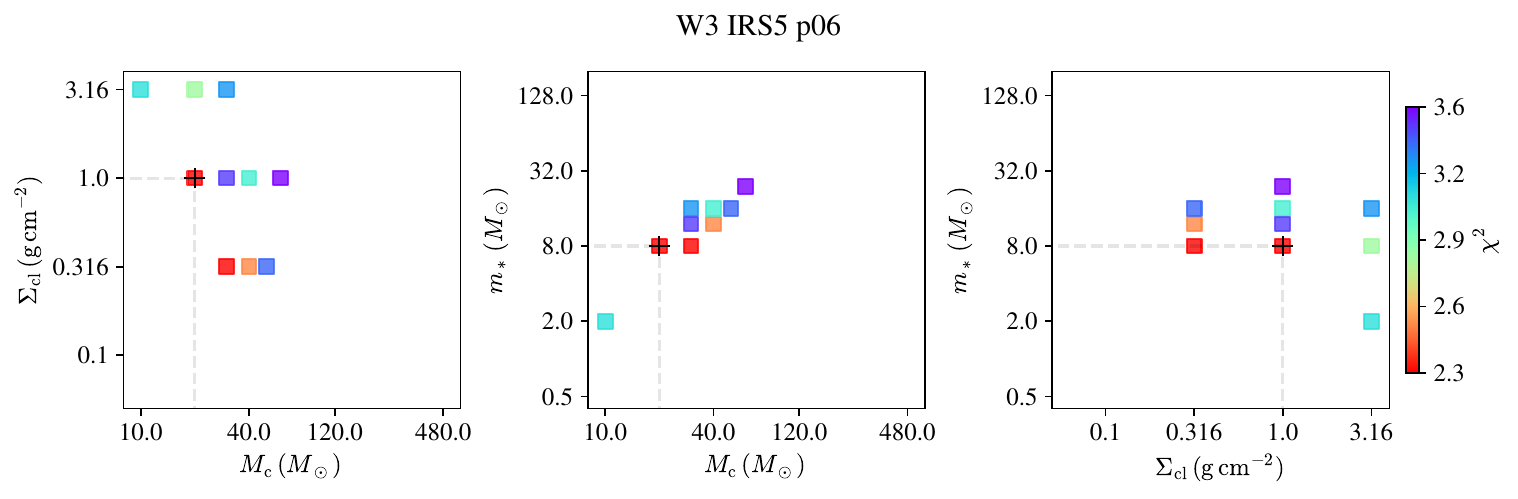}
\includegraphics[width=1.0\textwidth]{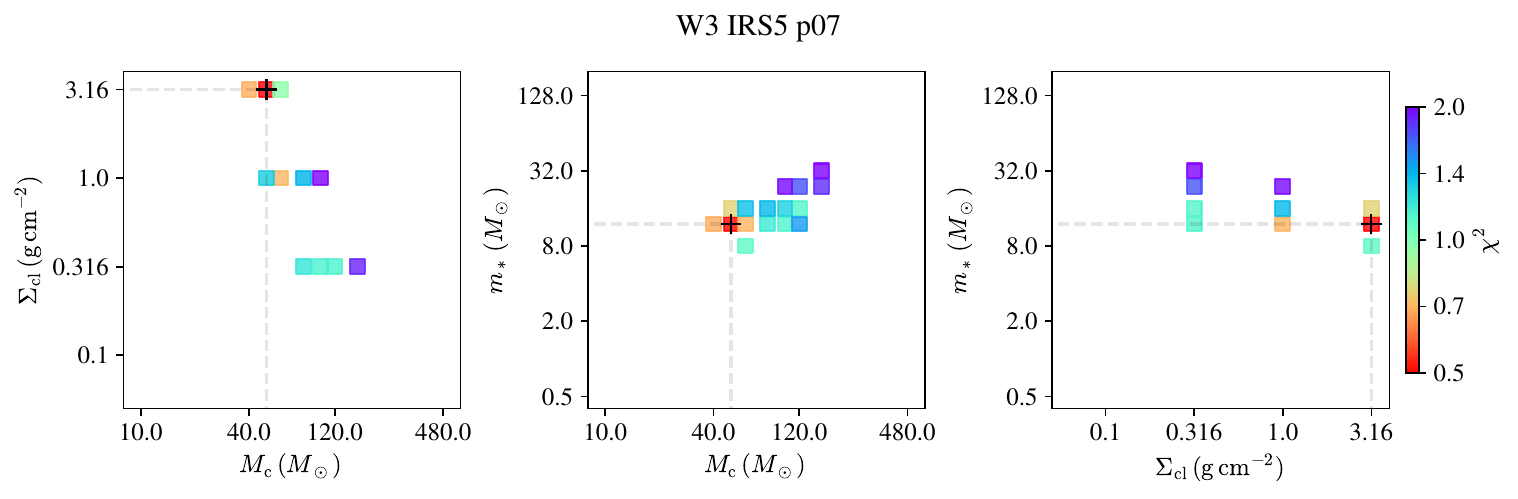}
\caption{(Continued.)}
\end{figure*}

\renewcommand{\thefigure}{A\arabic{figure}}
\addtocounter{figure}{-1}
\begin{figure*}[!htb]
\includegraphics[width=1.0\textwidth]{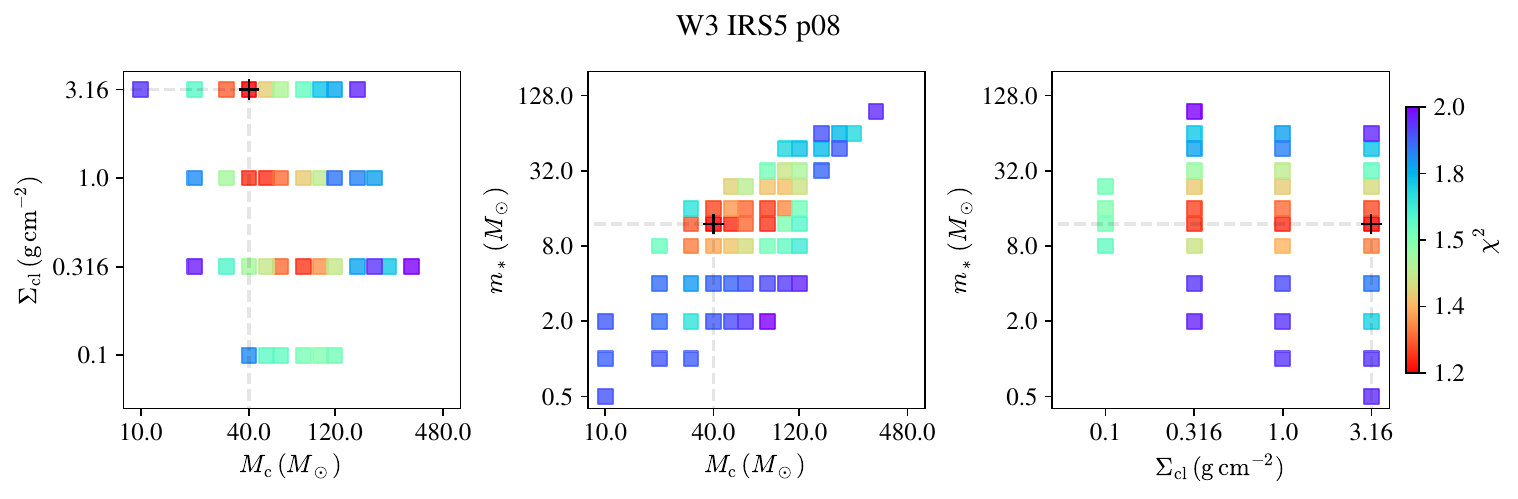}
\includegraphics[width=1.0\textwidth]{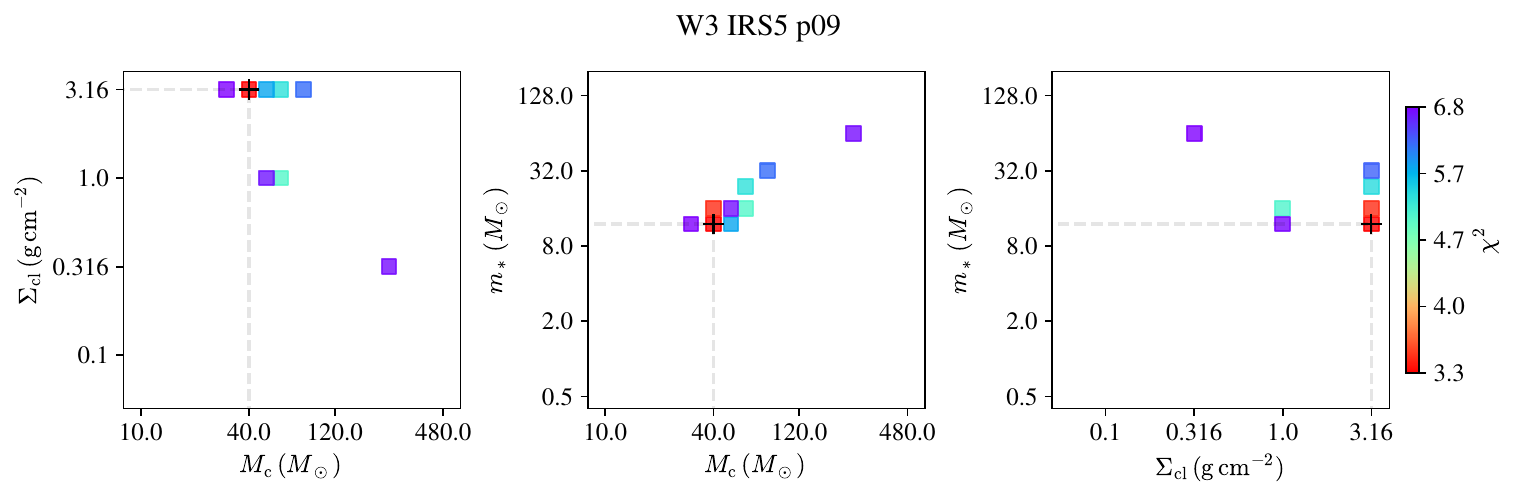}
\includegraphics[width=1.0\textwidth]{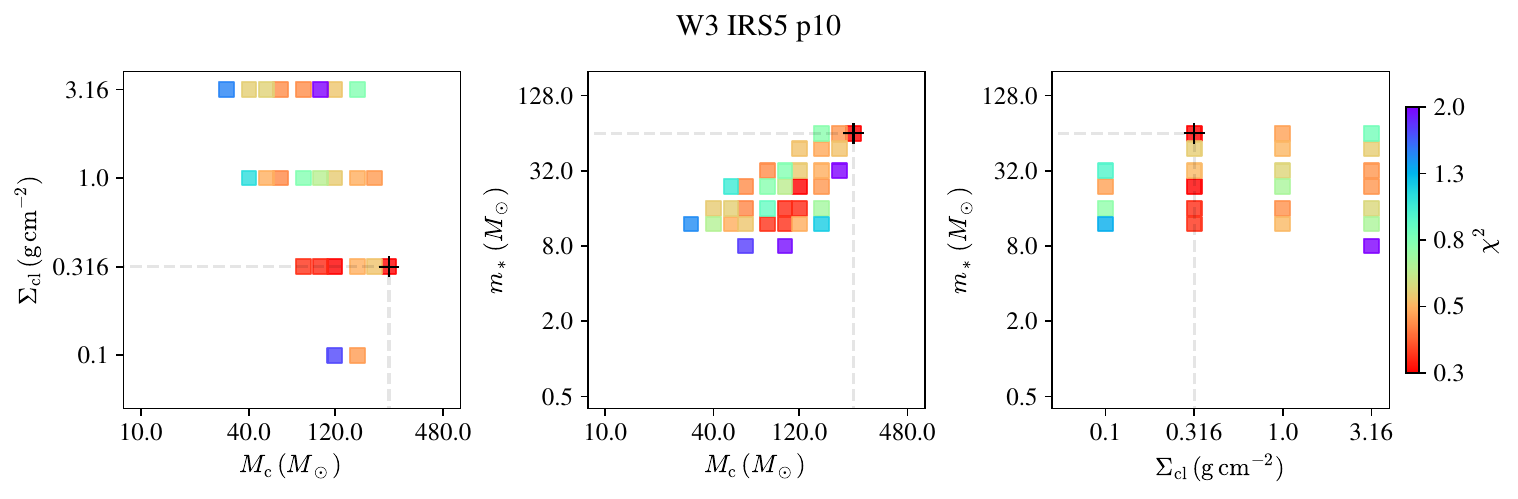}
\includegraphics[width=1.0\textwidth]{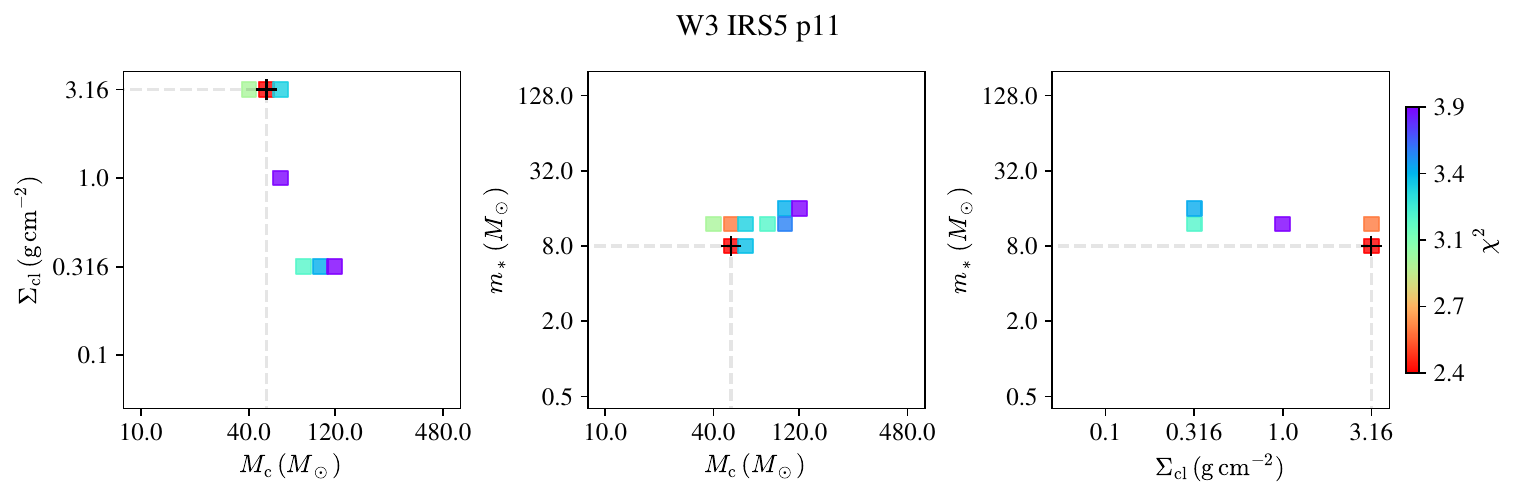}
\caption{(Continued.)}
\end{figure*}

\renewcommand{\thefigure}{A\arabic{figure}}
\addtocounter{figure}{-1}
\begin{figure*}[!htb]
\includegraphics[width=1.0\textwidth]{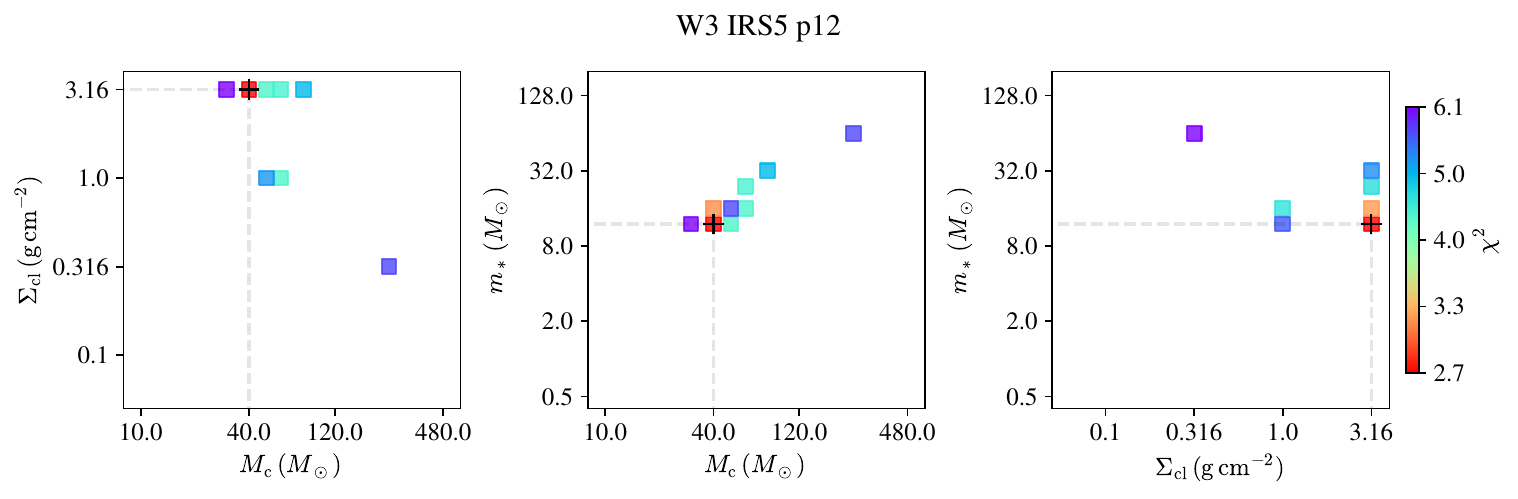}
\includegraphics[width=1.0\textwidth]{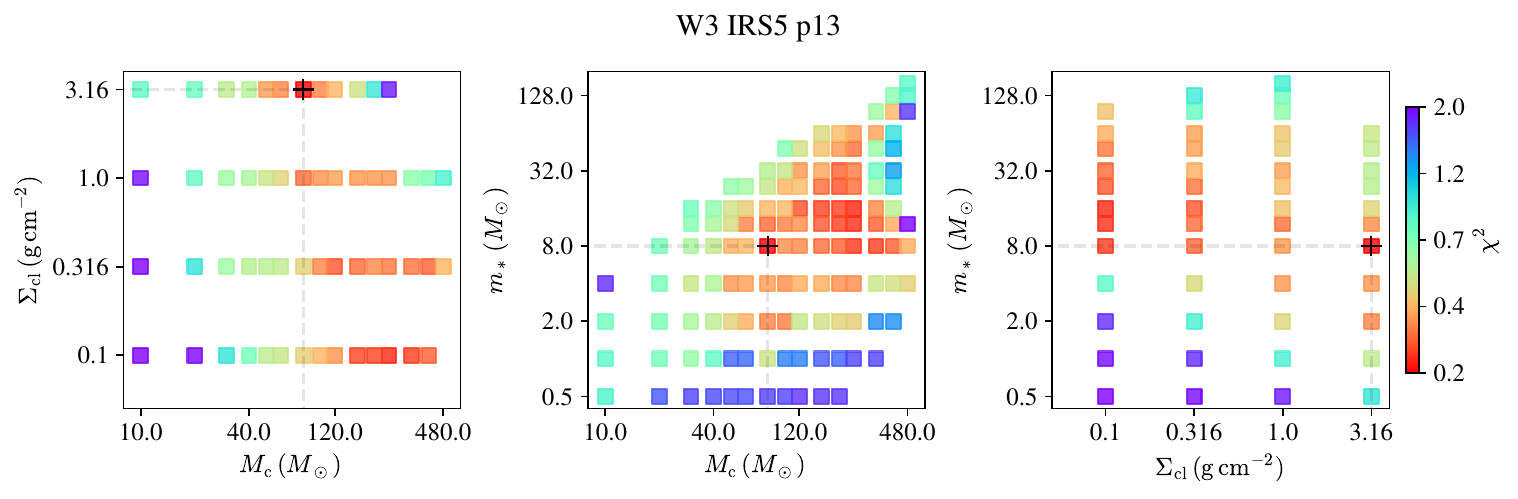}
\includegraphics[width=1.0\textwidth]{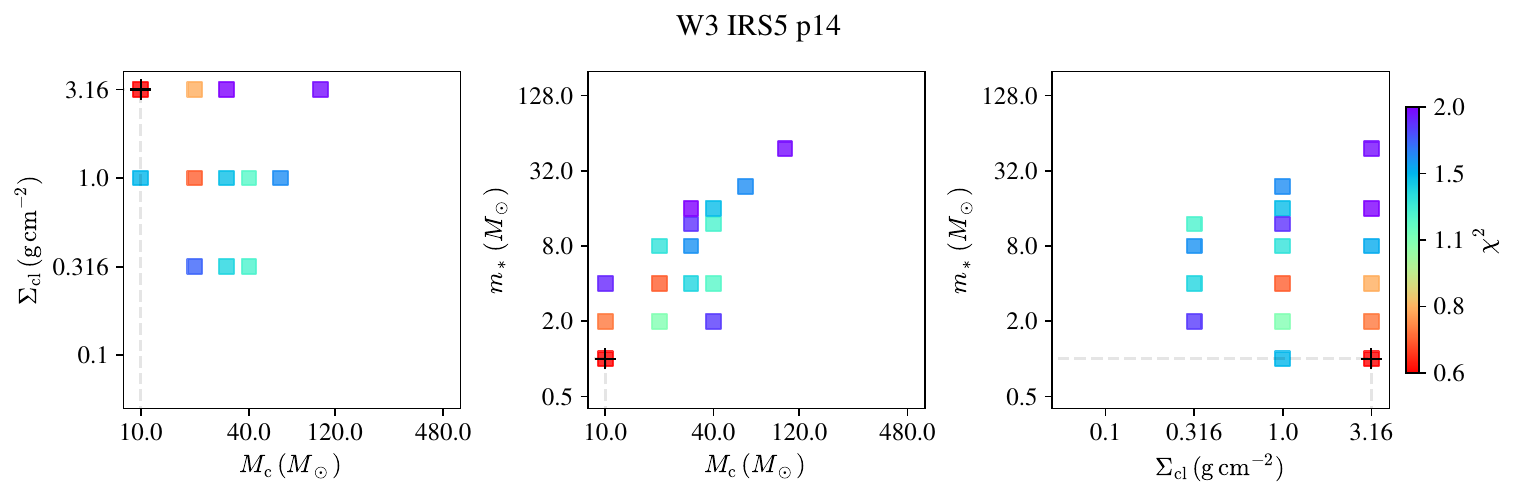}
\caption{(Continued.)}
\end{figure*}


\begin{longrotatetable}
\begin{deluxetable}{lccccccccccccccc}
\tabletypesize{\tiny}
\tablecaption{Integrated Flux Densities \label{tab:fluxes}}
\tablehead{
\colhead{Source} & \colhead{radius} & \colhead{$F_{3.6}$} & \colhead{$F_{4.5}$} & \colhead{$F_{5.8}$} & \colhead{$F_{7.7}$} & \colhead{$F_{8.0}$} & \colhead{$F_{19.7}$} & \colhead{$F_{24.2}$} & \colhead{$F_{31.5}$} & \colhead{$F_{37.1}$} & \colhead{$F_{70}$} & \colhead{$F_{160}$} & \colhead{$F_{250}$}  & \colhead{$F_{350}$} & \colhead{$F_{500}$} \\
\colhead{} & \colhead{(\arcsec/pc)} & \colhead{(Jy)} & \colhead{(Jy)} & \colhead{(Jy)} & \colhead{(Jy)} & \colhead{(Jy)} & \colhead{(Jy)} & \colhead{(Jy)} & \colhead{(Jy)} & \colhead{(Jy)} & \colhead{(Jy)} & \colhead{(Jy)} & \colhead{(Jy)} & \colhead{(Jy)} & \colhead{(Jy)}}

\startdata
\multirow{3}{*}{AFGL5180 p01} & \multirow{2}{*}{12.25} & $\cdots$ & $\cdots$ & $\cdots$ & 3.82 & $\cdots$ & $\cdots$ & $\cdots$ & $\cdots$ & 243.65 & 446.14 & 254.15 & 76.21 & 30.08 & 3.13 \\
& \multirow{2}{*}{0.10} & $\cdots$ & $\cdots$ & $\cdots$ & (3.14) & $\cdots$ & $\cdots$ & $\cdots$ & $\cdots$ & (248.27) & (518.46) & (379.70) & (133.99) & (52.59) & (9.16) \\
&  & $\cdots$ & $\cdots$ & $\cdots$ & [0.68] & $\cdots$ & $\cdots$ & $\cdots$ & $\cdots$ & [24.69] & [48.79] & [128.09] & [58.28] & [22.71] & [6.04] \\
\hline
\multirow{3}{*}{AFGL5180 p02} & \multirow{2}{*}{24.25} & $\cdots$ & $\cdots$ & $\cdots$ & 11.00 & $\cdots$ & $\cdots$ & $\cdots$ & $\cdots$ & 137.41 & 239.83 & 177.59 & 68.96 & 28.90 & 9.11 \\
& \multirow{2}{*}{0.21} & $\cdots$ & $\cdots$ & $\cdots$ & (9.06) & $\cdots$ & $\cdots$ & $\cdots$ & $\cdots$ & (133.23) & (253.28) & (223.10) & (101.28) & (45.48) & (17.32) \\
&  & $\cdots$ & $\cdots$ & $\cdots$ & [1.57] & $\cdots$ & $\cdots$ & $\cdots$ & $\cdots$ & [22.44] & [57.48] & [48.85] & [33.05] & [16.83] & [8.26] \\
\hline
\multirow{3}{*}{AFGL5180 p03} & \multirow{2}{*}{12.25} & $\cdots$ & $\cdots$ & $\cdots$ & 5.69 & $\cdots$ & $\cdots$ & $\cdots$ & $\cdots$ & 131.63 & 152.94 & 88.38 & 28.41 & 15.77 & 5.41 \\
& \multirow{2}{*}{0.10} & $\cdots$ & $\cdots$ & $\cdots$ & (6.62) & $\cdots$ & $\cdots$ & $\cdots$ & $\cdots$ & (144.09) & (206.44) & (166.81) & (64.48) & (35.20) & (13.31) \\
\enddata
\tablecomments{
{\tiny $F_{3.6}$, $F_{4.5}$, $F_{5.8}$, and $F_{8.0}$ refer to fluxes from Spitzer-IRAC at 3.6, 4.5, 5.8, and 8.0\,$\mu$m, respectively. $F_{7.7}$, $F_{19.1}$, $F_{31.5}$, and $F_{37.1}$ refer to fluxes from SOFIA-FORCAST at 7.7, 19.1, 31.5, and 37.5\,$\mu$m, respectively. $F_{70}$, $F_{160}$, $F_{250}$, $F_{350}$ and $F_{500}$ refer to fluxes from Herschel-PACS/SPIRE at 70, 160, 250, 350, and 500\,$\mu$m, respectively.  $F_{12.0}$, $F_{25.0}$, $F_{60.0}$, and $F_{100.0}$ refer to fluxes from IRAS-HIRES at 12.0, 25.0, 60.0, and 100.0\,$\mu$m, respectively. The three dots refer to either data not found or saturated at that wavelength. First row for each source refers to background subtracted flux. Second row for each source with fluxes in parenthesis refers to non background subtracted flux (in many FORCAST $7.7\:{\rm \mu m}$ images we find that the estimated background has a slight negative value, which results from a calibration offset problem; hence, background subtracted values appear larger in this table; however, we still expect these background subtracted values to be accurate). Third row for each source with square brackets refers to the associated error to the background subtracted fluxes. $\dagger$ No Herschel-PACS 70\,$\mu$m data is available and SOFIA-FORCAST 37\,$\mu$m was used to find the optimal aperture (\S\ref{sect:opt_rad}). For the radius column for each source, the first row refers to \arcsec and the second row to pc.}}
\end{deluxetable}
\end{longrotatetable}

\begin{longrotatetable}
\begin{deluxetable}{lcccccccccccc}
\tabletypesize{\tiny}
\tablecaption{Parameters of the Best Five Fitted Models and Average and Dispersion of Good Models\label{tab:best_models}} 
\tablewidth{18pt}
\tablehead{
\colhead{Source} &\colhead{$\chi^{2}$} & \colhead{$M_{c}$} & \colhead{$\Sigma_{\rm cl}$} & \colhead{$R_{c}$}  &\colhead{$m_{*}$} & \colhead{$\theta_{\rm view}$} &\colhead{$A_{V}$} & \colhead{$M_{\rm env}$} &\colhead{$\theta_{\rm w,esc}$} & \colhead{$\dot {M}_{\rm disk}$} & \colhead{$L_{\rm bol,iso}$} & \colhead{$L_{\rm bol}$} \\
\colhead{} & \colhead{} & \colhead{($M_\odot$)} & \colhead{(g $\rm cm^{-2}$)} & \colhead{(pc)} & \colhead{($M_{\odot}$)} & \colhead{(\arcdeg)} & \colhead{(mag)} & \colhead{($M_{\odot}$)} & \colhead{(deg)} &\colhead{($M_{\odot}$/yr)} & \colhead{($L_{\odot}$)} & \colhead{($L_{\odot}$)}
}
\startdata
AFGL5180 p01 & 0.21 & 60 & 0.100 & 0.18 & 8.0 & 88.6 & 30 & 40.2 & 33.2 & 4.4$\times10^{-5}$ & 4.1$\times10^{3}$ & 1.0$\times10^{4}$ \\
$d=1.8$ kpc & 0.27 & 60 & 0.316 & 0.10 & 16.0 & 64.8 & 42 & 19.0 & 55.6 & 1.1$\times10^{-4}$ & 4.4$\times10^{3}$ & 3.6$\times10^{4}$ \\
$R_\mathrm{ap}=12.25\arcsec$ & 0.28 & 80 & 0.100 & 0.21 & 16.0 & 79.9 & 38 & 31.2 & 53.8 & 5.0$\times10^{-5}$ & 4.4$\times10^{3}$ & 2.9$\times10^{4}$ \\
$R_\mathrm{ap}=0.10$ pc & 0.37 & 80 & 0.316 & 0.12 & 24.0 & 82.8 & 24 & 14.1 & 66.6 & 1.0$\times10^{-4}$ & 3.8$\times10^{3}$ & 8.1$\times10^{4}$ \\
 & 0.39 & 80 & 0.100 & 0.21 & 12.0 & 88.6 & 52 & 46.5 & 40.0 & 5.4$\times10^{-5}$ & 4.7$\times10^{3}$ & 1.6$\times10^{4}$ \\
Avg. model & \#161 &  $53^{+28}_{-18}$ & $0.309^{+0.700}_{-0.215}$ & $0.10^{+0.10}_{-0.05}$ & $12.0^{+7.1}_{-4.5}$ & $70.9 \pm 12.9$ & $58 \pm 51$ & $16.7^{+34.1}_{-11.2}$ & $45.9 \pm 12.7$ & $10.0^{+12.4}_{-5.5}\times10^{-5}$ & $4.9^{+2.9}_{-1.8}\times10^{3}$ & $2.5^{+4.4}_{-1.6}\times10^{4}$\\
Avg. model $\Sigma_\mathrm{cl,GB}=0.29$ & \#55 &  $52^{+24}_{-17}$ & $0.316$ & $0.09^{+0.02}_{-0.02}$ & $12.6^{+8.0}_{-4.9}$ & $70.9 \pm 12.7$ & $38 \pm 28$ & $18.6^{+4.5}_{-3.7}$ & $48.9 \pm 11.6$ & $9.7^{+1.0}_{-0.9}\times10^{-5}$ & $4.0^{+1.0}_{-0.8}\times10^{3}$ & $2.6^{+3.7}_{-1.5}\times10^{4}$\\
\hline
AFGL5180 p02 & 0.30 & 60 & 0.100 & 0.18 & 12.0 & 12.8 & 115 & 24.9 & 51.5 & 4.2$\times10^{-5}$ & 4.4$\times10^{4}$ & 1.5$\times10^{4}$ \\
\enddata
\tablecomments{{\small
For each source, the first five rows refer to the best five models taken from the 432 physical models for the non-restricted $\Sigma_\mathrm{cl}$ case. The sixth row shows the average and dispersion of good model fits (see the text) for the non-restricted $\Sigma_\mathrm{cl}$ case. The seventh row shows the average and dispersion of good models for the restricted $\Sigma_\mathrm{cl,GB}$ case. The number next to the symbol \# represents the number of models considered in the average of the good models. Upper and lower scripts in the average model rows refer to the upper and lower range of the dispersion of the good models.}}
\end{deluxetable}
\end{longrotatetable}

\section{SOMA I-IV SED Fits}
Here we present the results of the SED fitting for the sources in the SOMA I-IV sample using the updated definition of ``good'' models. Table \ref{tab:best_models_SOMA_I_IV} presents the best models including the $\Sigma_\mathrm{cl,GB}$ restricted models. Figure \ref{fig:sed_soma_I-IV} contains the source SEDs and ``good" model fits (see main text for details), and Figure \ref{fig:sed_2D_results_soma_I-IV} presents the ``good'' model distributions in the $\Sigma_{\rm cl}-M_c$, $m_*-M_c$, and $m_*-\Sigma_{\rm cl}$ planes.

\renewcommand{\thefigure}{C\arabic{figure}}
\addtocounter{figure}{-3}
\begin{figure*}[!htb]
\includegraphics[width=0.5\textwidth]{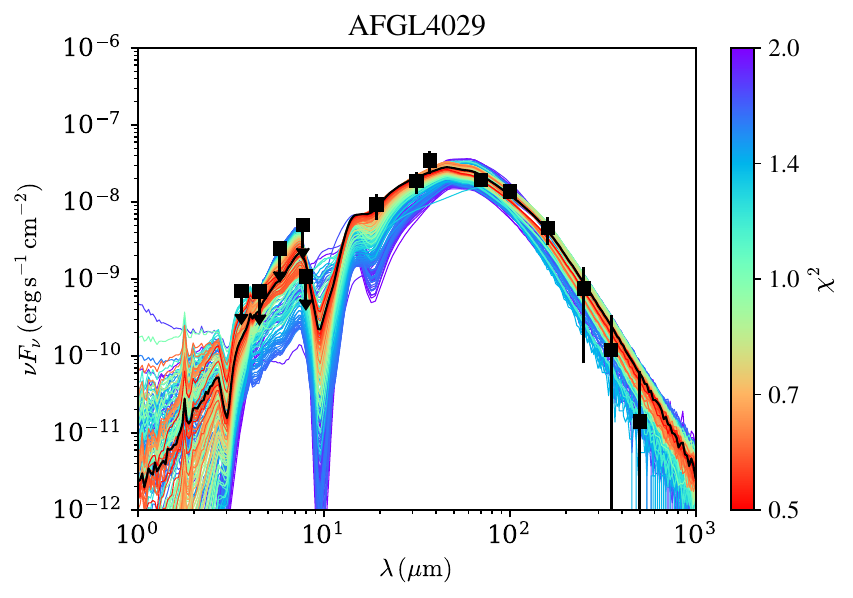}
\includegraphics[width=0.5\textwidth]{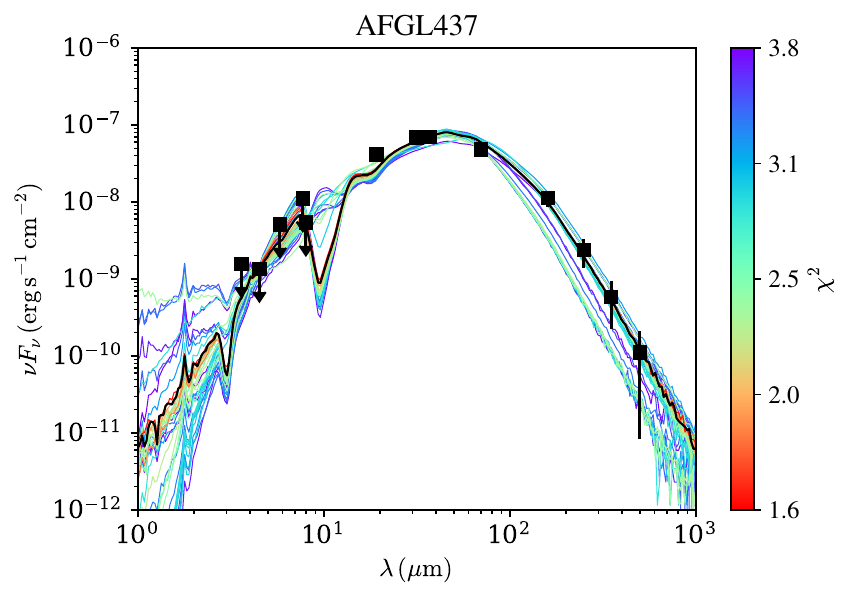}
\includegraphics[width=0.5\textwidth]{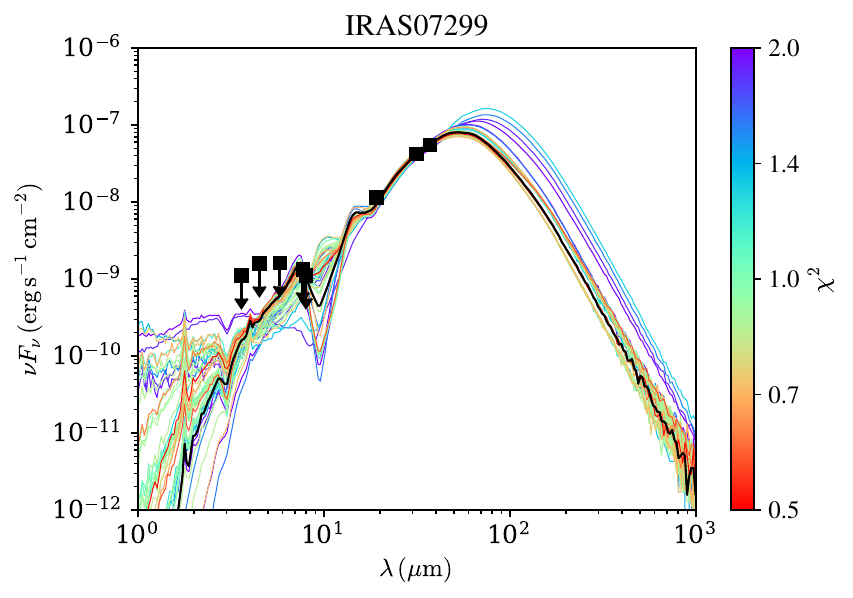}
\includegraphics[width=0.5\textwidth]{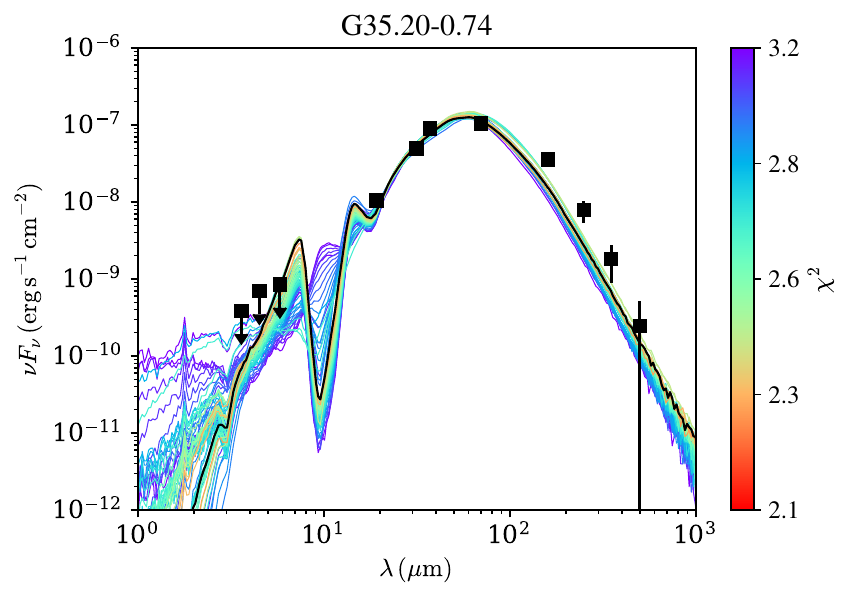}
\includegraphics[width=0.5\textwidth]{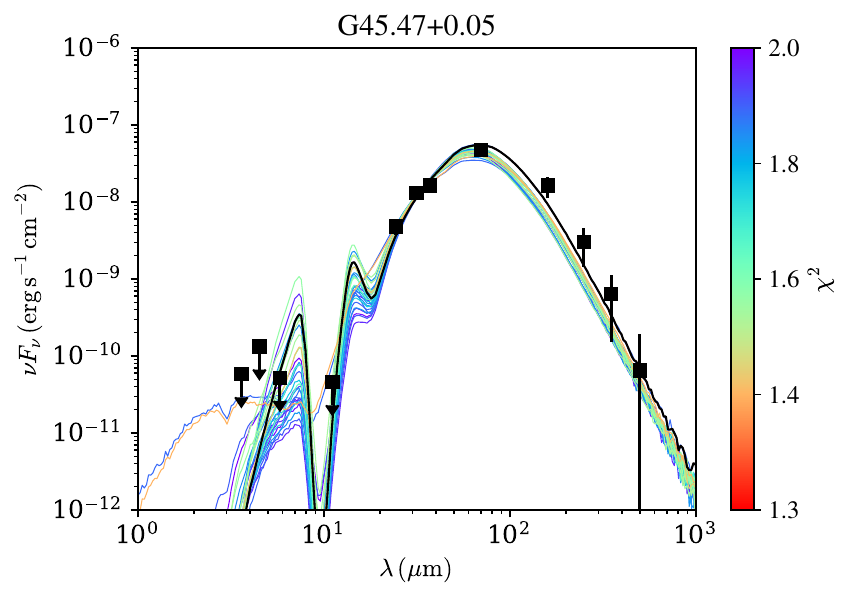}
\includegraphics[width=0.5\textwidth]{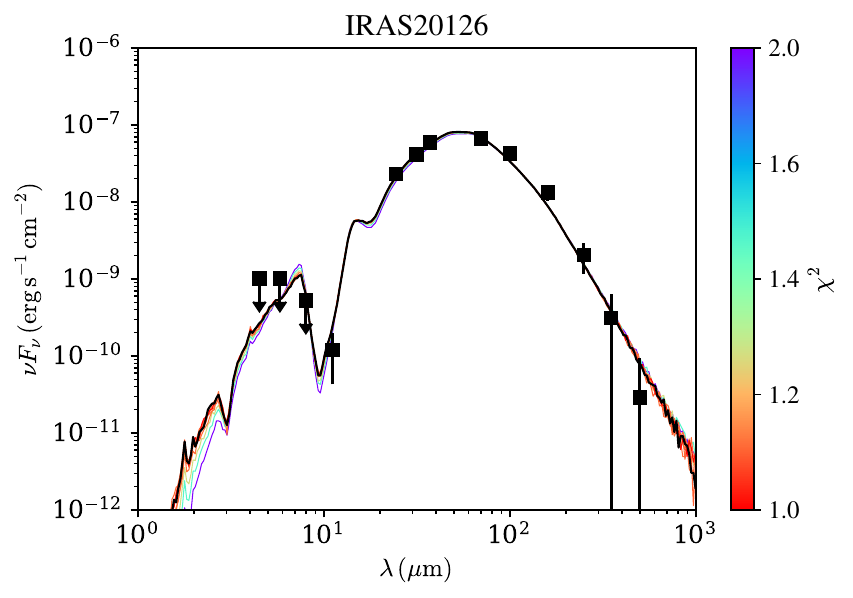}
\caption{SOMA I-IV sources reanalyzed with sedcreator an the new definition of ``good'' model. The resulting model parameters are listed in Table\,\ref{tab:best_models_SOMA_I_IV}.
\label{fig:sed_soma_I-IV}}
\end{figure*}

\renewcommand{\thefigure}{C\arabic{figure}}
\addtocounter{figure}{-1}
\begin{figure*}[!htb]
\includegraphics[width=0.5\textwidth]{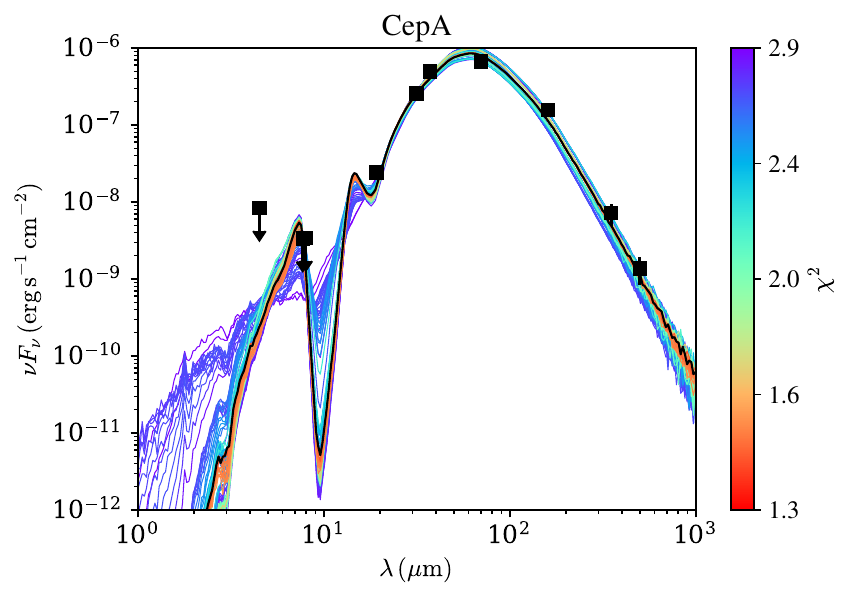}
\includegraphics[width=0.5\textwidth]{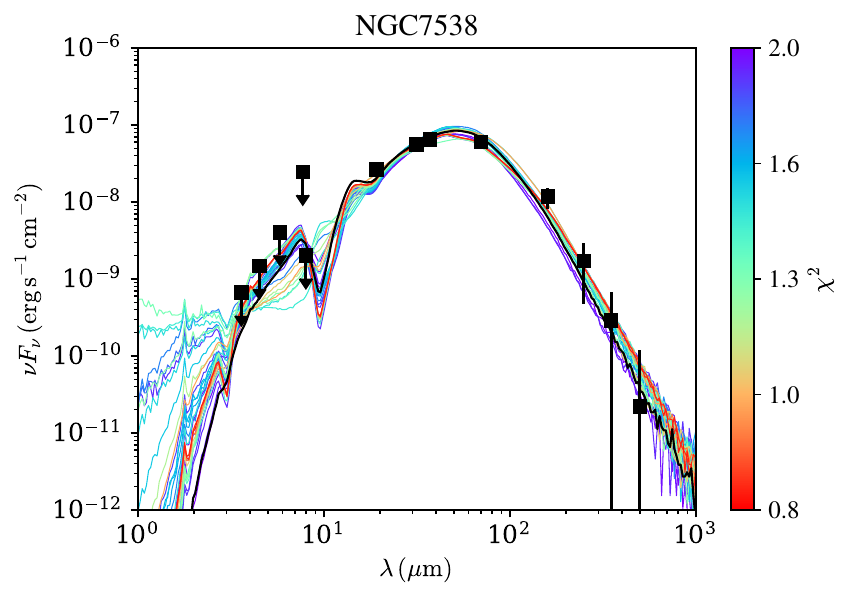}
\includegraphics[width=0.5\textwidth]{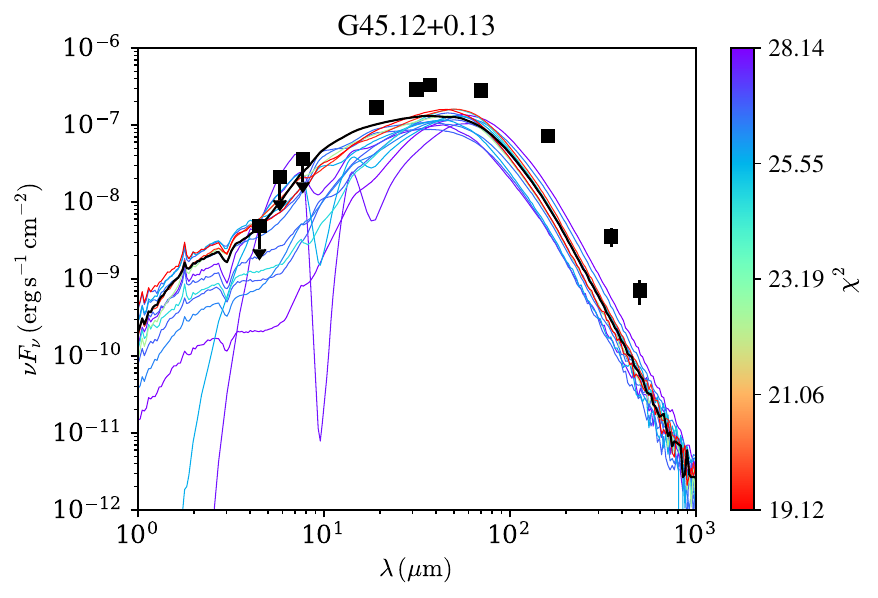}
\includegraphics[width=0.5\textwidth]{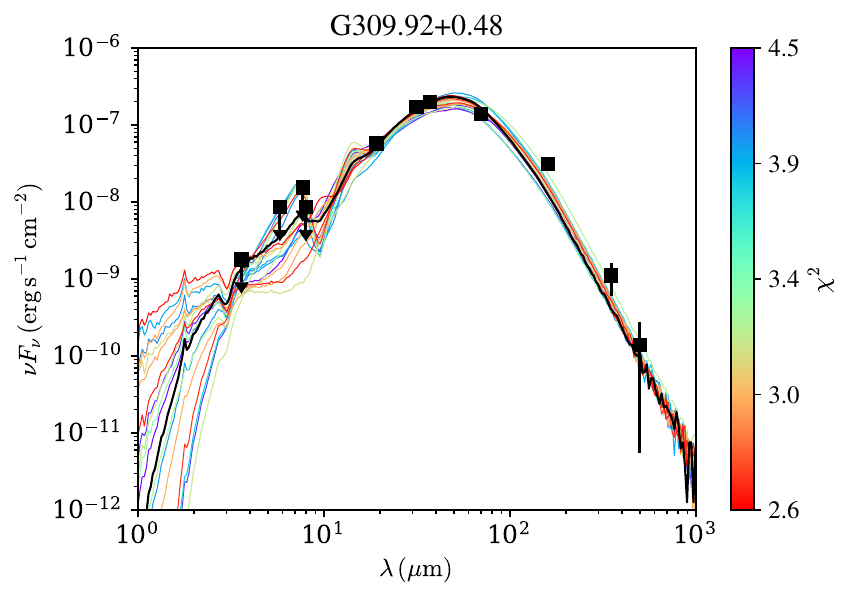}
\includegraphics[width=0.5\textwidth]{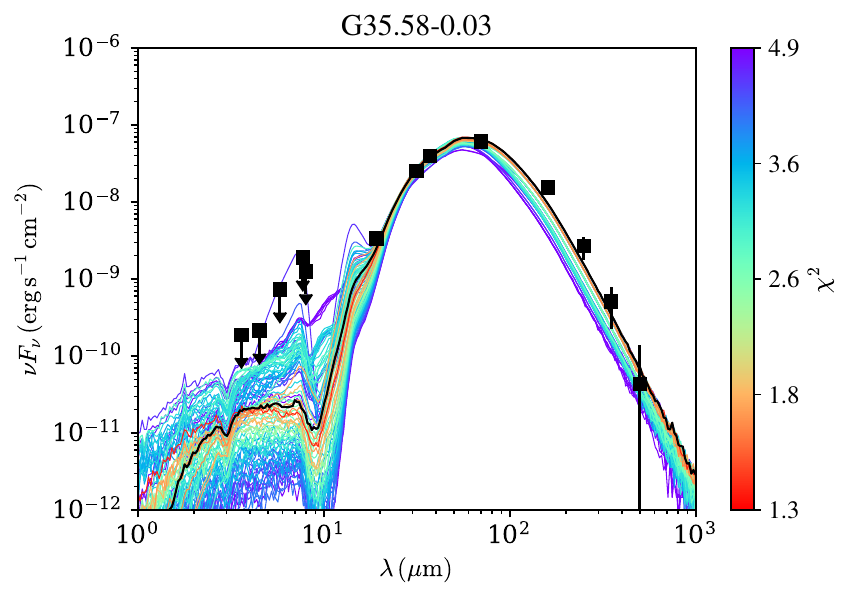}
\includegraphics[width=0.5\textwidth]{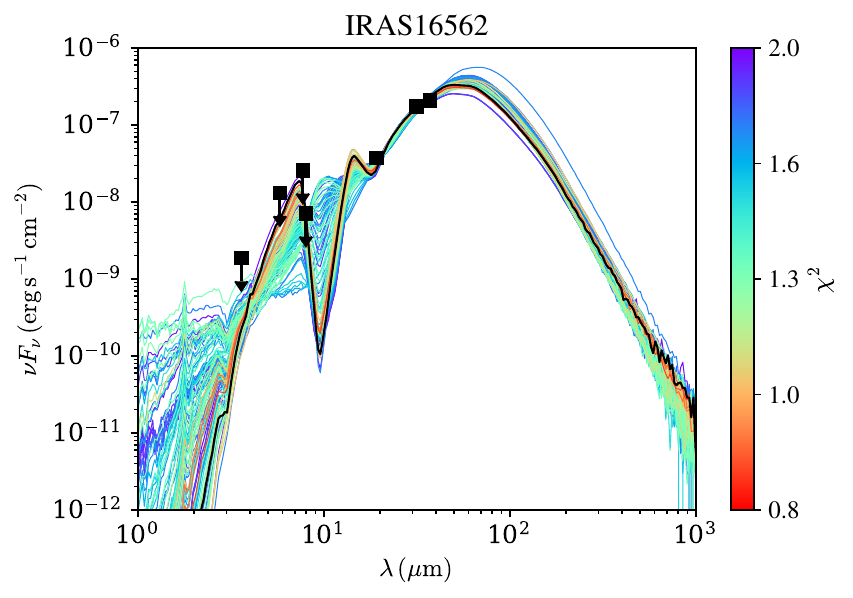}
\caption{(Continued.)}
\end{figure*}

\renewcommand{\thefigure}{C\arabic{figure}}
\addtocounter{figure}{-1}
\begin{figure*}[!htb]
\includegraphics[width=0.5\textwidth]{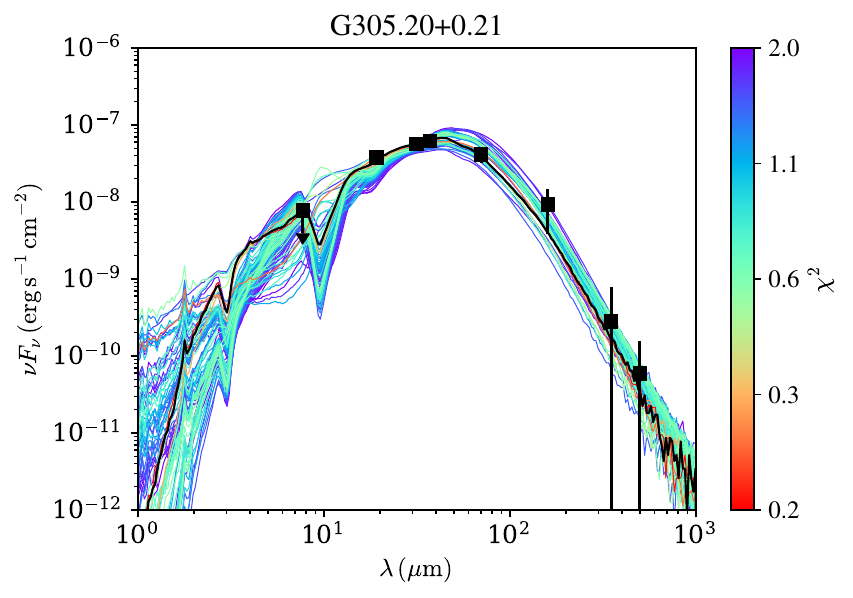}
\includegraphics[width=0.5\textwidth]{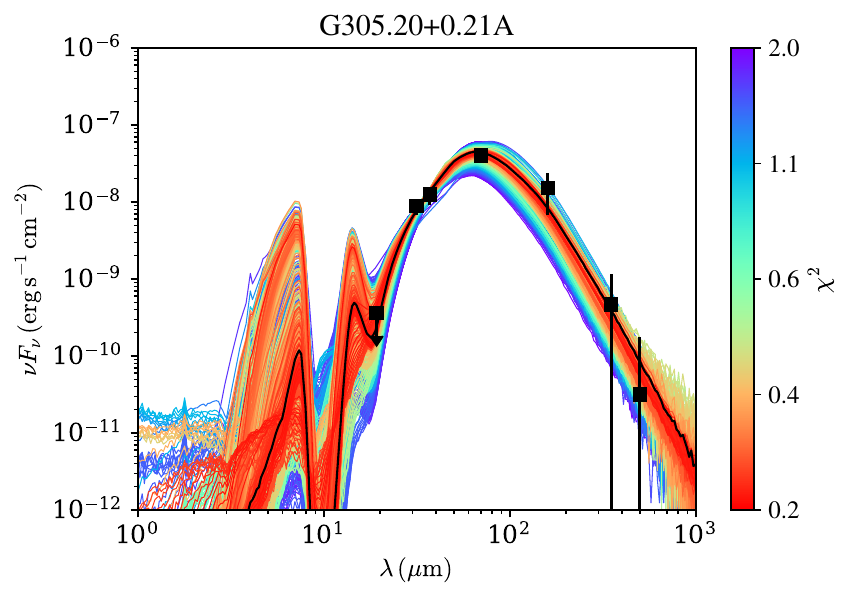}
\includegraphics[width=0.5\textwidth]{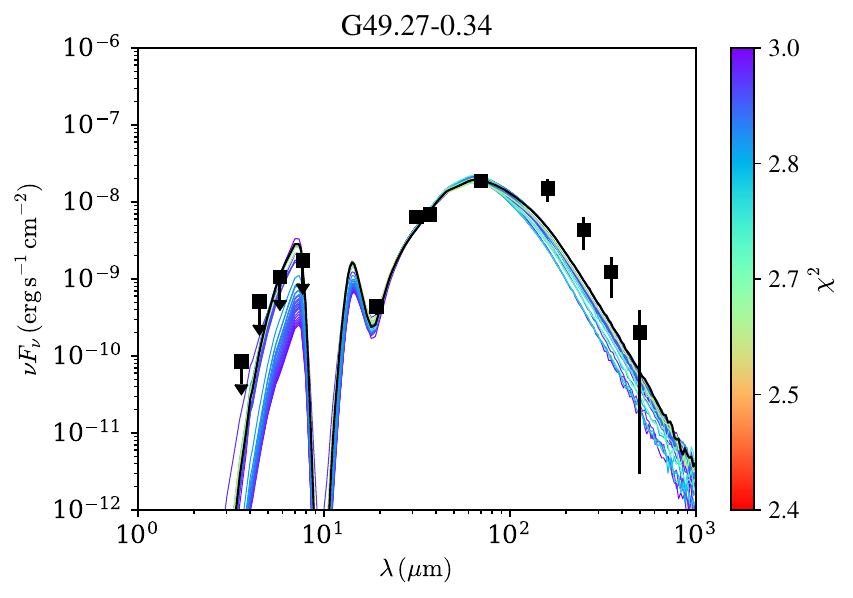}
\includegraphics[width=0.5\textwidth]{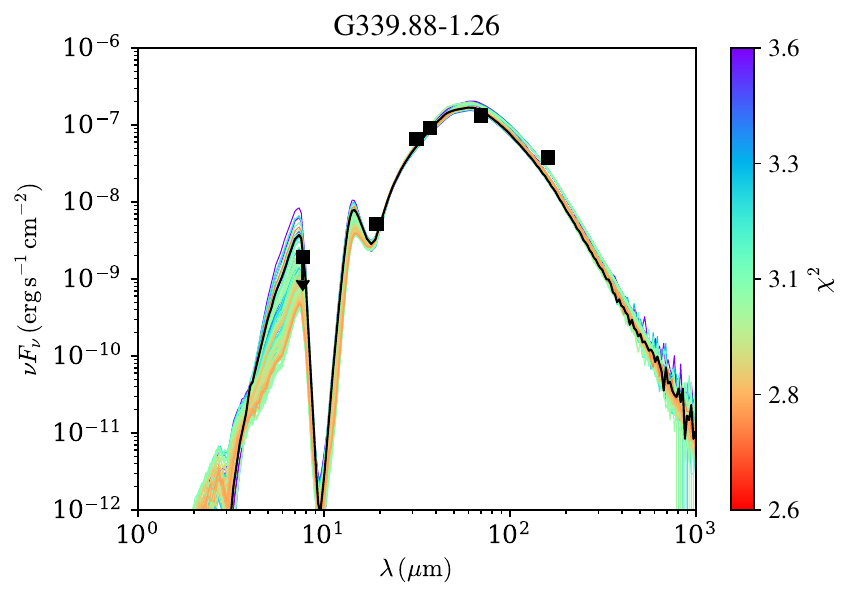}
\includegraphics[width=0.5\textwidth]{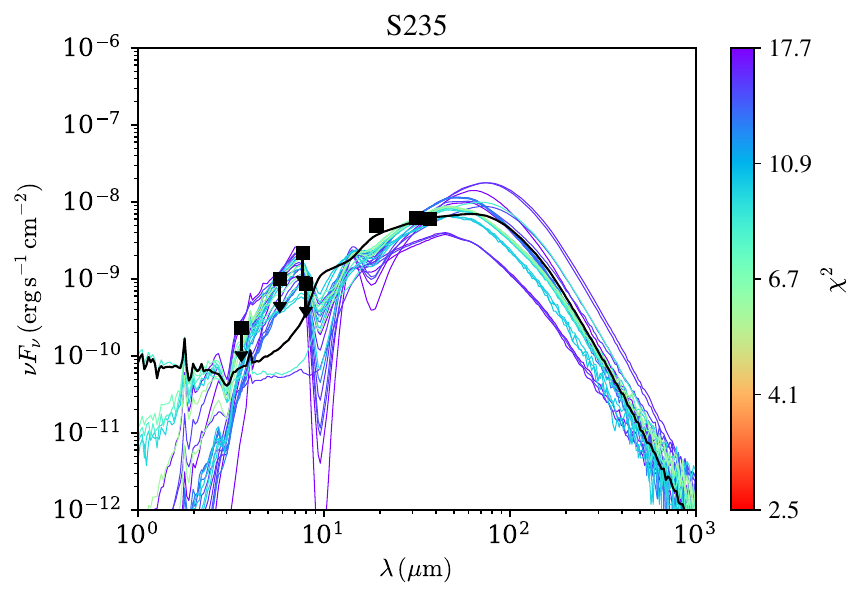}
\includegraphics[width=0.5\textwidth]{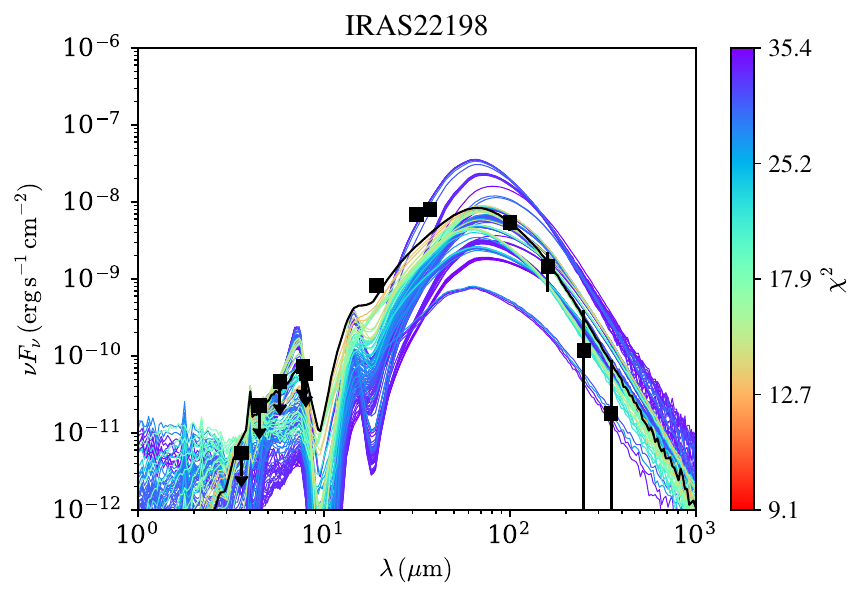}
\caption{(Continued.)}
\end{figure*}

\renewcommand{\thefigure}{C\arabic{figure}}
\addtocounter{figure}{-1}
\begin{figure*}[!htb]
\includegraphics[width=0.5\textwidth]{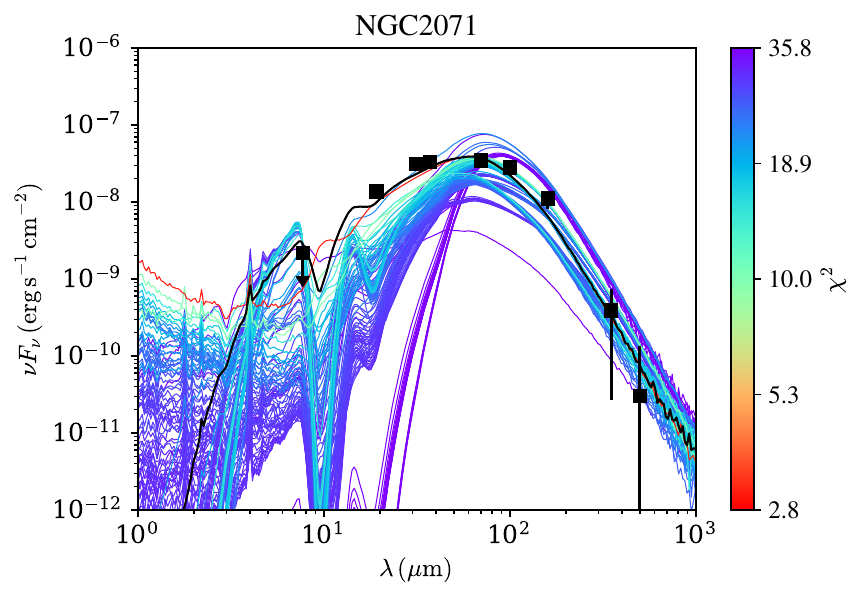}
\includegraphics[width=0.5\textwidth]{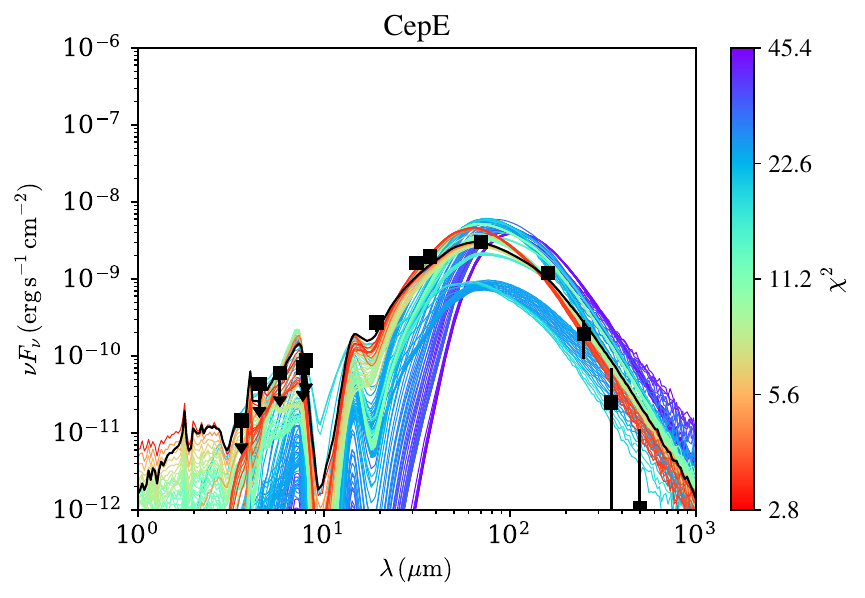}
\includegraphics[width=0.5\textwidth]{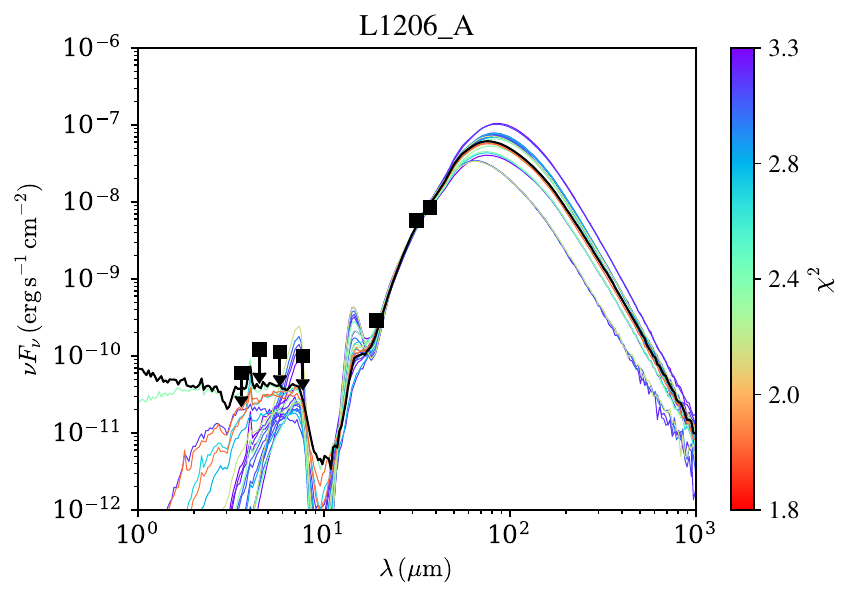}
\includegraphics[width=0.5\textwidth]{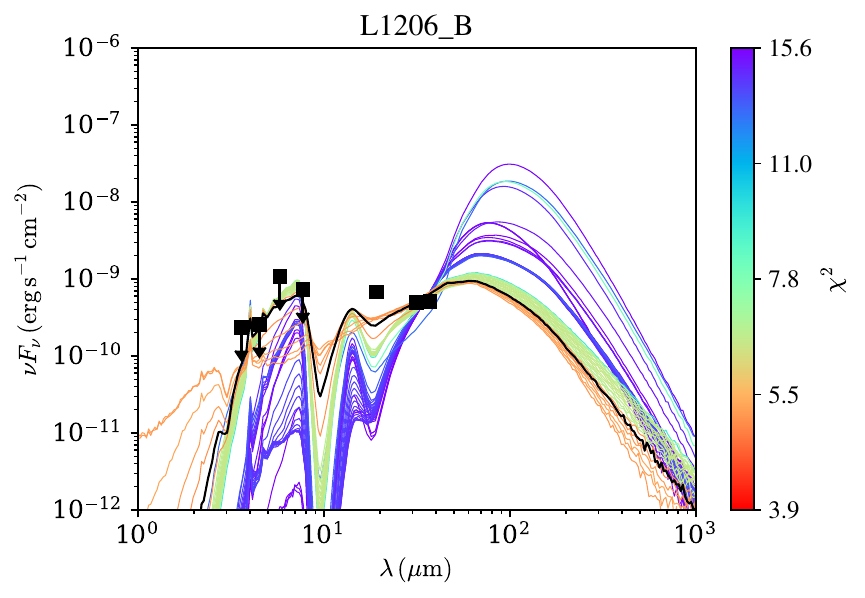}
\includegraphics[width=0.5\textwidth]{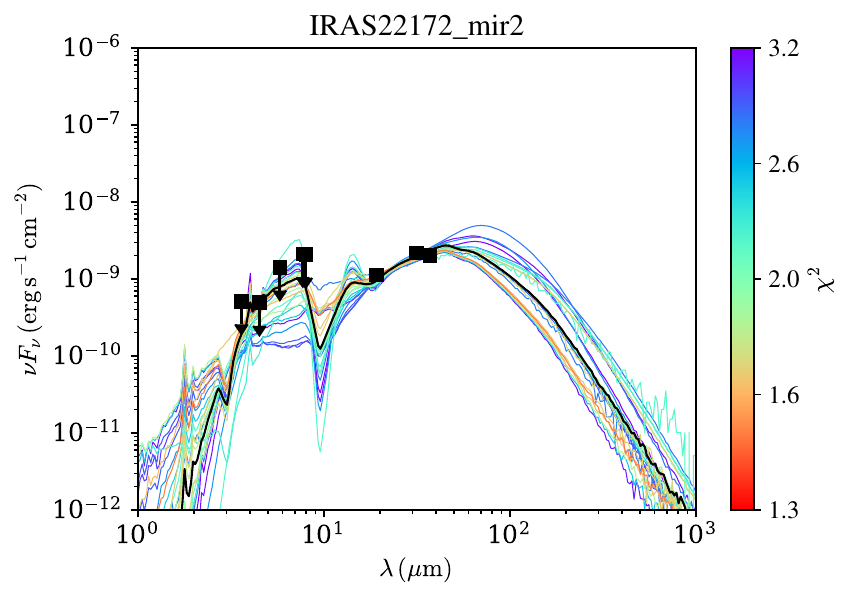}
\includegraphics[width=0.5\textwidth]{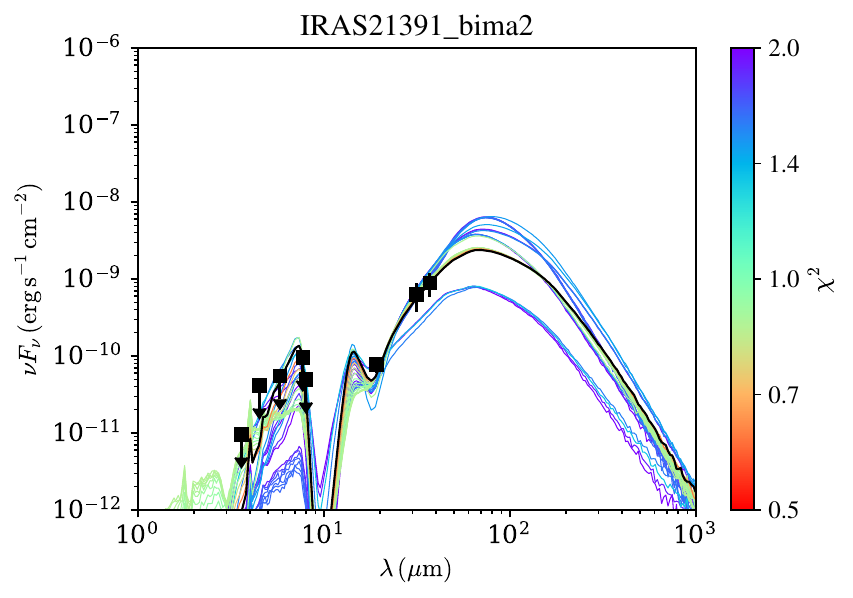}
\caption{(Continued.)}
\end{figure*}

\renewcommand{\thefigure}{C\arabic{figure}}
\addtocounter{figure}{-1}
\begin{figure*}[!htb]
\includegraphics[width=0.5\textwidth]{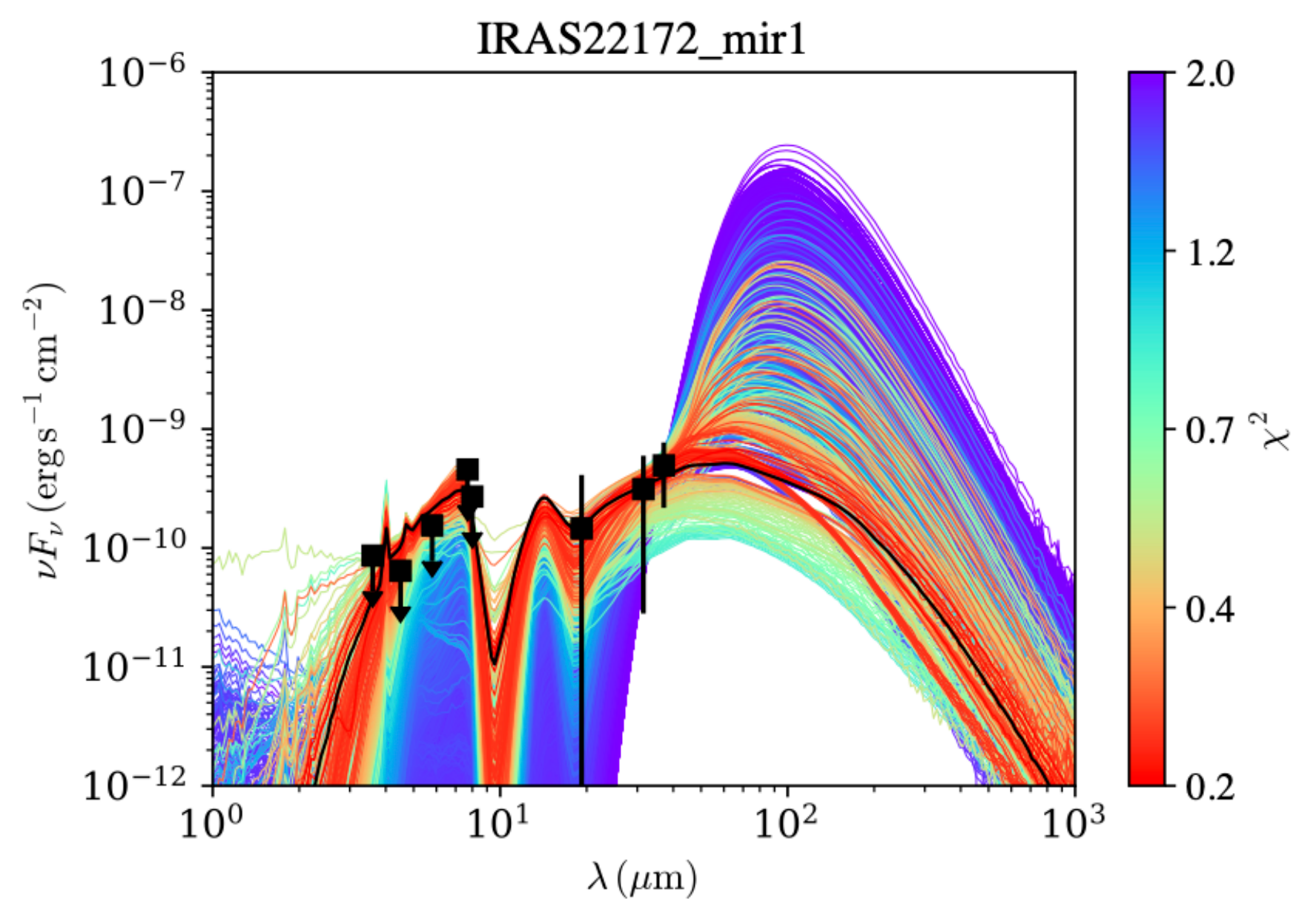}
\includegraphics[width=0.5\textwidth]{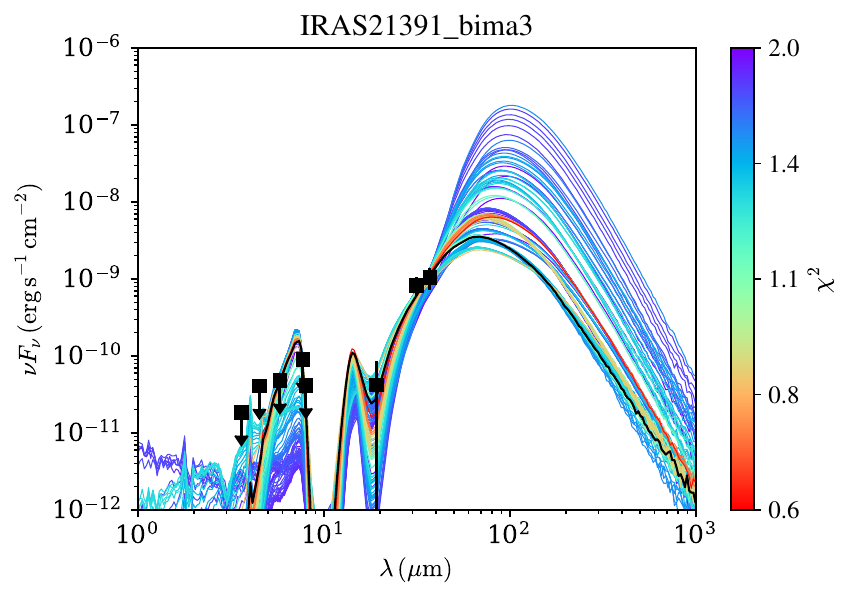}
\includegraphics[width=0.5\textwidth]{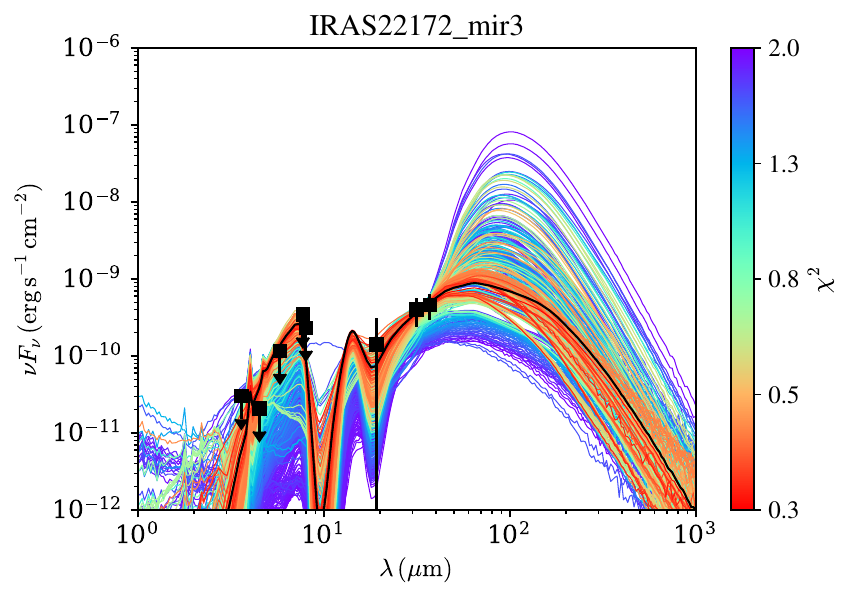}
\includegraphics[width=0.5\textwidth]{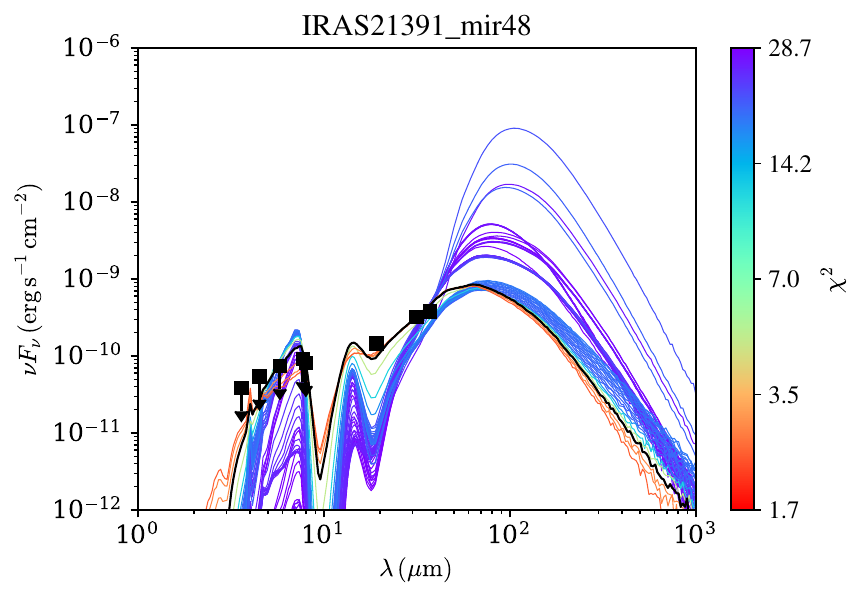}
\includegraphics[width=0.5\textwidth]{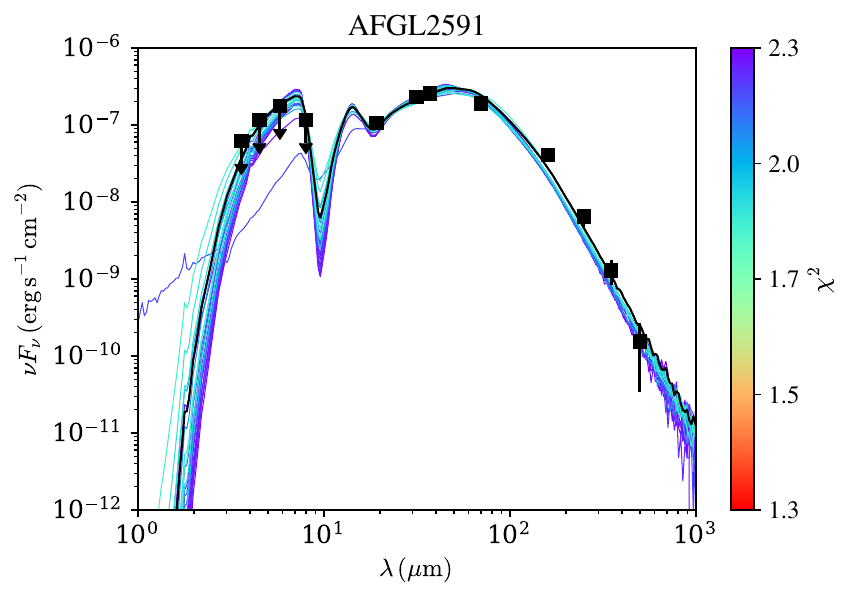}
\includegraphics[width=0.5\textwidth]{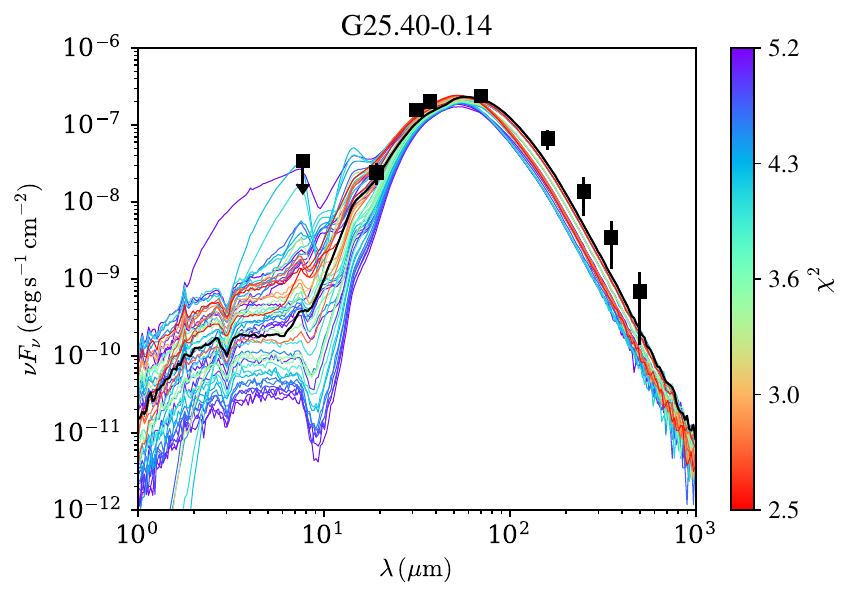}
\caption{(Continued.)}
\end{figure*}

\renewcommand{\thefigure}{C\arabic{figure}}
\addtocounter{figure}{-1}
\begin{figure*}[!htb]
\includegraphics[width=0.5\textwidth]{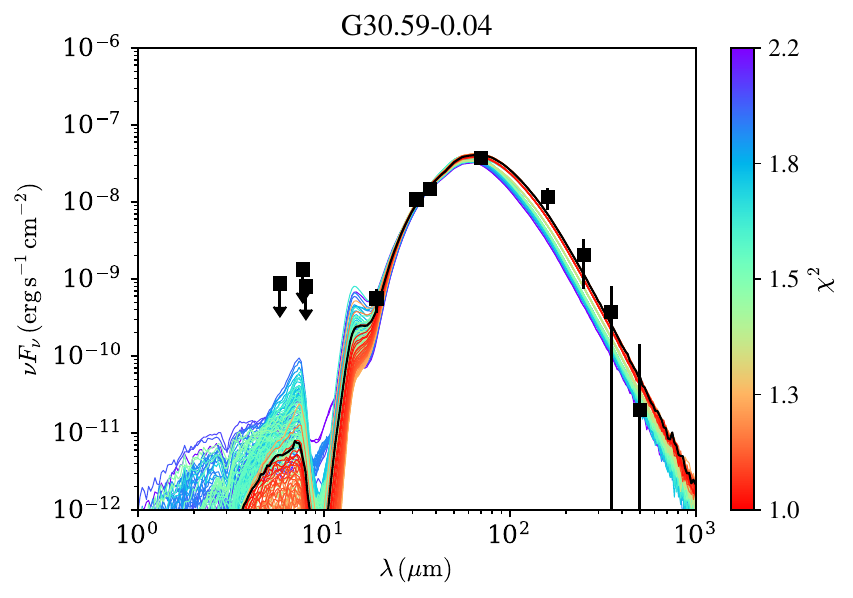}
\includegraphics[width=0.5\textwidth]{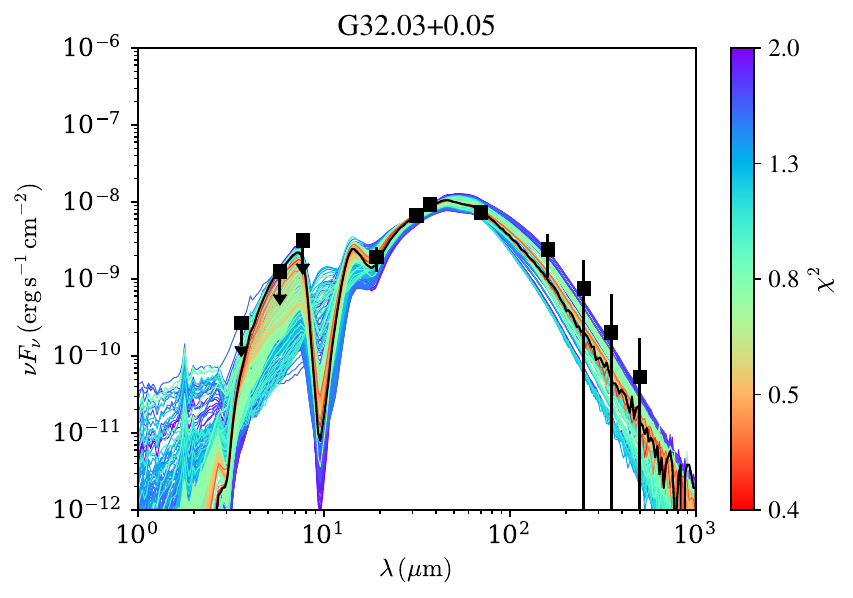}
\includegraphics[width=0.5\textwidth]{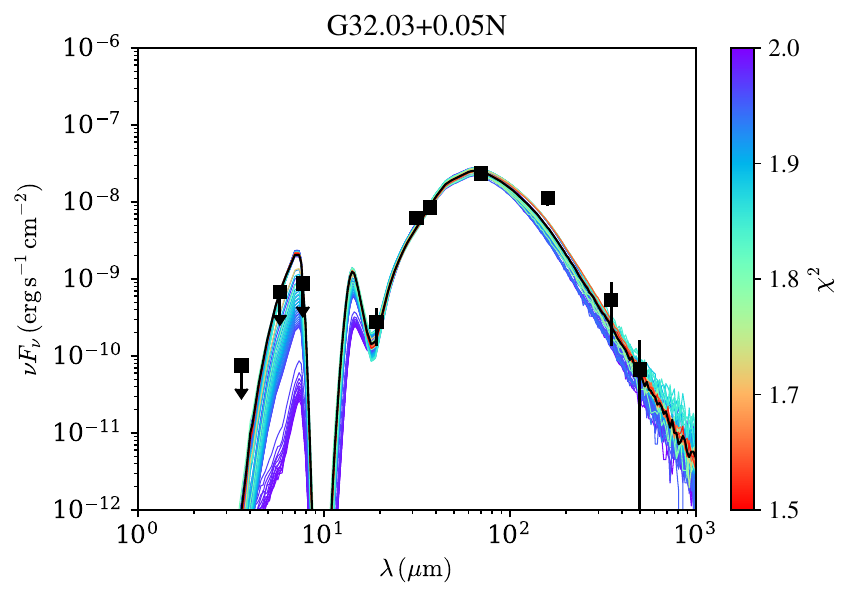}
\includegraphics[width=0.5\textwidth]{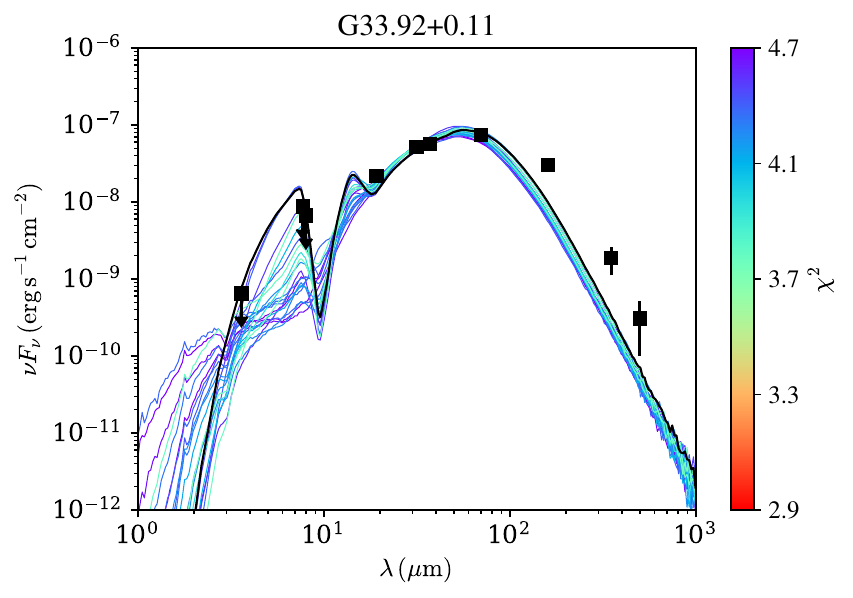}
\includegraphics[width=0.5\textwidth]{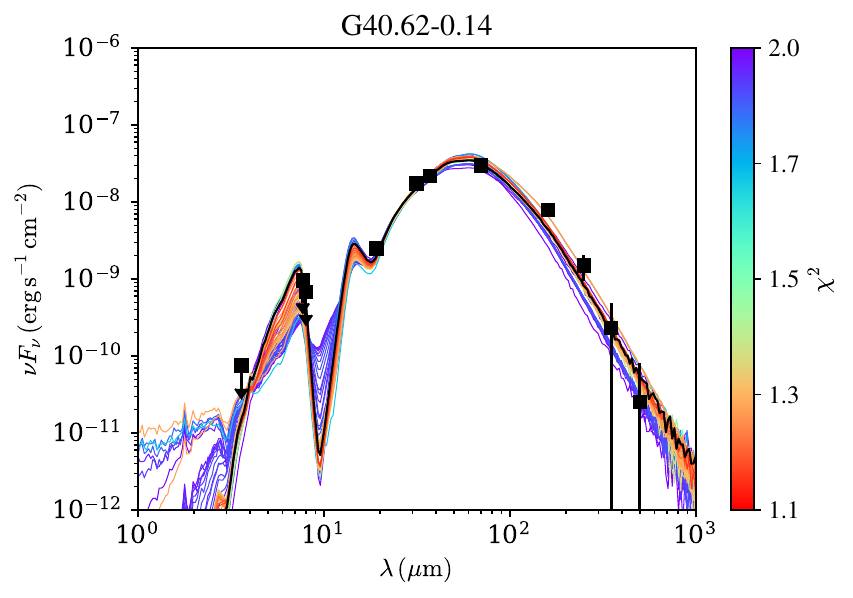}
\includegraphics[width=0.5\textwidth]{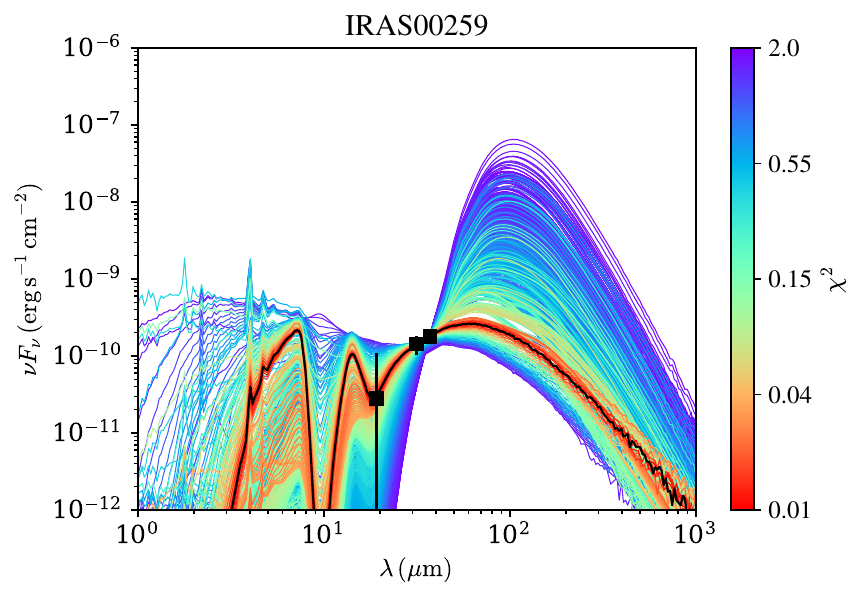}
\caption{(Continued.)}
\end{figure*}

\renewcommand{\thefigure}{C\arabic{figure}}
\addtocounter{figure}{-1}
\begin{figure*}[!htb]
\includegraphics[width=0.5\textwidth]{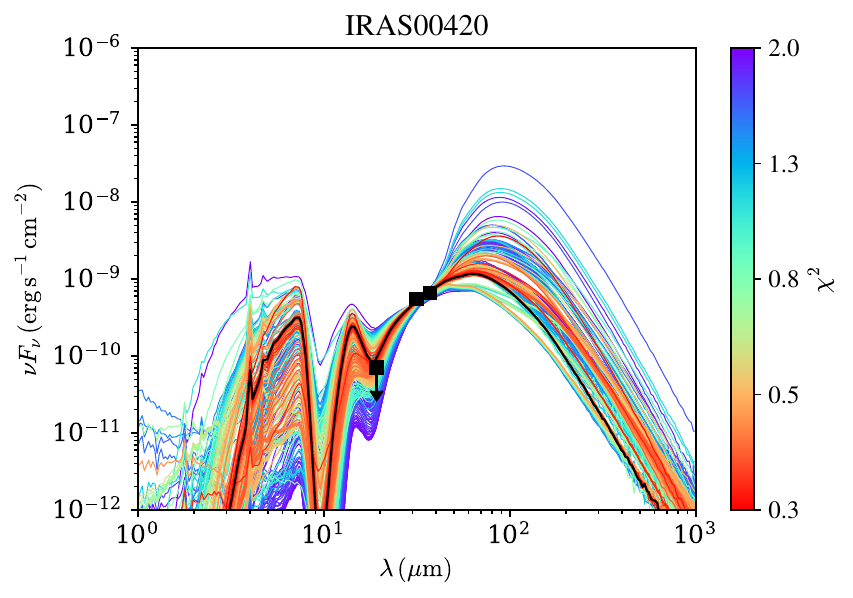}
\includegraphics[width=0.5\textwidth]{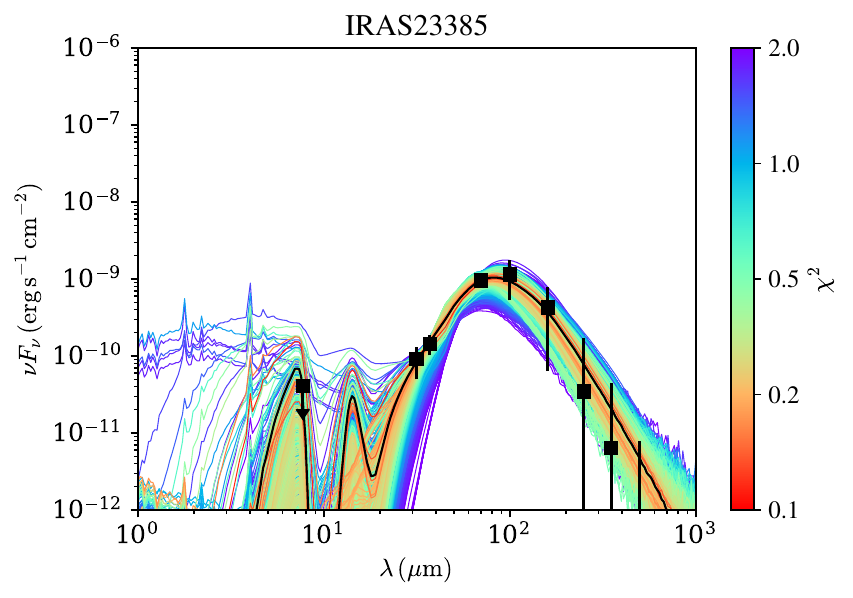}
\includegraphics[width=0.5\textwidth]{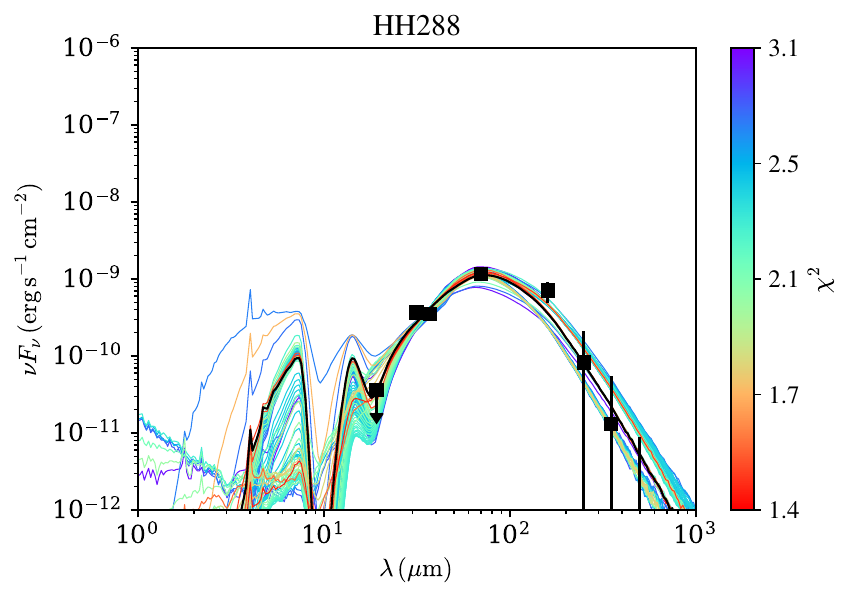}
\caption{(Continued.)}
\end{figure*}

\renewcommand{\thefigure}{C\arabic{figure}}
\begin{figure*}[!htb]
\includegraphics[width=1.0\textwidth]{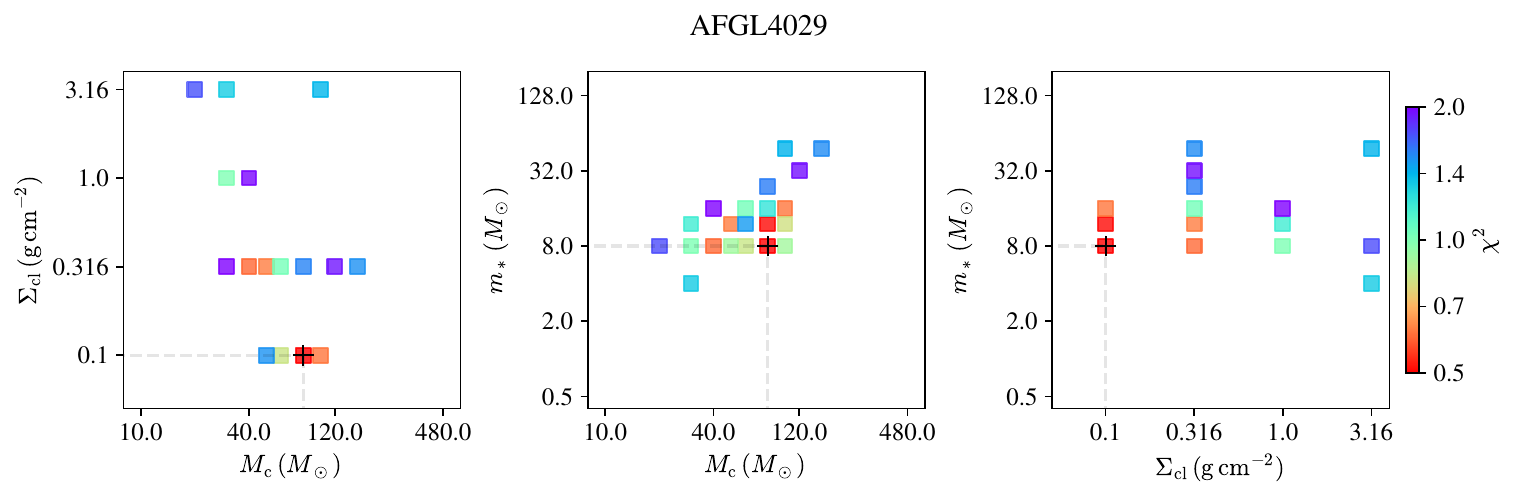}
\includegraphics[width=1.0\textwidth]{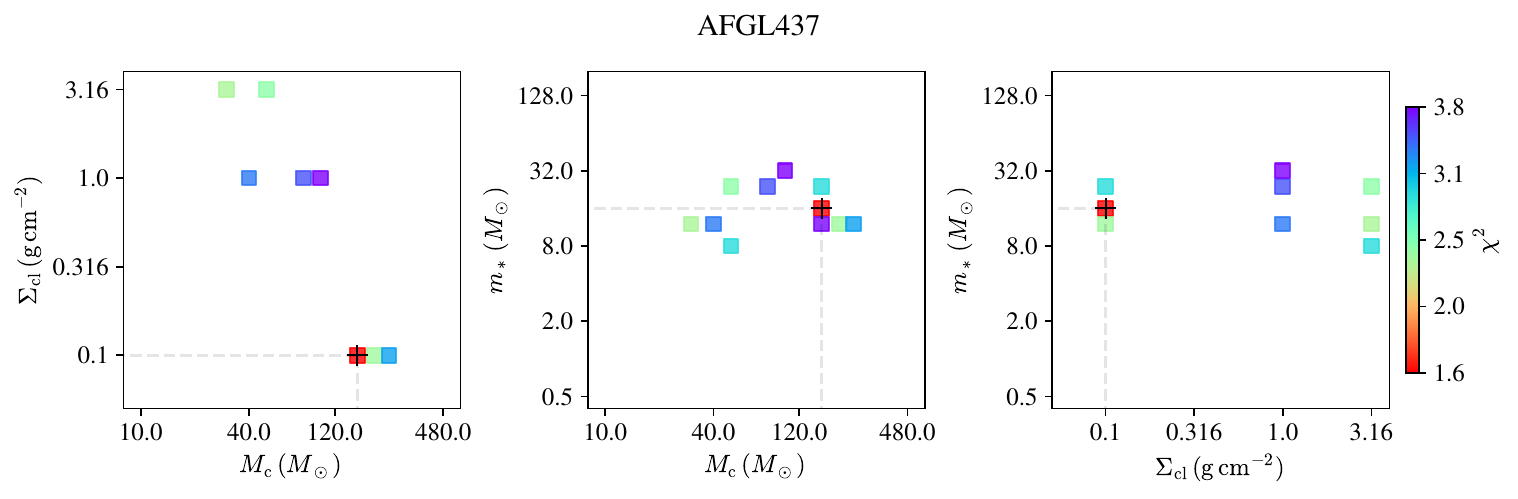}
\includegraphics[width=1.0\textwidth]{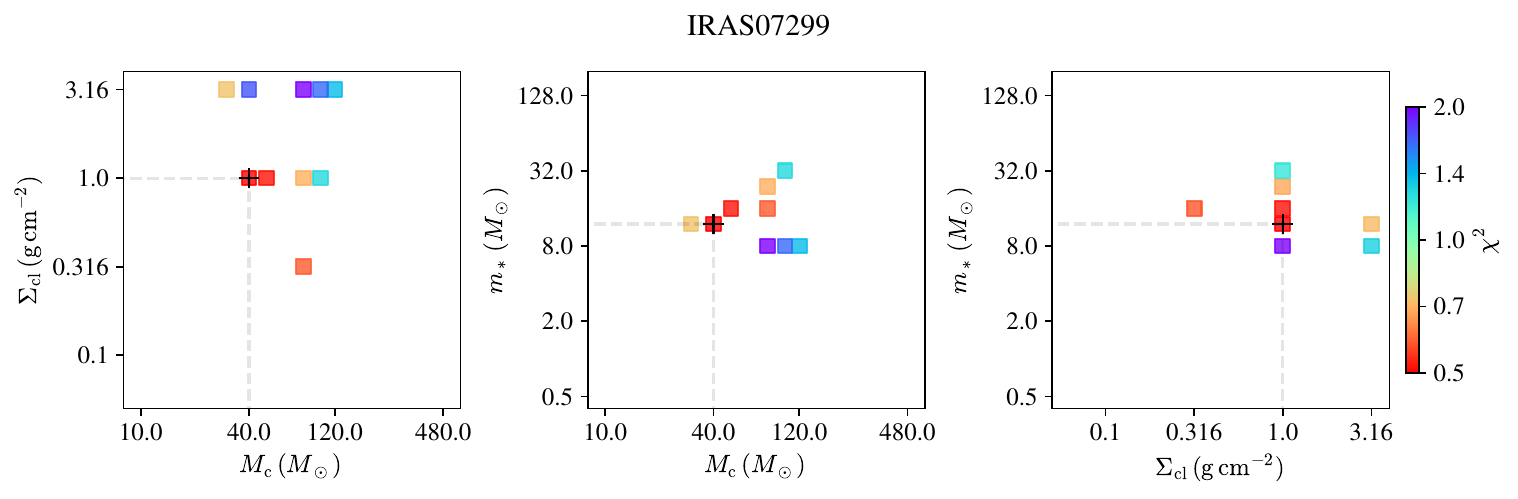}
\includegraphics[width=1.0\textwidth]{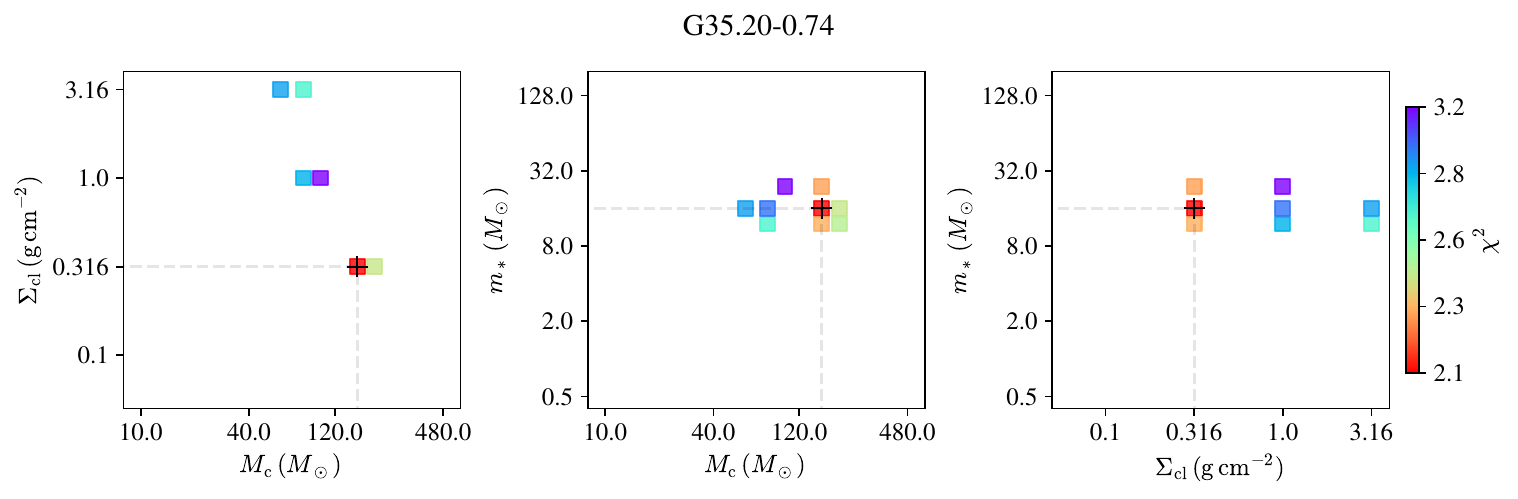}
\caption{Diagrams of $\chi^{2}$ distribution in $\Sigma_{\rm cl}$ - $M_{c}$ space (left), $m_{*}$ - $M_{\rm c}$ space (center) and $m_{*}$ - $\Sigma_{\rm  cl}$ space (right) for each source noted on top of each plot. The black cross is the best model.
\label{fig:sed_2D_results_soma_I-IV}}
\end{figure*}

\renewcommand{\thefigure}{C\arabic{figure}}
\addtocounter{figure}{-1}
\begin{figure*}[!htb]

\includegraphics[width=1.0\textwidth]{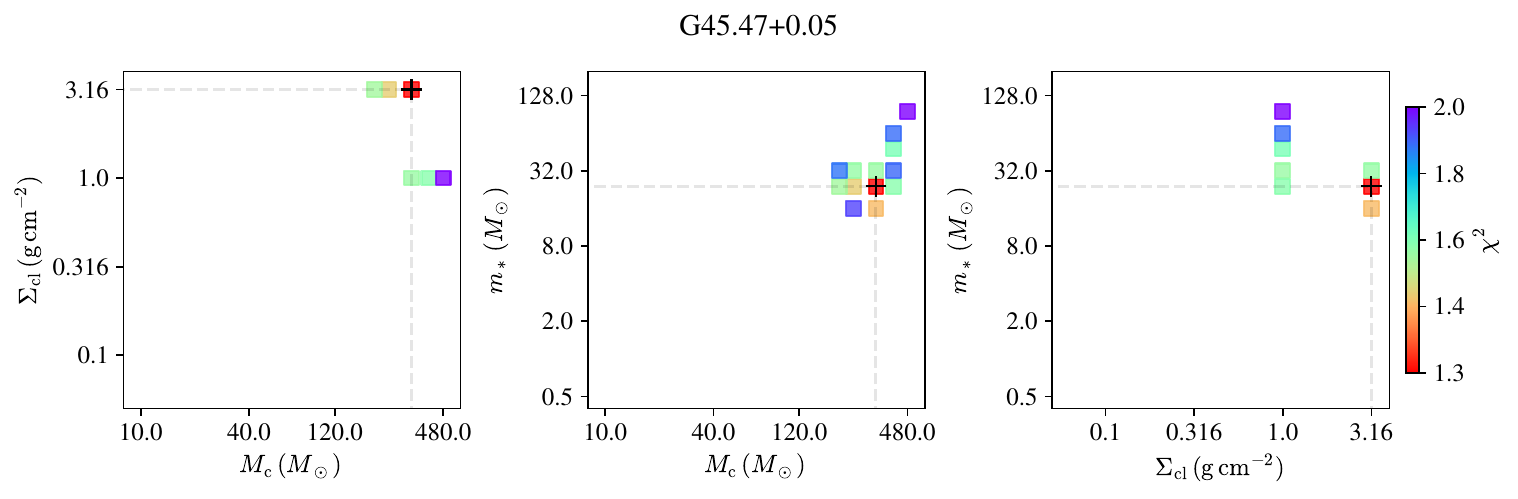}
\includegraphics[width=1.0\textwidth]{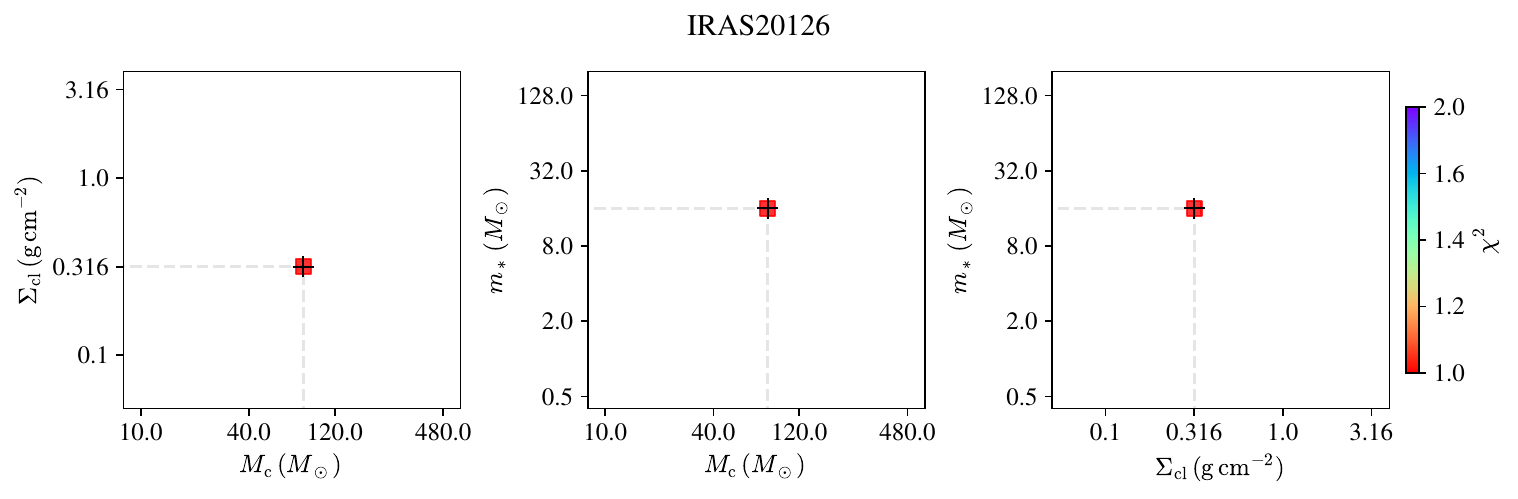}
\includegraphics[width=1.0\textwidth]{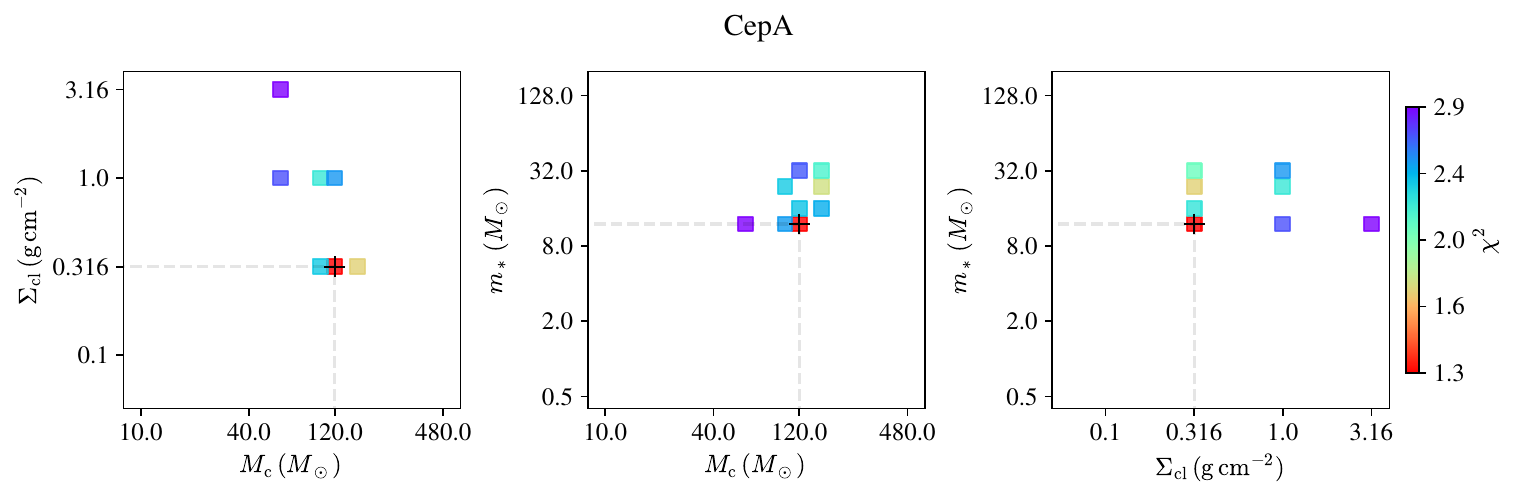}
\includegraphics[width=1.0\textwidth]{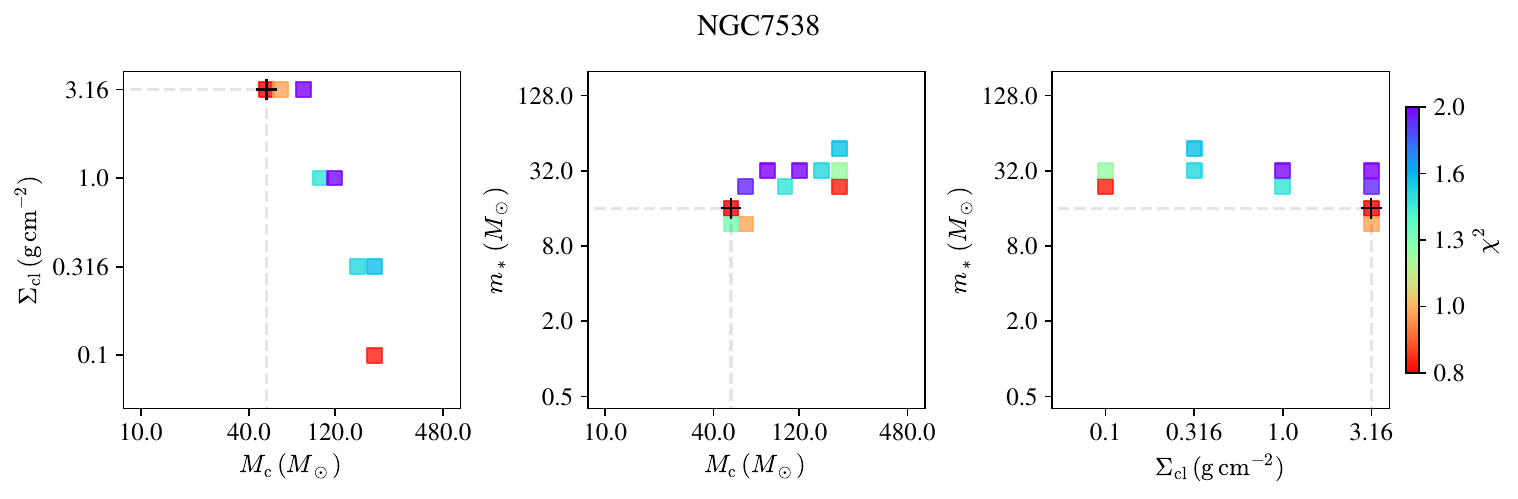}
\caption{(Continued.)}
\end{figure*}

\renewcommand{\thefigure}{C\arabic{figure}}
\addtocounter{figure}{-1}
\begin{figure*}[!htb]
\includegraphics[width=1.0\textwidth]{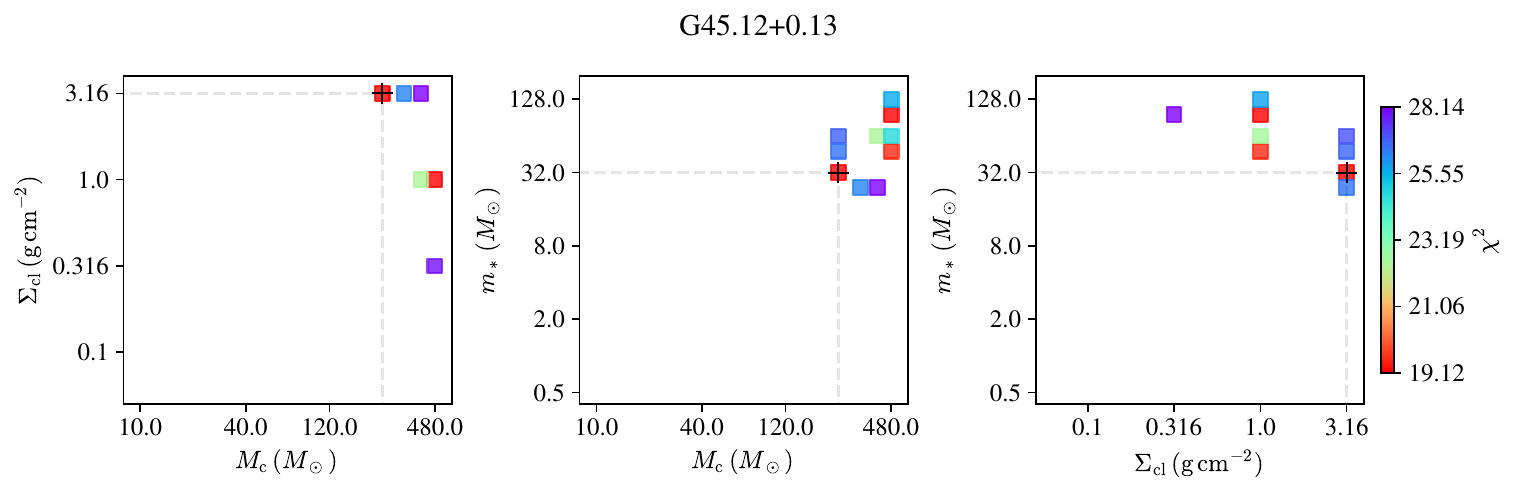}
\includegraphics[width=1.0\textwidth]{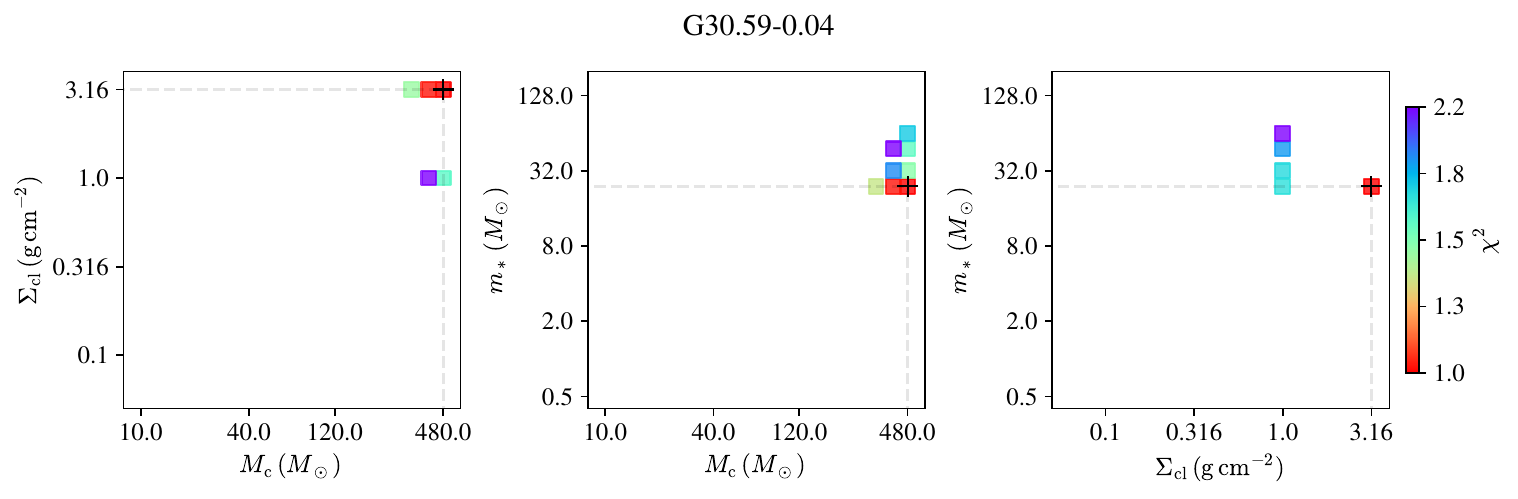}
\includegraphics[width=1.0\textwidth]{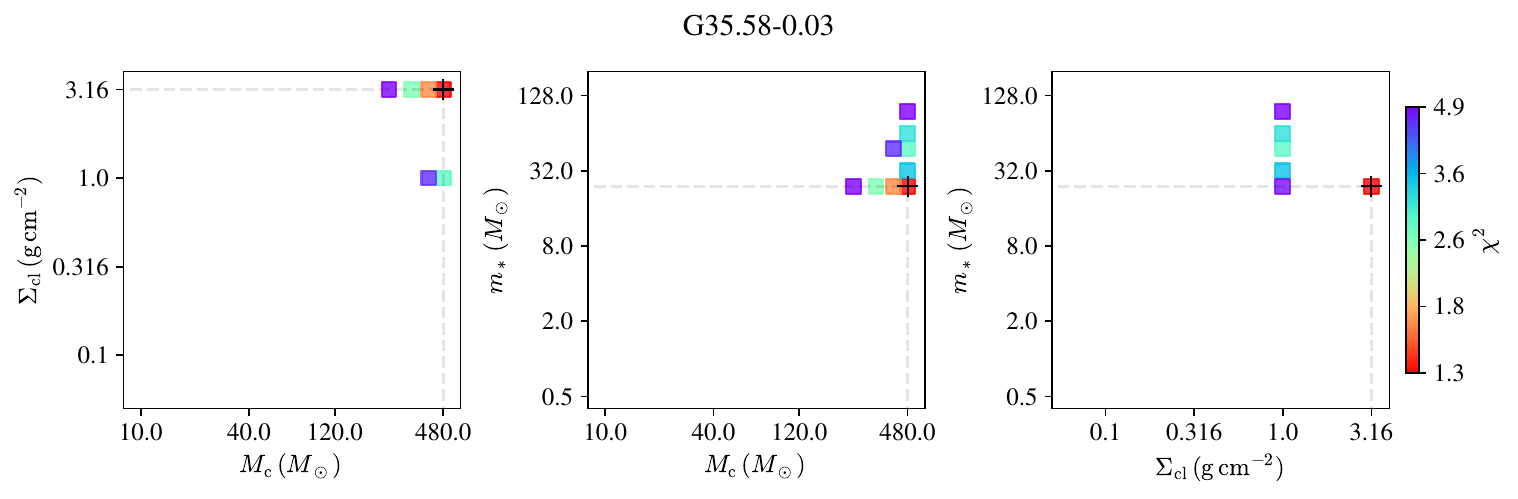}
\includegraphics[width=1.0\textwidth]{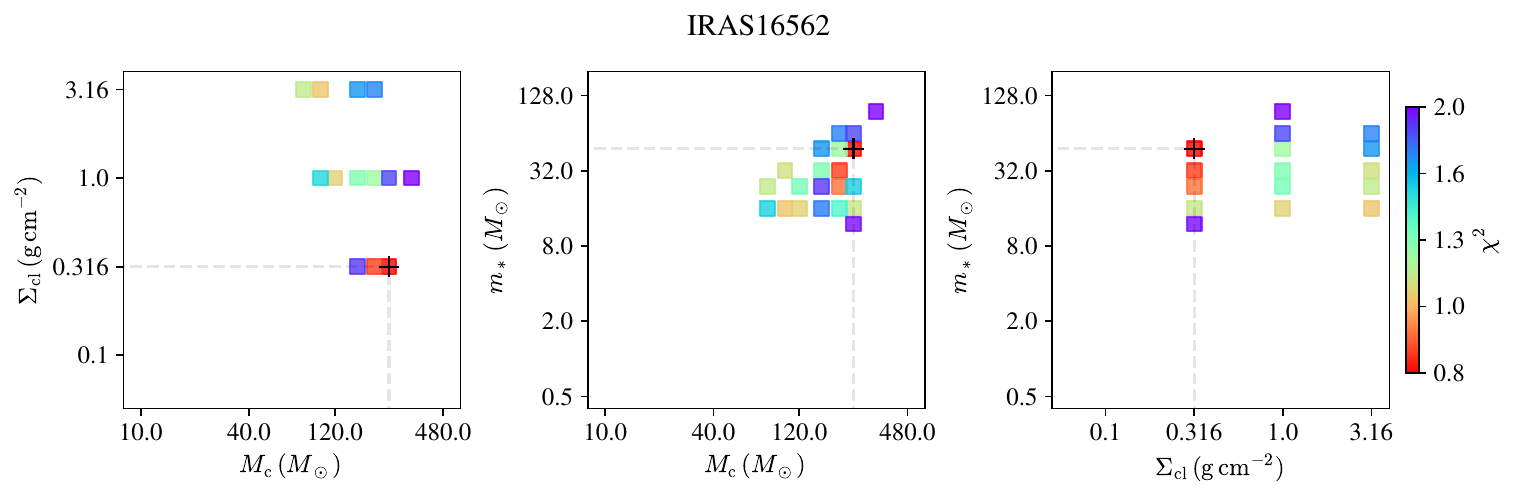}
\caption{(Continued.)}
\end{figure*}

\renewcommand{\thefigure}{C\arabic{figure}}
\addtocounter{figure}{-1}
\begin{figure*}[!htb]
\includegraphics[width=1.0\textwidth]{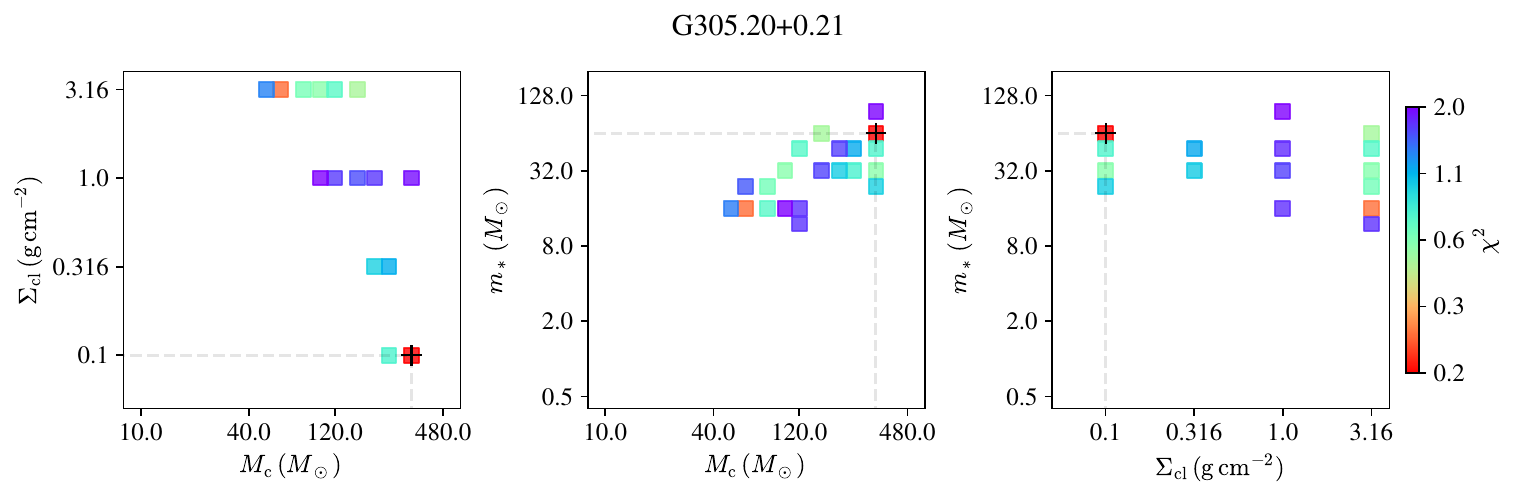}
\includegraphics[width=1.0\textwidth]{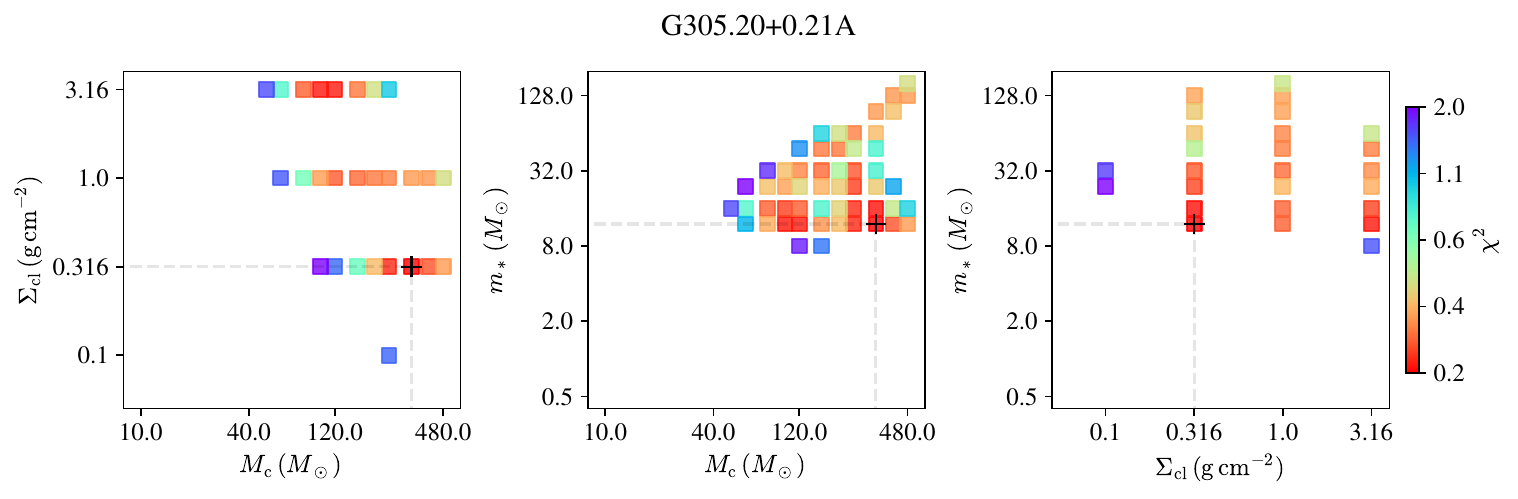}
\includegraphics[width=1.0\textwidth]{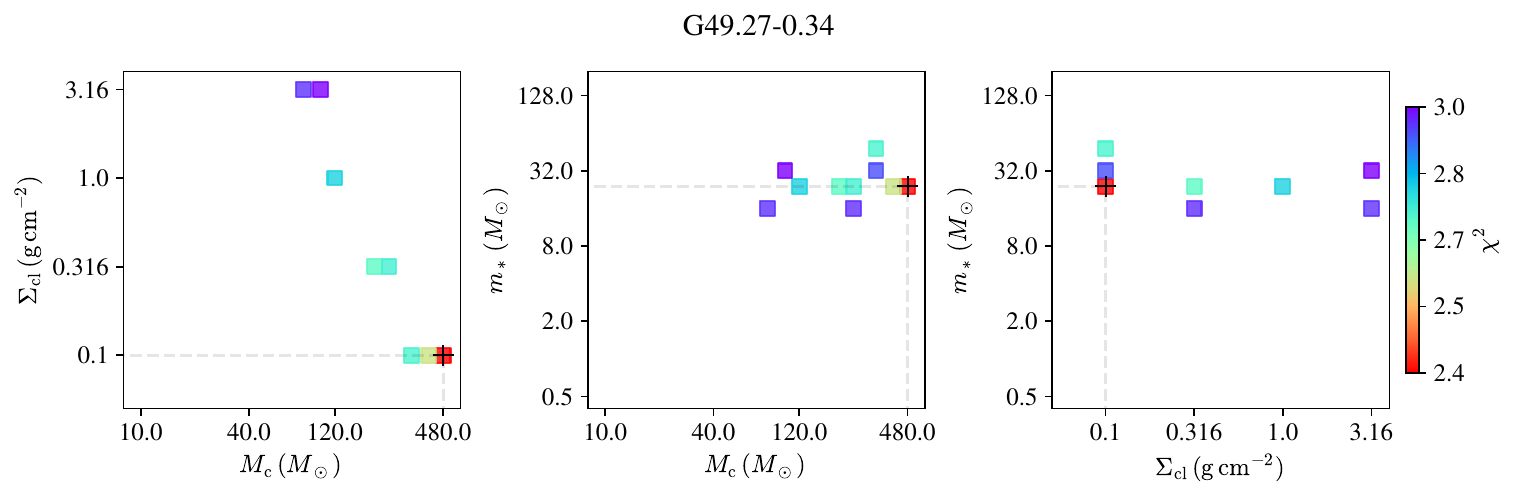}
\includegraphics[width=1.0\textwidth]{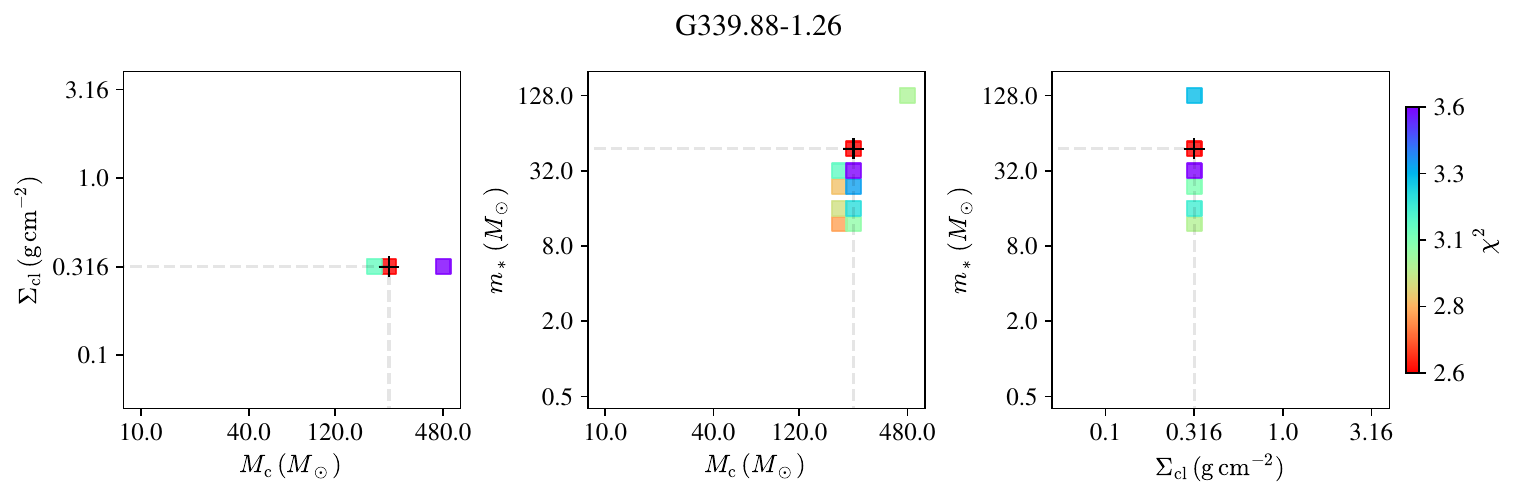}
\caption{(Continued.)}
\end{figure*}

\renewcommand{\thefigure}{C\arabic{figure}}
\addtocounter{figure}{-1}
\begin{figure*}[!htb]
\includegraphics[width=1.0\textwidth]{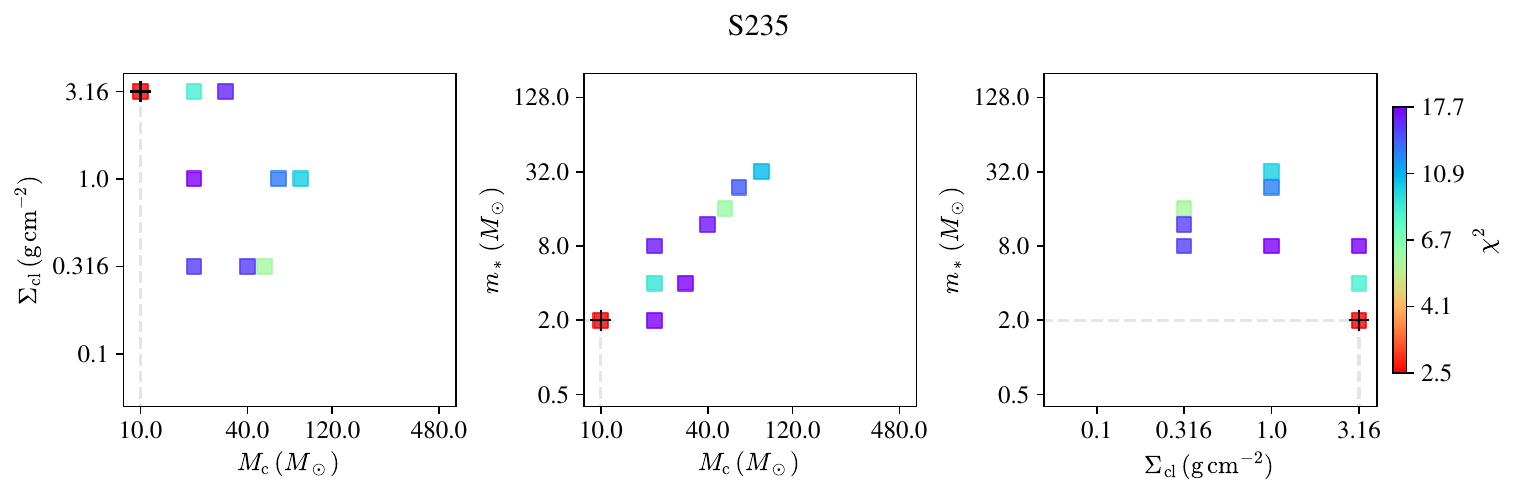}
\includegraphics[width=1.0\textwidth]{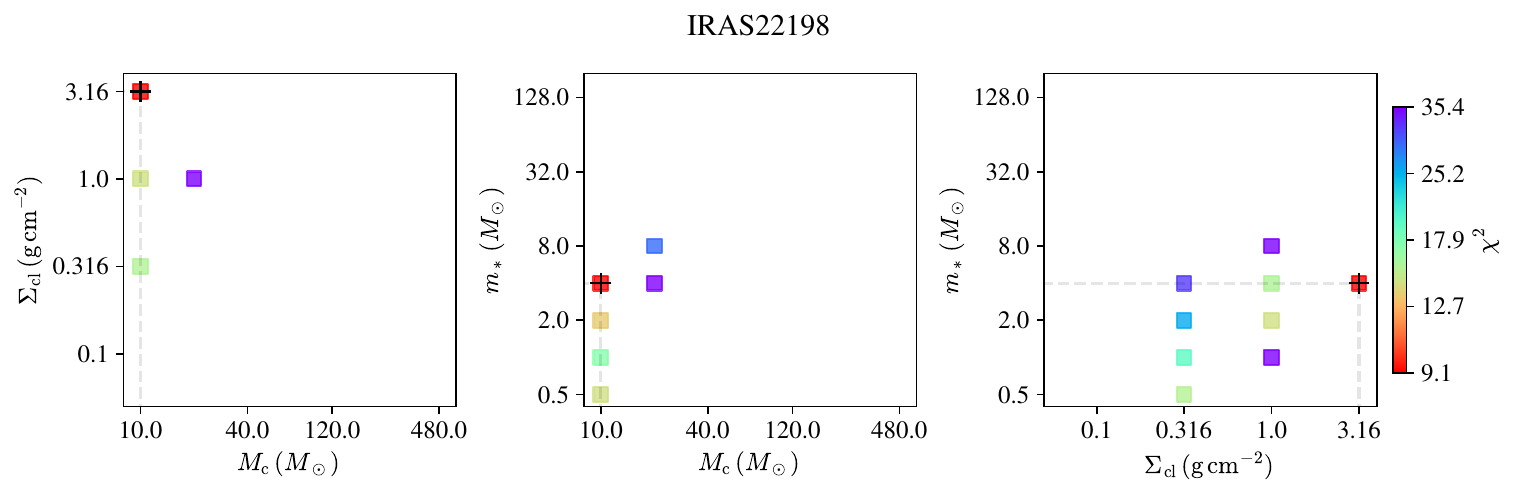}
\includegraphics[width=1.0\textwidth]{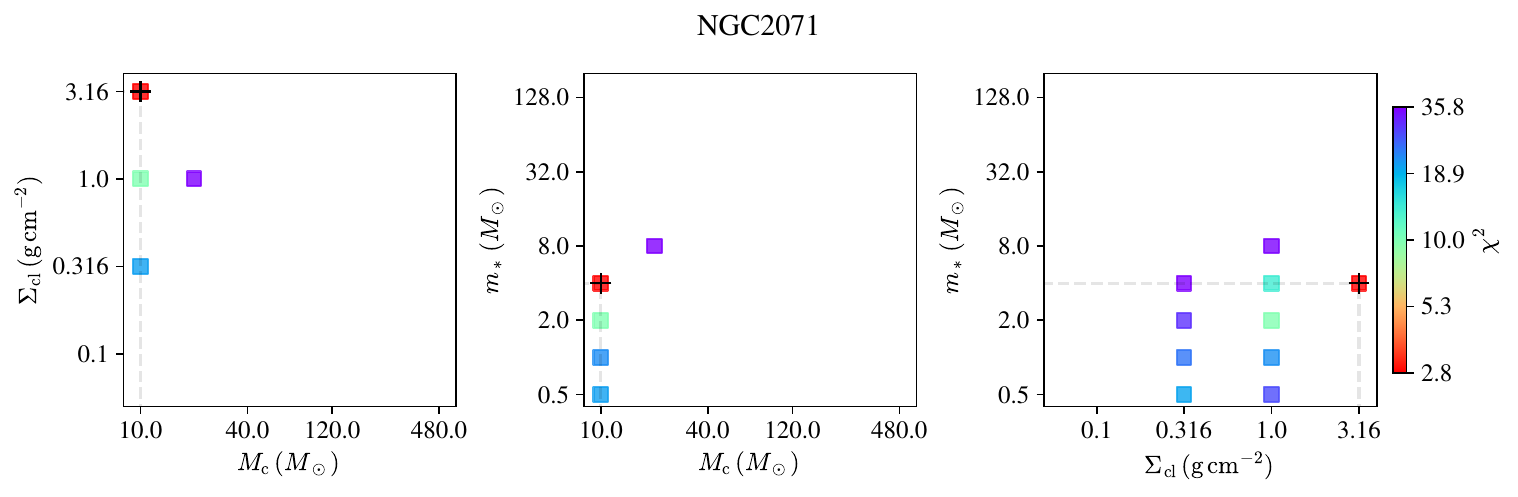}
\includegraphics[width=1.0\textwidth]{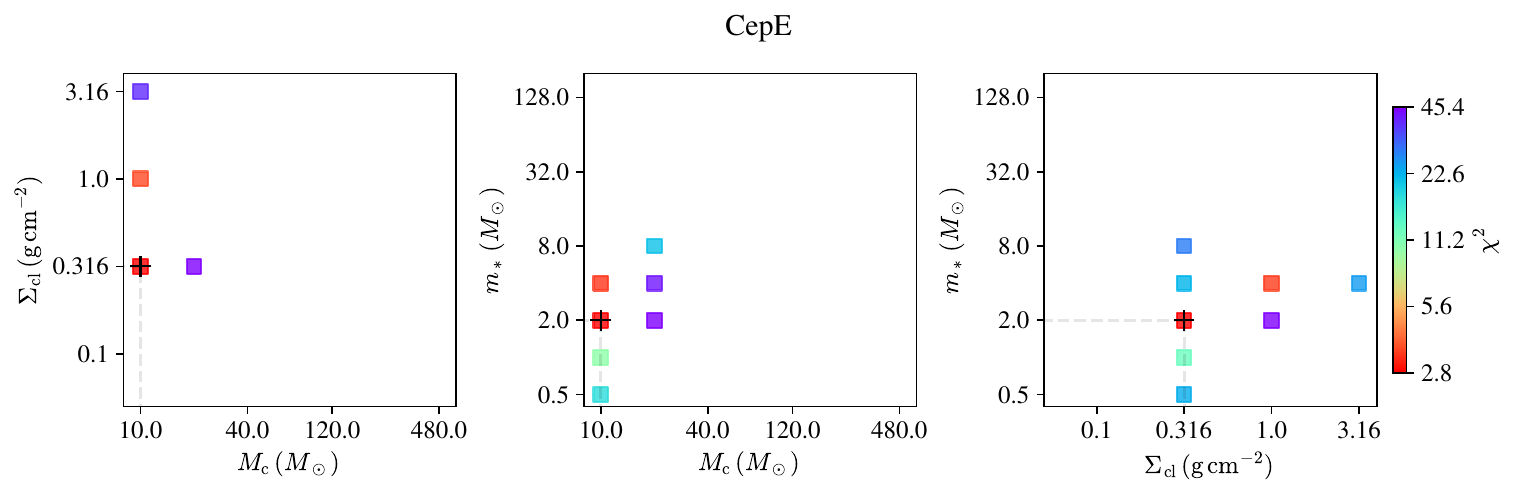}
\caption{(Continued.)}
\end{figure*}

\renewcommand{\thefigure}{C\arabic{figure}}
\addtocounter{figure}{-1}
\begin{figure*}[!htb]
\includegraphics[width=1.0\textwidth]{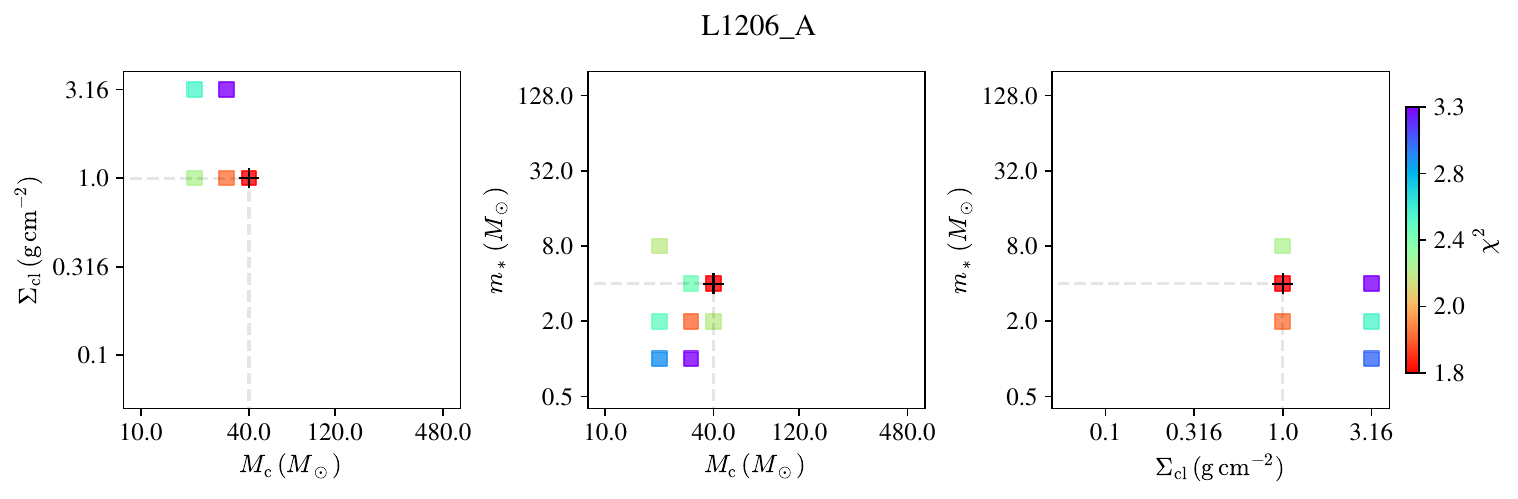}
\includegraphics[width=1.0\textwidth]{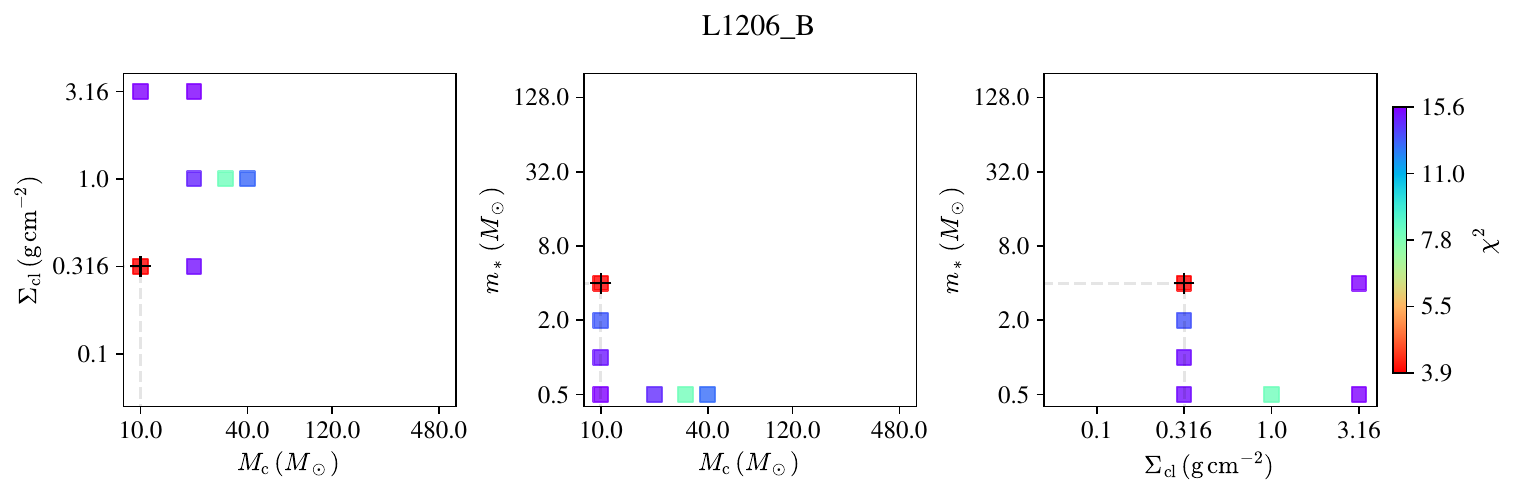}
\includegraphics[width=1.0\textwidth]{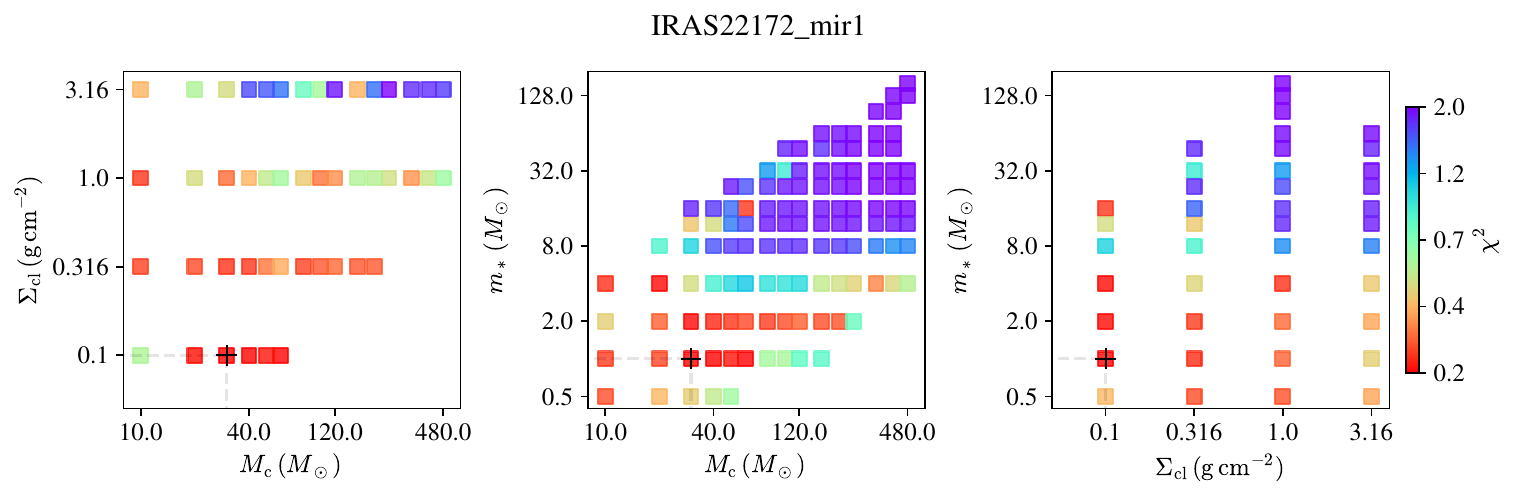}
\includegraphics[width=1.0\textwidth]{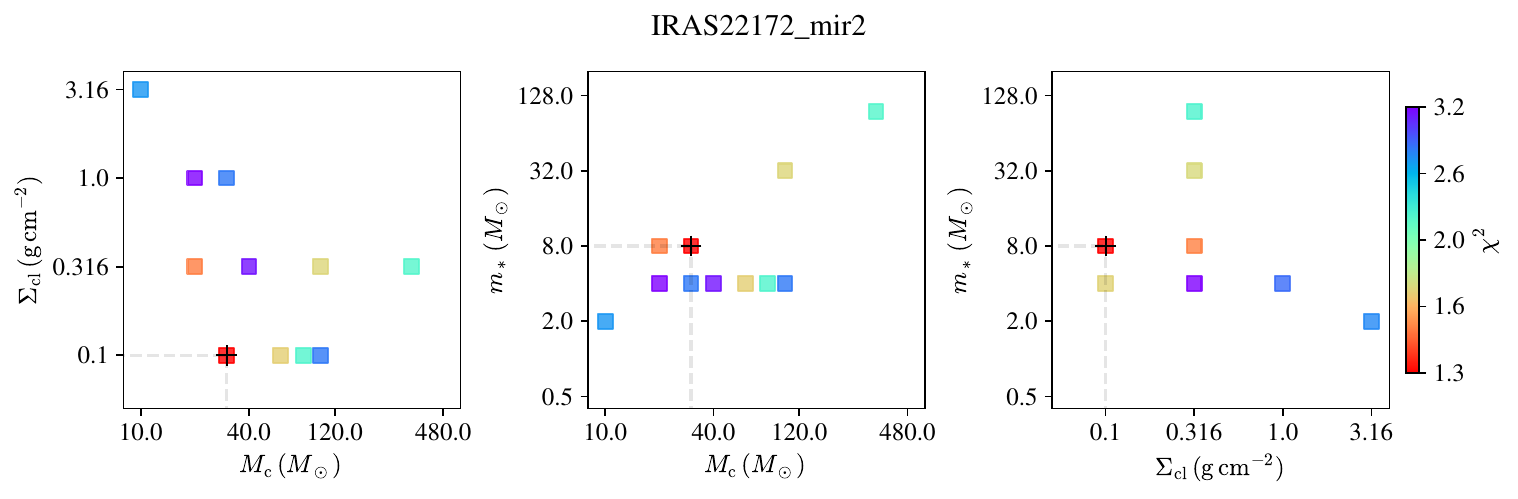}
\caption{(Continued.)}
\end{figure*}

\renewcommand{\thefigure}{C\arabic{figure}}
\addtocounter{figure}{-1}
\begin{figure*}[!htb]
\includegraphics[width=1.0\textwidth]{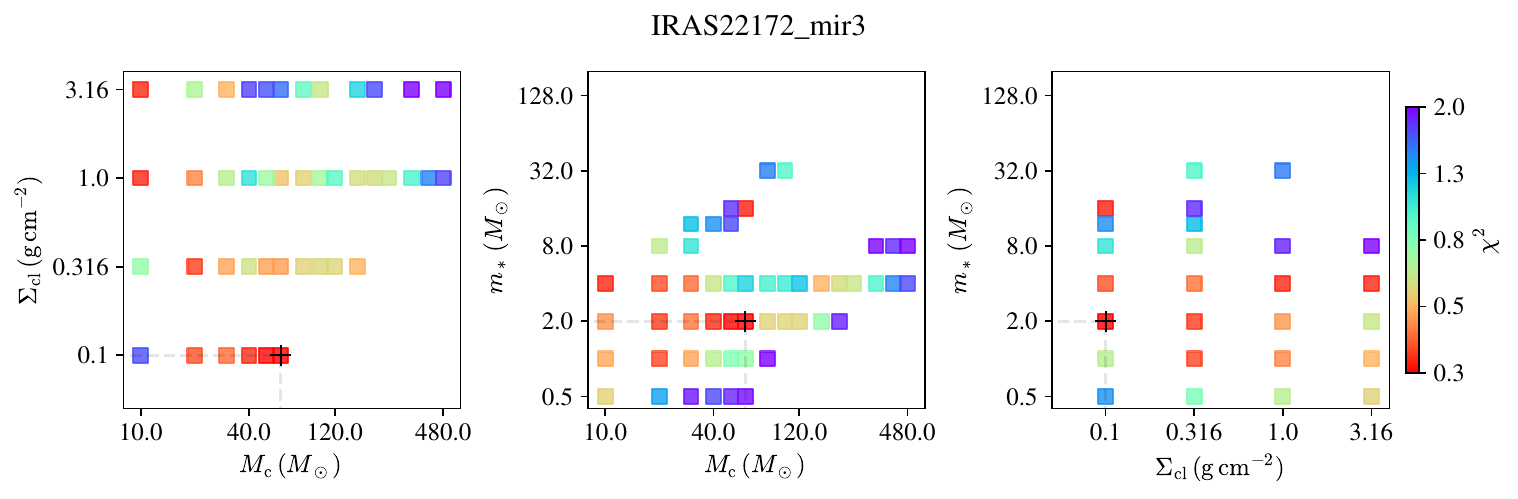}
\includegraphics[width=1.0\textwidth]{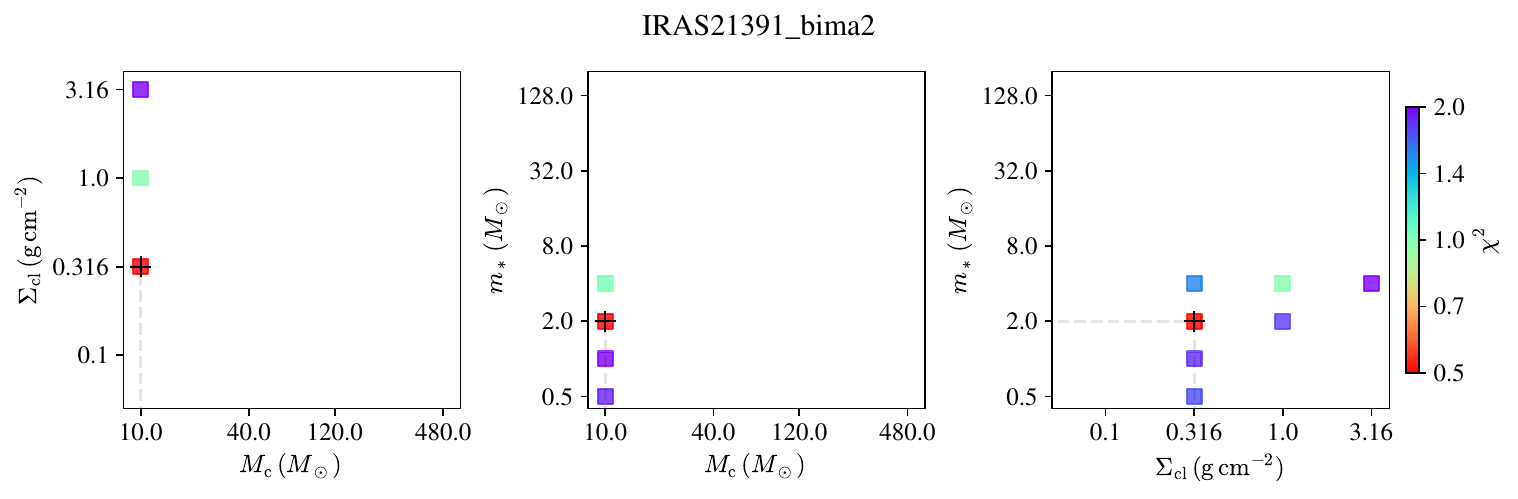}
\includegraphics[width=1.0\textwidth]{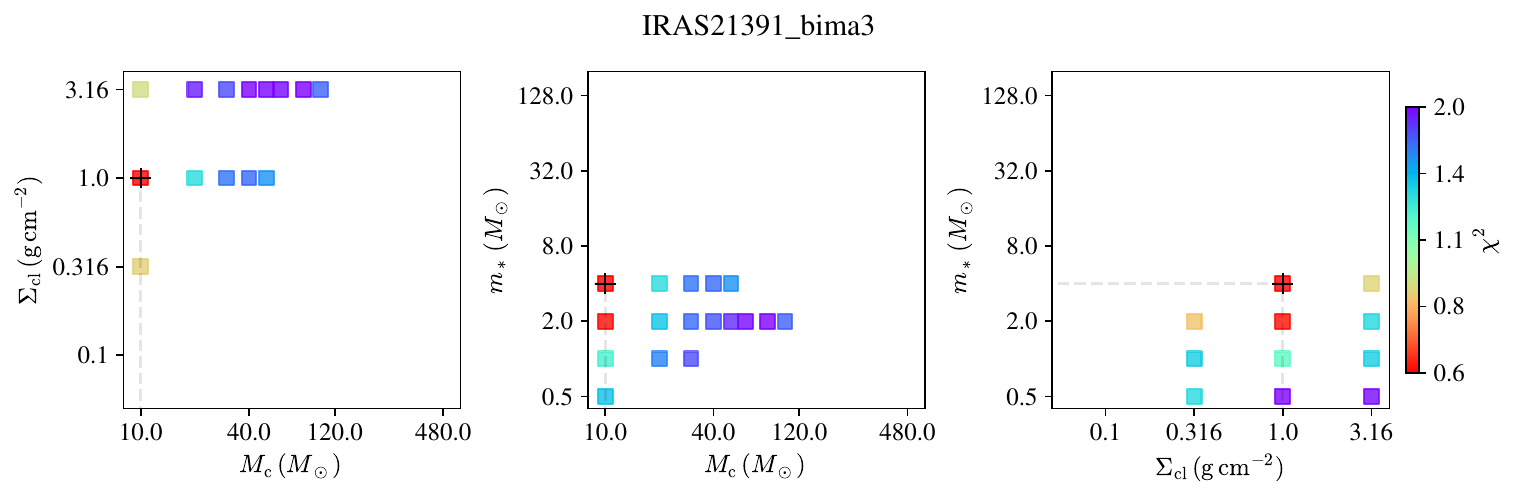}
\includegraphics[width=1.0\textwidth]{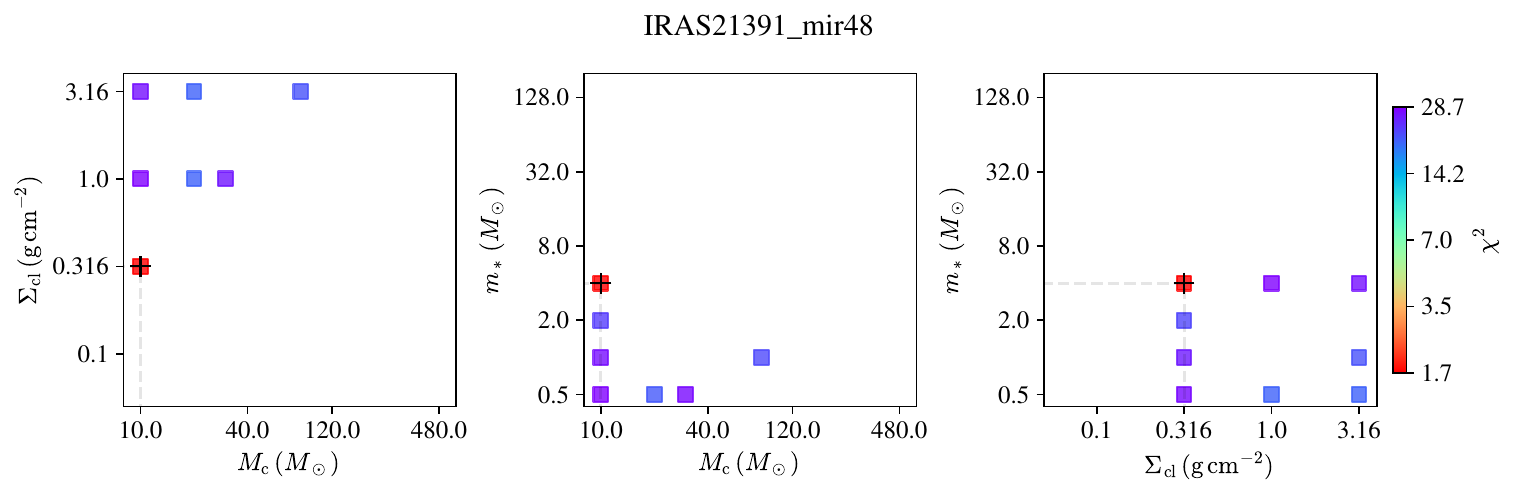}
\caption{(Continued.)}
\end{figure*}

\renewcommand{\thefigure}{C\arabic{figure}}
\addtocounter{figure}{-1}
\begin{figure*}[!htb]
\includegraphics[width=1.0\textwidth]{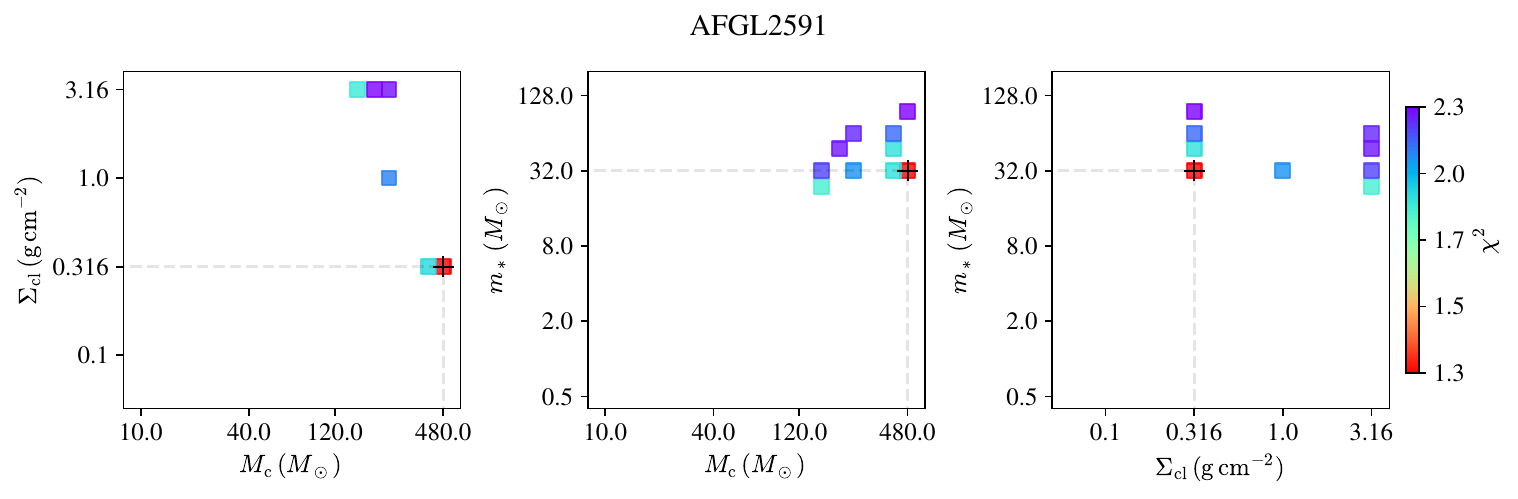}
\includegraphics[width=1.0\textwidth]{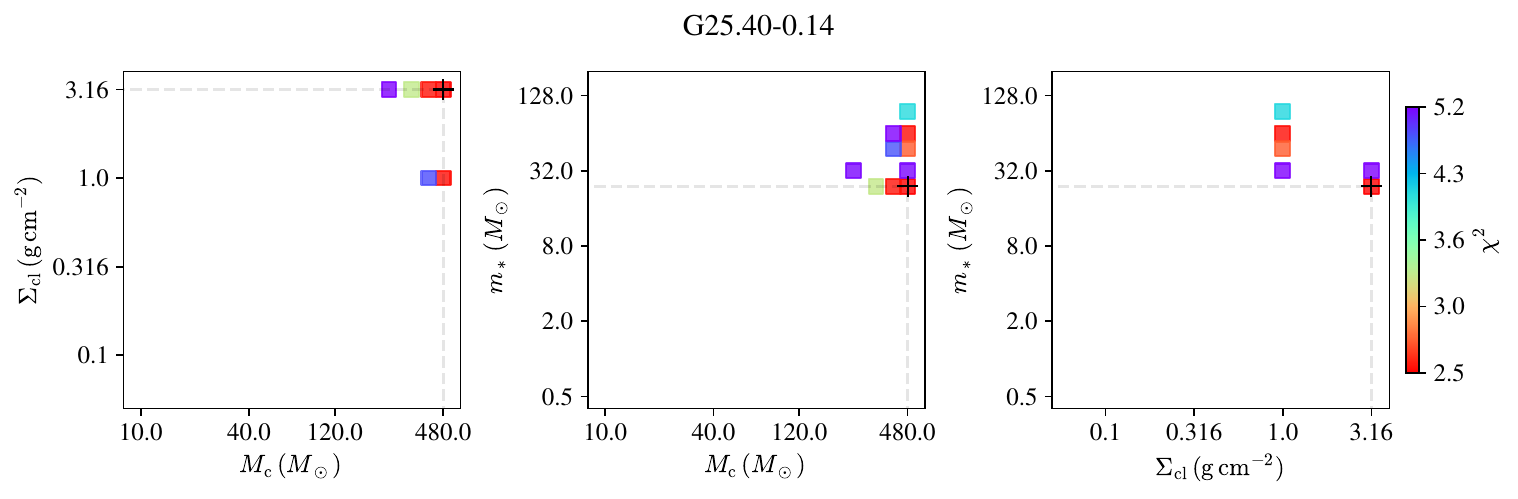}
\includegraphics[width=1.0\textwidth]{2D_plots_V2/SOMA_I_IV/G30.59-0.04_goodmodels_2d.pdf}
\includegraphics[width=1.0\textwidth]{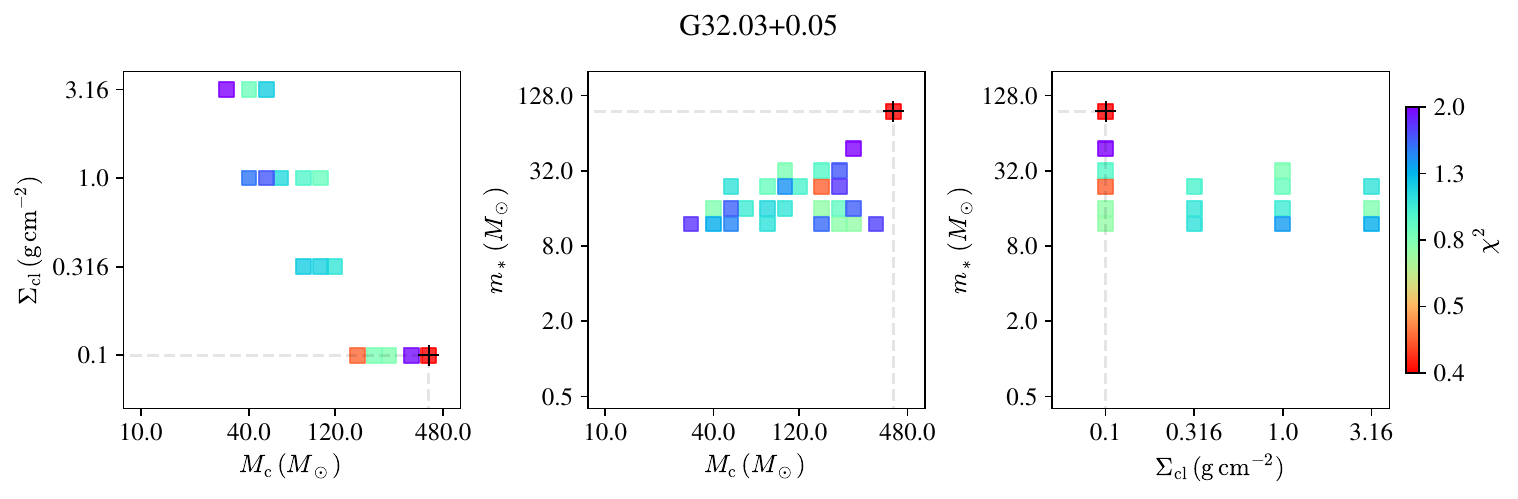}
\caption{(Continued.)}
\end{figure*}

\renewcommand{\thefigure}{C\arabic{figure}}
\addtocounter{figure}{-1}
\begin{figure*}[!htb]
\includegraphics[width=1.0\textwidth]{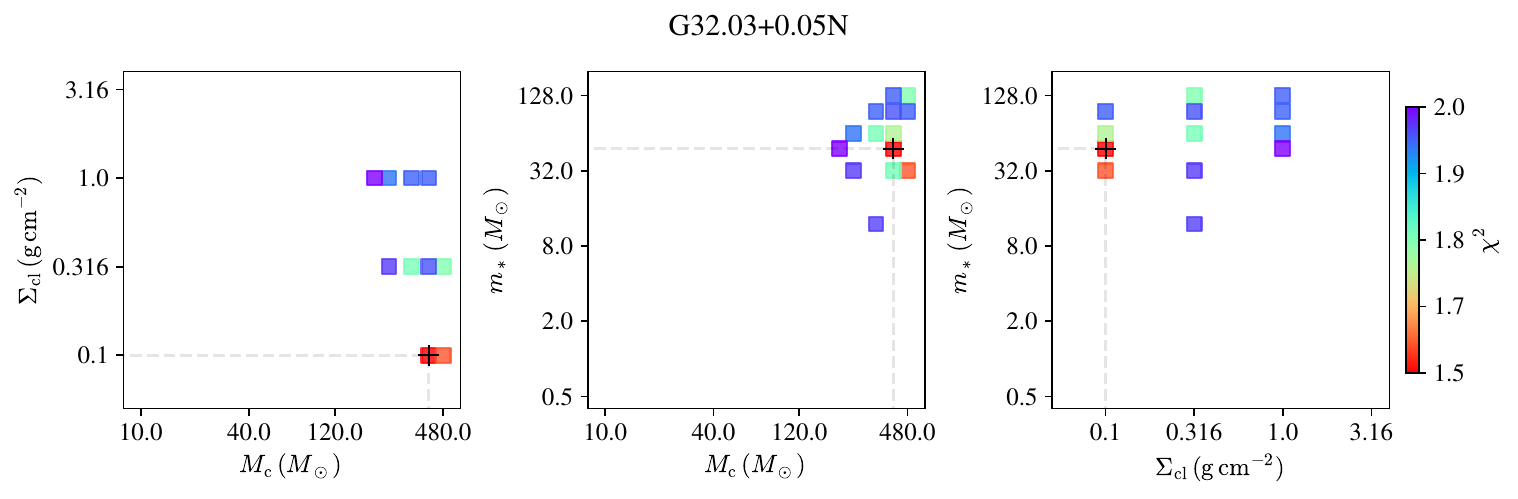}
\includegraphics[width=1.0\textwidth]{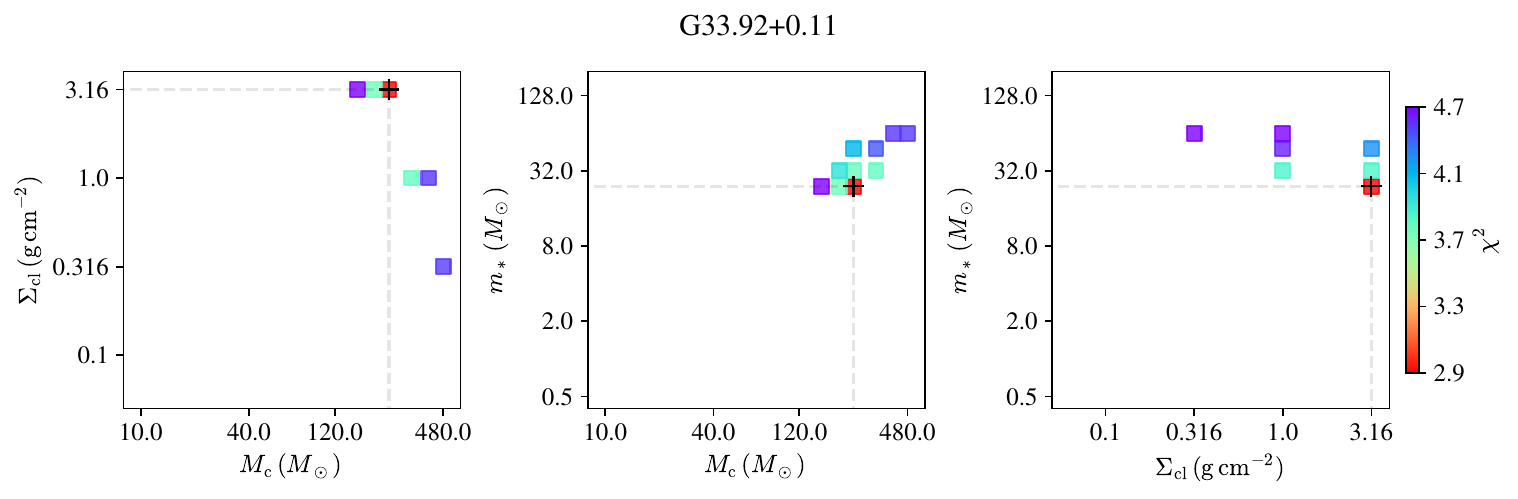}
\includegraphics[width=1.0\textwidth]{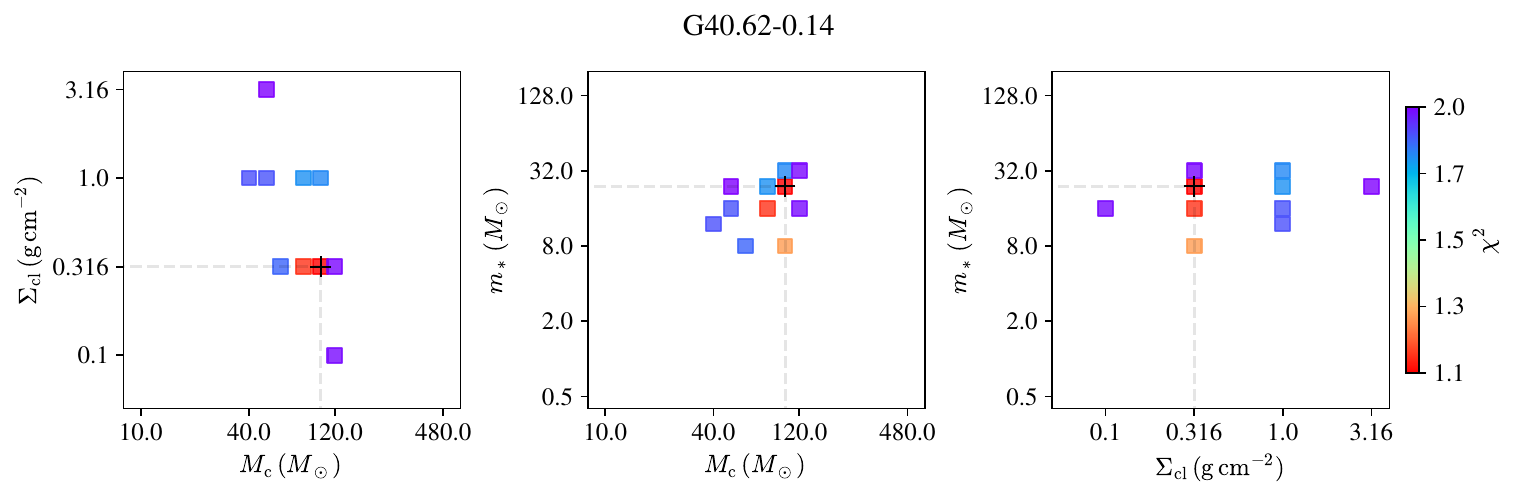}
\includegraphics[width=1.0\textwidth]{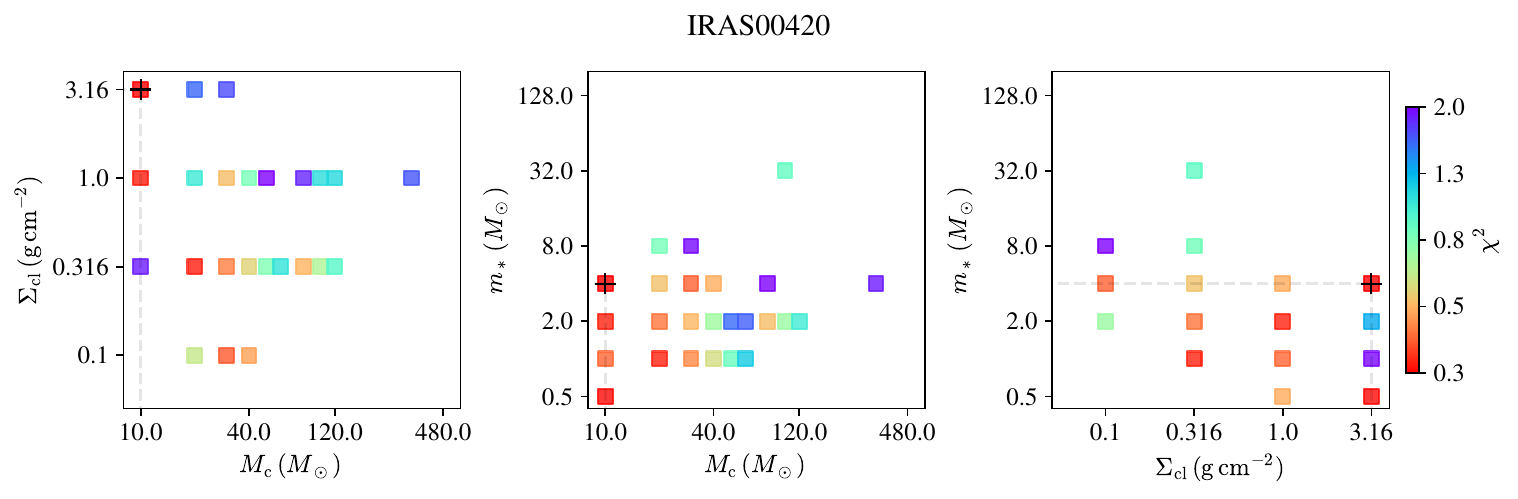}
\caption{(Continued.)}
\end{figure*}

\renewcommand{\thefigure}{C\arabic{figure}}
\addtocounter{figure}{-1}
\begin{figure*}[!htb]
\includegraphics[width=1.0\textwidth]{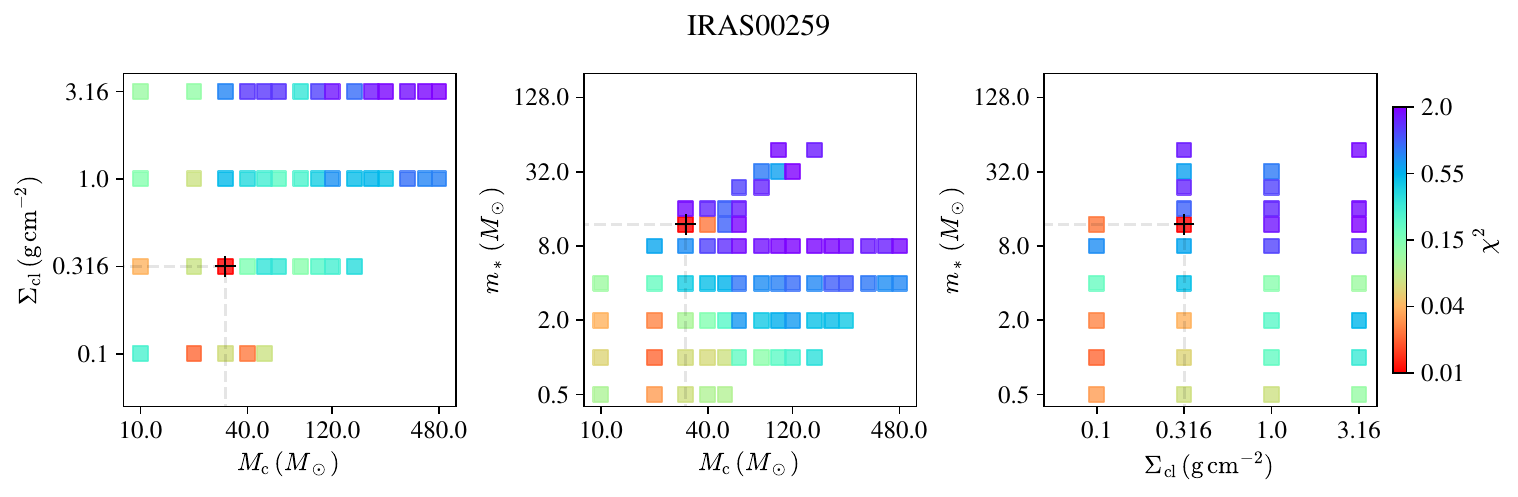}
\includegraphics[width=1.0\textwidth]{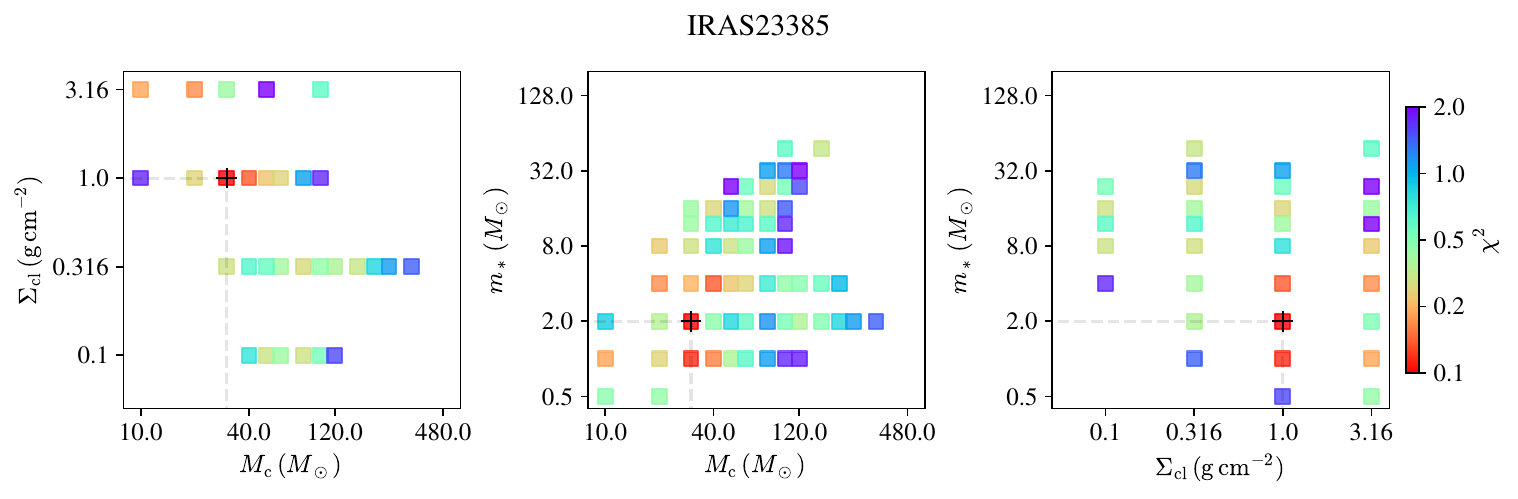}
\includegraphics[width=1.0\textwidth]{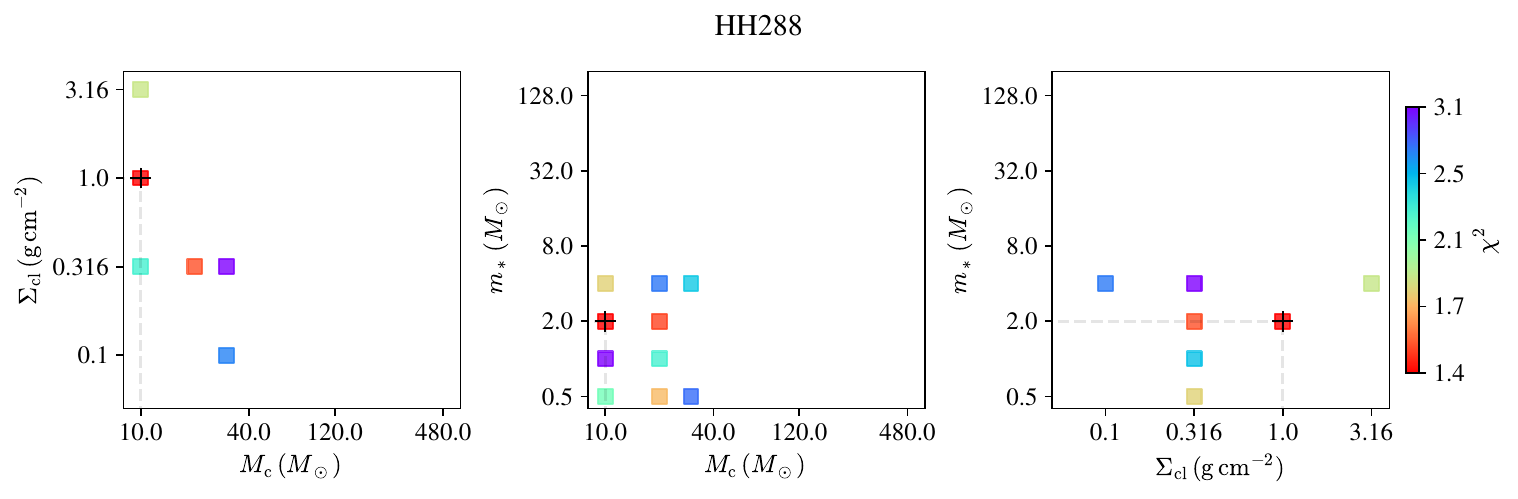}
\caption{(Continued.)}
\end{figure*}

\begin{longrotatetable}
\begin{deluxetable}{lcccccccccccc}
\tabletypesize{\tiny}
\tablecaption{Parameters of the Best Five Fitted Models and Average and Dispersion of Good Models for SOMA I-IV \label{tab:best_models_SOMA_I_IV}} 
\tablewidth{18pt}
\tablehead{
\colhead{Source} &\colhead{$\chi^{2}$} & \colhead{$M_{c}$} & \colhead{$\Sigma_{\rm cl}$} & \colhead{$R_{c}$}  &\colhead{$m_{*}$} & \colhead{$\theta_{\rm view}$} &\colhead{$A_{V}$} & \colhead{$M_{\rm env}$} &\colhead{$\theta_{\rm w,esc}$} & \colhead{$\dot {M}_{\rm disk}$} & \colhead{$L_{\rm bol,iso}$} & \colhead{$L_{\rm bol}$} \\
\colhead{} & \colhead{} & \colhead{($M_\odot$)} & \colhead{(g $\rm cm^{-2}$)} & \colhead{(pc)} & \colhead{($M_{\odot}$)} & \colhead{(\arcdeg)} & \colhead{(mag)} & \colhead{($M_{\odot}$)} & \colhead{(deg)} &\colhead{($M_{\odot}$/yr)} & \colhead{($L_{\odot}$)} & \colhead{($L_{\odot}$)}
}
\startdata
\multicolumn{13}{c}{{\it SOMA I}}\\
\hline
AFGL4029  & 0.46 & 80 & 0.100 & 0.21 & 8.0 & 64.8 & 1 & 61.5 & 26.8 & 5.0$\times10^{-5}$ & 5.4$\times10^{3}$ & 9.7$\times10^{3}$ \\
$d=2.0$ kpc & 0.47 & 80 & 0.100 & 0.21 & 12.0 & 71.0 & 4 & 46.5 & 40.0 & 5.4$\times10^{-5}$ & 5.0$\times10^{3}$ & 1.6$\times10^{4}$ \\
$R_\mathrm{ap}=12.75\arcsec$ & 0.56 & 40 & 0.316 & 0.08 & 8.0 & 43.5 & 0 & 22.3 & 36.3 & 9.2$\times10^{-5}$ & 5.0$\times10^{3}$ & 1.2$\times10^{4}$ \\
$R_\mathrm{ap}=0.12$ pc & 0.57 & 100 & 0.100 & 0.23 & 16.0 & 71.0 & 43 & 53.0 & 44.7 & 6.2$\times10^{-5}$ & 7.8$\times10^{3}$ & 3.0$\times10^{4}$ \\
 & 0.58 & 50 & 0.316 & 0.09 & 12.0 & 58.3 & 0 & 21.6 & 46.3 & 1.0$\times10^{-4}$ & 4.8$\times10^{3}$ & 2.4$\times10^{4}$ \\
Avg. model & \#270 &  $64^{+32}_{-22}$ & $0.214^{+0.321}_{-0.129}$ & $0.13^{+0.11}_{-0.06}$ & $11.3^{+6.0}_{-3.9}$ & $58.9 \pm 21.4$ & $55 \pm 69$ & $30.3^{+34.3}_{-16.1}$ & $40.7 \pm 12.1$ & $8.4^{+6.7}_{-3.7}\times10^{-5}$ & $8.8^{+15.0}_{-5.5}\times10^{3}$ & $2.0^{+2.5}_{-1.1}\times10^{4}$\\
Avg. model $\Sigma_\mathrm{cl,GB}=0.38$ & \#102 &  $55^{+20}_{-15}$ & $0.316$ & $0.10^{+0.02}_{-0.01}$ & $12.0^{+7.0}_{-4.4}$ & $57.7 \pm 22.8$ & $61 \pm 80$ & $23.4^{+7.6}_{-5.8}$ & $44.7 \pm 12.4$ & $10.4^{+1.1}_{-1.0}\times10^{-5}$ & $9.6^{+22.5}_{-6.7}\times10^{3}$ & $2.4^{+3.1}_{-1.4}\times10^{4}$\\
\enddata
\tablecomments{{\small
For each source, the first five rows refer to the best five models taken from the 432 physical models for the non-restricted $\Sigma_\mathrm{cl}$ case. The sixth row shows the average and dispersion of good model fits (see the text) for the non-restricted $\Sigma_\mathrm{cl}$ case. The seventh row shows the average and dispersion of good models for the restricted $\Sigma_\mathrm{cl,GB}$ case. The number next to the symbol \# represents the number of models considered in the average of the good models. Upper and lower scripts in the average model rows refer to the upper and lower range of the dispersion of the good models.}}
\end{deluxetable}
\end{longrotatetable}

\bibliography{phd_bibliography.bib}{}
\bibliographystyle{aasjournal}

\end{document}